\documentclass[11pt,a4paper]{article}
\pdfoutput=1
\usepackage{jheppub}
\usepackage{color}		
\usepackage{graphicx}	%
\usepackage{amsbsy}		
\usepackage{amsfonts}		
\usepackage{amsmath}		
\usepackage{amssymb}		
\usepackage{upgreek}		
\usepackage{wasysym}
\usepackage{multirow}
\usepackage{mciteplus}
\usepackage{psfrag}
\usepackage{url}
\usepackage{enumitem}
\usepackage{rotate}
\usepackage{float}
\usepackage{soul}
\usepackage{subfigure}
\usepackage{booktabs}
\usepackage{threeparttable}
\usepackage{afterpage}
\usepackage{array}
\usepackage{tabu}

\newcolumntype{x}[1]{>{\centering\let\newline\\\arraybackslash\hspace{0pt}}p{#1}}

\graphicspath{{figs/}}

\newcommand{\HBv}[1]{{\tt Higgs\-Bounds-{#1}}}
\newcommand{\HS}{{\tt Higgs\-Signals}}
\newcommand{\HSv}[1]{{\tt Higgs\-Signals-{#1}}}

\newcommand{\muobs}{\hat\mu}
\newcommand{\mobs}{\hat{m}}

\newcommand{\MeV}{~\mathrm{MeV}}
\newcommand{\pvalue}{$\mathcal{P}$-value}
\newcommand{\brinv}{\mathrm{BR}(H\to\mathrm{inv.})}
\newcommand{\brhnp}{\mathrm{BR}(H\to\mathrm{NP})}

\newcommand{\ku}{\kappa_u}

\newcommand{\kt}{\kappa_t}
\newcommand{\kd}{\kappa_d}
\newcommand{\kb}{\kappa_b}
\newcommand{\kg}{\kappa_g}
\newcommand{\kF}{\kappa_F}

\newcommand{\kga}{\kappa_\gamma}
\newcommand{\kZga}{\kappa_{Z\gamma}}
\newcommand{\kW}{\kappa_W}
\newcommand{\kZ}{\kappa_Z}
\newcommand{\kV}{\kappa_V}
\newcommand{\kl}{\kappa_\ell}
\newcommand{\ktau}{\kappa_\tau}
\newcommand{\kline}{\overline{\kappa}}

\newcommand{\invfb}{~\mathrm{fb}^{-1}}
\newcommand{\invab}{~\mathrm{ab}^{-1}}

\newcommand{\epem}{e^+e^-}
\newcommand{\epemtoZH}{e^+e^-\to ZH}

\newcommand{\htobb}{H\to b\bar{b}}
\newcommand{\htocc}{H\to c\bar{c}}
\newcommand{\htogg}{H\to gg}
\newcommand{\htoWW}{H\to WW^{(*)}}

\newcommand{\htoZZ}{H\to ZZ^{(*)}}
\newcommand{\htoZga}{H\to Z\gamma}
\newcommand{\htotautau}{H\to \tau^+\tau^-}

\newcommand{\htomumu}{H\to \mu^+\mu^-}

\newcommand{\htogaga}{H\to \gamma\gamma}
\newcommand{\brhbb}{\mathrm{BR}(\htobb)}
\newcommand{\brhcc}{\mathrm{BR}(\htocc)}
\newcommand{\brhgg}{\mathrm{BR}(\htogg)}
\newcommand{\brhWW}{\mathrm{BR}(\htoWW)}
\newcommand{\brhZZ}{\mathrm{BR}(\htoZZ)}
\newcommand{\brhtautau}{\mathrm{BR}(\htotautau)}
\newcommand{\brhgaga}{\mathrm{BR}(\htogaga)}
\newcommand{\brhZga}{\mathrm{BR}(\htoZga)}
\newcommand{\brhmumu}{\mathrm{BR}(\htomumu)}
\newcommand{\brat}{\mathrm{BR}}
\newcommand{\etmiss}{E_T^\mathrm{miss}}
\newcommand{\order}[1]{\ensuremath{{\cal O}(#1)}}
\newcommand{\MHexp}{125.7}
\newcommand{\pptoZH}{pp \to ZH}

\include{paperdef}

\newcommand{\citekappafits}{\cite{Lafaye:2009vr,Klute:2012pu,Plehn:2012iz,Klute:2013cx, Dobrescu:2012td,Espinosa:2012im,Cacciapaglia:2012wb,Belanger:2012gc,Ellis:2013lra,Djouadi:2013qya, Cheung:2013kla,Holdom:2013axa,Chpoi:2013wga,Bechtle:2013xfa,Belanger:2013xza}}
\newcommand{\citemodelfits}{\cite{Buckley:2012em,Arbey:2012na,Arbey:2012dq,Akula:2012kk,Cao:2012yn,Howe:2012xe,Drees:2012fb,Haisch:2012re,Bechtle:2012jw,Arbey:2012bp,Ke:2012zq, Chakraborty:2013si,Carmona:2013cq,Arbey:2013jla,Cao:2013wqa,Bhattacherjee:2013vga,Lopez-Val:2013yba,Belyaev:2013ida,Cao:2013gba,Bharucha:2013ela,Cheung:2013rva,Bechtle:2013mda,Cao:2013mqa,Djouadi:2013lra,Enberg:2013jba,Cao:2013cfa,Wang:2013sha,Fan:2014txa,Belanger:2014roa,Stal:2014sua}}
\newcommand{\citefits}{\cite{Lafaye:2009vr,Klute:2012pu,Plehn:2012iz,Klute:2013cx, Dobrescu:2012td,Espinosa:2012im,Cacciapaglia:2012wb,Belanger:2012gc,Ellis:2013lra,Djouadi:2013qya, Cheung:2013kla,Holdom:2013axa,Chpoi:2013wga,Bechtle:2013xfa,Belanger:2013xza,Buckley:2012em,Arbey:2012na,Arbey:2012dq,Akula:2012kk,Cao:2012yn,Howe:2012xe,Drees:2012fb,Haisch:2012re,Bechtle:2012jw,Arbey:2012bp,Ke:2012zq, Chakraborty:2013si,Carmona:2013cq,Arbey:2013jla,Cao:2013wqa,Bhattacherjee:2013vga,Lopez-Val:2013yba,Belyaev:2013ida,Cao:2013gba,Bharucha:2013ela,Cheung:2013rva,Bechtle:2013mda,Cao:2013mqa,Djouadi:2013lra,Enberg:2013jba,Cao:2013cfa,Wang:2013sha,Fan:2014txa,Belanger:2014roa,Stal:2014sua}}

\title{Probing the Standard Model with Higgs signal rates from the Tevatron, the LHC and a future ILC}
\author[a]{Philip Bechtle,}
\author[b]{Sven Heinemeyer,}
\author[c]{Oscar St{\aa}l,}
\author[a,d]{Tim Stefaniak}
\author[e]{and Georg Weiglein}

\affiliation[a]{Physikalisches Institut der Universit\"at Bonn,
 Nu{\ss}allee 12, D-53115 Bonn, Germany}
\affiliation[b]{Instituto de F\'isica de Cantabria (CSIC-UC), Santander,  Spain} 
\affiliation[c]{The Oskar Klein Centre, Department of Physics,
  Stockholm University,\\ SE-106 91 Stockholm, Sweden}
\affiliation[d]{Bethe Center for Theoretical Physics, Bonn University, Germany}  
\affiliation[e]{Deutsches Elektronen-Synchrotron DESY, Notkestra{\ss}e 85, D-22607 Hamburg, Germany}

\emailAdd{bechtle@physik.uni-bonn.de}
\emailAdd{Sven.Heinemeyer@cern.ch}
\emailAdd{oscar.stal@fysik.su.se}
\emailAdd{tim@th.physik.uni-bonn.de}
\emailAdd{Georg.Weiglein@desy.de}

\abstract{
We explore the room for possible deviations from the Standard Model
(SM) Higgs boson coupling structure in a systematic study of Higgs
coupling scale factor ($\kappa$) benchmark scenarios using the latest
signal rate measurements from the Tevatron and LHC experiments. We employ
$\chi^2$ fits performed with \HS,
which takes into account detailed information on signal efficiencies and
major correlations of theoretical and experimental uncertainties. All
considered scenarios allow for additional non-standard Higgs boson
decay modes, and various assumptions for constraining the total
decay width are discussed. No significant deviations from the SM Higgs boson 
coupling structure are found in any of the investigated benchmark
scenarios. We derive upper limits on an additional (undetectable)
Higgs decay mode under the assumption that the Higgs couplings to
weak gauge bosons do not exceed the SM prediction. We
furthermore discuss the capabilities of future facilities for probing
deviations from the SM Higgs couplings, comparing the high luminosity
upgrade of the LHC with a future International Linear Collider (ILC),
where for the latter various energy and luminosity scenarios are
considered. At the ILC model-independent
measurements of the coupling structure can be performed, and we provide
estimates of the precision that can be achieved.}


\arxivnumber{1403.1582}

\begin{document}

\maketitle
\flushbottom



\section{Introduction}

On July 4, 2012 the discovery of a narrow resonance, with a mass near 
$\MHexp\gev$, in the search for the Standard Model (SM) Higgs boson at the Large Hadron Collider (LHC) was announced at CERN by both the ATLAS and CMS experiments~\cite{Aad:2012tfa,Chatrchyan:2012ufa}. The initial discovery was based on the data collected at the LHC until June 2012, and these results have since been confirmed and refined using the full 2012 data
set~\cite{ATLAS:2013sla,ATLAS:2013mla,CMS:yva,CMS:2013wda}. Results from the Tevatron experiments~\cite{Aaltonen:2013kxa} support the findings.
Within the current experimental and theoretical uncertainties the properties of
the newly discovered particle are thus far in very good
agreement with the predictions for a SM Higgs boson, including the
measurements of signal rates as well as 
further properties such as spin.

In order to test the compatibility of the newly observed boson with the
predictions for the Higgs boson of the SM based on the data accumulated up to 2012,
the LHC Higgs Cross Section Working Group (LHCHXSWG)
proposed several benchmark scenarios within an ``interim framework'' 
employing \textit{Higgs coupling scale 
factors}~\cite{LHCHiggsCrossSectionWorkingGroup:2012nn,Heinemeyer:2013tqa}. 
This approach is based on earlier studies of the LHC sensitivity to Higgs couplings initiated in 
Refs.~\cite{Zeppenfeld:2000td,Duhrssen:2003tba,Duhrssen:2004cv,Duhrssen:2004uu} and has been influenced
by further analyses as described in
\citeres{LHCHiggsCrossSectionWorkingGroup:2012nn,Heinemeyer:2013tqa}, see also
\citere{LHCHiggsCrossSectionWorkingGroup:2012nn} concerning the discussion of a more general
framework based on suitable effective Lagrangians.
The coupling scale factors have been analyzed by the experimental
collaborations~\cite{ATLAS:2013sla,CMS:yva} 
as well as in further phenomenological studies where
Higgs coupling fits have been carried out~\citekappafits. 
The results of those analyses show no significant deviations from SM Higgs 
couplings.

The total Higgs decay width for a Higgs boson mass around $\MHexp\gev$ is
not expected to be directly observable at the LHC. In the SM, a
total width around $4\MeV$ is predicted, which is several orders of magnitude
smaller than the experimental mass resolution. Suggestions to achieve more
sensitive constraints on the total width other than the ones limited by the
experimental mass resolution have been made. These are based on the analysis of
off-shell contributions from above the Higgs resonance in Higgs decays to
$ZZ^*$ or $WW^*$ final states~\cite{Caola:2013yja,
Campbell:2011cu,Campbell:2013wga,Campbell:2013una} and of interference
effects between the $\htogaga$ signal and the background
continuum~\cite{Dixon:2013haa}. However, the ultimate sensitivities are
expected to remain about one order of magnitude above the level of the SM
width.
The limited access of the LHC to the Higgs width implies that only ratios of couplings can be
determined at the LHC, rather than couplings themselves, unless additional theory assumptions are made.

Looking beyond the SM, a generic property of many theories with extended Higgs sectors is that the
lightest scalar can have nearly identical properties to the SM Higgs boson. In this so-called decoupling limit, additional
states of the Higgs sector are heavy and may be difficult to detect in
collider searches. Deviations from the Higgs properties in the SM can arise from an
extended structure of the Higgs sector, for instance if there is more than one Higgs doublet. Another source of
possible deviations from the SM Higgs properties are loop effects from new
particles. The potential for deciphering the physics of electroweak symmetry breaking is directly related to the sensitivity for observing
deviations from the SM. Given the far-reaching consequences for our
understanding of the fundamental structure of matter and the basic laws
of nature, it is of the highest priority to probe the properties 
of the newly discovered particle with a comprehensive set of high-precision measurements.
In particular, the determination of its couplings to other particles with the
highest possible precision is crucial.

The aim of this paper is to investigate whether there are
hints of deviations from the SM Higgs couplings based on
a combined picture of all the latest results from the
Tevatron and LHC experiments. By investigating a complete selection of possible scale factor parametrizations of Higgs
coupling strengths ranging from highly constrained to very generic parametrizations,
we systematically study potential tendencies in the signal rates and correlations among the fit parameters. 
In all considered scenarios we allow for additional Higgs decay modes that are either assumed to be altogether \textit{invisible} Higgs decay modes, thus yielding a missing energy collider signature, or considered to be \textit{undetectable} decay modes.
In the latter case, additional model assumptions have to be imposed to
constrain the total width at the LHC. Based on those assumptions
an upper limit on the branching ratio of the undetectable decay mode can be derived for each
parametrization.

Going beyond the present status, we analyze the prospects
of Higgs coupling determination at future LHC runs with
$300~\ifb$ and $3000~\ifb$ of integrated luminosity, as well as with
a future $e^+e^-$ International Linear Collider (ILC). The estimated ILC capabilities are presented both
for a model-dependent and model-independent fit framework. In the first
case, the total width is constrained by imposing the same assumptions as
required for the LHC, and we compare the ILC capabilities directly with those
of the high-luminosity LHC (HL--LHC) with $3000~\ifb$. In the latter case,
the total width is only constrained by the total cross section measurement
of the $\epem \to ZH$ process at the ILC, thus enabling 
measurements of coupling scale factors free from theoretical prejudice.

Finding significant deviations in any Higgs coupling scale factors would 
provide a strong motivation for studying full models which exhibit a corresponding coupling
pattern. However, the fit results obtained within the framework of
coupling scale factors can in general \textit{not} be directly
translated into realistic new physics models, see Sect.~\ref{Sect:kappas}
for a discussion. Concerning the investigation of particular models of new physics, the 
most reliable and complete results are obtained by performing a dedicated
fit of the Higgs signal rates within the considered model.
Such model-dependent fits, see e.g.~\citemodelfits~for recent studies,
can easily be performed with the generic code \HS~\cite{Bechtle:2013xfa,Stal:2013hwa} that has been used to perform this work.
 
This paper is organized as follows. Sect.~\ref{Sect:methods}
  introduces the fit methodology and the statistical treatment employed
  in \HS. This section also contains a discussion of the experimental input and and the treatment of
  theoretical uncertainties, with further details given in the Appendix. In Sect.~\ref{Sect:currentfits} we present the fit results for
  the various benchmark parametrizations of Higgs coupling scale
  factors using all the currently available data
from the LHC and the Tevatron. Results for future
  expectations are presented in \refse{Sect:futurecouplings}. Here the
  current data is replaced by the projections for the future precision at the HL--LHC and the ILC, and we
  discuss the accuracy to which the Higgs coupling scale factors can be determined in
  the various scenarios. The conclusions are given in
  \refse{Sect:conclusions}. Additional 
  information can be found in the three appendices.
  Appendix A presents the
  experimental dataset that is used (and its validation). Appendix B
  contains a discussion of the statistical \pvalue\ derived from
  $\chi^2$ tests of model predictions against measurements of Higgs boson signal rates. Finally, Appendix C contains further details on how we
  treat the theoretical uncertainties of Higgs production and decay rates.


\section{Methodology}
\label{Sect:methods}

\subsection{Coupling scale factors}
\label{Sect:kappas}
The SM predicts the couplings of the Higgs boson
to all other known particles.
These couplings directly influence the rates and kinematic properties of
the production and decay of the Higgs boson. 
Therefore, measurements of the production and decay rates of the observed
state, as well as their angular correlations, yield information 
that can be used to probe whether data is compatible with the SM predictions. 

In the SM, once the numerical value of the Higgs
mass is specified, all the couplings of the Higgs boson to fermions, gauge
bosons and to itself are specified within the model.
It is therefore in
general not possible to perform a fit to experimental data within the
context of the SM where Higgs couplings are treated as free parameters \cite{Cornwall:1973tb,LlewellynSmith:1973ey}.
In order to test the compatibility of the newly observed boson with the
predictions for the SM Higgs boson and potentially to find evidence for
deviations in the 2012 data, the LHC Higgs Cross Section Working Group (LHCHXSWG)
proposed several benchmark scenarios containing \textit{Higgs coupling scale
factors} within an ``interim framework''
~\cite{LHCHiggsCrossSectionWorkingGroup:2012nn,Heinemeyer:2013tqa}.
This framework is based on several assumptions. In particular,
all deviations from the SM are computed assuming that there is only one underlying state at 
$\MHexp\gev$.
It is assumed that this state is a Higgs boson,
and that it is SM-like, in the sense that the experimental results so
far are compatible with the SM Higgs boson hypothesis.
Also the coupling tensor structures are assumed to be as in the
SM, meaning in particular that the state is \cp-even scalar.
Furthermore, the zero width approximation is assumed to be valid,
allowing for a clear separation and simple handling of production and decay of the Higgs particle. 

In order to take into account the currently best available SM
predictions for Higgs cross sections and partial widths, which include
higher-order QCD and EW
corrections~\cite{Dittmaier:2011ti,Dittmaier:2012vm,Heinemeyer:2013tqa},
while at the same time introducing possible deviations from the SM
values of the couplings, the  predicted SM Higgs cross sections and partial
decay widths are dressed with scale factors $\kappa_i$. 
The scale factors $\kappa_i$ are defined in such a way that the cross
sections $\sigma_{ii}$ or the partial decay widths $\Gamma_{ii}$
associated with the SM particle $i$ scale with the factor $\kappa_i^2$
when compared to the corresponding SM prediction.\footnote{Note, that in this interim framework, slight dependencies of the derived collider observables (cross sections $\sigma_{ii}$, partial widths $\Gamma_{ii}$) on the remaining Higgs coupling scale factors, $\kappa_j~(j\ne i)$, are often neglected. For instance, the cross section of the Higgs-strahlung process $\pptoZH$ features a small dependence on the top-Yukawa coupling scale factor entering via the NNLO process $gg\to Z^{*}\to HZ$~\cite{Harlander:2013mla}. However, for scale factor ranges, $\kappa_t \lesssim 3$, this effect is negligible. Hence, the $\pptoZH$ cross section can be simply rescaled by $\kappa_Z^2$.} 
The most relevant coupling scale factors are
$\kappa_t$, $\kappa_b$, $\kappa_\tau$, $\kappa_W$, $\kappa_Z$, \ldots
In the various benchmark scenarios defined in 
\citere{LHCHiggsCrossSectionWorkingGroup:2012nn,Heinemeyer:2013tqa}
several assumptions are made on the relations of these scale factors in order to investigate certain aspects of the Higgs
boson couplings, as will be discussed here in \refse{Sect:currentfits}.

One should keep in
mind that the inherent simplifications in the $\kappa$ framework make it rarely possible to directly
map the obtained results onto realistic models beyond the SM (BSM). The scale factor benchmark scenarios typically have more freedom to adjust the predicted signal rates to the measurements than realistic, renormalizable models.
The latter generally feature specific correlations among the predicted rates, which furthermore can depend non-trivially and non-linearly on the model parameters. 
Moreover, constraints from the electroweak precision data and possibly other sectors, such as dark matter, collider searches, vacuum stability, etc., can further restrict the allowed parameter space and thus the room for Higgs coupling deviations. Preferred values and confidence regions of the scale factors obtained from profiling over regions in the $\kappa$ parameter space, which are not covered by the allowed parameter space of the full model, cannot be transferred to the full model. The implications of the Higgs signal rate measurements for the full model can then only be investigated consistently in a dedicated, model-dependent analysis. In that sense, such
analyses of realistic BSM models are complementary to the approach followed here, and can easily be performed with the same tools and statistical methods as employed here.

One limitation at the LHC, but not at the ILC, is the fact that the
total width cannot be determined experimentally without additional
theory assumptions. In the absence of a total width measurement only
ratios of $\kappa$'s can be determined from experimental data.
In order to go beyond the measurement of ratios of coupling scale
factors to the determination of absolute coupling scale factors
$\kappa_i$ additional assumptions are necessary to remove one degree of
freedom. One possible and simple assumption is that there are no new Higgs decay modes
besides those with SM particles in the final state. 
Another possibility is to assume the final state of potentially present additional Higgs decay(s) to be purely invisible, leading to a $Z$ boson recoiling against missing
transverse energy in the Higgs-strahlung process at the LHC~\cite{Eboli:2000ze}. By employing constraints from dedicated
LHC searches for this signature the total width can be constrained.
In both cases, further assumptions need to be imposed on the partial widths of Higgs decays to SM particles
which are unobservable at the LHC, like for instance $H\to gg, cc, ss$. 
As a third possibility, an assumption can be made on the couplings of the Higgs to the
SM gauge bosons, $\kappa_{W,Z} \le 1$~\cite{Duhrssen:2004cv,Duhrssen:2004uu}. 
This assumption is theoretically well-motivated as it holds in a wide
class of models. For instance, they hold in any model with an arbitrary 
number of Higgs doublets, with and without additional Higgs singlets, or
in certain classes of composite Higgs models.
We will partly make use of these assumptions in our analysis below.
More details will be given in Sect.~\ref{Sect:currentfits}.


\subsection{The statistical analysis using \HS}
\label{Sect:HS}
We use the public computer program \HS~\cite{Bechtle:2013xfa,Stal:2013hwa}, based on the
\HBv{4}\ library~\cite{Bechtle:2008jh,Bechtle:2011sb,Bechtle:2013gu,Bechtle:2013wla}, which
is a dedicated tool to test model predictions of arbitrary Higgs sectors
against the mass and signal rate measurements from Higgs searches at the
LHC and the Tevatron. For both types of measurement a statistical $\chi^2$
value can be evaluated, denoted as $\chi^2_\mu$ for the signal rates
and $\chi^2_m$ for the Higgs mass. In this work we are only interested in the contribution from the signal rates and fix the Higgs mass to $m_H = \MHexp\gev$.

The Higgs signal rate measurement performed in an analysis $i$, denoted
by $\muobs_i$, is given by the experiments as a SM
normalized quantity. It contains all relevant Higgs collider processes, where each process is comprised of a production mode $P_j(H)$ and a decay mode $D_j(H)$, and features a specific efficiency $\epsilon_j$. The observed signal strength
modifier can thus be understood as a universal scale factor for the SM
predicted signal rates of all involved Higgs processes. The
corresponding model-predicted signal rates are calculated as 
\begin{equation}
\mu_i = \frac{\sum_j \epsilon_\mathrm{model}^{i,j} \,
  \sigma_\mathrm{model}(P_j(H)) \times \br_\mathrm{model}
    (D_j(H))}{\sum_j \epsilon^{i,j}_\mathrm{SM} \,
  \sigma_\mathrm{SM}(P_j(H)) \times \br_\mathrm{SM} (D_j(H))}. 
 \label{Eq:mu} 
\end{equation}
In general, the efficiencies $\epsilon^{i,j}$ can be different from the
SM for models where the influence of Higgs boson interaction terms with
a non-standard (higher-dimensional or \cp-odd) tensor-structure cannot
be
neglected~\cite{Banerjee:2013apa,Anderson:2013afp,Azatov:2013xha,Boos:2013mqa,Chen:2013waa,Buchalla:2013mpa,Dumont:2013wma}. 
In this paper, the efficiencies $\epsilon^{i,j}$ are assumed to be
identical for the SM and the unknown model predicting the rescaled
signal rates. This assumption is valid for \textit{small}
deviations from the SM Higgs couplings, where kinematic effects changing the efficiencies can be neglected. However, if
significant deviations from the SM are found from the analysis, 
a more careful investigation of anomalous Higgs couplings~\cite{Buchmuller:1985jz,Contino:2013kra,Pomarol:2013zra} becomes necessary, 
including a detailed study of their effects on the efficiencies. On the experimental side, the
publication of differential fiducial cross sections should then be
considered~\cite{Boudjema:2013qla}.

In this work we employ hybrid Bayesian-frequentist fits based on the $\chi^2$ value derived from \HS. In this approach the systematic uncertainties are parametrized using Gaussian probability density functions, while all model parameters are treated in a frequentist manner. In order to determine the uncertainties on individual model parameters in multi-dimensional fits we profile over the remaining model parameters. A ``naive'' \pvalue, i.e.~the
    probability of falsely rejecting a specific model assuming it is true, is quoted based on the
  agreement between the minimal $\chi^2$ value found in the fit and
  the number of degrees of freedom (ndf). However, the $\chi^2$
    value evaluated by \HS\ does not generically fulfill the
    prerequisite for this simple \pvalue\ estimation: Firstly, \HS\
  uses asymmetric uncertainties in order to take into account
    remaining non-Gaussian effects in the measurements. Secondly, the
    signal rate uncertainties are comprised of constant and relative
    parts. The latter include the theoretical uncertainties on the cross
    sections and branching ratios, which are proportional to the
    signal rate prediction, as well as the luminosity uncertainty,
    which is proportional to the measured signal rate. These features
    are necessary in order to effectively reproduce the properties of the full
  likelihood implementation as done by the experimental collaborations
  and ensure the correct scaling behavior when testing
    models different from the SM~\cite{Bechtle:2013xfa}.

These features potentially introduce deviations from the naive
  $\chi^2$ behavior, which could affect both the extraction of
 preferred parameter ranges at a certain
    confidence level (C.L.) as well as the
  calculation of the \pvalue. In order to estimate the impact of these effects, we performed a Monte Carlo (MC) toy study
  for a simple one-dimensional scale factor model, which is presented
  in Appendix~\ref{Sect:pvalue}. From this study two important
  conclusions can be drawn: Firstly, the central value and
  uncertainties of the estimated fit parameter extracted from
  the full toy study do indicate a small variation from the naive values
  extracted from profiling. However, these variations are each less
    than $2\%$. Hence, we are confident that the uncertainties and best fit values quoted later for the profile likelihood scans are
    valid to a good approximation. Secondly, the \pvalue\ obtained in the full MC toy study can be different to the naive $\chi^2$
  distribution. For an example of a change in the shape of the observed $\chi^2$ probability density function in
  toy experiments, see Fig.~\ref{fig:1Dfit_toy_chisq_dist} in
  Appendix~\ref{Sect:pvalue}, which indicates that the actual \pvalue\ may be higher than expected when assuming an ideal $\chi^2$ distribution. This effect could be significant and should be taken into account once this technique is used to exclude models, e.g.~once the $\chi^2$ probability comes close to 5\%. As will be discussed later in this work,
  we find naive ${\cal P}$-values for the best-fit points of all considered models in the range of 25 -- 35\%, which
  are far away from any critical border. Therefore, we are confident that the
  conclusions drawn from the naive ${\cal P}$-values in the remainder
  of the paper would not change in any significant way if a full toy
  study or, even better, a full likelihood analysis by the
  experimental collaborations, was done for every fit.

Within \HS\ the correlations of theoretical cross section, branching ratio and luminosity uncertainties among different observables are taken into account~\cite{Bechtle:2013xfa}. For this work, we further develop this implementation to also take into account major correlations of experimental systematic uncertainties for a few important analyses where the necessary information is provided. Specifically, this is the case for the CMS $\htogaga$ analysis~\cite{CMS:ril} and the ATLAS $\htotautau$ analysis~\cite{ATLAS-CONF-2013-108}. More details are given in Sect.~\ref{Sect:measurements} and Appendix~\ref{Sect:expdata}, including a comparison with official results.

We want to note that an alternative approach for transferring the experimental results into global Higgs coupling or model fits exists~\cite{Belanger:2013xza,Boudjema:2013qla}.
This approach suggests that the experiments provide combined higher-dimensional likelihood distributions for scale factors of the Higgs boson production modes. On first sight, an appealing feature is that correlations among the combined analyses of the experimental and theoretical uncertainties for the model investigated by the analysis (usually the SM) are automatically taken care of by the collaborations. However, when going beyond the combination of that specific selection of analyses, e.g.~when combining ATLAS and CMS results, or already when combining the likelihoods of different decay modes, detailed knowledge of the correlations of common uncertainty sources is again required. Moreover, a careful treatment of these correlations in a combination, as is done for the simple one-dimensional signal strength measurements in \HS, is far more complicated for higher-dimensional likelihoods. We therefore advise the collaborations to continue providing one-dimensional signal strength measurements, including detailed information on signal efficiencies and correlations, since then the amount of model-dependence in the experimental results is rather minimal~\cite{CorrSystDoc}. Nevertheless, we also support the suggestions made in Refs.~\cite{Boudjema:2013qla}, since these higher-dimensional likelihoods are still useful on their own and for validation. Note also recently proposed attempts to disentangle theoretical uncertainties from signal strength measurements~\cite{Cranmer:2013hia}.

The technical details of the profiled likelihood scans performed in this work are as follows.
For an efficient sampling of the parameter space the scans are performed with an adaptive Metropolis (AM) algorithm~\cite{haario2001ama} with flat prior probability distributions using the Markov-Chain Monte Carlo (MCMC) python package \texttt{PyMC}~\cite{Patil:Huard:Fonnesbeck:2010}. Appropriate initial values for the MCMC chains are found using the maximum a posteriori estimate (MAP) class of \texttt{PyMC}. The scans contain around $10^5$ -- $10^7$ points, depending on the dimensionality of the parameter space. For each scan, several Markov chains are run independently of each other, featuring a typical length of $\sim 10^5$ scan points. The results are presented in a purely frequentist's interpretation based on the global $\chi^2$ derived from \HS\ and, optionally, further $\chi^2$ contributions arising from constraints from invisible Higgs searches at the LHC. Hence, the Markov Chains merely serve as a sampling method of the parameter space and no Bayesian interpretation is employed. This resulting higher-dimensional $\chi^2$ distribution is then profiled in order to obtain one- and two-dimensional likelihoods for the fit parameters and related quantities. The $\{1, 2, 3\}\,\sigma$ parameter regions around the best-fit point are then obtained for values of the $\chi^2$ difference to the minimal value, $\Delta\chi^2 = \chi^2 - \chi^2_\text{min}$, of $\Delta \chi^2 \leq \{1.0, 4.0, 9.0\}$ for the one-dimensional, and $\Delta \chi^2 \leq \{2.30, 5.99, 11.90\}$ for the two-dimensional profiles, respectively. As discussed above, we also quote for each benchmark scenario the fit quality at the best-fit point, given by $\chi^2_\text{min}/\text{ndf}$, and the corresponding (naively estimated) \pvalue.

\subsection{Experimental input from the Tevatron and the LHC}
\label{Sect:measurements}

In the analysis of the present status of potential deviations in the Higgs couplings, presented in Sect.~\ref{Sect:currentfits}, we use the latest available signal strength measurements from the Tevatron and LHC experiments, which are included in \HSv{1.2.0}. Detailed information on these in total $80$ signal strength measurements and the assumed signal composition of the production modes is given in Appendix~\ref{Sect:expdata}. Notably, these measurements include the recently published results from ATLAS in the $\htotautau$ channel~\cite{ATLAS-CONF-2013-108}, for which we implement correlations of experimental systematic uncertainties in \HS, cf.~Appendix~\ref{Sect:expdata}. Based on the comparison of a six-dimensional scale factor fit to the official CMS results~\cite{CMS:yva}, we perform an approximate rescaling of the CMS $\htogaga$ measurements~\cite{CMS:ril} from the published Higgs mass value of $125.0\gev$ to the best-fit combined mass of $125.7\gev$. Using the rescaled measurements, we find very good agreement with the official CMS fit results, see Appendix~\ref{Sect:CMSvalidation} for details.

\subsection{Treatment of theoretical uncertainties}

We attempt to account for various correlations among the theoretical uncertainties of the cross section and branching ratio predictions. Correlations of theoretical uncertainties among different signal strength observables, as well as correlations among the theoretical uncertainties themselves induced by e.g.~common parametric dependencies, are taken into account in \HS~since version \texttt{1.1.0}~\cite{HSreleasenote}. Here we outline how the latter type of correlations is evaluated. More details are given in Appendix~\ref{Sect:AppTHU}.

The contributions of the major parametric and theoretical (higher-order) uncertainty sources to the
total uncertainties of the partial decay widths and production cross
sections are given separately by the LHCHXSWG in Refs.~\cite{Denner:2011mq,Heinemeyer:2013tqa}.
However, there is unfortunately no consensus on how these contributions can
be properly combined since the shapes of the underlying probability
distributions are unknown. Hence, thus far, the use of conservative
maximum error estimates is recommended. Nevertheless, such a
prescription is needed in order to account for the correlations. In this work we employ covariance
matrices evaluated by a Monte Carlo (MC) simulation, which combines the parametric
and theoretical uncertainties in a correlated way. The importance of a combination prescription for
precision Higgs coupling determination in the future ILC era is briefly discussed in Appendix~\ref{Sect:AppTHU}.

The relative parametric uncertainties (PU) on the partial Higgs decay widths, $\Delta\Gamma^i_\mathrm{PU} (H\to X_k)$, from the strong coupling, $\alpha_s$, and the charm, bottom and top quark mass, $m_c$, $m_b$ and $m_t$, respectively, as well as the theoretical uncertainties (THU) from missing higher order corrections, $\Delta \Gamma_\mathrm{THU}(H\to X_k)$, are given in Tab.~1 of Ref.~\cite{Heinemeyer:2013tqa}. The PUs are given for each decay mode for both positive and negative variation of the parameter. From this response to the parameter variation we can deduce the correlations among the various decay modes resulting from the PUs. More importantly, correlations between the branching ratio uncertainties are introduced by the total decay width, $\Gamma^\mathrm{tot} = \sum_k \Gamma (H\to X_k)$.

The covariance matrix for the Higgs branching ratios is then evaluated with a toy MC: all PUs are smeared by a Gaussian of width $\Delta\Gamma^i_\mathrm{PU} (H\to X_k)$, where the derived correlations are taken into account. Similarly, the THUs are smeared by a Gaussian or a uniform distribution within their uncertainties. Hereby, we treat all THUs as uncorrelated, except for those of $WH$ and $ZH$ production. These are treated as fully correlated due to their common higher-order QCD effects. We find that both probability distributions give approximately the same covariance matrix. A detailed description of our procedure is given in Appendix~\ref{Sect:AppTHU}, including a comparison of different implementations and assumptions on the theoretical uncertainties in the light of future data from the high luminosity LHC and ILC.
Overall, we find slightly smaller estimates for the uncertainties than those advocated by the LHCHXSWG, cf.~Appendix~\ref{Sect:AppTHU}. This is not surprising, since the (very conservative) recommendation is to combine the uncertainties linearly. 

Using the present uncertainty estimates~\cite{Heinemeyer:2013tqa}, the correlation matrix for the branching ratios in the basis ($H\to \gamma\gamma, WW, ZZ, \tau\tau, bb, Z\gamma, cc, \mu\mu, gg$)  is given by
\begin{equation}
\footnotesize
(\rho_{\brat,ij}^\mathrm{SM}) = 
\begin{pmatrix}
  1 & 0.91	& 0.91 & 0.71 & -0.88 & 0.41 & -0.13 & 0.72 & 0.60 \\
  0.91 & 1 & 0.96 & 0.75 & -0.94 & 0.43 & -0.14 & 0.76 & 0.64 \\
 0.91 & 0.96 & 1  &0.75 &  -0.93 & 0.43 & -0.13 & 0.76 & 0.64 \\
 0.71 & 0.75 & 0.75 & 1  & -0.79 & 0.34 & -0.12 & 0.59 & 0.50 \\
 -0.88 & -0.94 & -0.93 & -0.79 & 1 & -0.42 & 0.11 & -0.73 & -0.79 \\
 0.41 & 0.43 & 0.43 & 0.34 & -0.42 & 1 & -0.05 & 0.34 & 0.29 \\
 -0.13 & -0.14 & -0.13 & -0.12 & 0.11 & -0.05 & 1 & -0.12 & -0.50 \\
 0.72 & 0.76 & 0.76 & 0.59 & -0.73 & 0.34 & -0.12 & 1 & 0.50 \\
 0.60 & 0.64 & 0.64 & 0.50 & -0.79 & 0.29 & -0.50 & 0.50 & 1\\
 \end{pmatrix}.
 \label{Eq:corrBR}
\end{equation}
As can be seen, strong correlations are introduced via the total width. As a result, the $\htobb$ channel, which dominates the total width, as well as the $\htocc$ channel are anti-correlated with the remaining decay modes.

For the production modes at the LHC with a center-of-mass energy of $8\tev$ the correlation matrix in the basis (ggH, VBF, $WH$, $ZH$, $t\bar{t}H$) is given by
\begin{equation}
(\rho_{\sigma,ij}^\mathrm{SM}) = 
\begin{pmatrix}
1		&	 -2.0\cdot 10^{-4} &	3.7\cdot 10^{-4} &	9.0\cdot 10^{-4} &  0.524 \\
-2.0\cdot 10^{-4} & 	 1 		&	0.658		& 	0.439     &		2.5\cdot 10^{-4} \\
3.7\cdot 10^{-4}  & 	0.658 	&		1		&	 0.866   & 	-9.8\cdot10^{-5} \\
9.0\cdot 10^{-4}  & 	0.439	&	0.866		& 	1		&	2.8\cdot10^{-4}\\
 0.524		& 	2.5\cdot 10^{-4} & 		-9.8\cdot10^{-5} & 	2.8\cdot10^{-4} &  	1\\
 \end{pmatrix}.
 \label{Eq:corrXS}
\end{equation}
Significant correlations appear between the gluon fusion (ggH) and $t\bar{t}H$ production processes due to common uncertainties from the parton distributions and QCD-scale dependencies, as well as among the vector boson fusion (VBF) and associate Higgs-vector boson production ($WH$, $ZH$) channels. 

These correlations are taken into account in all fits presented in this work. The numerical values presented in Eqs.~\eqref{Eq:corrBR} and \eqref{Eq:corrXS} are evaluated for the setting used in the fits to current measurements, cf.~Sect.~\ref{Sect:currentfits}, as well as in the conservative future LHC scenario (S1), see Sect.~\ref{Sect:futureLHC}. For the other future scenarios discussed in Sect.~\ref{Sect:futurecouplings}, we re-evaluate the covariance matrices based on the assumptions on future improvements of parametric and theoretical (higher-order) uncertainties. However, while the magnitude of the uncertainties changes in the various scenarios discussed later, we find that the correlations encoded in Eqs.~\eqref{Eq:corrBR} and \eqref{Eq:corrXS} are rather universal. A comparison of uncertainty estimates among all future scenarios we discuss, as well as with the recommended values from the LHCHXSWG, can be found in Appendix~\ref{Sect:AppTHU}.

\section{Current status of Higgs boson couplings}
\label{Sect:currentfits}

In this section we explore the room for possible deviations from
the SM Higgs boson couplings for various benchmark models, each
targeting slightly different aspects of the Higgs sector. We follow the
LHCHXSWG interim framework~\cite{LHCHiggsCrossSectionWorkingGroup:2012nn, Heinemeyer:2013tqa} for
probing (small) deviations from the SM Higgs boson couplings by
employing simple scale factors for the production and decay
rates. Hereby, we assume that the LHC Higgs signal is due to a single
narrow resonance with a mass of $\sim \MHexp\gev$. The experimental
signal efficiencies of the various analyses are assumed to be unchanged
with respect to the SM Higgs signal. This corresponds to the underlying
assumption that the tensor structure of the couplings is the same as in
the SM, i.e.~we investigate the coupling structure of a $\cp$-even,
scalar boson. 

The LHC signal rate measurements, i.e.~measurements of the product of a
production cross section times the branching ratio to a certain final
state, do not provide direct information about the total width of the
Higgs boson. Hence, the LHC is regarded as being insensitive to probe the
total Higgs width, $\Gamma_H$, unless it features a very broad
resonance, $\Gamma_H \sim \mathcal{O}(\mathrm{few}\gev)$. The
current best limit, $\Gamma_H < 3.4\gev$ at $95\%~\mathrm{C.L.}$, is
obtained by CMS using the $\htoZZ\to 4 \ell$
channel~\cite{Chatrchyan:2013mxa}. An even more recent proposal is to
exploit the $ZZ$ invariant mass spectrum in the process $gg \to
ZZ^{(*)}\to 4\ell$, where the total Higgs decay width can be constrained
due to strongly enhanced contributions from off-shell Higgs
production. This has been projected to yield a $95\%~\mathrm{C.L.}$
upper limit of $\Gamma_H \lesssim 40 \times~\Gamma_H^\mathrm{SM}$ using
current data, and a potential future sensitivity of $\sim 10 \times
\Gamma_H^\mathrm{SM}$ is claimed for increased integrated
luminosity~\cite{Caola:2013yja,Campbell:2013una}. A total width of that
order still allows for a significant branching fraction to
undetectable/invisible final states and sizable coupling
modifications. SM-like signal rates for a Higgs boson with an increased
total width can always be obtained by a simultaneous increase of the
branching fraction to undetectable particles and the Higgs couplings to
SM particles, if both are allowed to vary and no further assumptions are
imposed~\cite{Duhrssen:2004cv,Duhrssen:2004uu,Belanger:2013kya}.\footnote{Although the
  $\kappa$ scale factor framework technically features a perfect
  degeneracy between an increasing $\brhnp$ and increasing scale factors
  of the Higgs couplings to SM particles if no additional constraints
  are imposed, the validity of the underlying model assumptions --- in
  particular the assumption of identical signal efficiencies as in the
  SM --- need to be scrutinized carefully in parameter regions with
  significant deviations from the SM Higgs couplings. In general,
  effects leading to such large coupling deviations within the
  underlying (unknown) model may potentially also lead to different
  kinematical distributions and hence to changed signal
  efficiencies. Furthermore, the narrow width approximation will become
  worse for an increasing total width.} 
Given the signal rate measurements
from the experiments at the Tevatron and the LHC, this degeneracy can
only be overcome by additional model assumptions and constraints.

In our analysis, we generally allow for an additional branching fraction to new physics, $\brhnp$. Concerning the assumptions needed to constrain the total width, we
distinguish the two cases of the additional branching fraction being comprised of either {\em invisible} or {\em undetectable}
Higgs decays. The {\em invisible} decays are considered to measurable/detectable
via e.g.~the Higgs-strahlung process, leading to a $Z$ boson recoiling
against missing transverse energy at the LHC~\cite{Eboli:2000ze}. Invisible Higgs decays can appear in models where the Higgs boson couples to a light dark matter (DM) candidate, as for instance in light singlet DM models~\cite{Cline:2013gha} or supersymmetry with a stable neutralino as the lightest supersymmetric particle (LSP).
In contrast, the {\em undetectable} decays cannot be constrained by any present LHC analysis. Possible examples are $H\to gg, cc, ss$ or other light flavored hadronic Higgs decays as these signatures are considered indistinguishable from the background. Other examples can be found in theories beyond the SM, like for instance, decays to supersymmetric particles that further decay via SUSY cascades or via $\mathcal{R}$-parity violating interactions~\cite{Dreiner:1997uz,Barbier:2004ez}, which also potentially leading to detached vertices. In this work we investigate the following two options to overcome this degeneracy:
\begin{itemize}
\item[\textit{(i)}] All additional Higgs decay modes yield an \textit{invisible} final state, i.e.
\begin{align}
\brhnp \equiv \brinv .
\end{align} 
Hence, results from ATLAS and CMS searches measuring the recoil of a $Z$ boson against missing transverse energy in the $\pptoZH$ production can be used to constrain $\kappa_Z^2 \, \brhnp$.
\item[\textit{(ii)}] The Higgs-vector boson coupling scale factor is required to be $\kappa_V \le 1~(V=W,Z)$. The Higgs production in the $VH$ and VBF channels is then constrained from above~\cite{Duhrssen:2004cv,Duhrssen:2004uu}. In this case, no assumption on additional Higgs decay modes needs to be imposed. Hence, an upper limit on $\brhnp$ can be derived from the fit result~\cite{Belanger:2013xza}. This assumption is valid for models that contain only singlet and doublet Higgs fields. However, in models with higher Higgs field representations~\cite{Georgi1985,Chanowitz1985,Gunion1990, oai:arXiv.org:hep-ph/0306034, oai:arXiv.org:1202.1532, Chang2012,Hisano:2013sn,Kanemura:2013mc} this assumption does generally not hold.
\end{itemize}
As will be discussed in Sect.~\ref{Sect:ILC}, both assumptions become obsolete once the direct cross section measurement for $\epem\to HZ$ becomes available from the ILC. 

In the following Sections \ref{Sect:universal}--\ref{Sect:7dim} we discuss several fits to benchmark parametrizations of Higgs coupling deviations, where we also allow for an additional Higgs boson decay mode $\brhnp$ leading to an invisible final state (\textit{i}). In this case, we further constrain the product $\kappa_Z^2 \, \brhnp$ by adding the profile likelihood, $-2\log \Lambda$, from the ATLAS search $\pptoZH\to Z(\mathrm{inv.})$~\cite{ATLAS:2013pma} to the global $\chi^2$ obtained from \HS.\footnote{CMS carried out similar searches for the $\pptoZH \to Z(\mathrm{inv.})$ process and obtained $95\%~\mathrm{C.L.}$ upper limits corresponding to $\kappa_Z^2 \, \brinv\leq 0.75$ (for $Z\to \ell^+\ell^-$)~\cite{CMS:2013yda} and $\leq 1.82$ (for $Z\to b\bar{b}$)~\cite{CMS-PAS-HIG-13-028}. However, unlike ATLAS, CMS does not provide a profile likelihood that can be incorporated into our fit.}
In Sect.~\ref{Sect:kvle1} we instead employ the theoretical constraint $\kV \le 1$ (\textit{ii}) to constrain the total width. Under this condition we derive for each benchmark parametrization upper limits on a new Higgs decay mode, which apply irrespectively of whether the final state is truly invisible or just undetectable.

If the individual scale factors for the loop-induced Higgs couplings to gluons and photons, $\kg$ and $\kga$, respectively, are not treated as individual free parameters in the fit, they can be derived from the fundamental Higgs coupling scale factors. We generally denote such \textit{derived} scale factors as $\kline$. Additional genuine loop contributions from BSM particles to these effective couplings are then assumed to be absent. The Higgs-gluon scale factor is then given in terms of $\kt$ and $\kb$ as~\cite{LHCHiggsCrossSectionWorkingGroup:2012nn,Heinemeyer:2013tqa}
\begin{equation}
\kline_g^2 (\kb, \kt, m_H) = \frac{\kt^2 \cdot \sigma_\mathrm{ggH}^{tt}(m_H) +\kb^2 \cdot \sigma_\mathrm{ggH}^{bb}(m_H) + \kt\kb \cdot \sigma_\mathrm{ggH}^{tb}(m_H)}{\sigma_\mathrm{ggH}^{tt}(m_H) + \sigma_\mathrm{ggH}^{bb}(m_H) + \sigma_\mathrm{ggH}^{tb}(m_H)}.
\label{Eq:kg}
\end{equation}
Here, $\sigma_\mathrm{ggH}^{tt}(m_H)$, $\sigma_\mathrm{ggH}^{bb}(m_H)$ and $\sigma_\mathrm{ggH}^{tb}(m_H)$ denote the contributions to the cross section from the top-quark loop, the bottom-quark loop and the top-bottom interference, respectively. For a Higgs mass around $\MHexp\gev$ the interference term is negative for positive scale factors. Details about state-of-the-art calculations have been summarized in Refs.~\cite{Dittmaier:2011ti,Dittmaier:2012vm,Heinemeyer:2013tqa}. We use numerical values for the different contributions to Eq.~\eqref{Eq:kg} extracted from {\tt FeynHiggs-2.9.4}~\cite{Heinemeyer:1998yj,Hahn:2009zz} for a center-of-mass energy of $8 \tev$. These evaluations are based on the calculations presented in Ref.~\cite{Aglietti:2006tp,Bonciani:2007ex}. The top Yukawa contributions are calculated up to NNLO, whereas the bottom Yukawa contributions are evaluated up to NLO. These calculations agree well with the numbers used so far by the experimental collaborations~\cite{Dittmaier:2011ti}.

Similarly to $\kline_g$, the scale factor for the loop-induced Higgs-photon coupling, $\kline_\gamma$, is derived from the coupling scale factors and contributions to the partial width of the involved particles in the loop,
\begin{equation}
\kline_{\gamma}^2 (\kb, \kt, \ktau, \kW, m_H) = \frac{\sum_{i,j} \kappa_i \kappa_j \cdot \Gamma_{\gamma\gamma}^{ij}(m_H)}{\sum_{i,j} \Gamma_{\gamma\gamma}^{ij}(m_H)},
\label{Eq:kga}
\end{equation}
where $(i,j)$ loops over the particles $tt$, $bb$, $\tau\tau$, $WW$, $tb$, $t\tau$,$tW$, $b\tau$, $bW$, $\tau W$. The $\Gamma^{ij}$ have been evaluated with \texttt{HDECAY}~\cite{Djouadi:1997yw,Spira:1997dg}. The partial widths $\Gamma_{\gamma\gamma}^{ii}$ are derived by setting $\kappa_i =1$, $\kappa_j = 0$ ($i\ne j$). Then the cross terms are derived by first calculating $\Gamma_{\gamma\gamma}$ with $\kappa_i = \kappa_j = 1$ and $\kappa_k = 0~(k\ne i,j)$, and then subtracting $\Gamma_{\gamma\gamma}^{ii}$ and $\Gamma_{\gamma\gamma}^{jj}$.
Despite the absence of a sensitive observable probing the Higgs coupling to $Z\gamma$ directly, we also derive the coupling scale factor $\kline_{Z\gamma}$ in order to infer indirect constraints on this quantity and to evaluate its contribution to the total decay width. This scale factor coupling is derived in complete analogy to $\kline_\gamma$.

In the following benchmark fits, we choose to parametrize our results in terms of the absolute scale factors, $\kappa_i$, and an additional branching ratio to new particles, $\brhnp$. These parameters can be transformed into the total width scale factor $\kappa_H^2$ used in the benchmark model proposals from the LHCHXSWG~\cite{LHCHiggsCrossSectionWorkingGroup:2012nn,Heinemeyer:2013tqa},
\begin{align}
\kappa_H^2 = \frac{\kline_H^2 (\kappa_i)}{1-\brhnp},
\label{Eq:kappaH}
\end{align}
where $\kline_H^2(\kappa_i)$ is the derived scale factor for the SM total width as induced by the modified Higgs couplings to SM particles, $\kappa_i$ (both including the fundamental and loop-induced couplings). For an allowed range $\brhnp \in [0,1]$, the total width scale factor, $\kappa_H^2$, thus ranges from $\kline_H^2$ to infinity.

Before we study potential deviations from the SM Higgs couplings it is worthwhile to look at the fit quality of the SM itself: Tested against the 80 signal rate measurements we find $\chi^2/\text{ndf} = 84.3/80$ which corresponds to a (naive) \pvalue\ of $\sim 35.0\%$.\footnote{For the SM, where we have no additional invisible or undetectable decay modes, we do not count the ATLAS $\brinv$ limit into the ndfs.} Thus, the the measurements are in good agreement with the SM predictions. However, coupling variations may be able to improve the fit quality if the signal rates actually feature systematic under- or over-fluctuations, indicating possible deviations in some of the Higgs couplings from their SM values. It is the goal of the next sections to systematically search for such tendencies as well as to determine the viable parameter space of possible deviations. Note that, if we slightly modify the SM by only adding new Higgs decay modes while keeping the couplings at their SM predictions ($\kappa_i = 1$), we obtain $95\%~\mathrm{C.L.}$ upper limits of $\brinv \le 17\%$ in the case of purely invisible final states of the additional decay modes, and $\brhnp \le 20\%$ in the case of undetectable decay modes.

\subsection{Universal coupling modification}
\label{Sect:universal}

The first benchmark model that we consider contains only one universal Higgs coupling scale factor,
$\kappa$, in addition to the invisible Higgs decay mode. Hence, all Higgs production cross sections and
partial widths to SM particles are universally scaled by $\kappa^2$. Although this scenario seems to be
overly simplistic it actually represents realistic physics models, such as the extension of the SM Higgs
sector by a real or complex scalar
singlet~\cite{Cline:2013gha,Schabinger:2005ei,Patt:2006fw,Barger:2007im,Barger:2008jx,Bhattacharyya:2007pb,Bock:2010nz,Englert:2011yb,Englert:2011aa,Pruna:2013bma}.
In the presence of singlet-doublet mixing $\kappa$ can be identified with the mixing angle. Another
example arises from strongly interacting Higgs sectors, which typically lead to a universal form factor correction to all Higgs couplings \cite{Giudice:2007fh}.
Both undetectable and invisible Higgs decays are potentially present in these models. 

\begin{figure}
\centering
\includegraphics[width=0.65\textwidth]{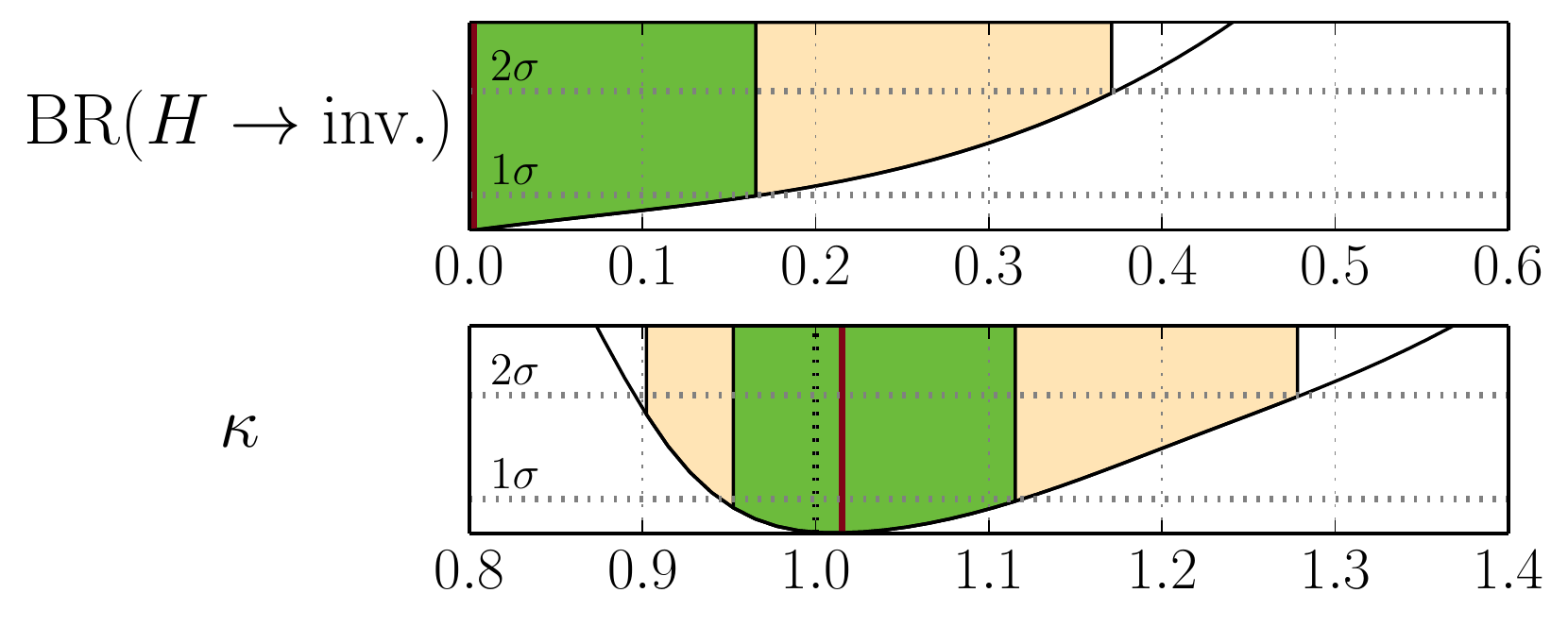}
\caption{One-dimensional $\Delta\chi^2$ profiles for the parameters in the ($\kappa, \brinv$) fit. The best-fit point is indicated by the red line. The $68\%$ ($95\%$) C.L.~regions are illustrated by the green (pale yellow) bands.}
\label{Fig:universal1D}
\end{figure}

\begin{table}
\renewcommand{\arraystretch}{1.3}
\centering
\begin{tabular}{cccc}
\toprule
Fit parameter & best-fit value & $68\%~\mathrm{C.L.}$ range (1D) & $95\%~\mathrm{C.L.}$ range (1D)\\
\midrule
$ \mathrm{BR}(H \to \mathrm{inv.}) $ & $ 0.00 $ & $\substack{+  0.17 \\ - 0.00 }$ & $\substack{+  0.37 \\ - 0.00 }$ \\ 
$ \kappa $ & $ 1.01 $ & $\substack{+  0.10 \\ - 0.08 }$ & $\substack{+  0.26 \\ - 0.13 }$ \\ 
\midrule
$\kappa_H^2 $ &$ 1.03 $ & $\substack{+  0.43 \\ - 0.13 }$ & $\substack{+  1.55 \\ - 0.23 }$ \\ 
\bottomrule
\end{tabular}
\caption{Best-fit values and $68\%$ and $95\%~\mathrm{C.L.}$ ranges for the fit parameters obtained from the one-dimensional $\Delta\chi^2$ profiles in the ($\kappa, \brinv$) fit.}
\label{Tab:universal1D}
\end{table}

We show the fit results obtained under the assumption of a fully invisible additional Higgs decay mode as one- and two-dimensional profiled $\Delta\chi^2$ distributions in Figs.~\ref{Fig:universal1D} and \ref{Fig:universal2D}(a), respectively.
The best fit point is found at $\kappa = 1.01 \substack{+0.10\\ - 0.08}$ with a $\chi_\text{min}^2/\mathrm{ndf} = 84.3/79$, which corresponds to a \pvalue\ of $\sim32.2\%$. The $68\%$ and $95\%~\mathrm{C.L.}$ ranges are also listed in Tab.~\ref{Tab:universal1D}, along with the corresponding range for the total width scale factor $\kappa_H^2$. The two-dimensional $\Delta\chi^2$ distribution in Fig.~\ref{Fig:universal2D}(a) shows a strong positive correlation between $\kappa$ and $\brinv$. This reflects the fact that a suppression of the branching ratios to SM particles introduced by an additional invisible decay mode needs to be compensated by an increase of the production rates. The allowed region is however bounded at increasing $\brinv$ by the limit from the invisible Higgs search from ATLAS. 
\begin{figure}
\centering
\subfigure[Assuming $\brhnp \equiv \brinv$]{\includegraphics[width=0.48\textwidth]{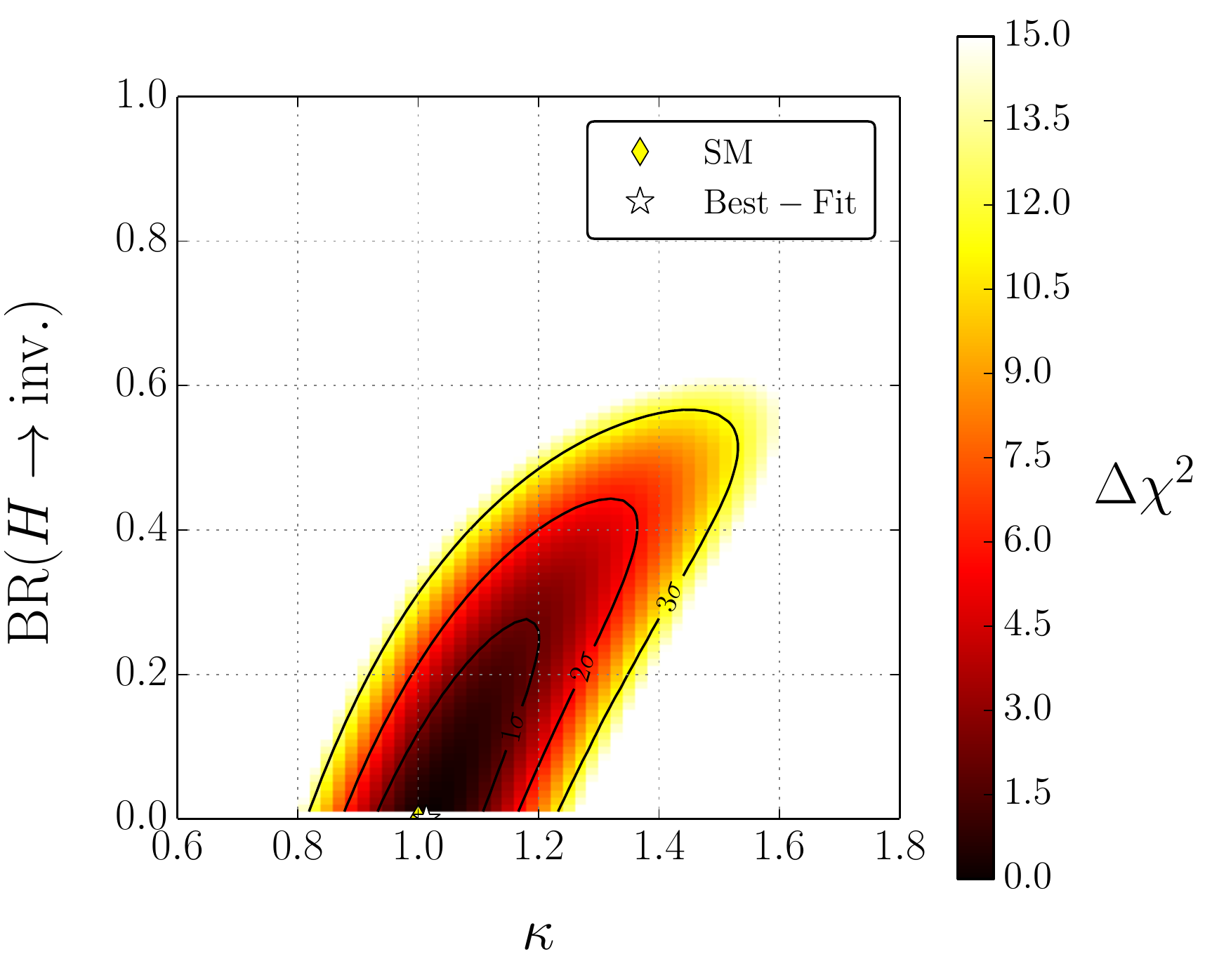}}\hfill
\subfigure[No assumptions. The blue and red contours indicate current and prospective limits, respectively, on the total width from off-shell Higgs production at the LHC~\cite{Caola:2013yja,Campbell:2013una}.]{\includegraphics[width=0.48\textwidth]{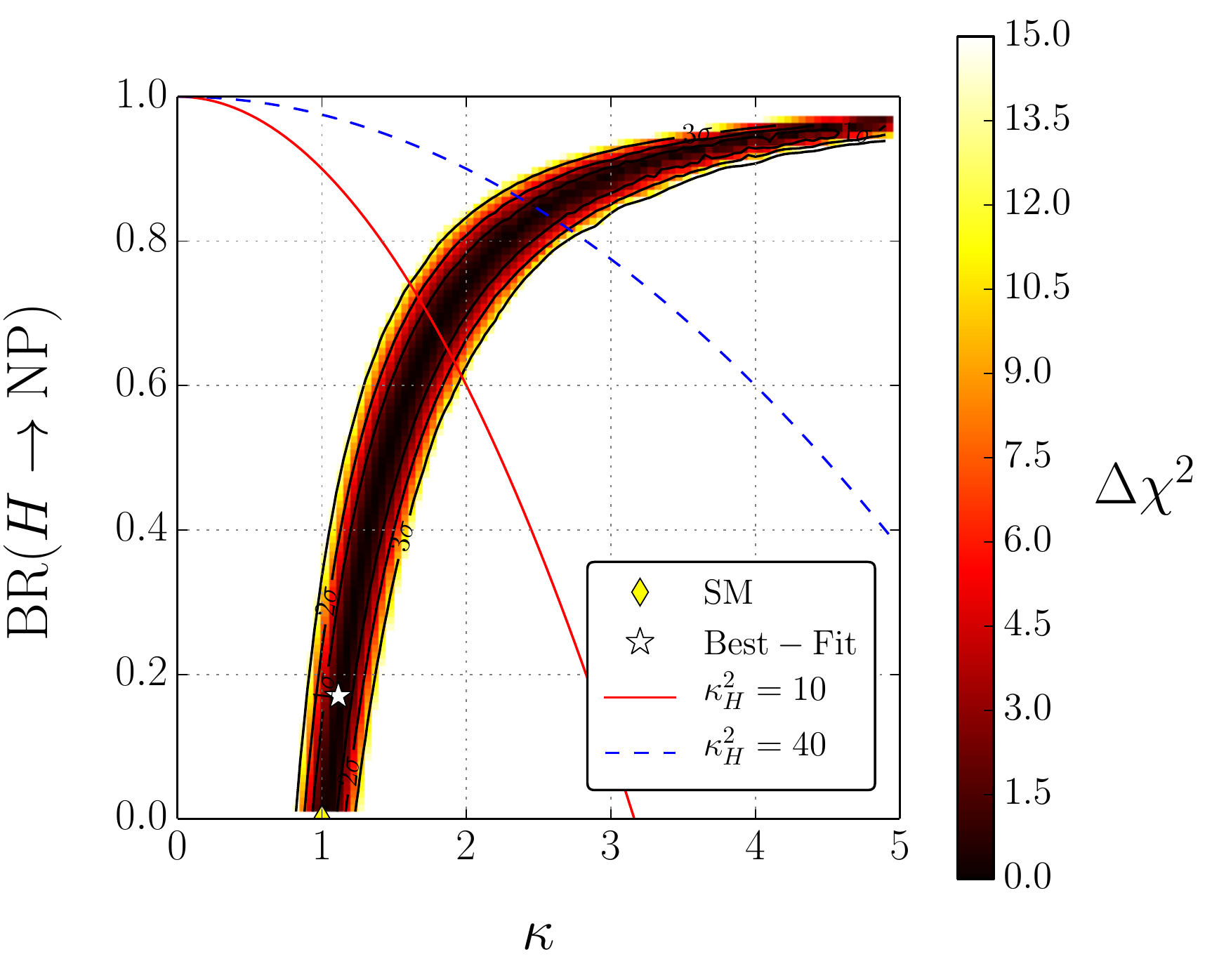}}
\caption{Two-dimensional $\Delta\chi^2$ profiles for the fit parameters in the ($\kappa, \brhnp$) fit.}
\label{Fig:universal2D}
\end{figure}

In Fig.~\ref{Fig:universal2D}(b) we illustrate what happens if this constraint is absent, i.e.~if no assumptions on the additional Higgs decay mode or model parameters, such as $\kV \le 1$, are imposed. The allowed parameter range then extends towards arbitrarily large values of $\kappa$, and $\brhnp\to 1$ due to the perfect degeneracy mentioned above. In the same figure we indicate present ($\kappa_H^2 \le 40$) and potential future ($\kappa_H^2 \le 10$) LHC constraints on the total width that could be derived from off-shell Higgs production in $gg \to ZZ^{(*)}\to 4\ell$~\cite{Caola:2013yja,Campbell:2013una}. 
Such upper limits on the total width scale factor, $\kappa_{H,\text{limit}}^2$, can be used to infer indirect bounds on $\brhnp$ and the coupling scale factor $\kappa$.\footnote{Note, that this argument applies also for more general, higher-dimensional scale factor models since all scale factors $\kappa_i$ are identical in the degenerate case.} Using \refeq{Eq:kappaH}, the limit can be parametrized by 
\begin{align}
\frac{\kappa^2}{1-\brhnp} \le \kappa_{H,\text{limit}}^2,
\end{align}
while SM signal rates are obtained for 
\begin{align}
\kappa^2\cdot[1-\brhnp] =1.
\end{align}
For a given upper limit of the total width scale factor, $\kappa_{H,\text{limit}}^2$, we can thus infer the indirect bounds
\begin{align}
\kappa \le \sqrt{\kappa_{H,\text{limit}}}, \qquad \brhnp = 1 - \kappa_{H,\text{limit}}^{-1}.
\end{align}
For a current (prospective) upper limit of $\kappa_{H,\text{limit}}^2=40~(10)$ at the (high-luminosity) LHC, this would translate into $ \kappa \le 2.51~(1.78)$ and $\brhnp \le 84\%~(68\%)$. However, even when taking these constraints into account there remains a quite large parameter space with possibly sizable $\brhnp$. Hence, the LHC will not be capable to determine absolute values of the Higgs couplings in a model-independent way. This is reserved for future $e^+e^-$ experiments like the ILC, cf.~Sect.~\ref{Sect:ILC}.

Returning to the current fit results displayed in Fig.~\ref{Fig:universal2D}, we can also infer from this fit a lower limit on the total signal strength into known final states (normalized to the SM):
\begin{align}
\kappa^2\cdot [ 1 - \brhnp] \ge 0.81 \quad (\mbox{at}~95\%~\mathrm{C.L.}).
\end{align}
Note, that this limit is irrespective of the final state(s) of the additional Higgs decay mode(s).

\subsection{Couplings to gauge bosons and fermions}
\label{Sect:VF}

The next benchmark model contains one universal scale factor for all Higgs couplings to fermions, $\kF$, and one for the $SU(2)$ gauge bosons, $\kV$ ($V=W,Z$). This coupling pattern occurs, for example, in minimal composite Higgs models \cite{Agashe:2004rs,Contino:2006qr,Giudice:2007fh}, where the Higgs couplings to fermions and vector bosons can be suppressed with different factors.
The loop-induced coupling scale factors are scaled as expected from the SM structure, Eqs.~\eqref{Eq:kg} and \eqref{Eq:kga}. Note that $\kline_g$ scales trivially like $\kF$ in this case, whereas $\kline_\gamma$ depends on the relative sign of $\kV$ and $\kF$ due to the $W$ boson-top quark interference term, giving a negative contribution for equal signs of the fundamental scale factors. Due to this sign dependence we allow for negative values of $\kappa_F$ in the fit, while we restrict $\kappa_V \ge 0$. The assumption of universality of the Higgs-gauge boson couplings, $\kW = \kZ$, corresponds to the (approximately fulfilled) custodial global $SU(2)$ symmetry of the SM Higgs sector. We will explore the possibility of non-universal Higgs-gauge boson couplings in the next section.

\begin{figure}
\centering
\includegraphics[width=0.65\textwidth]{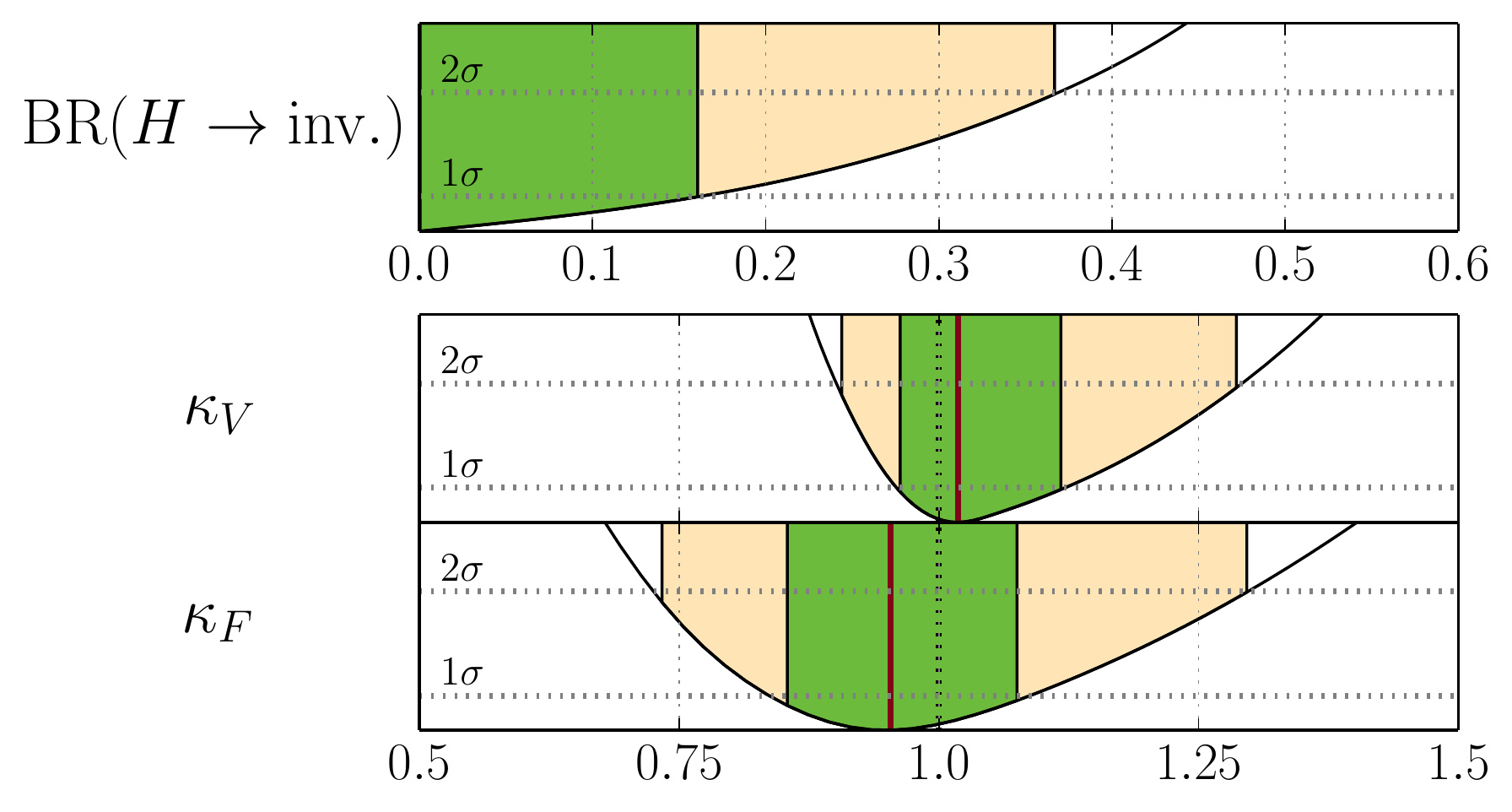}
\caption{One-dimensional $\Delta\chi^2$ profiles for the parameters in the ($\kV, \kF, \brinv$) fit.}
\label{Fig:VF1D}
\end{figure}
\begin{table}
\renewcommand{\arraystretch}{1.3}
\centering
\begin{tabular}{cccc}
\toprule
Fit parameter & best-fit value & $68\%~\mathrm{C.L.}$ range (1D) & $95\%~\mathrm{C.L.}$ range (1D)\\
\midrule
$ \mathrm{BR}(H \to \mathrm{inv.}) $ & $ 0.00 $ & $\substack{+  0.16 \\ - 0.00 }$ & $\substack{+  0.37 \\ - 0.00 }$ \\ 
$ \kappa_V $ & $ 1.02 $ & $\substack{+  0.11 \\ - 0.06 }$ & $\substack{+  0.27 \\ - 0.12 }$ \\ 
$ \kappa_F $ & $ 0.95 $ & $\substack{+  0.14 \\ - 0.12 }$ & $\substack{+  0.34 \\ - 0.22 }$ \\ 
\midrule
$ \kappa_H^2 $ & $ 0.95 $ & $\substack{+  0.40 \\ - 0.20 }$ & $\substack{+  1.51 \\ - 0.30 }$ \\ 
$ \kline_g $ & $ 0.95 $ & $\substack{+  0.14 \\ - 0.12 }$ & $\substack{+  0.34 \\ - 0.23 }$ \\ 
$ \kline_\gamma $ & $ 1.04 $ & $\substack{+  0.11 \\ - 0.07 }$ & $\substack{+  0.28 \\ - 0.14 }$ \\ 
$ \kline_{Z\gamma} $ & $ 1.03 $ & $\substack{+  0.10 \\ - 0.06 }$ & $\substack{+  0.27 \\ - 0.12 }$ \\ 
\bottomrule
\end{tabular}
\caption{Best-fit values and $68\%$ and $95\%~\mathrm{C.L.}$ regions for the fit parameters obtained from the one-dimensional $\Delta\chi^2$ profiles in the ($\kV, \kF, \brinv$) fit.}
\label{Tab:VF1D}
\end{table}
\begin{figure}
\centering
\includegraphics[width=0.7\textwidth]{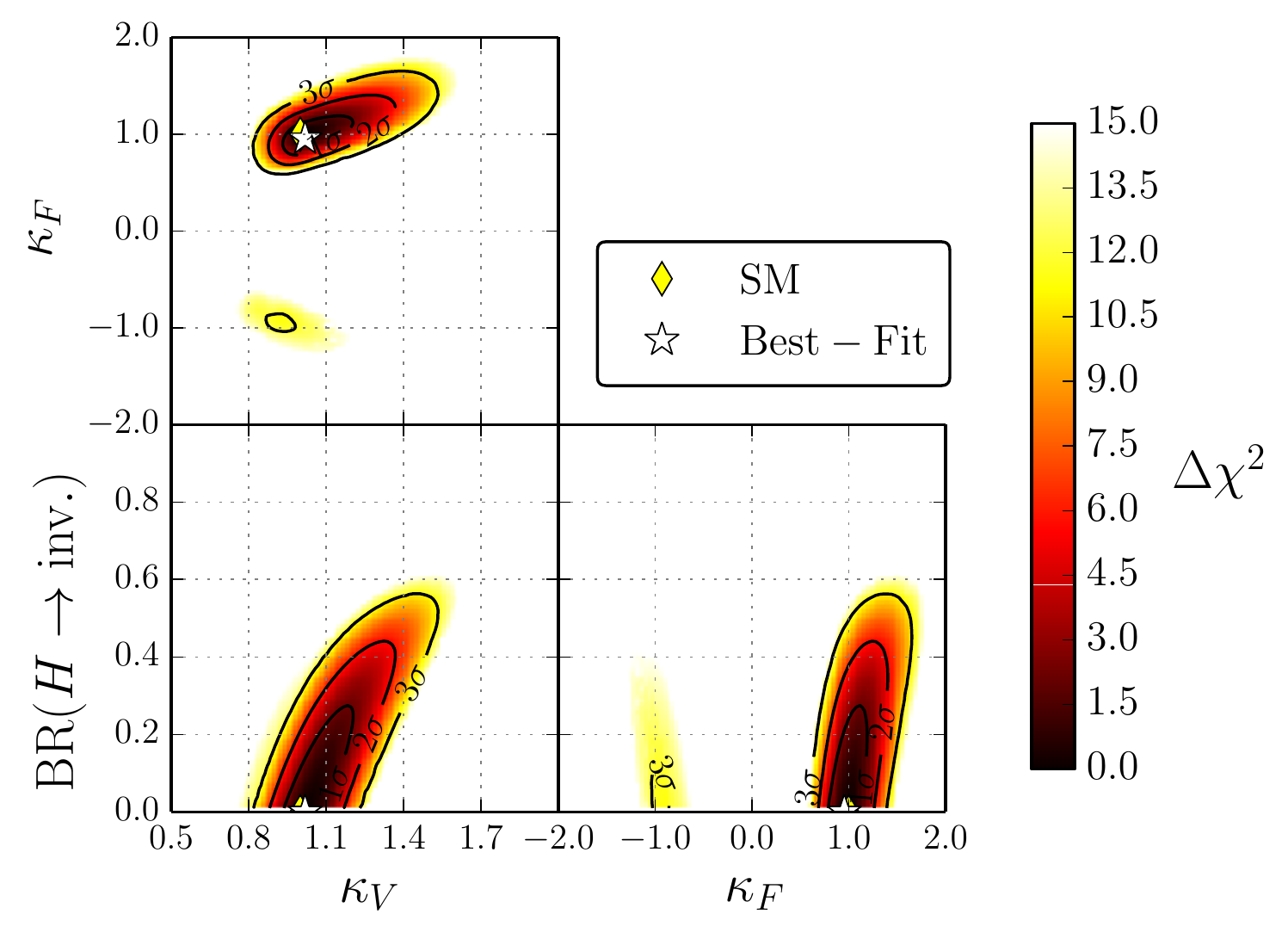}
\caption{Two-dimensional $\Delta\chi^2$ profiles for the parameters in the ($\kV, \kF, \brinv$) fit.}
\label{Fig:VF2D}
\end{figure}

We show the one- and two-dimensional profiled $\Delta\chi^2$ distributions in Figs.~\ref{Fig:VF1D} and \ref{Fig:VF2D}, respectively. 
At the best-fit point we have $\chi^2_\mathrm{min}/\mathrm{ndf} = 84.0 / 78$, corresponding to a \pvalue\ of $\sim30.1\%$. The best-fit values of the fit parameters and the (derived) scale factors for the total width and loop-induced couplings are listed in Tab.~\ref{Tab:VF1D} including the one-dimensional $68\%$ and $95\%~\mathrm{C.L.}$ ranges. Both the Higgs-fermion couplings and Higgs-gauge boson couplings are very close to their SM values. At most, $\kappa_F$ indicates a very weak tendency to a slight suppression. We can obtain $95\%~\mathrm{C.L.}$ upper limits on the branching ratio to invisible final states, $\brinv \le 37\%$, and the total decay width $ \Gamma^\mathrm{tot} \le 2.46 \cdot\Gamma^\mathrm{tot}_\mathrm{SM} \approx 10.3\MeV$.

From the two-dimensional $\chi^2$ profiles, shown in Fig.~\ref{Fig:VF2D}, we see that the sector with negative $\kF$ is disfavored by more than $2\sigma$. In the positive $\kF$ sector, $\kV$ and $\kF$ show a strong positive correlation to preserve SM-like relations among the production cross sections and branching ratios. At this stage, due to the assumed scaling universality of all Higgs couplings to fermions and gauge bosons, the fit does not have enough freedom to resolve small potentially present tendencies in the Higgs signal rates, but rather reflects the overall global picture. Hence, we expect the correlation of $\kappa_V$ with the Higgs fermion coupling scale factor(s) to diminish once more freedom is introduced in the Yukawa coupling sector. This will be discussed in Sect.~\ref{Sect:Vudl}. Furthermore, Fig.~\ref{Fig:VF2D} shows that both $\kV$ and $\kF$ are positively correlated with $\brinv$, similarly to the case with an overall coupling scale factor (cf.~Fig.~\ref{Fig:universal2D}). 

\subsection{Probing custodial symmetry}
\label{Sect:WZF}

\begin{figure}[b]
\centering
\includegraphics[width=0.65\textwidth]{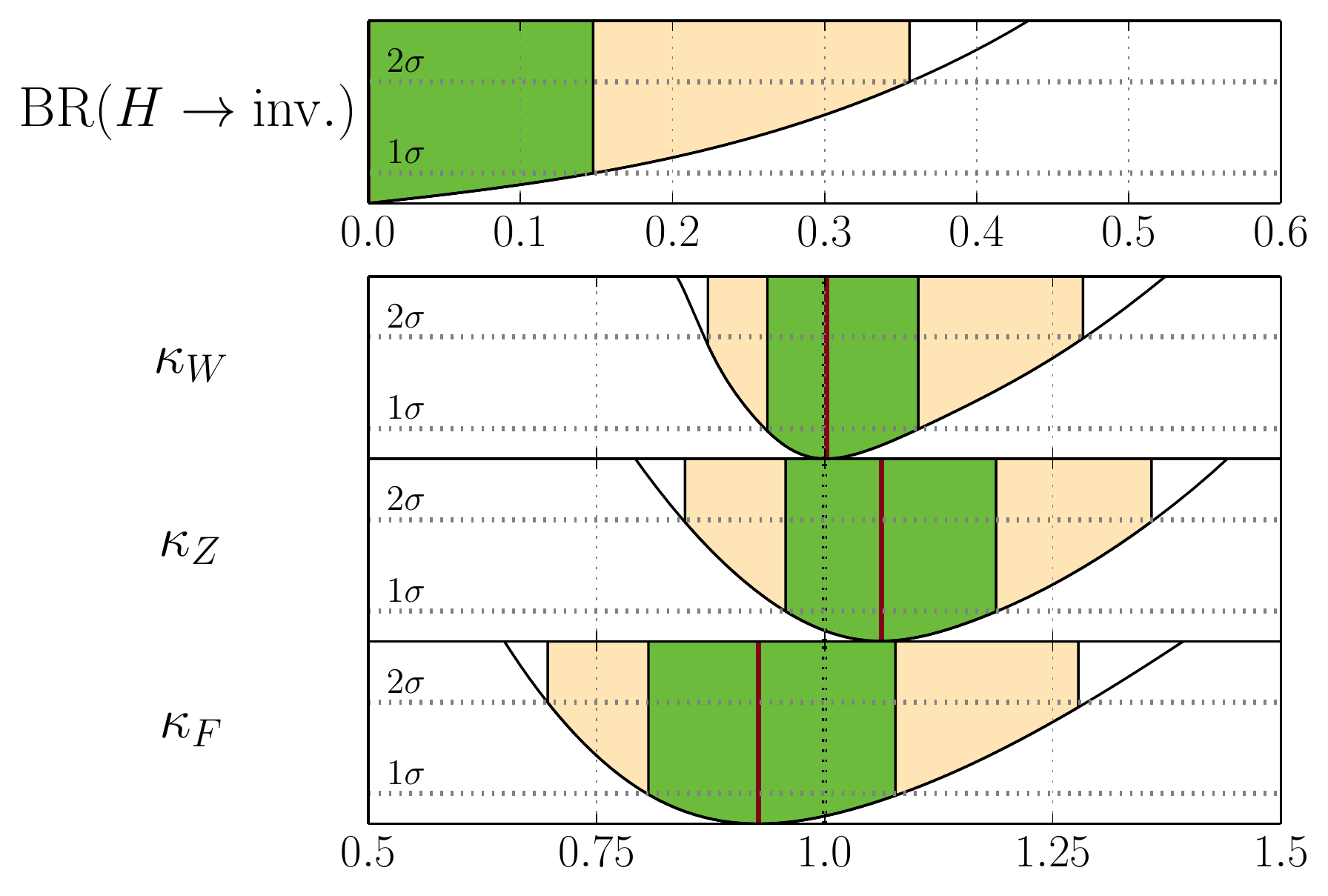}
\caption{One-dimensional $\Delta\chi^2$ profiles for the parameters in the ($\kW, \kZ, \kF, \brinv$) fit.}
\label{Fig:WZF1D}
\end{figure}

\begin{table}
\renewcommand{\arraystretch}{1.3}
\centering
\begin{tabular}{cccc}
\toprule
Fit parameter & best-fit value & $68\%~\mathrm{C.L.}$ range (1D) & $95\%~\mathrm{C.L.}$ range (1D)\\
\midrule
$ \mathrm{BR}(H \to \mathrm{inv.}) $ & $ 0.00 $ & $\substack{+  0.15 \\ - 0.00 }$ & $\substack{+  0.36 \\ - 0.00 }$ \\ 
$ \kappa_W $ & $ 1.00 $ & $\substack{+  0.10 \\ - 0.07 }$ & $\substack{+  0.28 \\ - 0.14 }$ \\ 
$ \kappa_Z $ & $ 1.06 $ & $\substack{+  0.13 \\ - 0.11 }$ & $\substack{+  0.30 \\ - 0.22 }$ \\ 
$ \kappa_F $ & $ 0.93 $ & $\substack{+  0.16 \\ - 0.12 }$ & $\substack{+  0.36 \\ - 0.23 }$ \\ 
\midrule
$ \kappa_H^2 $ & $ 0.90 $ & $\substack{+  0.41 \\ - 0.18 }$ & $\substack{+  1.45 \\ - 0.31 }$ \\ 
$ \kline_g $ & $ 0.93 $ & $\substack{+  0.15 \\ - 0.12 }$ & $\substack{+  0.35 \\ - 0.23 }$ \\ 
$ \kline_\gamma $ & $ 1.02 $ & $\substack{+  0.11 \\ - 0.08 }$ & $\substack{+  0.29 \\ - 0.16 }$ \\ 
$ \kline_{Z\gamma} $ & $ 1.00 $ & $\substack{+  0.11 \\ - 0.07 }$ & $\substack{+  0.29 \\ - 0.13 }$ \\ 
\bottomrule
\end{tabular}
\caption{Best-fit parameter values and $68\%$ and $95\%~\mathrm{C.L.}$ regions obtained from the one-dimensional $\Delta\chi^2$ profiles in the ($\kW, \kZ, \kF, \brinv$) fit.}
\label{Tab:WZF1D}
\end{table}

Experimentally, deviations from the custodial global $SU(2)$ symmetry are strongly constrained by the oblique (Peskin-Takeuchi) $T$ parameter~\cite{Altarelli:1990zd,Peskin:1991sw} obtained in global electroweak fits~\cite{Baak:2012kk,Ciuchini:2013pca}. Nevertheless, as an independent and complementary test, it is important to investigate the universality of the Higgs-gauge boson couplings directly using the signal rate measurements.

Here, we restrict the analysis to the simplest benchmark model probing the custodial symmetry, consisting of individual scale factors for the Higgs couplings to $W$ and $Z$-bosons, $\kappa_W$ and $\kappa_Z$, respectively, and a universal scale factor for the Higgs-fermion couplings, $\kF$. Again, we also allow for an additional invisible decay mode, $\brinv$. Note that, besides the direct signal rate measurements in the channels $\htoWW$ and $\htoZZ$, different constraints apply to the scale factors $\kappa_W$ and $\kappa_Z$: The loop-induced coupling scale factors $\kline{\ga}$ and $\kline_{Z\ga}$ are only affected by $\kappa_W$ and $\kappa_F$, hence $\kappa_Z$ plays a subdominant role in the important channel $\htogaga$ by only affecting the less important production modes $HZ$ and VBF. In contrast, the invisible Higgs search does not constrain $\kW$ at all, but only the product $\kZ^2\,\brinv$. Since the $W$--$Z$ boson interference term in the vector boson fusion channel is neglected, we can impose $\kappa_Z \ge 0$ without loss of information. As in Sect.~\ref{Sect:VF}, we furthermore impose $\kappa_W \ge 0$ and accommodate the sign dependence in the loop-induced couplings by allowing $\kappa_F$ to take on negative values.

The results of the fit are shown in Figs.~\ref{Fig:WZF1D} and~\ref{Fig:WZF2D} as one- and two-dimensional $\chi^2$ profiles in the fit parameters. The best-fit values and the (1D) $68\%$ and $95\%$ C.L.~intervals of the fit parameters and derived scale factors are listed in Tab.~\ref{Tab:WZF1D}. The best fit point features $\chi^2_\mathrm{min}/\mathrm{ndf}= 83.7/77$, corresponding to a \pvalue\ of $\sim28.2\%$. Similar as in the previous fit, a very small non-significant suppression of the Higgs-fermion coupling scale factor $\kF \sim 0.93$ can be observed. Furthermore, the fit has a small tendency towards slightly enhanced $\kZ \sim 1.06$, whereas $\kW$ is very close to the SM value.\footnote{A stronger tendency like this was also seen in the official ATLAS result~\cite{ATLAS:2013sla}. Due to the combination with measurements from other experiments, the tendency observed in our fit is much weaker.} The Higgs-gauge boson coupling scale factors both agree well with the SM predictions, and also with being equal to each other. Since the fit shows excellent agreement of the data with the assumption of custodial symmetry, we will assume $\kW = \kZ \equiv \kV$ in the following.

As can be seen from the two-dimensional $\Delta\chi^2$ profiles, Fig.~\ref{Fig:WZF2D}, the sector with negative $\kF$ is less disfavored than in the previous fit, albeit still by more than $2\sigma$. Since the connection between $\kW$ and $\kZ$ is dissolved, the signal rates of $\htogaga$ can be accommodated more easily in the negative $\kF$ sector than before. It can be seen in Fig.~\ref{Fig:WZF2D} that the least constrained region for negative $\kF$ favors values of $\kW \sim 0.70 - 0.85$ and $\kZ \sim 0.95 - 1.20$, i.e.~a much larger discrepancy between $\kW$ and $\kZ$ than in the positive $\kF$ sector, where we found the overall best fit. 

\begin{figure}
\centering
\includegraphics[width=0.6\textwidth]{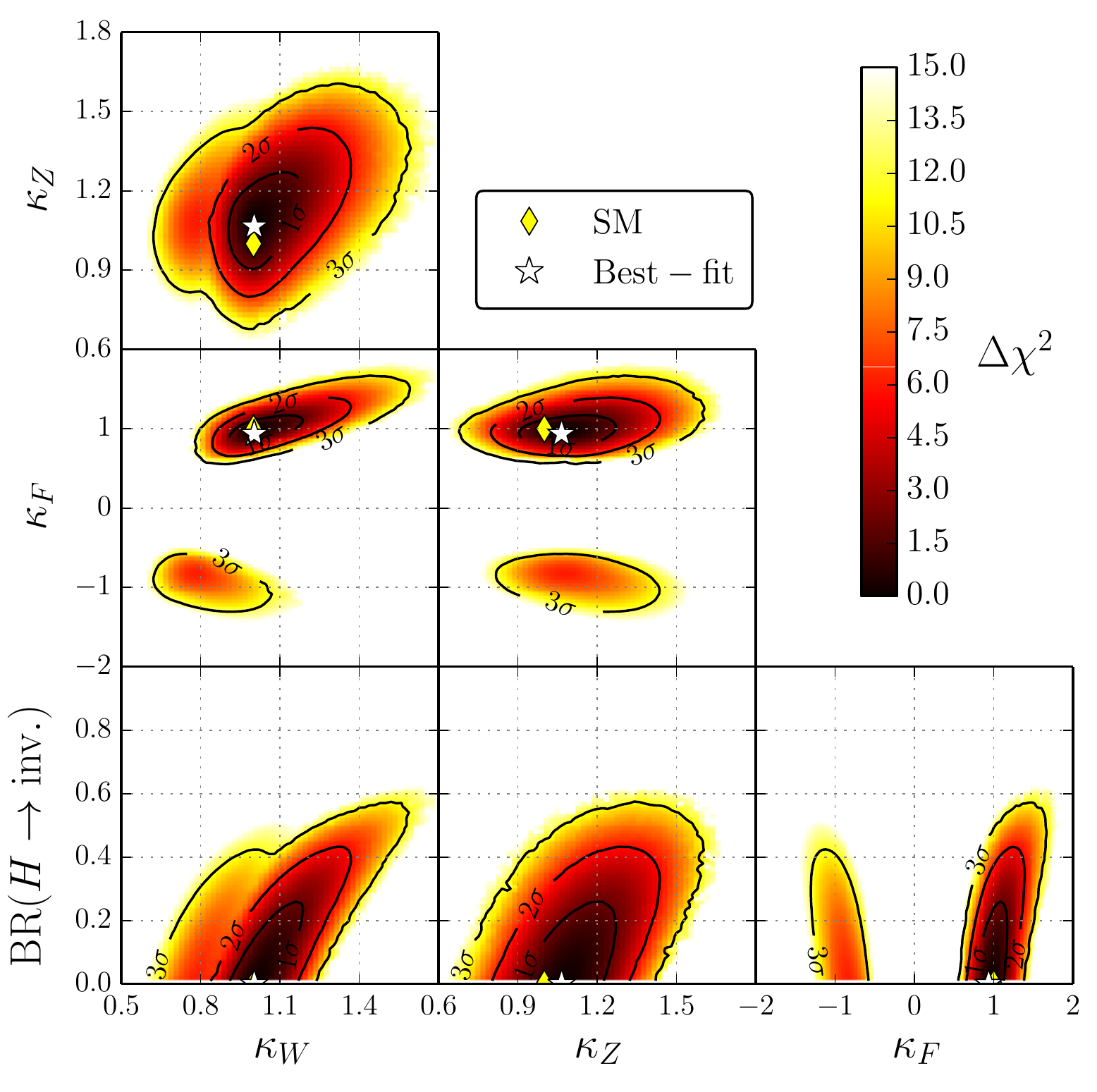}
\caption{Two-dimensional $\Delta\chi^2$ profiles for the parameters in the ($\kW, \kZ, \kF, \brinv$) fit.}
\label{Fig:WZF2D}
\end{figure}


\subsection{Probing the Yukawa structure}
\label{Sect:Vudl}

\begin{figure}
\centering
\includegraphics[width=0.65\textwidth]{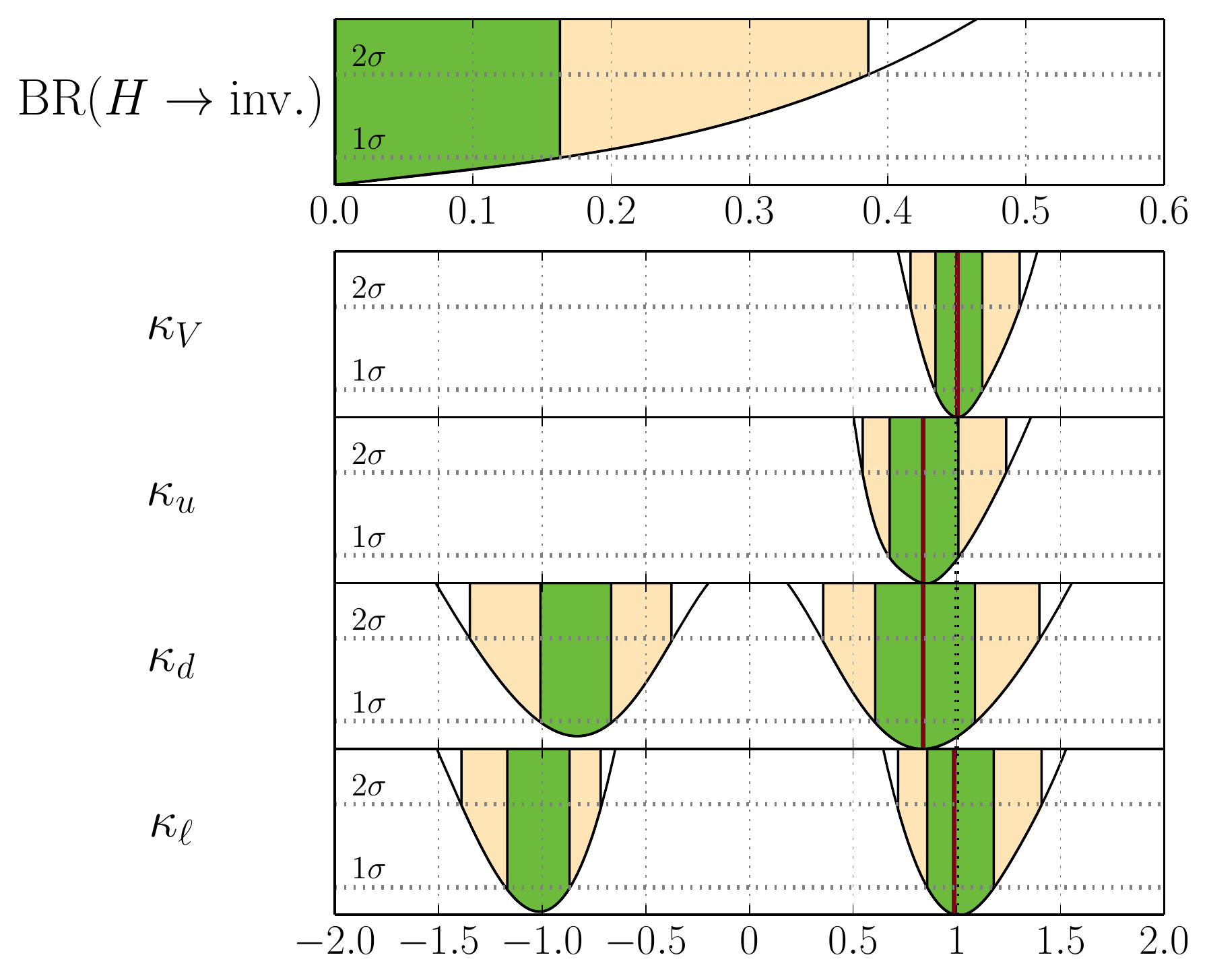}
\caption{One-dimensional $\Delta\chi^2$ profiles for the parameters in the ($\kV, \ku, \kd, \kl, \brinv$) fit.}
\label{Fig:Vudl1D}
\end{figure}

\begin{table}
\renewcommand{\arraystretch}{1.3}
\centering
\begin{tabular}{cccc}
\toprule
Fit parameter & best-fit value & $68\%~\mathrm{C.L.}$ range (1D) & $95\%~\mathrm{C.L.}$ range (1D)\\
\midrule
$ \mathrm{BR}(H \to \mathrm{inv.}) $ & $ 0.00 $ & $\substack{+  0.17 \\ - 0.00 }$ & $\substack{+  0.39 \\ - 0.00 }$ \\ 
$ \kappa_V $ & $ 1.00 $ & $\substack{+  0.13 \\ - 0.11 }$ & $\substack{+  0.31 \\ - 0.23 }$ \\ 
$ \kappa_u $ & $ 0.84 $ & $\substack{+  0.18 \\ - 0.17 }$ & $\substack{+  0.40 \\ - 0.29 }$ \\ 
$ \kappa_d $ & $ 0.84 $ & $\substack{+  0.26 \\ - 0.24 }$ & $\substack{+  0.56 \\ - 0.49 }$ \\ 
$ \kappa_\ell $ & $ 0.99 $ & $\substack{+  0.19 \\ - 0.13 }$ & $\substack{+  0.42 \\ - 0.28 }$ \\ 
\midrule
$ \kappa_H^2 $ & $ 0.80 $ & $\substack{+  0.45 \\ - 0.28 }$ & $\substack{+  1.53 \\ - 0.50 }$ \\ 
$ \kline_g $ & $ 0.84 $ & $\substack{+  0.16 \\ - 0.12 }$ & $\substack{+  0.38 \\ - 0.24 }$ \\ 
$ \kline_\gamma $ & $ 1.04 $ & $\substack{+  0.15 \\ - 0.11 }$ & $\substack{+  0.33 \\ - 0.24 }$ \\ 
$ \kline_{Z\gamma} $ & $ 1.01 $ & $\substack{+  0.13 \\ - 0.10 }$ & $\substack{+  0.31 \\ - 0.22 }$ \\
\bottomrule
\end{tabular}
\caption{Best-fit values and $68\%$ and $95\%~\mathrm{C.L.}$ regions for the fit parameters around the best fit point (positive sector only) obtained from the one-dimensional $\Delta\chi^2$ profiles in the ($\kV, \ku, \kd, \kl, \brinv$) fit.}
\label{Tab:Vudl1D}
\end{table}

We will now have a closer look at the Higgs-fermion coupling structure. In fact, assuming that all Higgs-fermion couplings can be described by one common scale factor --- as we have done until now --- is motivated in only a few special BSM realizations. A splitting of up- and down-type Yukawa couplings appears in many BSM models, e.g.~Two-Higgs-Doublet Models (2HDM)~\cite{Lee:1973iz,Glashow:1976nt,Deshpande:1977rw,Donoghue:1978cj,Haber:1978jt,Hall:1981bc,Gunion:2002zf,Branco:2011iw,Cheng:1987rs} of Type II or in the minimal supersymmetric extension of the SM (MSSM)~\cite{Haber:1984rc,Gunion:1984yn,Gunion:1986nh,Gunion:1989we}. Moreover, realistic 2HDMs with more generic Yukawa couplings featuring additional freedom for the Higgs-charged lepton coupling can be constructed to be consistent with constraints from flavor-changing neutral currents (FCNCs)~\cite{Cheng:1987rs,Mahmoudi:2009zx,Crivellin:2012ye,Crivellin:2013wna}. Also in the MSSM, the degeneracy of bottom-type quarks and leptons can be abrogated by radiative SUSY QCD corrections (so-called $\Delta_b$ corrections)~\cite{Hempfling:1993kv,Hall:1993gn,Carena:1994bv}. Therefore, we now relax the assumption of a universal Higgs-fermion coupling scale factor and introduce common scale factors for all up-type quarks, $\ku$, all down-type quarks, $\kd$, and all charged leptons, $\kl$. All Higgs-fermion coupling scale factors are allowed to take positive and negative values. The parameters $\kV$ and $\brinv$ remain from before.

\begin{figure}[t]
\centering
\includegraphics[width=0.9\textwidth]{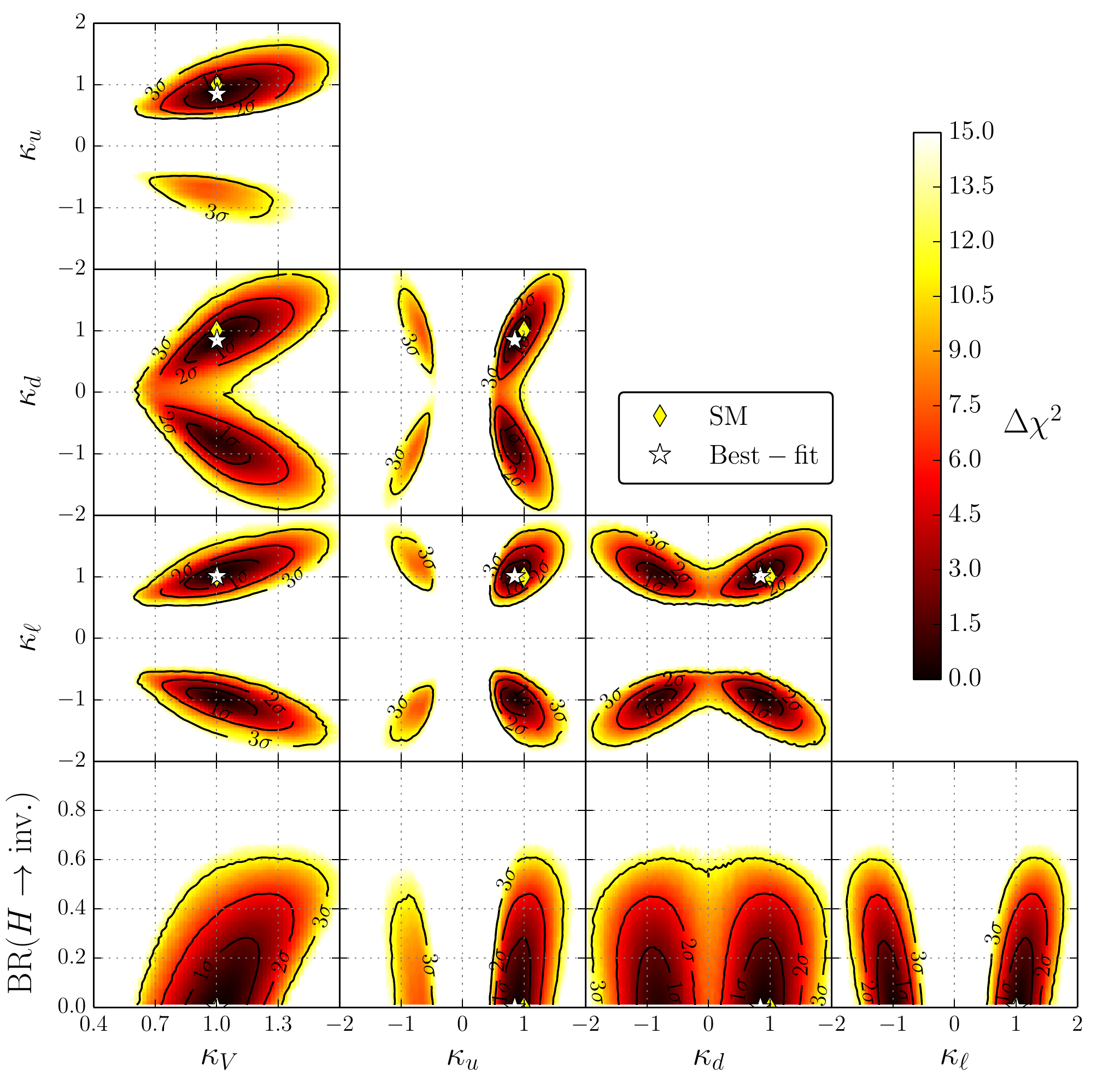}
\caption{Two-dimensional $\Delta\chi^2$ profiles for the parameters in the ($\kV, \ku, \kd, \kl, \brinv$) fit.}
\label{Fig:Vudl2D}
\end{figure}

The one-dimensional $\Delta\chi^2$ profiles of the fit parameters are shown in Fig.~\ref{Fig:Vudl1D}. The parameter values of the best-fit point, which is found in the sector with all scale factors being positive, are given in Tab.~\ref{Tab:Vudl1D} along with the (1D) $68\%$ and $95\%$~C.L.~intervals. The fit quality is $\chi^2_\mathrm{min}/\mathrm{ndf} = 82.8/76$ corresponding to a \pvalue\ of $\sim27.8\%$. As can be clearly seen in Fig.~\ref{Fig:Vudl1D}, negative values of $\kd$ and $\kl$ are still consistent with the measurements within $68\%~\mathrm{C.L.}$ due to their small influence on the loop-induced Higgs couplings to gluons and/or photons. In particular the sign discrimination of $\kl$ is very weak. In contrast, negative values of $\ku$ are disfavored by more than $2\sigma$ due to the influence on the Higgs-photon effective coupling in the convention $\kappa_V \ge 0$. The fit prefers slightly suppressed values of $\ku\sim 0.84$ since $\kg \simeq \ku$, Eq.~\eqref{Eq:kg}, which is sensitively probed by the LHC measurements via the gluon fusion production mode. Due to the recent $\htotautau$ results from ATLAS~\cite{ATLAS-CONF-2013-108} and CMS~\cite{CMS:utj,Chatrchyan:2014nva,Chatrchyan:2014vua}, $\kl$ is determined to be very close to its SM value with a precision of $\sim15\%$. We observe a slight but non-signifcant suppression of the Higgs-down type quark coupling, $\kd\sim 0.84$. This scale factor has the worst precision of the fitted parameters, about $\sim 30\%$. The sign degeneracy of $\kd$ is slightly broken via the sensitivity of the Higgs-gluon coupling scale factor to the relative sign of $\kt$ and $\kb$, cf.~\refeq{Eq:kg}.

The correlation of the fitted coupling scale factors can be seen from the shape of the ellipses in the two-dimensional $\chi^2$ profiles, shown in Fig.~\ref{Fig:Vudl2D}. The slope of the major axis of the ellipse in the positive sector of the ($\kV,~\ku$) plane is $\sim 0.6-0.7$ and thus much shallower than the slopes in the ($\kV,~\kd$) and ($\kV,~\kl$) planes, approximately given by $\sim 1.7$ and $\sim 1.3$, respectively. Therefore, this parameterization exhibits more freedom to adjust the predicted signal rates to the Tevatron and LHC measurements. Nevertheless, the best-fit point and favored region is in perfect agreement with the SM and thus the additional freedom does not improve the fit quality. Once more precise measurements of the $\htotautau$ and $\htobb$ channels become available, this parametrization can be expected to provide a good test of the SM due to the different correlations among $\kV$ and the Higgs-fermion coupling scale factors.

\subsection{Probing new physics in loop-induced couplings}
\label{Sect:gga}

\begin{figure}[t]
\centering
\includegraphics[width=0.65\textwidth]{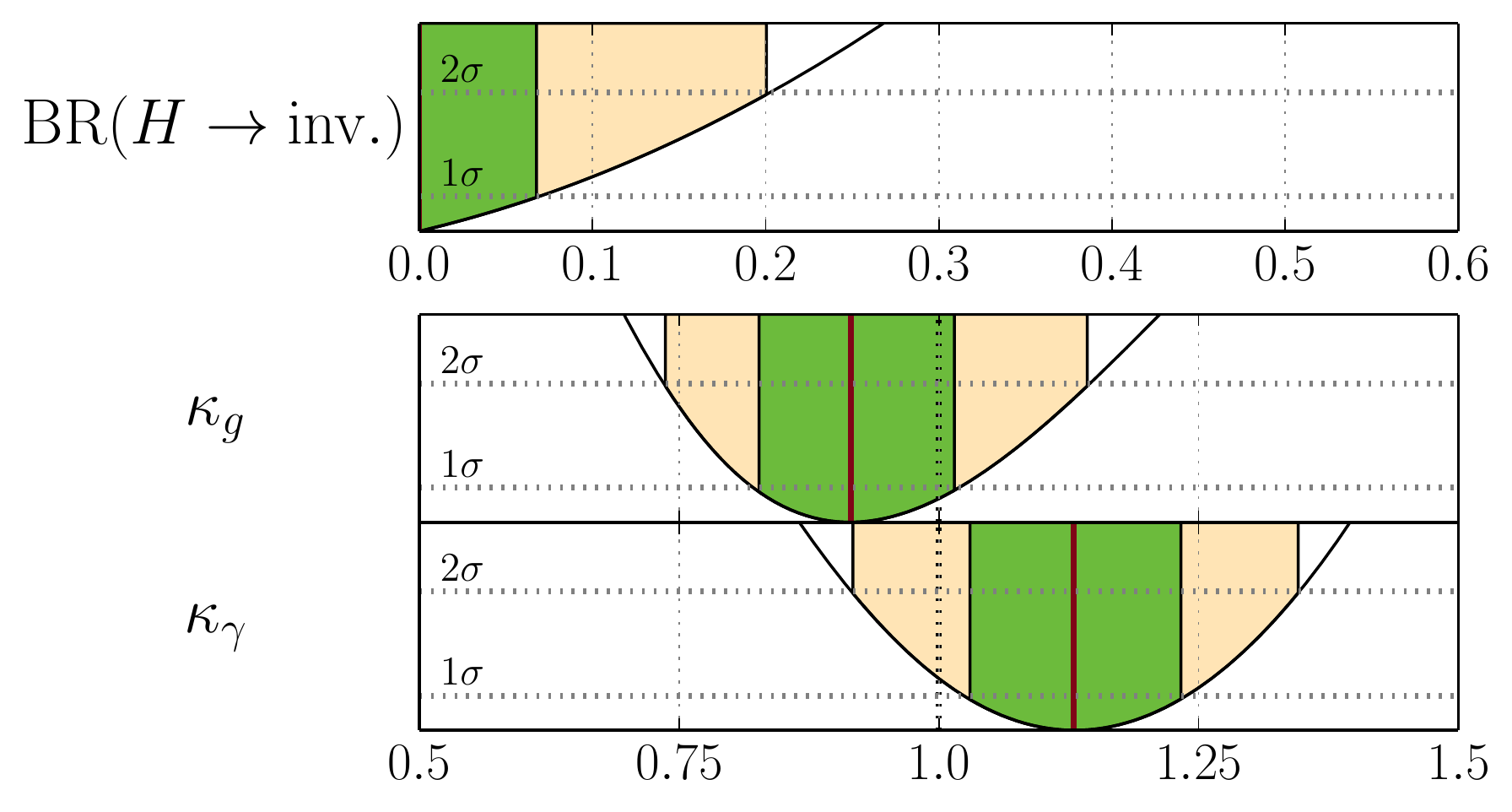}
\caption{One-dimensional $\Delta\chi^2$ profiles for the parameters in the ($\kg, \kga, \brinv$) fit. The best-fit point is indicated by the red line, $68\%$ and $95\%$ C.L.~regions are illustrated by the green and pale yellow bands.}
\label{Fig:gga1D}
\end{figure}

Up to now we have investigated possible modifications of the fundamental tree-level Higgs boson couplings to SM particles and derived the loop-induced couplings to gluons and photons using Eq.~\eqref{Eq:kg} and~\eqref{Eq:kga}, respectively. In this section, we modify these coupling scale factors, $\kg$ and $\kga$, directly. Such modifications could be introduced by unknown new physics loop contributions, while the tree-level Higgs boson couplings are unaffected. Triggered by the hints in the experimental data for a possible $\htogaga$ enhancement, new physics sources for modifications of the Higgs-photon coupling have been subject to many recent studies. For instance, charged supersymmetric particles such as light staus~\cite{Carena:2011aa,Carena:2012gp,Basso:2012tr,Buckley:2012em,Bechtle:2012jw,Carena:2013qia} and charginos~\cite{SchmidtHoberg:2012yy,Hemeda:2013hha} could give potentially substantial contributions. In 2HDMs the Higgs-photon coupling can be altered due to contributions from the charged Higgs boson~\cite{Posch:2010hx,Drozd:2012vf,Cordero-Cid:2013sxa}, and in the special case of the Inert Doublet Model~\cite{Arhrib:2012ia,Goudelis:2013uca,Krawczyk:2013jta,Krawczyk:2013pea}, modifications of $\kga$ and $\kZga$ are indeed the only possible change to the Higgs coupling structure. In addition, many of these models can also feature invisible or undetectable Higgs decays. The effective Higgs-gluon coupling can be modified in supersymmetric models by stop contributions, where one can easily find rate predictions for Higgs production in gluon fusion corresponding to $\kappa_g < 1$~\cite{Carena:2013qia,oai:arXiv.org:hep-ph/9806315,oai:arXiv.org:hep-ph/0202167}.

Our fit parametrization represents the case where indirect new physics effects may be visible only in the loop-induced Higgs-gluon and Higgs-photon couplings. Direct modifications to the tree-level couplings, as introduced e.g.~if the observed Higgs boson is a mixed state, are neglected. The more general case where all couplings are allowed to vary will be discussed in the next section. Due to the very small branching ratio $\mathrm{BR}(H\to Z\gamma) \times \mathrm{BR}(Z\to \ell\ell)$ in the SM, the LHC is not yet sensitive to probe $\kZga$. We therefore set $\kZga = \kga$. In addition, here we assume that any additional Higgs decays lead to  invisible final states. Undetectable Higgs decays are discussed in Sect.~\ref{Sect:kvle1}.

\begin{table}
\renewcommand{\arraystretch}{1.3}
\centering
\begin{tabular}{cccc}
\toprule
Fit parameter & best-fit value & $68\%~\mathrm{C.L.}$ range (1D) & $95\%~\mathrm{C.L.}$ range (1D)\\
\midrule
$ \mathrm{BR}(H \to \mathrm{NP}) $ & $ 0.00 $ & $\substack{+  0.07 \\ - 0.00 }$ & $\substack{+  0.20 \\ - 0.00 }$ \\ 
$ \kappa_g $ & $ 0.92 $ & $\substack{+  0.11 \\ - 0.10 }$ & $\substack{+  0.23 \\ - 0.18 }$ \\ 
$ \kappa_\gamma $ & $ 1.14 $ & $\substack{+  0.11 \\ - 0.11 }$ & $\substack{+  0.21 \\ - 0.22 }$ \\ 
\midrule
$ \kappa_H^2 $ & $ 1.01 $ & $\substack{+  0.08 \\ - 0.03 }$ & $\substack{+  0.28 \\ - 0.03 }$ \\ 
\bottomrule
\end{tabular}
\caption{Best-fit values and $68\%$ and $95\%~\mathrm{C.L.}$ regions for the fit parameters around the best fit point obtained from the one-dimensional $\Delta\chi^2$ profiles in the ($\kg, \kga, \brinv$) fit.}
\label{Tab:gga1D}
\end{table}

\begin{figure}[b]
\centering
\includegraphics[width=0.65\textwidth]{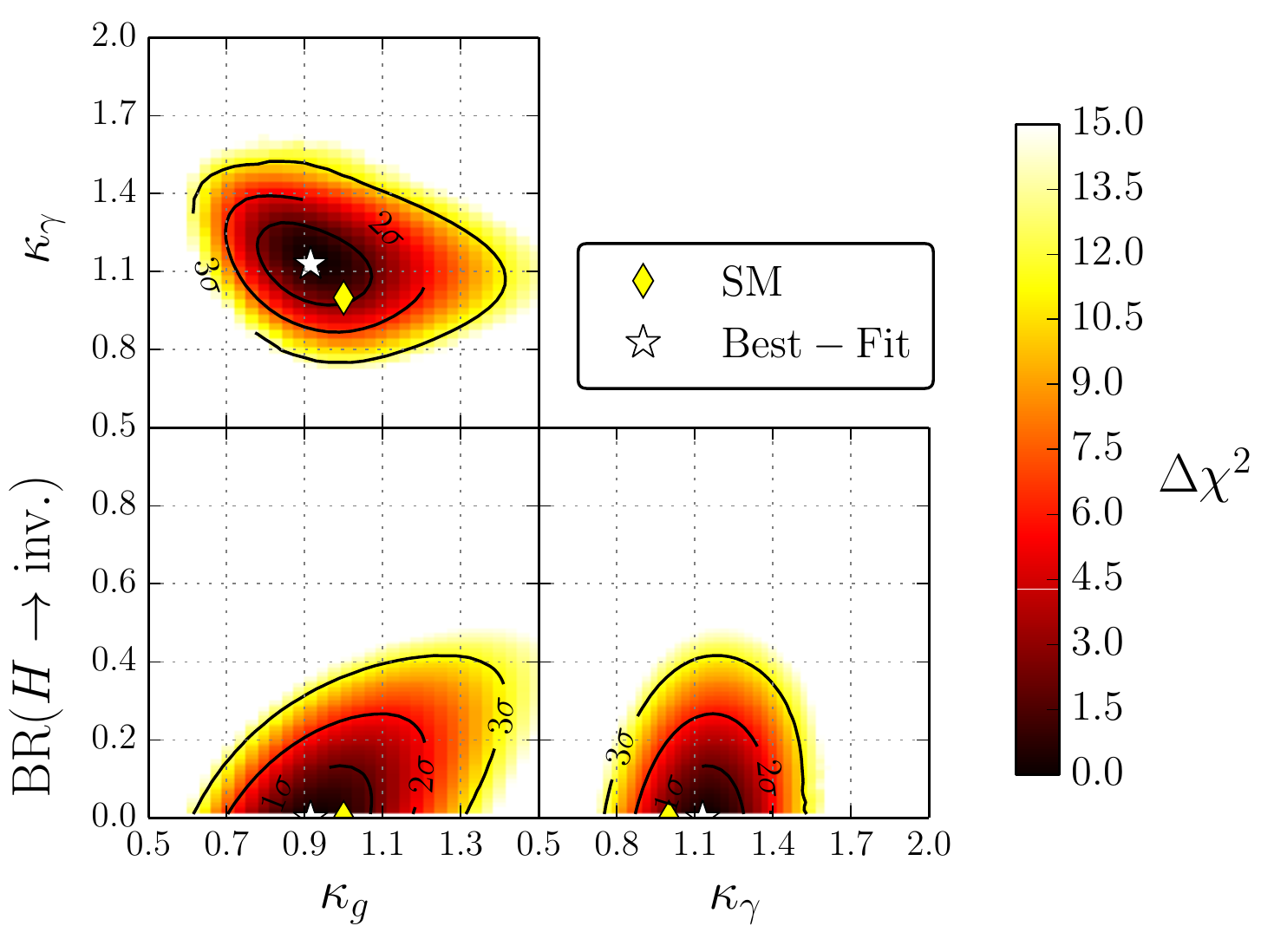}
\caption{Two-dimensional $\chi^2$ profiles for the fit parameters in the ($\kg, \kga, \brinv$) fit.}
\label{Fig:gga2D}
\end{figure}

The fit results are shown as one- and two-dimensional $\Delta\chi^2$ profiles in the fit parameters in Fig.~\ref{Fig:gga1D} and~\ref{Fig:gga2D}, respectively. The (1D) preferred parameter values are also provided in Tab.~\ref{Tab:gga1D}. In this scenario, the best fit indicates a slight suppression of the Higgs-gluon coupling, $\kg= 0.92$, with a simultaneous enhancement in the Higgs-photon coupling, $\kga= 1.14$. The anti-correlation of these two parameters can be seen in Fig.~\ref{Fig:gga2D}. It is generated by the necessity of having roughly SM-like $gg\to\htogaga$ signal rates. The best fit point, which has $\chi^2_\mathrm{min}/\mathrm{ndf} = 82.6/78$, is compatible with the SM expectation at the $1\sigma$ level, as can be seen in Fig.~\ref{Fig:gga2D}. The estimated \pvalue\ is $\sim33.9\%$. Note that $\brinv$ is much stronger constrained to $\le 20\%$ (at $95\%~\mathrm{C.L.}$) in this parametrization than in the previous fits. The reason being that the suppression of the SM decay modes with an increasing $\brinv$ cannot be fully compensated by an increasing production cross sections since the tree-level Higgs couplings are fixed. The partial compensation that is possible with an increased gluon fusion cross section is reflected in the strong correlation between $\kg$ and $\brinv$, which can be seen in Fig.~\ref{Fig:gga2D}.

\subsection{General Higgs couplings}
\label{Sect:7dim}

\begin{figure}
\centering
\includegraphics[width=0.65\textwidth]{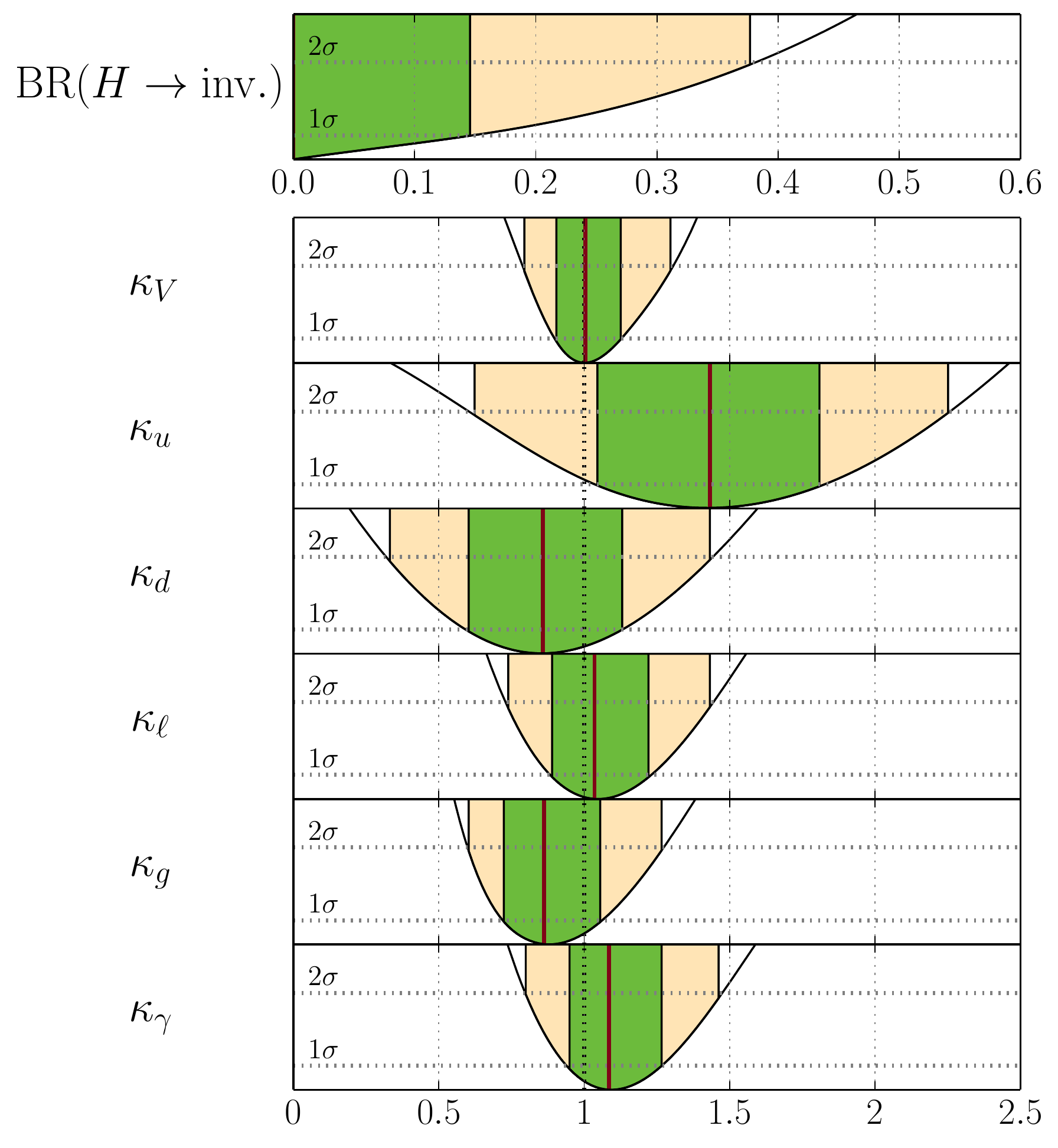}
\caption{One-dimensional $\Delta\chi^2$ profiles for the parameters in the ($\kV, \ku, \kd, \kl, \kg, \kga, \brinv$) fit.}
\label{Fig:7dim1D}
\end{figure}

We now allow for genuine new physics contributions to the loop-induced couplings by treating $\kg$ and $\kga$ as free fit parameters in addition to a general parametrization of the Yukawa sector as employed in Sect.~\ref{Sect:Vudl}. This gives in total seven free fit parameters, $\kV$, $\ku$, $\kd$, $\kl$, $\kg$, $\kga$ and $\brinv$. Note, that this parametrization features a sign degeneracy in \textit{all} coupling scale factors, since the only derived scale factor, $\kappa_H^2$, depends only on the squared coupling scale factors. For practical purposes, we thus restrict ourselves to the sector where all scale factors are positive. Furthermore, it can be illustrative to decompose $\kg$ and $\kga$ into scale factors $\kline_i$ for the calculable contributions from SM particles, with rescaled couplings, appearing in the loop, as described by Eqs.~\eqref{Eq:kg}--\eqref{Eq:kga}, and a scale factor $\Delta\kappa_i$ for the genuine new physics contributions:
\begin{align}
\kappa_g &= \kline_g + \Delta\kappa_g,\\
\kappa_\gamma &= \kline_\gamma + \Delta\kappa_\gamma.
\end{align}
This decomposition assumes that the unknown new physics does not alter the loop contri\-butions from SM particles, Eqs.~\eqref{Eq:kg}--\eqref{Eq:kga}. 
Technically, the $\kappa_{g,\ga}$ are used as fit parameters, and the derived $\kline$ are evaluated for each scan point. $\De\kappa = \kappa - \kline$ can thus be extracted for each point, and by this construction, the likelihood distribution in $\De\kappa$ includes the full correlations.

\begin{table}
\renewcommand{\arraystretch}{1.3}
\centering
\begin{tabular}{cccc}
\toprule
Fit parameter & best-fit value & $68\%~\mathrm{C.L.}$ range (1D) & $95\%~\mathrm{C.L.}$ range (1D)\\
\midrule
$ \mathrm{BR}(H \to \mathrm{inv.}) $ & $ 0.00 $ & $\substack{+  0.15 \\ - 0.00 }$ & $\substack{+  0.39 \\ - 0.00 }$ \\ 
$ \kappa_V $ & $ 1.00 $ & $\substack{+  0.13 \\ - 0.11 }$ & $\substack{+  0.31 \\ - 0.22 }$ \\ 
$ \kappa_u $ & $ 1.42 $ & $\substack{+  0.40 \\ - 0.39 }$ & $\substack{+  0.83 \\ - 0.82 }$ \\ 
$ \kappa_d $ & $ 0.86 $ & $\substack{+  0.28 \\ - 0.27 }$ & $\substack{+  0.59 \\ - 0.54 }$ \\ 
$ \kappa_\ell $ & $ 1.05 $ & $\substack{+  0.19 \\ - 0.17 }$ & $\substack{+  0.40 \\ - 0.32 }$ \\ 
$ \kappa_g $ & $ 0.88 $ & $\substack{+  0.18 \\ - 0.16 }$ & $\substack{+  0.39 \\ - 0.28 }$ \\ 
$ \kappa_\gamma $ & $ 1.09 $ & $\substack{+  0.18 \\ - 0.15 }$ & $\substack{+  0.38 \\ - 0.29 }$ \\ 
\midrule
$ \kline_H^2 $ & $ 0.86 $ & $\substack{+  0.36 \\ - 0.27 }$ & $\substack{+  0.90 \\ - 0.48 }$ \\ 
$ \kappa_H^2 $ & $ 0.88 $ & $\substack{+  0.43 \\ - 0.28 }$ & $\substack{+  1.56 \\ - 0.50 }$ \\ 
$ \Delta\kappa_\gamma $ & $ 0.19 $ & $\substack{+  0.14 \\ - 0.14 }$ & $\substack{+  0.30 \\ - 0.28 }$ \\ 
$ \Delta\kappa_g $ & $ -0.63 $ & $\substack{+  0.36 \\ - 0.32 }$ & $\substack{+  0.90 \\ - 0.62 }$ \\ 
$ \kline_{Z\gamma} $ & $ 0.98 $ & $\substack{+  0.13 \\ - 0.13 }$ & $\substack{+  0.29 \\ - 0.25 }$ \\ 
\bottomrule
\end{tabular}
\caption{Best-fit values and $68\%$ and $95\%~\mathrm{C.L.}$ regions for the fit parameters (above) and derived scale factors (below) obtained from the one-dimensional $\Delta\chi^2$ profiles in the ($\kV, \ku, \kd, \kl, \kg, \kga, \brinv$) fit.}
\label{Tab:7dim1D}
\end{table}

The one-dimensional $\Delta\chi^2$ profiles in the fit parameters are shown in Fig.~\ref{Fig:7dim1D}. Their best-fit values and preferred parameter ranges are listed together with those of derived scale factors in Tab.~\ref{Tab:7dim1D}. The best-fit point features a fit quality of $\chi^2_\mathrm{min}/\mathrm{ndf} = 79.9/74$ and thus a \pvalue\ of $\sim29.9\%$. Due to the dissolved dependence between the Yukawa couplings and the effective Higgs-gluon and Higgs-photon couplings, $\ku$ is significantly less accurately determined than in previous more constrained fits. In fact, it is now dominantly influenced by the recent CMS measurements targeting $t\bar{t}H$ production~\cite{CMS:2013tfa,CMS:2013sea,CMS:2013fda}, which give a combined signal strength of $\muobs_\mathrm{CMS}^{t\bar{t}H} = 2.5\substack{+1.1\\-1.0}$~\cite{CMStthcombination}. Hence, the fit prefers slightly enhanced values, $\ku\sim 1.42$, albeit with very large uncertainties. The scale factors $\kg$ and $\kga$ can now be freely adjusted to match the combined rates of Higgs production in gluon fusion and $\br(\htogaga)$, respectively. Here we observe the same tendencies as in the previous fit, cf.~Sect.~\ref{Sect:gga}. Due to the slight preference for enhanced $\ku$ and suppressed $\kg$, the fitted new physics contribution to the Higgs-gluon coupling is quite sizable and negative, $\Delta\kg \sim - 0.63$. In contrast, the Higgs-photon coupling is fairly well described by the rescaled contributions from SM particles alone because the enhanced $\ku$ also enhances $\kline_\gamma$ slightly. The favored magnitude for the genuine new physics contribution to the Higgs-photon coupling is $\Delta\kga \sim 0.19$.

\begin{figure}
\centering
\includegraphics[width=0.95\textwidth]{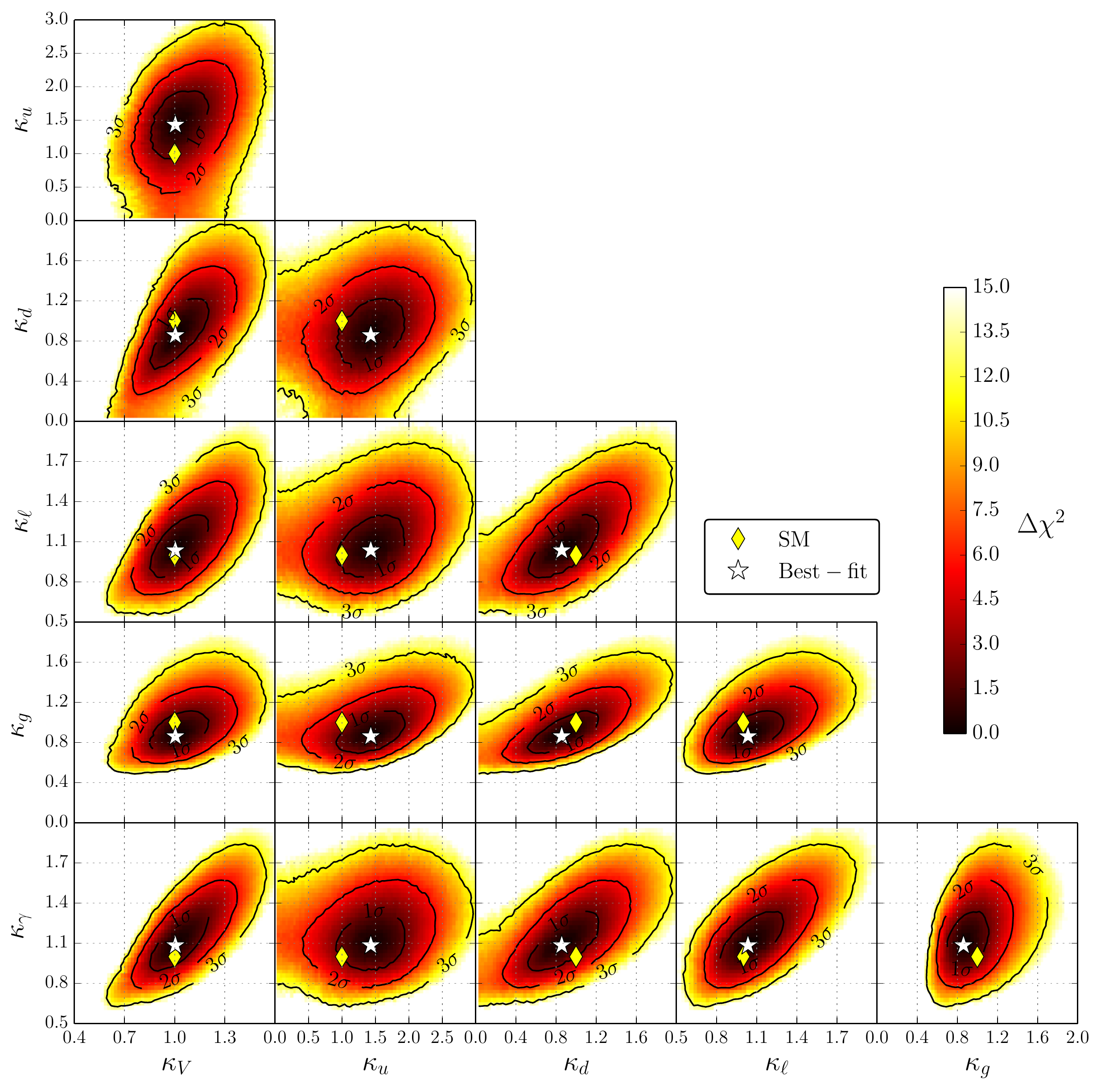}
\caption{Two-dimensional $\Delta\chi^2$ profiles for the fitted Higgs coupling scale factors in the ($\kV, \ku, \kd, \kl, \kg,\kga, \brinv$) fit.}
\label{Fig:7dim2D}
\end{figure}

\begin{figure}
\centering
\includegraphics[width=0.7\textwidth]{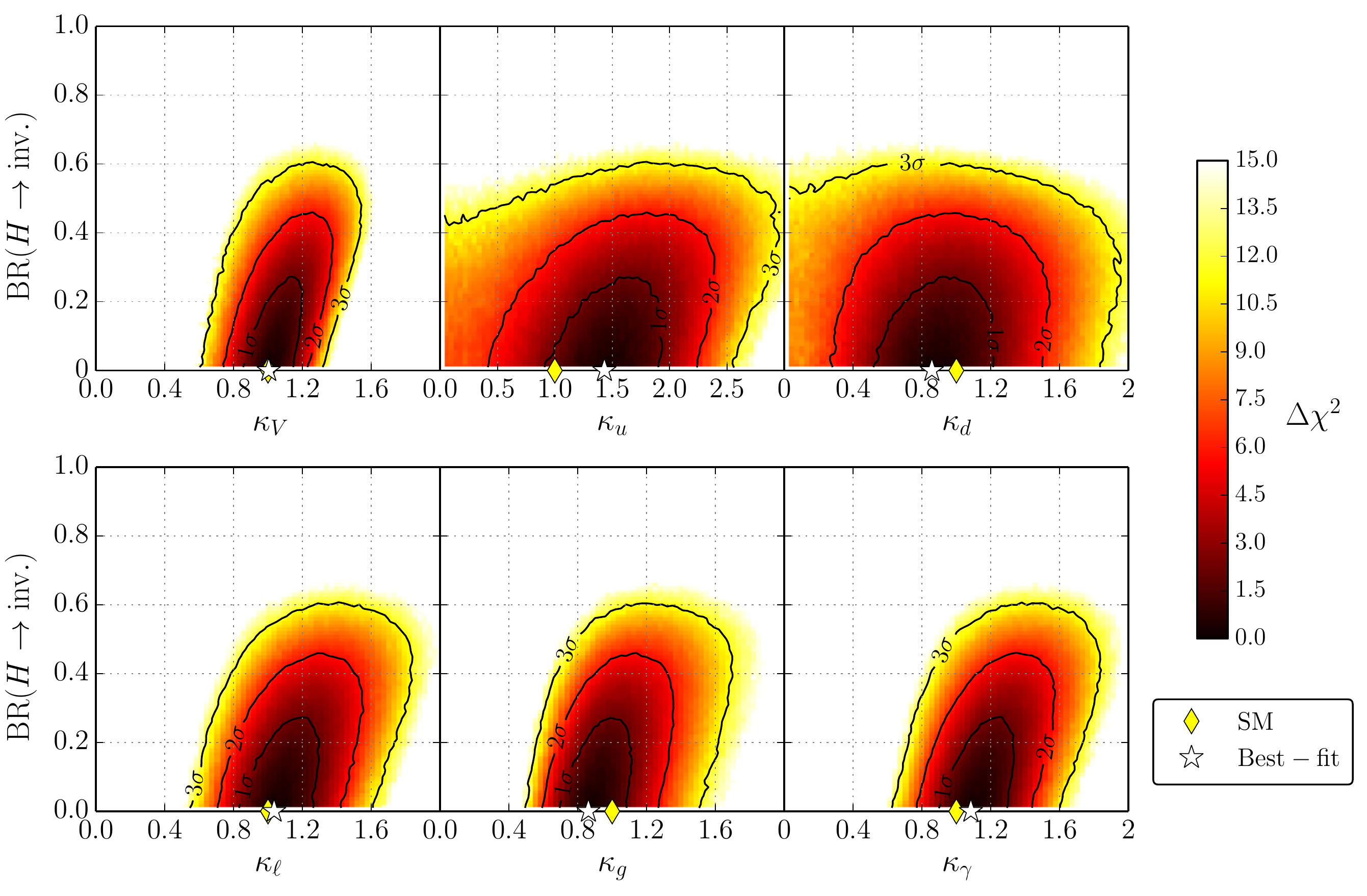}
\caption{Two-dimensional $\Delta\chi^2$ profiles of the fitted Higgs coupling scale factors with the invisible Higgs decay mode, $\brinv$, in the ($\kV, \ku, \kd, \kl, \kg,\kga, \brinv$) fit.}
\label{Fig:7dim2D_brinv}
\end{figure}

The two-dimensional $\chi^2$ profiles of the fitted Higgs coupling scale factors are shown Fig.~\ref{Fig:7dim2D} and their correlations with $\brinv$ are given in Fig.~\ref{Fig:7dim2D_brinv}. Similarly as in the fit to the Yukawa structure in Sect.~\ref{Sect:Vudl}, all fundamental coupling scale factors are positively correlated. However, the correlations here are much weaker due to the additional freedom introduced for the loop-induced Higgs couplings. In the projection planes for $\kV$ and the Higgs-fermion coupling scale factors, the ellipses are tilted in comparison to the previous fit in Sect.~\ref{Sect:Vudl}, now featuring  larger slopes of the major axes, which are roughly given by $7.5$, $2.5$ and $2.8$ for the ($\kV, \ku$), ($\kV, \kd$) and ($\kV, \kl$) planes, respectively. This represents the fact that $\ku$, $\kd$ and $\kl$ are less accurately determined since they are now only probed by the poorly measured $t\bar{t}H$, $\htobb$ and $\htotautau$ rates, respectively, while $\kV$ is still strongly constrained by both the VBF and $VH$ production modes and the decay modes $\htoWW$ and $\htoZZ$.

The correlations of the fundamental coupling scale factors to the loop-induced couplings scale factors $\kg$ and $\kga$ also turn out to be positive. Here the strongest correlation is observed among $\kg$ and $\kd$, which govern the dominant production and decay modes, respectively. Since the decay $\htobb$ is not yet probed to any reasonable accuracy at the LHC, the fit allows for an enhanced decay rate if at the same time the dominant production cross section is also increased in order to compensate for the reduced branching ratios of the remaining decay modes.\footnote{A similar correlation was found in the fit presented in Sect.~\ref{Sect:Vudl} for $\ku$ and $\kd$, because there $\ku$ was dominantly influencing the derived Higgs-gluon coupling.} Nevertheless, the preferred fit region is found for slightly suppressed values of both $\kg$ and $\kd$. A strong positive correlation is also found between $\kV$ and $\kga$.

It should be noted that the correlation of the loop-induced couplings scale factors $\kg$ and $\kga$ has changed with respect to the previous fit, \refse{Sect:gga}. They now show a weak positive correlation. This is because the general parametrization features again the degeneracy of increasing scale factors and the additional decay mode, which is only broken by the $\brinv$ constraint. This leads to a positive correlation among all $\kappa_i$ which dominates over the small anti-correlations needed to adjust the small tendencies in the observed signal rates. This is also reflected in Fig.~\ref{Fig:7dim2D_brinv}, where all scale factors show a positive correlation with $\brinv$.

\begin{table}
\centering
\renewcommand{\arraystretch}{1.3}
\scalebox{0.92}{
\begin{tabular}{l|x{1.4cm}x{1.4cm}x{1.4cm}x{1.4cm}x{1.4cm}x{1.4cm}}
\hline
Fit				& \multicolumn{6}{c}{$68\%~\mathrm{C.L.}$ precision of the  Higgs coupling scale factors [in $\%$]}	\\
				&	$\kV$	&	$\kg$	& 	$\kga$	&	$\ku$	&	$\kd$	&	$\kl$	\\
\hline
CMS Moriond 2013	&	$20\%$	&	$28\%$	&	$25\%$	&	$100\%$	&	$55\%$	&	$30\%$	\\
\HS\ (LHC $\oplus$ Tev.) 	&	$12\%$	&	$20\%$	&	$15\%$	&	$30\%$	&	$35\%$	&	$18\%$	\\
\hline
\end{tabular}
}
\caption{Comparison of the relative $68\%~\mathrm{C.L.}$ precision of the Higgs coupling scale factors obtained by the CMS combination presented at the Moriond 2013 conference~\cite{CMS:yva} and our results from the seven-dimensional scale factor fit using both LHC and Tevatron measurements. The quoted numbers are rough estimates from the (sometimes asymmetric) likelihood shapes, cf.~Ref.~\cite{CMS:yva} and~\reffi{Fig:7dim1D}.}
\label{Tab:currentprecision}
\end{table}

Comparing the relative ($1\sigma$) precision on the individual scale factors obtained here with the results of an official CMS fit analysis\footnote{The CMS fit parametrizes the Higgs couplings via the same scale factors as used here, however, the fit does not allow for an additional Higgs decay mode. We furthermore used the CMS fit results to validate our fit procedure, see Appendix~\ref{Sect:CMSvalidation}.} presented at the Moriond 2013 conference~\cite{CMS:yva}, we assert the improvements listed in Tab.~\ref{Tab:currentprecision}. Here only rough symmetrical estimates of the sometimes quite asymmetrical uncertainties are given.
With a common interpretation of the latest data from ATLAS, CMS and the Tevatron experiments, a significant improvement of the scale factor determination is achieved. Moreover, the strong improvement in the precision of $\ku$ is due to the dedicated CMS $t\bar{t}H$ tagged analyses~\cite{CMS:2013tfa,CMS:2013sea,CMS:2013fda} which had not been included in the CMS fit. With the latest $\htotautau$ measurement by ATLAS the precision of $\kl$ has also improved significantly. Nevertheless, for all scale factors potential deviations within $\sim 10\%$ or even more are still allowed at the $1\sigma$ level within this benchmark model.

\begin{figure}[t]
\centering
\includegraphics[width=0.65\textwidth]{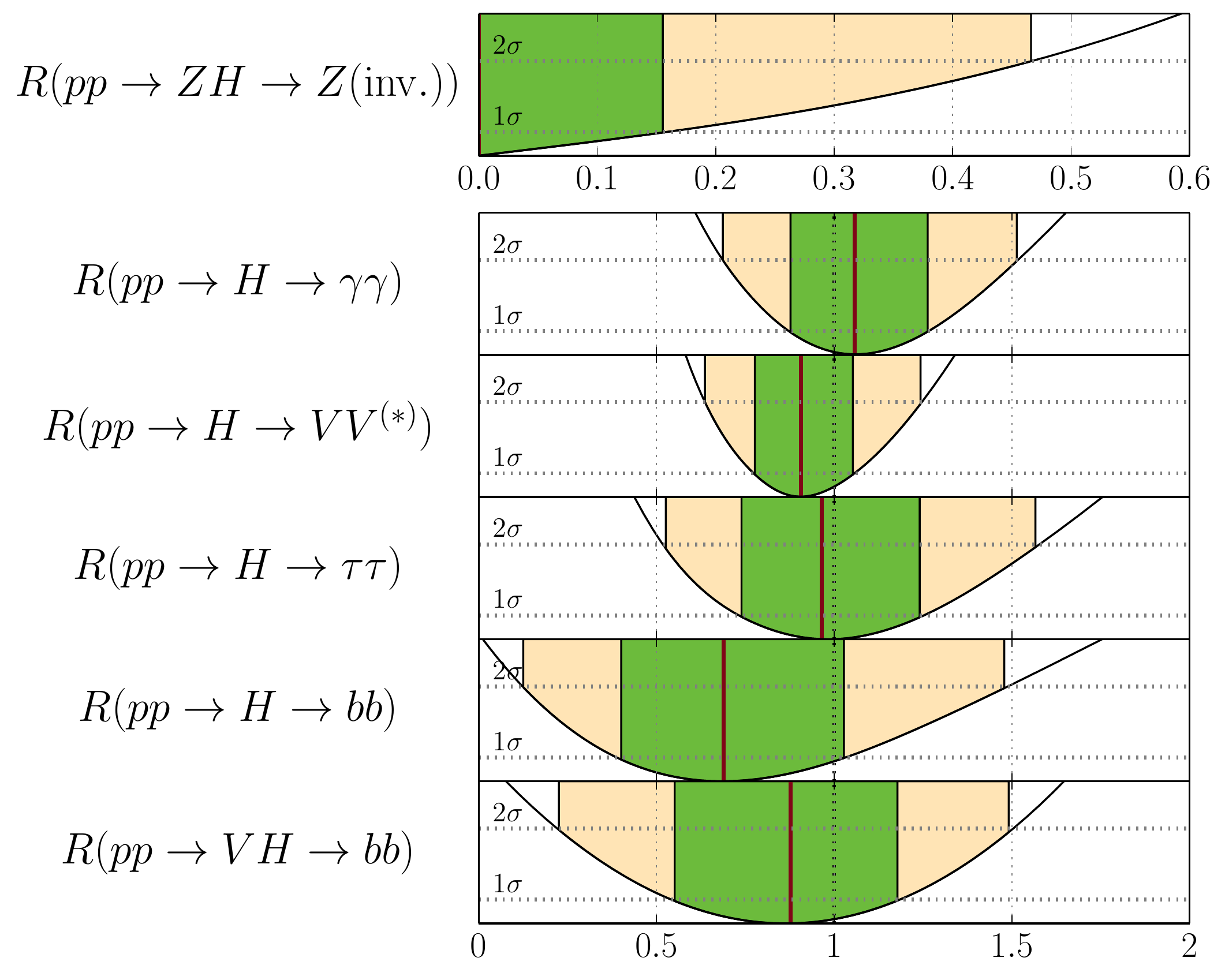}
\caption{One-dimensional $\Delta\chi^2$ profiles from the ($\kV, \ku, \kd, \kl, \kg, \kga, \brinv$) fit for the (idealized, SM normalized) signal rates at $8\tev$ for the main LHC channels.}
\label{Fig:7dim1D_rates}
\end{figure}

For this most general fit we also show the predicted signal rates for the preferred parameter space in Fig.~\ref{Fig:7dim1D_rates}. The rates $R(pp\to H\dots \to XX)$ are idealized LHC $8\tev$ signal rates where all included channels $j$ contribute with the same efficiency $\epsilon_j$, i.e.,
\begin{align}
R(pp\to H\dots\to XX) &\equiv \left.\mu(pp\to H\dots\to XX)\right|_{\epsilon_j=1},
\label{Eq:Rvalues}
\end{align}
where $\mu$ is defined in \refeq{Eq:mu}. The production mode $pp\to H$
denotes inclusive production, i.e.~we include all five LHC Higgs
production modes at their (rescaled) SM values, whereas the rates
denoted by $\pptoZH~[VH]$ include only production through
Higgs-strahlung [and $WH$ production]. It can be seen from the figure that all rates agree
with the SM expectation at $68\%~\mathrm{C.L.}$ A very weak enhancement
of the $pp\to\htogaga$ rate is observed, while the remaining channels
with fermionic or weak gauge boson final states are slightly suppressed.

\begin{figure}
\centering
\includegraphics[width=\textwidth]{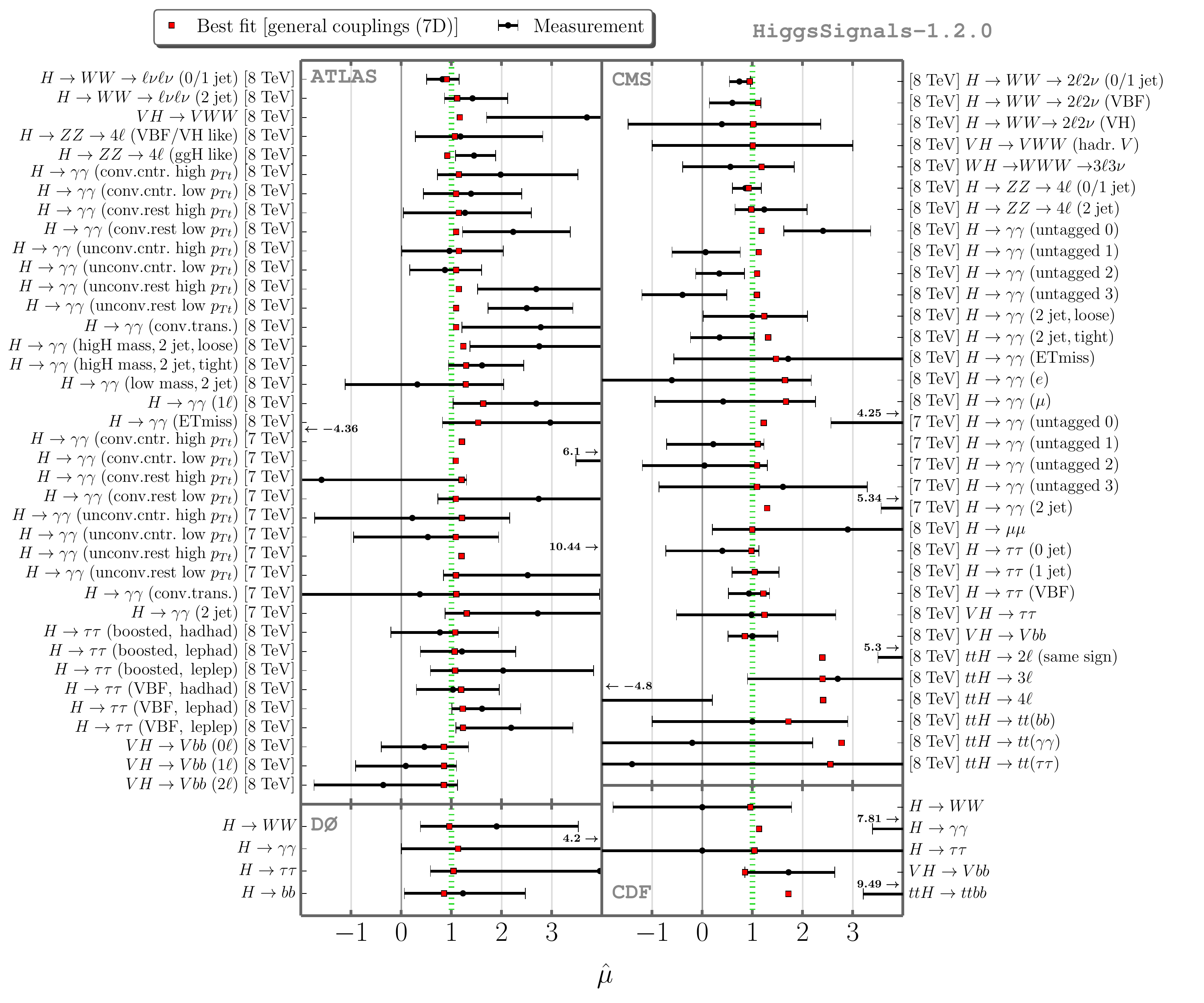}
\vspace{-0.8cm}
\caption{Comparison of the predicted signal rates of the best fit point in the general (seven-dimensional) Higgs couplings scale factor benchmark fit with the measurements from the ATLAS, CMS, CDF and D\O~collaborations. The green line indicates the prediction for the SM.}
\label{Fig:7dim_bestfit}
\end{figure}

Finally, in \reffi{Fig:7dim_bestfit} we show the actual signal
rates $\muobs$ predicted by the best fit point, depicted as red squares,
compared to all 80 measurements from the Tevatron and LHC experiments
that went into our analysis. The latter are given by the black dots and the error bars indicate the $68\%~\mathrm{C.L.}$ uncertainty.
 In the left column we show the ATLAS and D\O\ results, whereas in the
right column the CMS and CDF observables are given. The SM, located at
$\muobs = 1$, is marked as a green dashed line. 
It can be seen that most signal rates are predicted to be very close to
the SM, note however the relatively large range shown
for $\muobs$. An exception can be observed for the channels which
comprise a substantial $t\bar{t}H$ component. Moreover, we find a
slight enhancement in $\htogaga$ channels with a significant
contribution from vector boson fusion and/or associated Higgs-weak
gauge boson production.
Overall, \reffi{Fig:7dim_bestfit} demonstrates again that despite
the large available freedom to adjust the signal rates 
in this very general parametrization, the preferred region agrees
remarkably well with the SM. No significant improvement of the fit
quality is gained by allowing the additional freedom. This implies
that no significant, genuine
tendencies of deviations in the SM Higgs coupling structure can be
found.

\subsection{Upper limits on additional undetectable Higgs decay modes}
\label{Sect:kvle1}

We now discuss the case where the additional Higgs decay modes are not
detectable with the current Higgs analyses, i.e.~their final states do not lead
to the missing transverse energy signature, as discussed in the
  beginning of \refse{Sect:currentfits}.
As discussed earlier, SM-like Higgs signal rates can be achieved even with a sizable branching fraction to
undetectable final states, if at the same time the Higgs boson production rates are enhanced.  
In the absence of direct measurements of the Higgs total width or absolute cross sections the degeneracy between simultaneously increasing $\brhnp$ and coupling scale factors $\kappa_i$ can only be ameliorated with further model assumptions. Recall that $\brhnp$ in general refers to any Higgs decays that are {\em undetectable} at present collider experiments, and can in general have both SM or BSM particles in the final state. Requiring that $\kV \le 1$ (or $\kW \le 1$ and $\kZ \le 1$), an upper limit on $\brhnp$ can be derived for each investigated benchmark model without assuming that the additional decay modes leads to a missing energy signature.

Remarkably, we find that some of the six benchmark parametrizations discussed in Sect.~\ref{Sect:universal}--\ref{Sect:7dim} yield very similar limits on $\brhnp$. We therefore categorize them in three Types:
\begin{itemize}[leftmargin=0.6in]
\item[Type 1:] Benchmark models with universal Yukawa couplings and no additional freedom in the loop-induced couplings. This comprises the fits in Sect.~\ref{Sect:universal}--\ref{Sect:WZF}.
\item[Type 2:] Benchmark models with fixed tree-level couplings but free loop-induced couplings, cf.~Sect.~\ref{Sect:gga}.
\item[Type 3:] Benchmark models with non-universal Yukawa couplings, as discussed in Sect.~\ref{Sect:Vudl} and~\ref{Sect:7dim}.
\end{itemize}

\begin{table}[b]
\centering
\scalebox{0.88}{
\begin{tabular}{l  x{1.9cm}x{2.4cm}x{2.4cm}x{3.4cm}}
\toprule
category					&	SM		&		Type 1		&	Type 2			&		Type 3 		\\	
\midrule
						&			&$\kappa$		&					&	$\kV,\ku,\kd,\kl$			\\
Fitted coupling scale factors	&	-		&$\kV, \kF$		&	$\kg,\kga$			&	$\kV,\ku,\kd,\kl,\kg,\kga$	\\
						&			&$\kW, \kZ, \kF$	&					&								\\
\midrule
$\brhnp$ ($68\%~\mathrm{C.L.}$) 	&	$\le 9\%$	&	$\le 9\%$			&	$\le 10\%$	& $\le 20\%$	\\
$\brhnp$ ($95\%~\mathrm{C.L.}$) 	&	$\le 20\%$	&	$\le 20\%$			&	$\le 26\%$	& $\le 40\%$	\\
\bottomrule
\end{tabular}
}
\caption{Upper limits at $68\%$ and $95\%~\mathrm{C.L.}$ on the undetectable Higgs decay mode, $\brhnp$, obtained under the assumption $\kV\le 1$ ($V=W,Z$). All considered benchmark scenarios can be categorized into three types. The fitted coupling scale factors are given in the middle row.}
\label{Tab:BRhnp_limits}
\end{table}

The resulting upper limits on $\brhnp$ are given in
\refta{Tab:BRhnp_limits}. The corresponding profiled $\Delta\chi^2$ distributions
are displayed in \reffi{Fig:BRhnp_limits}. The most stringent limits are obtained for Type~1, where the limit is nearly identical to what is obtained with fixed SM Higgs couplings. The weakest limits are obtained for Type~3. But even in the latter, least restricted case a $\brhnp \le 40\%$
at the $95\%~\mathrm{C.L.}$ is found. 

For some of the benchmark fits considered here a similar study has been performed in Ref.~\cite{Belanger:2013xza} using Moriond 2013 results. The limits presented there are in good agreement with our results.

\begin{figure}
\centering
\includegraphics[width=0.65\textwidth]{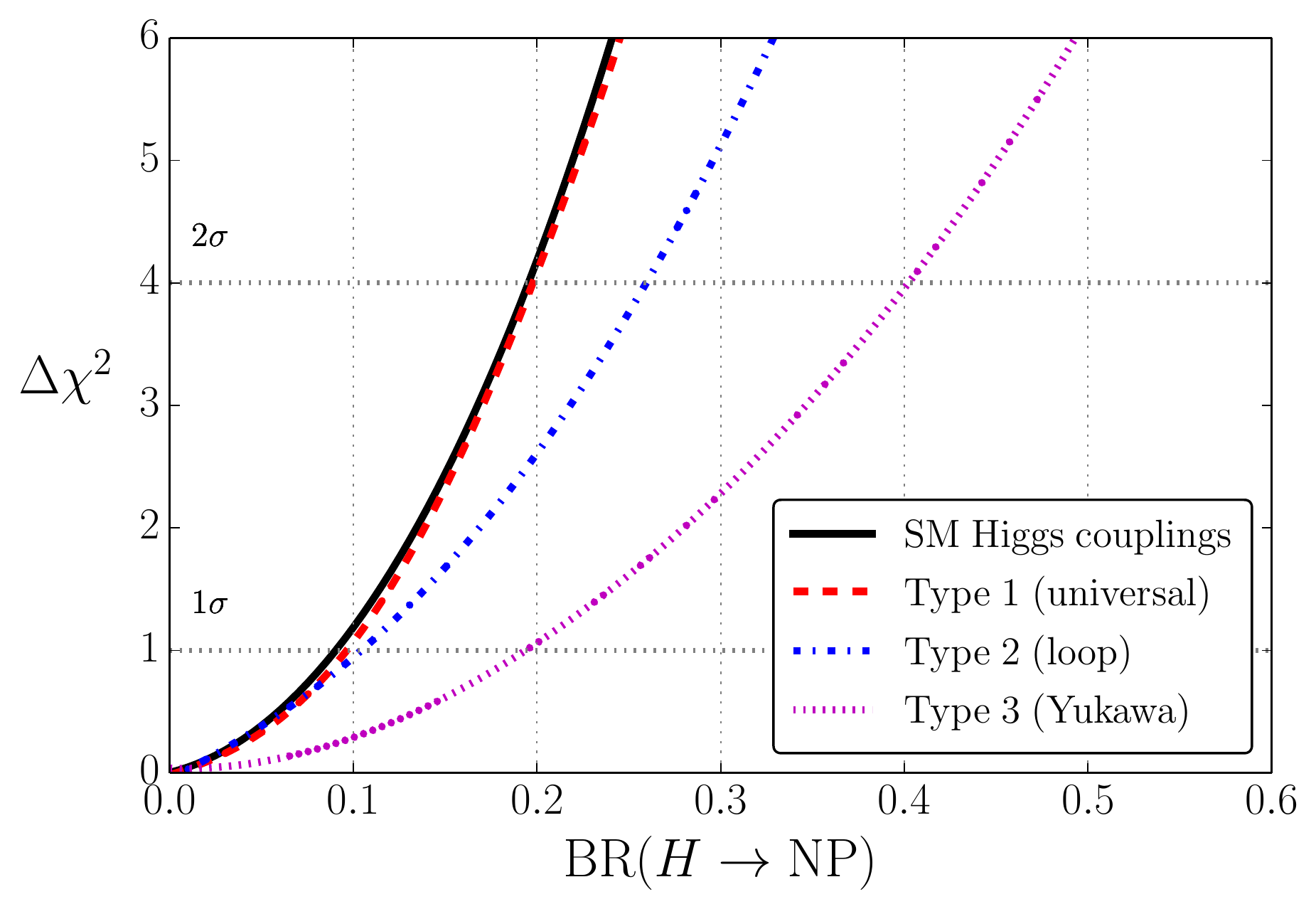}
\caption{One-dimensional $\chi^2$ profiles of $\brhnp$ in all benchmark scenarios with the assumption $\kV\le 1$ ($V=W,Z$). The three scenario types are defined in the text.}
\label{Fig:BRhnp_limits}
\end{figure}

\section{Future precision of Higgs coupling determinations}
\label{Sect:futurecouplings}
\subsection{Prospective Higgs coupling determination at the LHC}
\label{Sect:futureLHC}

The LHC experiments ATLAS and CMS have provided estimates of the future
precision for the Higgs signal rate measurements in most of the relevant
channels for integrated luminosities of $300~\ifb$ and $3000~\ifb$ at
$\sqrt{s}=14\tev$
\cite{EuropeanStrategyforParticlePhysicsPreparatoryGroup:2013fia}. The
first numbers (from 2012) have recently been
updated~\cite{ATLAS:2013hta,ATL-PHYS-PUB-2013-014,CMS:2013xfa}. In this
section we use these updated projections to determine the accuracy of
future Higgs coupling determination at the LHC. Similar studies based
on the updated projections were recently performed in Refs.~\cite{Dawson:2013bba,Peskin:2013xra},
using a slightly different methodology and parametrization of the Higgs
couplings. For earlier studies see also \citeres{Zeppenfeld:2000td,Duhrssen:2004cv,Duhrssen:2004uu,Lafaye:2009vr,Klute:2012pu,Klute:2013cx}.

Concerning the projected sensitivities for rate measurements from
ATLAS, a detailed compilation has been provided in 
\citeres{ATLAS:2013hta,ATL-PHYS-PUB-2013-014} which in most cases contains
information on the signal composition, and the projections are given with and without
theoretical uncertainties. ATLAS has also provided projections for
sub-channels including tags for the different production modes.
Unfortunately, a projection for the important channel $\htobb$ is not yet
available. This channel plays an important role in any global fit, since
the partial decay width for $\htobb$ dominates the total width in the
SM. Moreover, the ATLAS $\htotautau$ projection is based on an older analysis, and
one could expect a potential improvement from an updated study.

CMS has provided estimates for the capabilities to measure the Higgs signal
rates only for inclusive channels~\cite{CMS:2013xfa}. Unfortunately,
detailed information about the 
signal composition is missing. We are therefore
forced here to assume typical values for the signal efficiencies guided by
present LHC measurements. Moreover, the treatment of
theoretical uncertainties in the CMS projections is not very
transparent.\footnote{
See also \citere{Peskin:2013xra} for a discussion of this issue.}
CMS discusses two scenarios:
Scenario 1 uses current systematic and theoretical
uncertainties.\footnote{Note that improvements of systematical
uncertainties that can be reduced with increasing statistics in the data
control regions are however taken into account. Furthermore, even the
assumption that the same systematical uncertainties as at present 
can be reached for the harsher experimental conditions in future is based
on a projection involving a certain degree of improvement.} In Scenario 2 the
theoretical uncertainties are reduced by $1/2$, whereas the experimental
systematic uncertainties are decreased with the square root of the
integrated luminosity. No projections without theoretical uncertainties
are provided by CMS. However, the Scenario 2 projections appear quite
aggressive since they are of the same order as --- or even more precise
than --- the purely experimental projections from ATLAS. 
Furthermore, our estimates of theoretical uncertainties, rescaled under
the assumptions of Scenario 2, yield in some cases, e.g.~in the $H \to
\ga\ga, ZZ^{(*)}$ and $WW^{(*)}$ channels with $3000~\ifb$, values that are larger than the CMS estimates of the total (i.e., theoretical and experimental) uncertainties of the measurements, at least when assuming that the main production mechanism for the signal
is gluon fusion. Following a conservative approach, we therefore use the
projected CMS rate measurements given for Scenario 2, but interpret the
uncertainties as being purely experimental.\footnote{Another way to
circumvent this problem is discussed in \citere{Peskin:2013xra}, where an
alternative set of projected CMS measurements is proposed.} However, it
should be noted that the dominant effect leading to differences between
our results and the official CMS estimates of prospective Higgs coupling
determination is the absence of publicly available CMS projections of
the category measurements. Using only the inclusive measurements generally
leads to lower precision estimates in higher-dimensional scale factor
fits.

\begin{table}
\centering
\begin{threeparttable}[b]
\footnotesize
 \renewcommand{\arraystretch}{1.2}
\begin{tabular}{ l | cccc | c }
\toprule
Future scenario		&	PDF		&	$\alpha_s$	& 	$m_c$, $m_b$, $m_t$	&	THU\tnote{$\dagger$}	& $\brinv$ constraint	\\
\midrule
LHC300 (S1)		&	$100\%$		&	$100\%$	&	all $100\%$			& $100\%$ 	&	conservative, Eq.~\eqref{Eq:LHC300_inv_conservative}  \\
LHC300 (S2, csv.)	&	$50\%$	&	$100\%$		&	all $100\%$			& $50\%$ 	&	conservative, Eq.~\eqref{Eq:LHC300_inv_conservative} \\
LHC300 (S2, opt.)	&	$50\%$	&	$100\%$		&	all $100\%$			& $50\%$ 	&	optimistic, Eq.~\eqref{Eq:LHC300_inv_optimistic} \\
HL--LHC (S1)		&	$100\%$		&	$100\%$	&	all $100\%$			& $100\%$ 	&	conservative, Eq.~\eqref{Eq:LHC3000_inv_conservative} \\
HL--LHC (S2, csv.)		&	$50\%$	&	$100\%$		&	all $100\%$			& $50\%$		&	conservative, Eq.~\eqref{Eq:LHC3000_inv_conservative} \\
HL--LHC (S2, opt.)		&	$50\%$	&	$100\%$		&	all $100\%$			& $50\%$		&	optimistic, Eq.~\eqref{Eq:LHC3000_inv_optimistic} \\
\midrule
ILC250			&	-			&	$50\%$	&	all $50\%$			& $50\%$	& $\le 0.9\%$ (cf.~Tab.~\ref{Tab:ILCobs})	\\
ILC500			&	-			&	$50\%$	&	all $50\%$			& $50\%$	& $\le 0.9\%$ (cf.~Tab.~\ref{Tab:ILCobs})	\\
ILC1000			&	-			&	$50\%$	&	all $50\%$			& $50\%$	& $\le 0.9\%$ (cf.~Tab.~\ref{Tab:ILCobs})	\\
ILC1000 (LumiUp)	&	-			&	$50\%$	&	all $50\%$			& $50\%$	& $\le 0.4\%$ (cf.~Tab.~\ref{Tab:ILCobs})	\\
\midrule
HL--LHC $\oplus$ ILC250 ($\sigma_{ZH}^\mathrm{total}$)\tnote{$\ddag$}&		$50\%$		&	$50\%$		&	all $50\%$				& $50\%$ & \tnote{$*$}	\\
HL--LHC $\oplus$ ILC250	&		$50\%$		&	$50\%$		&	all $50\%$				& $50\%$ & 	\tnote{$*$}\\
HL--LHC $\oplus$ ILC500&		$50\%$		&	$50\%$		&	all $50\%$				& $50\%$ & 	\tnote{$*$}\\
HL--LHC $\oplus$ ILC1000&		$50\%$		&	$50\%$		&	all $50\%$				& $50\%$ & 	\tnote{$*$}\\
HL--LHC $\oplus$ ILC1000 (LumiUp)&	$50\%$		&	$50\%$		&	all $50\%$				& $50\%$ & 	\tnote{$*$}\\
\bottomrule
\end{tabular}
 \begin{tablenotes}
 \footnotesize
 \item[$\dagger$] Affects the theoretical uncertainties (THU) of all partial widths except for the decay modes $\htoWW$ and $\htoZZ$ (kept unchanged) as well as the uncertainties from missing higher-order QCD corrections, often
     estimated via a scale variation, and missing higher-order EW corrections for all LHC production modes.
\item[$\ddag$] In this scenario only the direct ILC measurement of
$\sigma(\epemtoZH)$ with $250~\ifb$ at $\sqrt{s}=250\gev$ is added to the
HL--LHC projections to constrain the total width.
\item[*] For the HL--LHC $\oplus$ ILC combinations we do not use the assumption $\brhnp \equiv \brinv$.
 \end{tablenotes}
 \end{threeparttable}
\caption{List of all future scenarios considered. Given are for each
scenario the assumptions on uncertainties (relative to the current
values, i.e.\ the entry ``100\%'' denotes the current value, while
the entry ``50\%'' denotes an improvement by a
factor of two)
from parton distribution functions (PDF), the strong coupling $\alpha_s$,
the quark masses ($m_c, m_b, m_t$), and theoretical uncertainties (THU) on
the predictions for the LHC Higgs cross sections and partial decay widths.
The last column gives for each scenario the constraint that is employed \textit{if} the additional Higgs decay(s) are assumed to be invisible.
The considered integrated luminosities for the three energy stages $250\gev$, $500\gev$ and $1\tev$
of the ILC for a baseline scenario and for a luminosity upgrade (LumiUp) are
specified in Sect.~\ref{Sect:ILC}, based on \citere{Asner:2013psa}.
The various ILC scenarios include the projected measurements from the preceding stages. 
}
\label{Tab:scenarios}
\end{table}

The ATLAS and CMS estimates of the experimental precision used in our analysis are listed in Tab.~\ref{Tab:futureLHCmeasurements} in Appendix~\ref{Sect:Appprojections}, which also gives the assumed signal composition for each channel. For both experiments we assume that the experimental precision includes a $3\%$ systematic uncertainty on the integrated luminosity, which is treated as fully correlated among each experiment.

On top of these experimental precisions we add theoretical rate
uncertainties within \HS. We discuss two future scenarios for the LHC-only
projections: In the first scenario (S1) we take the current theoretical
uncertainties as already used in the previous fits in
Sect.~\ref{Sect:currentfits}. This scenario thus represents the rather
pessimistic --- or conservative --- case that no improvement in the
theoretical uncertainties can be achieved. With increasing integrated
luminosity, however, the uncertainty from the parton density functions
(PDF) can be expected to decrease~\cite{Campbell:2013qaa}. Future progress can
also be expected in calculations of higher-order corrections to the Higgs
production cross sections and decay widths, which may further decrease the
theoretical uncertainties, in particular the QCD scale dependence and
remaining uncertainties from unknown electroweak (EW) corrections. Hence, in the second scenario (S2) we assume that
uncertainties from the PDFs, as well as most\footnote{This includes
uncertainties from missing higher-order QCD corrections, often estimated by scale variation, and unknown EW corrections for the LHC
Higgs production modes, as well as the uncertainties of all partial decay
widths except the decays to $W$ and $Z$ bosons where higher-order EW
corrections are already known with high accuracy.} theoretical
uncertainties, are halved. In both scenarios, the parametric uncertainties
from the strong coupling constant, $\alpha_s$, and the heavy quark masses,
$m_c$, $m_b$ and $m_t$, are unchanged. 
The different future scenarios considered in our analysis together
with the respective assumptions on the future uncertainties and constraints are 
summarized in Tab.~\ref{Tab:scenarios}. The entry ``100\%'' in
Tab.~\ref{Tab:scenarios} corresponds to the present value of the
considered quantity, and accordingly, ``50\%'' denotes an improvement by a
factor of two. More details and estimates of the cross section and branching ratio uncertainties for these scenarios are given in Appendix~\ref{Sect:AppTHU}.

ATLAS and CMS also provide projections for the $95\%~\mathrm{C.L.}$ upper
limit on the rate of an invisibly decaying Higgs boson in the
Higgs-strahlung process, $\pptoZH$. Assuming, like we have done in
Sect.~\ref{Sect:universal}--\ref{Sect:7dim}, that additional Higgs
decay modes give rise to purely invisible final states\footnote{We state
explicitly in Tab.~\ref{Tab:scenarios} which constraint on the additional decay modes is applied, if purely invisible final states are assumed.}, these constraints are incorporated in our fit as ideal $\chi^2$ likelihoods of the form
\begin{equation}
\chi^2 = 4 \cdot \tilde\sigma^2/\tilde\sigma_{95\%\mathrm{C.L.}}^2.
\end{equation}
The quantity $\tilde\sigma$ corresponds to the product $\kZ^2 \brinv$, i.e~the cross section of $\pptoZH\to Z(\mathrm{inv.})$ normalized to the SM cross section for $\pptoZH$. Both ATLAS and CMS consider two scenarios for the projected limits~\cite{ATL-PHYS-PUB-2013-014,CMS:2013xfa}: The conservative (csv.) scenario,
\begin{align}
\mbox{LHC}~300~\ifb:\qquad &\tilde\sigma_{95\%\mathrm{C.L.}} = 0.32~\mbox{(ATLAS)} \qquad \tilde\sigma_{95\%\mathrm{C.L.}} = 0.28~\mbox{(CMS)} \label{Eq:LHC300_inv_conservative}  \\
\mbox{LHC}~3000~\ifb:\qquad &\tilde\sigma_{95\%\mathrm{C.L.}} = 0.16~\mbox{(ATLAS)} \qquad \tilde\sigma_{95\%\mathrm{C.L.}} = 0.17~\mbox{(CMS)} \label{Eq:LHC3000_inv_conservative}
\end{align}
and the optimistic (opt.) scenario,
\begin{align}
\mbox{LHC}~300~\ifb:\qquad &\tilde\sigma_{95\%\mathrm{C.L.}} = 0.23~\mbox{(ATLAS)} \qquad \tilde\sigma_{95\%\mathrm{C.L.}} = 0.17~\mbox{(CMS)} \label{Eq:LHC300_inv_optimistic} \\
\mbox{LHC}~3000~\ifb:\qquad &\tilde\sigma_{95\%\mathrm{C.L.}} = 0.08~\mbox{(ATLAS)} \qquad \tilde\sigma_{95\%\mathrm{C.L.}} = 0.06~\mbox{(CMS)}. \label{Eq:LHC3000_inv_optimistic}
\end{align}
We combine the projected ATLAS and CMS limits by adding their respective
$\chi^2$ contributions. For the scenario S1 we only employ the
conservative constraints, Eqs.~\eqref{Eq:LHC300_inv_conservative} and
\eqref{Eq:LHC3000_inv_conservative}, whereas for the scenario S2 with
reduced uncertainties we compare fits using either the conservative or the
optimistic constraint. These cases are denoted by (S2, csv.) and (S2,
opt.), respectively. 

\begin{figure}
\centering
\subfigure[Assuming $\brhnp \equiv \brinv$.]{\includegraphics[width=0.49\textwidth]{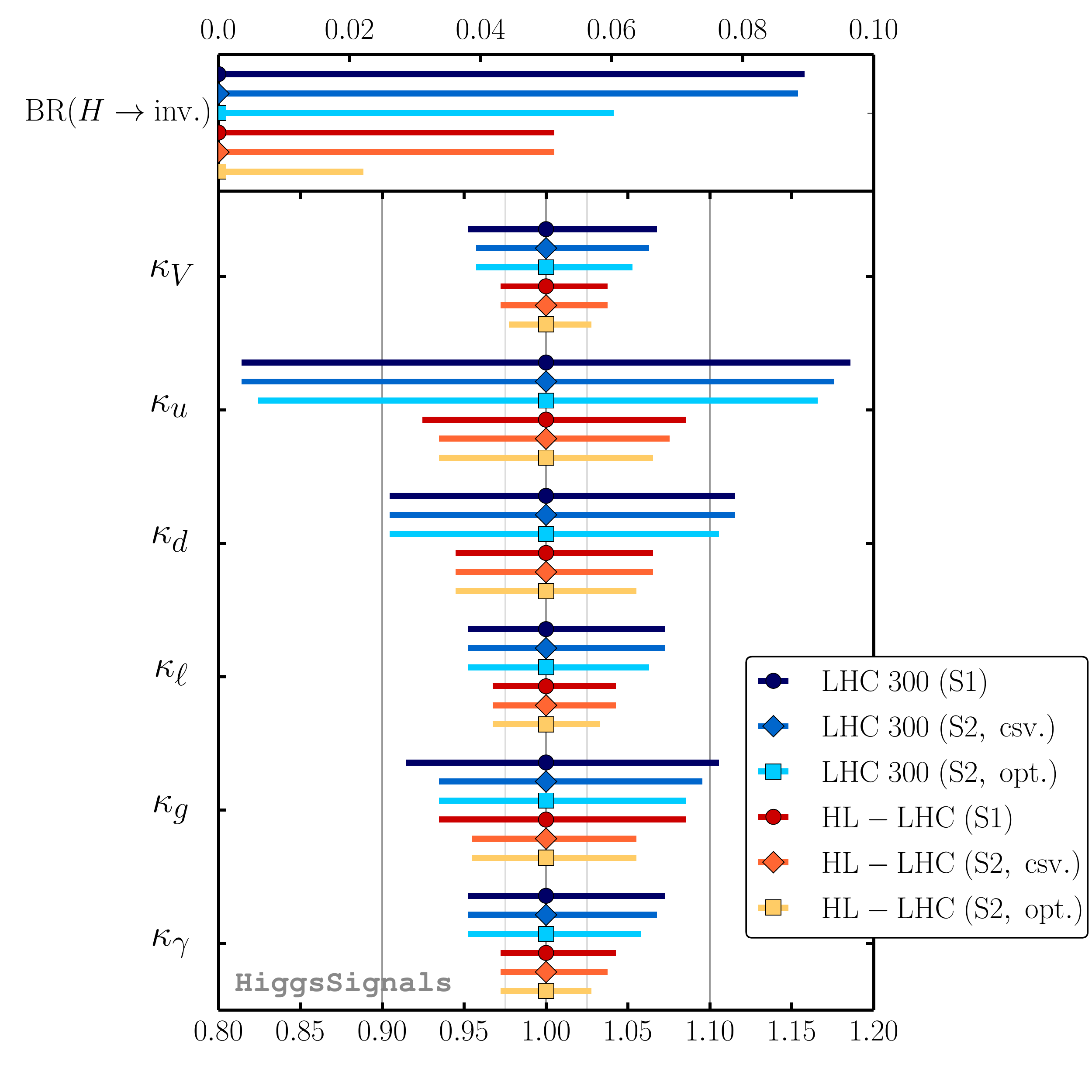}}
\subfigure[Assuming $\kV \le 1$.]{\includegraphics[width=0.49\textwidth]{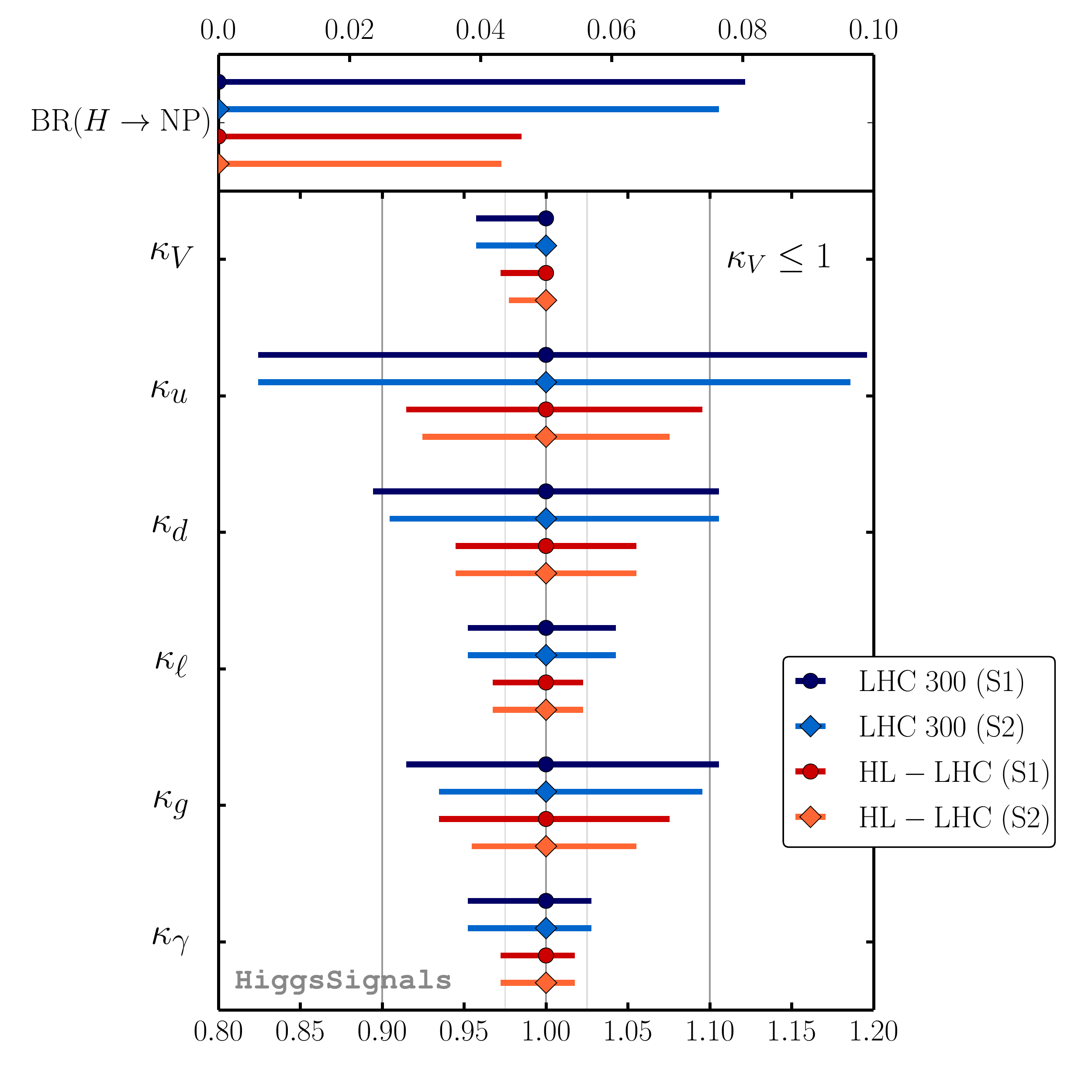}}
\caption{Projected future precision for the determination of Higgs
coupling scale factors at the LHC with integrated luminosities of
$300~\ifb$ and $3000~\ifb$ (HL--LHC).}
\label{Fig:LHCprospects}
\end{figure}

For the LHC projections we employ the same seven-dimensional scale factor parametrization as discussed in Sect.~\ref{Sect:7dim}. The resulting $68\%~\mathrm{C.L}$ precision estimates obtained under the assumption that the additional decay mode $\brhnp\equiv \brinv$ are displayed in Fig.~\ref{Fig:LHCprospects}(a) and listed in Tab.~\ref{Tab:LHCprospects_7dim_brinv}. The plot includes all six LHC-only scenarios as listed in Tab.~\ref{Tab:scenarios}.

In general the obtained $68\%~\mathrm{C.L.}$ limit on $\brinv$ is weaker
than the limit obtained from a Gaussian combination of the limits in
Eqs.~\eqref{Eq:LHC300_inv_conservative}--\eqref{Eq:LHC3000_inv_optimistic},
because the fit has the freedom to adjust $\kZ (\equiv \kV)$ to values $<
1$. Improvements in the theoretical uncertainties will mostly affect the
effective Higgs-gluon coupling. At an integrated luminosity of
$300~\ifb$ we obtain a precision estimate for the scale factor of the
effective Higgs-gluon coupling of $\delta\kg \sim 9.5\%$ in the more
conservative scenario S1,\footnote{Here and in the following the Higgs
coupling precision at $68\%~\mathrm{C.L.}$ is denoted by $\delta\kappa$. The values quoted in
the text usually correspond to symmetric averages. For the exact
asymmetric values see the corresponding tables, e.g.~here 
Tab.~\ref{Tab:LHCprospects_7dim_brinv}.}
which is improved to $\delta\kg \sim 7.5\%$ in the most optimistic scenario
S2. At the high luminosity LHC with $3000~\ifb$ the corresponding
projections are $\delta\kg \sim 7.5\%$ for the scenario S1 and 
$\delta\kg \sim 5\%$ for the scenario S2, irrespective of the
assumed precision of the $\brinv$ constraint. The assumed improvements of
the theoretical uncertainties hence lead to a significant increase of the
$\kg$ precision at the HL--LHC, while the precision at $300~\ifb$ is still mostly limited by statistics.

The impact of more optimistic limits on the invisible Higgs decays,
Eqs.~\eqref{Eq:LHC300_inv_optimistic}--\eqref{Eq:LHC3000_inv_optimistic},
can directly be seen in the projected upper $68\%~\mathrm{C.L.}$ limit on
$\brinv$ in Fig.~\ref{Fig:LHCprospects}(a). Since this improved constraint
also applies to the Higgs--$Z$ boson coupling the precision of the
Higgs--vector-boson coupling scale factor, $\delta \kV$, also improves
from $\sim 5.3\%$ [$3.3\%$] to $\sim 4.8\%$ [$2.6\%$] at $300~\ifb$
[$3000~\ifb$], assuming the improved theoretical uncertainties of Scenario
S2. The impact on the remaining scale factors is rather insignificant and
results mostly from their positive correlation with $\kV$ and $\brinv$.
Hence, these are slightly more constrained from above if a more optimistic
limit on the invisible Higgs decays can be achieved.

\begin{table}
\renewcommand{\arraystretch}{1.3}
\centering
\small
\begin{tabu}{c | x{1.2cm}x{1.2cm}x{1.2cm} | x{1.2cm}x{1.2cm}x{1.2cm}}
\toprule
		     	& \multicolumn{6}{c}{$68\%~\mathrm{C.L.}$ Higgs
coupling scale factor precision [in $\%$]} \\
\midrule			
			& \multicolumn{3}{c |}{LHC 300} &
\multicolumn{3}{c}{HL--LHC} \\
\rowfont{\footnotesize}
Scenario 	& S1 & S2, csv. & S2, opt. &  S1 & S2, csv. & S2, opt. \\
\midrule
\rowfont{\footnotesize}
$ \mathrm{BR}(H \to \mathrm{inv.}) $ & $ \le 8.9 $ & $ \le 8.8 $ & $ \le 6.0 $ & $ \le 5.1 $ & $ \le 5.1 $ & $ \le 2.2 $ \\ 
\midrule
$ \kappa_V $ & $\substack{+ 6.8 \\ - 4.8 }$& $\substack{+ 6.3 \\ - 4.3 }$ & $\substack{+ 5.3 \\ - 4.3 }$ & $\substack{+ 3.8 \\ - 2.8 }$ & $\substack{+ 3.8 \\ - 2.8 }$ & $\substack{+ 2.8 \\ - 2.3 }$ \\ 
$ \kappa_u $ & $\substack{+ 18.6 \\ - 18.6 }$ & $\substack{+ 17.6 \\ - 18.6 }$ & $\substack{+ 16.6 \\ - 17.6 }$ & $\substack{+ 8.5 \\ - 7.5 }$ & $\substack{+ 7.5 \\ - 6.5 }$ & $\substack{+ 6.5 \\ - 6.5 }$ \\  
$ \kappa_d $ & $\substack{+ 11.6 \\ - 9.5 }$ & $\substack{+ 11.6 \\ - 9.5 }$ & $\substack{+ 10.6 \\ - 9.5 }$ & $\substack{+ 6.5 \\ - 5.5 }$ & $\substack{+ 6.5 \\ - 5.5 }$ & $\substack{+ 5.5 \\ - 5.5 }$ \\ 
$ \kappa_\ell $ & $\substack{+ 7.3 \\ - 4.8 }$ & $\substack{+ 7.3 \\ - 4.8 }$ & $\substack{+ 6.3 \\ - 4.8 }$ & $\substack{+ 4.3 \\ - 3.3 }$ & $\substack{+ 4.3 \\ - 3.3 }$ & $\substack{+ 3.3 \\ - 3.3 }$ \\ 
$ \kappa_g $ & $\substack{+ 10.6 \\ - 8.5 }$ & $\substack{+ 9.5 \\ - 6.5 }$ & $\substack{+ 8.5 \\ - 6.5 }$ & $\substack{+ 8.5 \\ - 6.5 }$ & $\substack{+ 5.5 \\ - 4.5 }$  & $\substack{+ 5.5 \\ - 4.5 }$ \\ 
$ \kappa_\gamma $ & $\substack{+ 7.3 \\ - 4.8 }$ & $\substack{+ 6.8 \\ - 4.8 }$ & $\substack{+ 5.8 \\ - 4.8 }$ & $\substack{+ 4.3 \\ - 2.8 }$ & $\substack{+ 3.8 \\ - 2.8 }$ & $\substack{+ 2.8 \\ - 2.8 }$ \\  
\bottomrule
\end{tabu}
\caption{Estimates of the future $68\%~\mathrm{C.L.}$ precision of Higgs
coupling scale factors at the LHC under the assumption $\brhnp \equiv \brinv$. The values correspond to those in Fig.~\ref{Fig:LHCprospects}(a).}
\label{Tab:LHCprospects_7dim_brinv}
\end{table}

\begin{table}
\renewcommand{\arraystretch}{1.3}
\centering
\small
\begin{tabu}{c | x{1.9cm}x{1.9cm} | x{1.9cm}x{1.9cm}}
\toprule
		     	& \multicolumn{4}{c}{$68\%~\mathrm{C.L.}$ Higgs coupling scale factor precision [in $\%$]} \\
\midrule			
			& \multicolumn{2}{c |}{LHC 300} &
\multicolumn{2}{c}{HL--LHC} \\
\rowfont{\footnotesize}
Scenario 	& S1 &  S2 &  S1 & S2 \\
\midrule
\rowfont{\footnotesize}
$ \mathrm{BR}(H \to \mathrm{NP}) $ & $ \le 8.0 $ & $ \le 7.6 $ & $ \le 4.6 $ & $ \le 4.3 $ \\ 
\midrule
$ \kappa_V $ & $\substack{+ 0.0 \\ - 4.3 }$ & $\substack{+ 0.0 \\ - 4.3 }$ & $\substack{+ 0.0 \\ - 2.8 }$ & $\substack{+ 0.0 \\ - 2.3 }$ \\ 
$ \kappa_u $ & $\substack{+ 19.6 \\ - 17.6 }$ & $\substack{+ 18.6 \\ - 17.6 }$ & $\substack{+ 9.5 \\ - 8.5 }$  & $\substack{+ 7.5 \\ - 7.5 }$ \\ 
$ \kappa_d $ & $\substack{+ 10.6 \\ - 10.6 }$ & $\substack{+ 10.6 \\ - 9.5 }$ & $\substack{+ 5.5 \\ - 5.5 }$ & $\substack{+ 5.5 \\ - 5.5 }$ \\ 
$ \kappa_\ell $ & $\substack{+ 4.3 \\ - 4.8 }$ & $\substack{+ 4.3 \\ - 4.8 }$ & $\substack{+ 2.3 \\ - 3.3 }$ & $\substack{+ 2.3 \\ - 3.3 }$ \\ 
$ \kappa_g $ & $\substack{+ 10.6 \\ - 8.5 }$ & $\substack{+ 9.5 \\ - 6.5 }$ & $\substack{+ 7.5 \\ - 6.5 }$ & $\substack{+ 5.5 \\ - 4.5 }$ \\ 
$ \kappa_\gamma $ & $\substack{+ 2.8 \\ - 4.8 }$ & $\substack{+ 2.8 \\ - 4.8 }$ & $\substack{+ 1.8 \\ - 2.8 }$ & $\substack{+ 1.8 \\ - 2.8 }$ \\
\bottomrule
\end{tabu}
\caption{Estimates of the future $68\%~\mathrm{C.L.}$ precision of Higgs coupling
scale factors at the LHC under the assumption $\kV \le 1$. The values correspond to those in Fig.~\ref{Fig:LHCprospects}(b).}
\label{Tab:LHCILC7dim_kVle1}
\end{table}

Taking into account the possibility that an
additional Higgs decay mode may result in an undetectable final state, we
show the fit results obtained under the assumption $\kV \le 1$ in
Fig.~\ref{Fig:LHCprospects}(b) and Tab.~\ref{Tab:LHCILC7dim_kVle1}.
Overall, the achievable precision in the Higgs coupling scale factors
with this assumption on the Higgs coupling to gauge bosons is
very similar to what was obtained with the assumption of allowing
only additional Higgs decays into invisible final states,
cf.~Fig.~\ref{Fig:LHCprospects}(a). A notable difference is, however, that
in particular the scale factors $\kl$ and $\kga$ are more strongly
constrained from above due to their positive correlation with $\kV$, which
is forced to be $\le 1$ by assumption in this case. 
The obtained $68\%~\mathrm{C.L.}$ limit
projection on $\brhnp$ can be regarded as an independent limit projection
inferred from the model assumption on $\kV$ and the chosen parametrization,
see also the discussion in Sect.~\ref{Sect:kvle1}. Remarkably, the limit
projections obtained here are stronger than the allowed range for $\brinv$
in the previous fits in Fig.~\ref{Fig:LHCprospects}(a) where the
constraints from searches for an invisibly decaying Higgs boson have
been applied.

Overall, we find estimates of Higgs coupling scale factor precisions
within $\sim 5-18\%$ at $300~\ifb$ and $\sim 3-10\%$ at $3000~\ifb$
obtained under the assumption $\brhnp\equiv\brinv$. These estimates 
improve slightly if one assumes $\kV \le 1$ instead. 
Comparisons with results in the literature based on the same
projections of the future capabilities provided by ATLAS and CMS,
our results agree quite well with those presented in Ref.~\cite{Dawson:2013bba}. A comparison of
our results with Ref.~\cite{Peskin:2013xra} would need to take into account
the different approaches of implementing the CMS projections. In view of
this fact, we also find reasonable agreement with the results presented in
Ref.~\cite{Peskin:2013xra}.

It should be noted that this seven-parameter fit within the 
``interim framework'' of Higgs-coupling scale factors still contains important simplifying assumptions and 
restrictions, which one would want to avoid as much as possible 
in a realistic analysis at the time when $300~\ifb$ or $3000~\ifb$ of integrated luminosity will have been collected, see the
discussion in~\citeres{LHCHiggsCrossSectionWorkingGroup:2012nn,Heinemeyer:2013tqa}.


\subsection{Prospective Higgs coupling determination at the ILC}
\label{Sect:ILC}

Looking beyond the LHC, an $e^+e^-$ linear collider (LC) with a
center-of-mass energy that can be raised at least up to 
$\sqrt{s} \sim 500\gev$ is widely regarded to be ideally suited for
studying the properties of the discovered new particle with high precision. 
The Technical Design Report for the International Linear Collider, ILC, 
has recently been submitted~\cite{Baer:2013cma}, and there are encouraging
signs that a timely realisation of this project may become possible due to
the strong interest of the Japanese scientific community and the Japanese
government to host the ILC.

The ILC offers a clean experimental environment enabling precision
measurements of the Higgs boson mass, width, its quantum numbers and 
$\cp$-properties as well as the signal rates of a variety of production 
and decay channels, including a high-precision measurement of the decay
rate into invisible final states. The highest statistics can be
accumulated at the highest energy, $\sqrt{s} \sim 1 \tev$, from the
$t$-channel process where a Higgs boson is produced in $WW$ fusion
($\epem\to \nu\nu H$). At $\sqrt{s} \sim 250 \gev$ an absolute measurement of the production 
cross section can be performed from the Higgs-strahlung
process ($\epemtoZH$) near threshold using the recoil of the Higgs boson
against the $Z$~boson, decaying via
$Z \to \mu^+\mu^-$ or $Z \to e^+e^-$, without having to consider the
actual pattern of the Higgs decay. The absolute measurement of the
production cross section can be exploited to obtain absolute measurements
of the decay branching ratios and of the total width of the decaying
particle. Consequently, no additional model
assumptions are necessary to constrain the total width
and thus the Higgs boson couplings.
For $\sqrt{s} \sim 250 \gev$ an integrated luminosity of $250~\ifb$ 
will result in \order{10^5} Higgs bosons.
The ILC will provide high-precision measurements of channels that are known 
to be difficult (such as $H \to b \bar b$) or may even be impossible 
(such as $H \to c \bar c, gg$) at the LHC. 
At $\sqrt{s} \sim 500 \gev$ the weak boson fusion process already dominates over
the Higgs-strahlung process 
for a $126\gev$ SM-like Higgs boson, and the two production channels
together provide data with very high statistics.
Starting from this energy, the top Yukawa coupling and, for sufficiently
high luminosity,
the trilinear self-coupling will become accessible.

In this section we study the capabilities of Higgs coupling determinations at the ILC. Similar studies have been performed in Ref.~\cite{Klute:2013cx,Han:2013kya,Dawson:2013bba,Peskin:2013xra}.
We discuss fit results using prospective ILC measurements both alone and
in combination with measurements from the HL--LHC. Since the two major Higgs production modes,
Higgs-strahlung and $WW$ fusion, are governed by the Higgs-$Z$-$Z$ and Higgs-$W$-$W$ couplings,
respectively, from now on we abandon the assumption of custodial symmetry. In comparison to the LHC, the sensitivity to probe custodial symmetry is greatly enhanced at the ILC, since clean measurements of the two main production modes can be performed individually at high precision.
Instead of the previous parametrization, we fit individual scale factors for the Higgs-$Z$-$Z$ and Higgs-$W$-$W$ couplings from now on. Thus, we employ an eight-dimensional fit in the parameters $\kW$, $\kZ$, $\ku$, $\kd$, $\kl$, $\kg$, $\kga$ and $\brhnp$.

The projected ILC measurements have been presented in Ref.~\cite{Baer:2013cma} and recently updated in a Snowmass White paper~\cite{Asner:2013psa}. These updated numbers, which we use in our fits, are summarized in Tab.~\ref{Tab:ILCobs} in Appendix~\ref{Sect:Appprojections}. In particular, we include the measurements of the total $ZH$ cross section, cf.~\refta{Tab:ILCobs}, which constrain the total width and enable a \textit{model-independent} determination of the Higgs couplings. An assumed luminosity uncertainty of $0.1\%$ and theoretical uncertainties of the $\epemtoZH$, $\epem \to \nu\nu H$ and $\epem\to t\bar{t}H$ cross section predictions of $0.5\%$, $1\%$ and $1\%$, respectively, are treated as fully correlated in our fit. We assume the same improvements of the theoretical uncertainties for the Higgs decay modes as in Scenario S2 of the LHC projections. In addition, we assume that the parametric uncertainties from dependences on $\alpha_s$ and the heavy quark masses $m_c$, $m_b$ and $m_t$ can also be reduced by $50\%$ with prospective ILC measurements and
lattice calculations~\cite{Campbell:2013qaa}. A further reduction of the top quark mass uncertainty --- anticipated to improve by a factor of $\sim 10$ with respect to the current precision~\cite{Baer:2013cma} --- has negligible impact on the partial width uncertainties is and therefore not further considered here.
 A summary of all future scenarios that we consider in our analysis is given in Tab.~\ref{Tab:scenarios}. Estimates of the theoretical
uncertainties on the Higgs branching ratios that we apply for the ILC scenarios are provided in Appendix~\ref{Sect:AppTHU}.

In our analysis of the ILC projections we consider three stages of center-of-mass energies, namely $250\gev$ (stage 1), $500\gev$
(stage 2) and $1\tev$ (stage 3). For the integrated luminosities at those energy stages
we investigate both a baseline program with integrated luminosities of 
$250~\ifb$ at stage 1, $500~\ifb$ at stage 2 and $1~\iab$ at stage 3, as
well as a scenario corresponding to a luminosity upgrade (LumiUp). For the
latter the integrated luminosities of $1150~\ifb$ at stage 1, $1600~\ifb$ at stage 2 and $2.5~\iab$ at stage 3 are
assumed, see \citere{Asner:2013psa}.

\begin{figure}
\centering
\subfigure[Assuming $\brhnp\equiv\brinv$.]{\includegraphics[width=0.49\textwidth]{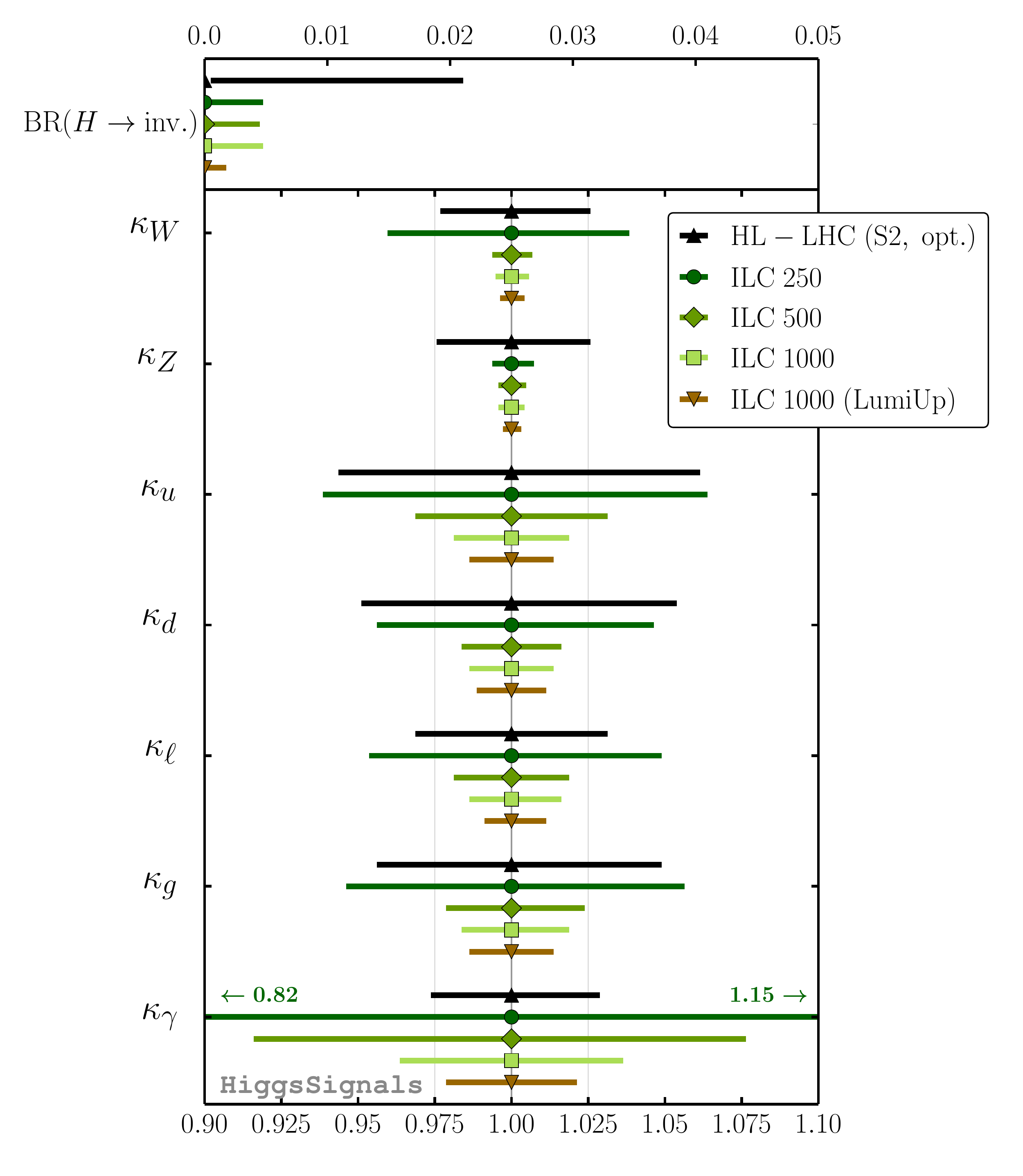}}
\subfigure[Assuming $\kW, \kZ \le 1$.]{\includegraphics[width=0.49\textwidth]{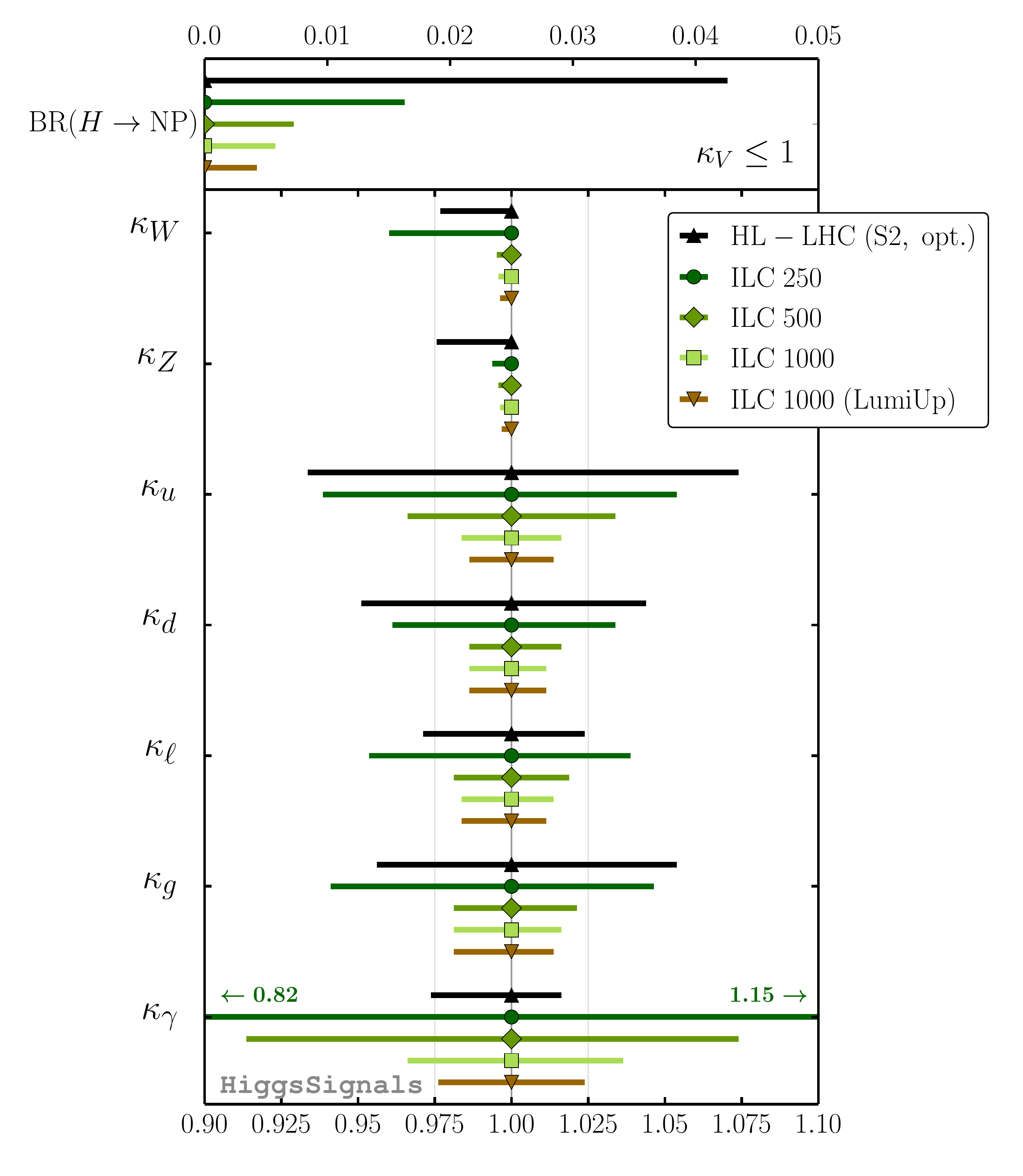}}
\caption{Prospective Higgs coupling scale factor determination at the ILC in comparison
with the (optimistic) HL--LHC scenario under the same model assumptions
as in \reffi{Fig:LHCprospects}.}
\label{Fig:LHC_vs_ILCprospects_8dim_modeldependent}
\end{figure}

In Fig.~\ref{Fig:LHC_vs_ILCprospects_8dim_modeldependent} we show the
estimated accuracies of the Higgs coupling scale factors at the ILC
obtained under \textit{model-dependent} assumptions, in analogy to
the analyses performed above for the projections of future accuracies at
the LHC: In~\reffi{Fig:LHC_vs_ILCprospects_8dim_modeldependent}(a) we
assume that any additional Higgs decay results in invisible final
states; accordingly we also take into account the projected ILC upper
limit on $\brinv$, cf.~\refta{Tab:ILCobs} (or~\refta{Tab:scenarios}). In
\reffi{Fig:LHC_vs_ILCprospects_8dim_modeldependent}(b) we apply the
theoretical constraint $\kW,\kZ \le 1$. For comparison we also show the
fit results for the optimistic HL--LHC scenario (S2, opt) obtained under these assumptions.

Overall, the scale factor precisions achieved under those two
assumptions are very similar to each other. Comparing
the results of the first ILC stage, where just a `baseline' value for
the integrated luminosity of $250~\ifb$ is assumed
(ILC250), with the ultimate precision that can be reached at the LHC, we 
see already at this stage a substantial improvement in the precision of
the scale factor $\kZ$ (from $\sim 2.5\%$ to $\sim 0.7\%$). 
This is already a crucial improvement since this coupling is of central importance in the experimental test
of the electroweak symmetry breaking mechanism. 
Furthermore, the ILC provides at this stage important measurements that are complementary 
to the HL--LHC measurements. For instance, the independent determination
of the Higgs coupling to gluons via the decay $\htogg$ is advantageous 
in order to eliminate the dependence of this quantity on the remaining PDF uncertainties of
the LHC gluon fusion process. In addition, the measurement of the rate $\sigma(\epem\to ZH) \times {\rm BR}(H \to b \bar b)$ with 
1.2\% accuracy, see \refta{Tab:ILCobs}, together with the absolute cross section measurement of the $ZH$ production process with a precision of $2.6\%$, give important constraints on the $\htobb$ decay mode, which dominantly contributes to the total width of a SM-like Higgs boson. However, the corresponding scale factors $\kZ$ and $\kd$ are still strongly correlated. Another independent measurement of the $\htobb$ mode with similar precision --- as it is provided e.g.~at the ILC stage 2 with $\sqrt{s}=500\gev$ in $WW$ fusion (see below) --- is required to abrogate this correlation, thus allowing for a precise determination of $\kappa_d$.

The most striking improvement that the ILC already provides at the first stage with $\sqrt{s}=250\gev$,
however, is the \textit{model-independent} measurement of the $ZH$
production process and correspondingly \textit{model-independent}
determinations of Higgs branching ratios. Combining this input from the
ILC with the measurements performed at the HL--LHC leads to a significant
improvement of the latter, as will be discussed below (see 
\reffi{Fig:ILCprospects_8dim_modelindependent}).

While $\kZ$  can be probed already quite accurately at the early ILC stage
at $250\gev$ due to the dominant Higgs-strahlung process, the $\kW$
determination is less precise, $\delta \kW \sim 4.0\%$. This picture
changes at the later stages of the ILC with higher center-of-mass energies,
denoted as ILC500 and ILC1000, where the `baseline' integrated
luminosities of  $500~\ifb$ and $1~\iab$, respectively, have been
assumed. At ILC500 and ILC1000 the $WW$ fusion becomes the
dominant production mode.
Here, all scale factors in this parametrization except $\kga$ can be
determined to a precision of better than $2.5\%$ using only ILC
measurements. With the ultimate ILC integrated luminosity,
denoted as ILC1000 (LumiUp), even the $\kga$ coupling can be probed with an accuracy of $\lesssim
2.5\%$, and the remaining couplings are determined at the $\lesssim 1\%$
level, again using ILC measurements only. In the case where $\kV \le 1$ is
imposed instead of assuming non-standard Higgs decays to result in
invisible final states, the sensitivity for setting an upper limit on
$\brhnp$ inferred from the fit improves significantly at the ILC from
$4.3~(8.5)\%$ to $1.6~(3.3)\%$ at the $68~(95)\%~\mathrm{C.L.}$.

\begin{figure}
\centering
\subfigure[ILC only.
]{\includegraphics[width=0.49\textwidth]{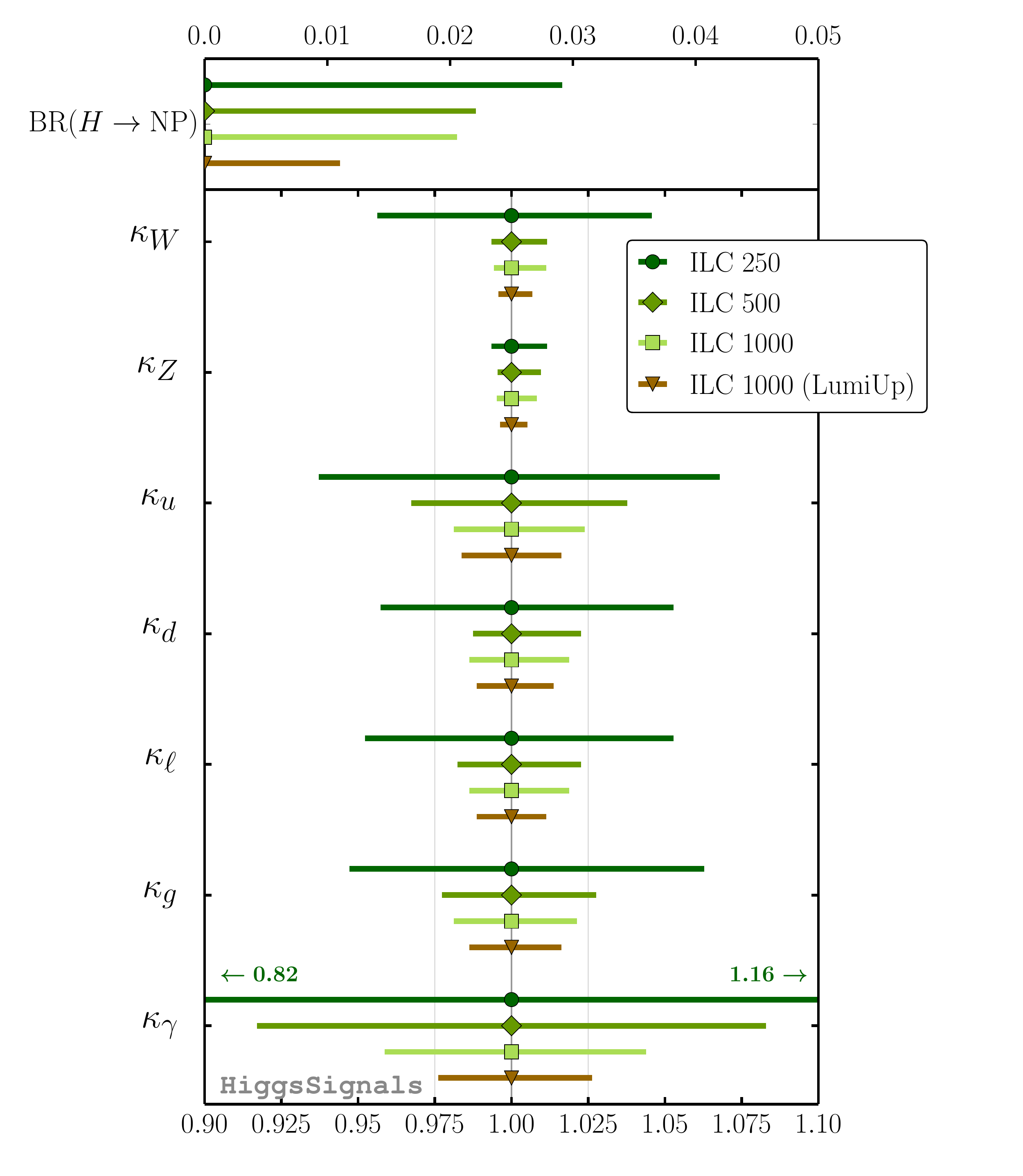}}
\subfigure[HL--LHC and combination of HL--LHC and ILC.]{\includegraphics[width=0.49\textwidth]{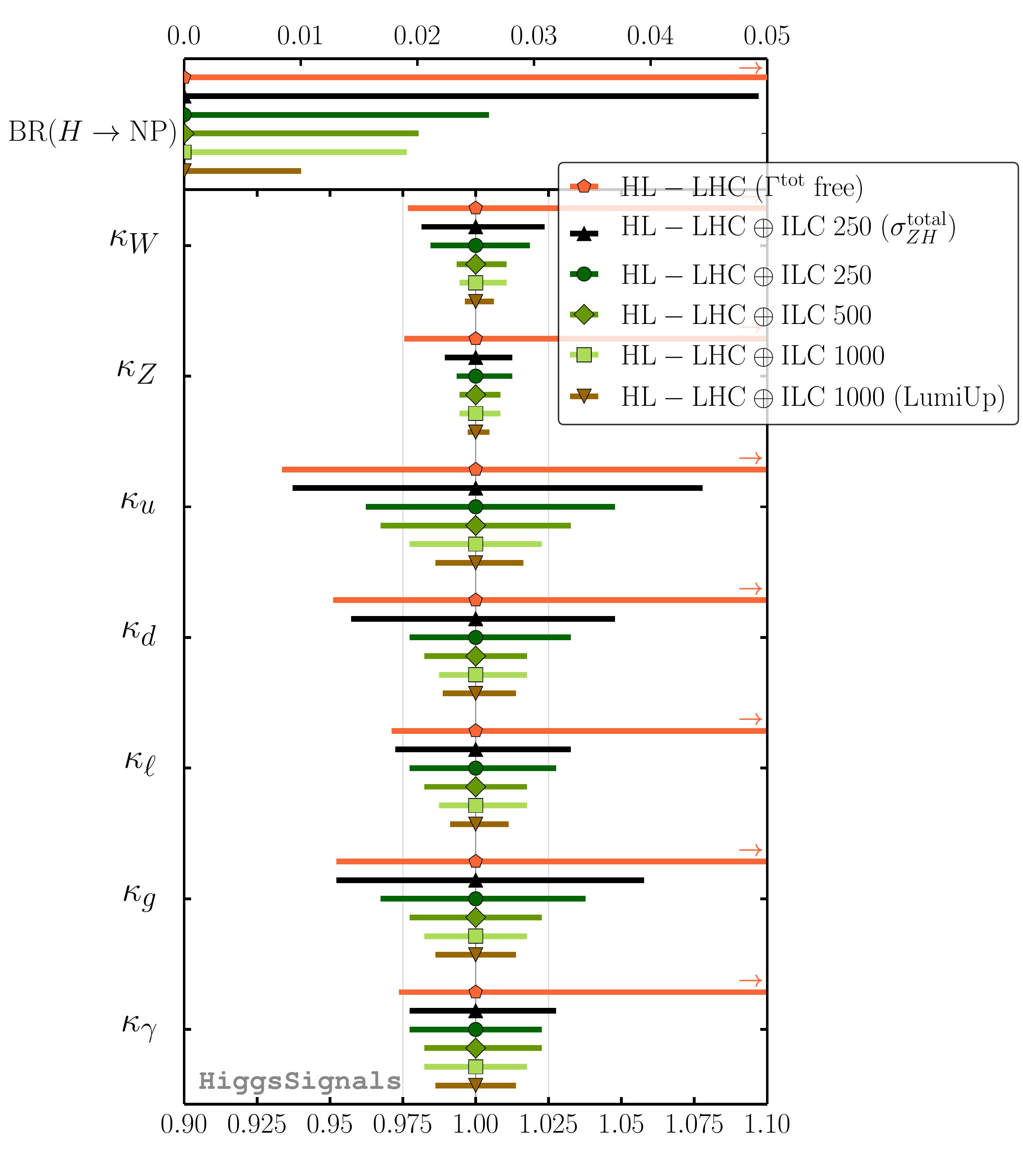}}
\caption{Future prospects of \textit{model-independent} Higgs coupling scale factor
determinations at the ILC alone (\textit{a}) and in combination with the
HL--LHC (\textit{b}). For comparison, we also show the results obtained at the
HL--LHC if the total width is not constrained by any assumptions on 
additional non-standard Higgs decay modes or limited scale factor
ranges (like~$\kV\le 1$).}
\label{Fig:ILCprospects_8dim_modelindependent}
\end{figure}

As stated earlier, the assumptions made in the previous fits are
actually unnecessary at the ILC once the total cross section measurement of the
$\epemtoZH$ process is taken into account. Therefore,
\textit{model-independent} estimates of the Higgs coupling accuracies can
be obtained, which are shown in
\reffi{Fig:ILCprospects_8dim_modelindependent}(a) and (b) for the ILC only
and HL--LHC $\oplus$ ILC combined measurements, respectively. The values are also listed in Tab.~\ref{Tab:LHCILC8dim}.
The estimated accuracies obtained for the ILC-only measurements in
this model-independent approach are only slightly weaker than the
ones obtained above under additional model assumptions,
cf.~\reffi{Fig:LHC_vs_ILCprospects_8dim_modeldependent}. 
At the early ILC stage (ILC250) the sensitivity for setting a 
\textit{model-independent} $95\%~\mathrm{C.L.}$ upper limit on $\brhnp$ of
$\lesssim 5.8\%$ is obtained from the fit. This sensitivity
improves to $\lesssim 4.1-4.4\%$ at the later baseline ILC stages. The
more precise measurement of the $\epem \to ZH$ cross section  at $250\gev$ with 
the ILC luminosity upgrade improves the sensitivity
further, such that $\brhnp \lesssim 2.2\%$ at
$95\%~\mathrm{C.L.}$ can be reached 
at the ultimate ILC stage at $\sqrt{s}=1\tev$.

For the combination of HL--LHC and ILC measurements for a
model-independent Higgs coupling determination, as shown in
\reffi{Fig:ILCprospects_8dim_modelindependent}(b), it is illustrative to
consider first the results obtained using the HL--LHC only or with a
minimal amount of ILC input, i.e.\ by only adding the total cross section
measurement of the $\epem \to ZH$ process. In the first case, as already
demonstrated in Sect.~\ref{Sect:universal}, the unconstrained fit (HL--LHC
($\Gamma^\text{tot}$ free) in
\reffi{Fig:ILCprospects_8dim_modelindependent}(b)) features a degeneracy
of increasing $\brhnp$ and increasing scale factors $\kappa_i$, until the
LHC is finally capable to observe broad width effects via off-shell Higgs
production. As a result, there is virtually no sensitivity in determining an upper limit for very large values of the scale factors.\footnote{The fact that the error bars for the scenario HL--LHC ($\Gamma^\text{tot}$ free) extend to values far outside of the right-hand side
of \reffi{Fig:ILCprospects_8dim_modelindependent}(b) is indicated by little arrows in the plot.}

\begin{table}
\renewcommand{\arraystretch}{1.3}
\centering
\small
\scalebox{0.89}{
\begin{tabu}{c | cccc | x{1.6cm}cccc}
\toprule
		     	& \multicolumn{9}{c}{$68\%~\mathrm{C.L.}$ Higgs
coupling scale factor precision [in $\%$]} \\
\midrule			
			& \multicolumn{4}{c |}{ILC only} &
\multicolumn{5}{c}{HL--LHC $\oplus$ ILC} \\
\rowfont{\footnotesize}
Scenario	 	& 250 & 500 & 1000 & 1000 (LumiUp) & 250 ($\sigma_{ZH}^\mathrm{total}$) & 250 & 500 & 1000 & 1000 (LumiUp)\\
\midrule
\rowfont{\footnotesize}
$ \mathrm{BR}(H \to \mathrm{NP}) $ & $ \le 2.9 $ &	$\le 2.2$& $ \le 2.1 $& $ \le 1.1 $& $ \le 4.9 $& $ \le 2.6 $ & $ \le 2.0 $ & $ \le 1.9 $ & $ \le 1.0 $ \\ 
\midrule
$ \kappa_W $ & $\substack{+ 4.6 \\ - 4.4 }$ &  $\substack{+ 1.2 \\ - 0.7 }$ & $\substack{+ 1.2 \\ - 0.6 }$&$\substack{+ 0.7 \\ - 0.4 }$ & $\substack{+ 2.4 \\ - 1.9 }$& $\substack{+ 1.9 \\ - 1.6 }$ & $\substack{+ 1.1 \\ - 0.7 }$  & $\substack{+ 1.1 \\ - 0.6 }$ & $\substack{+ 0.6 \\ - 0.4 }$ \\ 
$ \kappa_Z $ & $\substack{+ 1.3 \\ - 0.7 }$  &$\substack{+ 1.0 \\ - 0.6 }$ & $\substack{+ 0.9 \\ - 0.6 }$ &$\substack{+ 0.5 \\ - 0.4 }$ & $\substack{+ 1.3 \\ - 1.1 }$ & $\substack{+ 1.3 \\ - 0.7 }$ & $\substack{+ 0.9 \\ - 0.6 }$ & $\substack{+ 0.9 \\ - 0.6 }$ & $\substack{+ 0.5 \\ - 0.3 }$ \\ 
$ \kappa_u $ & $\substack{+ 6.8 \\ - 6.3 }$ & $\substack{+ 3.8 \\ - 3.3 }$ & $\substack{+ 2.3 \\ - 2.3 }$ &$\substack{+ 1.6 \\ - 1.6 }$ & $\substack{+ 7.8 \\ - 6.3 }$ & $\substack{+ 4.8 \\ - 3.8 }$ & $\substack{+ 3.3 \\ - 3.3 }$ & $\substack{+ 2.3 \\ - 2.3 }$ & $\substack{+ 1.6 \\ - 1.4 }$ \\ 
$ \kappa_d $ & $\substack{+ 5.3 \\ - 4.3 }$ & $\substack{+ 2.3 \\ - 1.8 }$ & $\substack{+ 1.8 \\ - 1.3 }$ &$\substack{+ 1.4 \\ - 1.1 }$  &$\substack{+ 4.8 \\ - 4.3 }$ & $\substack{+ 3.3 \\ - 2.3 }$ & $\substack{+ 1.8 \\ - 1.8 }$ & $\substack{+ 1.8 \\ - 1.3 }$ & $\substack{+ 1.4 \\ - 1.1 }$ \\ 
$ \kappa_\ell $ & $\substack{+ 5.3 \\ - 4.8 }$ & $\substack{+ 2.3 \\ - 1.8 }$ & $\substack{+ 1.8 \\ - 1.3 }$ &$\substack{+ 1.9 \\ - 1.6 }$  &$\substack{+ 3.3 \\ - 2.8 }$ & $\substack{+ 2.8 \\ - 2.3 }$ & $\substack{+ 1.8 \\ - 1.8 }$ & $\substack{+ 1.8 \\ - 1.3 }$ & $\substack{+ 1.1 \\ - 0.9 }$ \\ 
$ \kappa_g $ & $\substack{+ 6.3 \\ - 5.3 }$ & $\substack{+ 2.8 \\ - 2.3 }$ & $\substack{+ 2.3 \\ - 1.8 }$&$\substack{+ 1.9 \\ - 1.6 }$ & $\substack{+ 5.8 \\ - 4.8 }$& $\substack{+ 3.8 \\ - 3.3 }$ & $\substack{+ 2.3 \\ - 2.3 }$ & $\substack{+ 1.8 \\ - 1.8 }$ & $\substack{+ 1.4 \\ - 1.4 }$ \\ 
$ \kappa_\gamma $ & $\substack{+ 15.8 \\ - 17.8 }$ & $\substack{+ 8.3 \\ - 8.3 }$ & $\substack{+ 3.8 \\ - 3.8 }$& $\substack{+ 2.6 \\ - 2.6 }$& $\substack{+ 2.8 \\ - 2.3 }$& $\substack{+ 2.3 \\ - 2.3 }$ & $\substack{+ 2.3 \\ - 1.8 }$ & $\substack{+ 1.8 \\ - 1.8 }$ & $\substack{+ 1.4 \\ - 1.4 }$ \\ 
\bottomrule
\end{tabu}
}
\caption{$68\%~\mathrm{C.L.}$ precision estimates and upper limits for the \textit{model-independent} determination of Higgs coupling scale factors  and $\brhnp$, respectively, using only ILC measurements or in combination with HL--LHC measurements. The ultimate ILC scenario at $\sqrt{s}=1\tev$ after a full luminosity upgrade (LumiUp) is denoted as ILC 1000 (LU) here. These values correspond to those depicted in Fig.~\ref{Fig:ILCprospects_8dim_modelindependent}.}
\label{Tab:LHCILC8dim}
\end{table}

By adding only the total $\epemtoZH$ cross section measurement from the baseline
ILC250 run to the HL--LHC observables the degeneracy is broken. This
leads to a very significant improvement in the determination
of all Higgs coupling scale factors. Besides this effect one can see that
the combination with this single input value from the ILC leads to further
significant improvements affecting also the lower limits on the scale
factors. In particular, the precision on the lower limit of $\kZ$ improves
from $\sim2.5\%$ to $\sim1.1\%$.
Moreover, the
$95\%~\mathrm{C.L.}$ upper limit on $\brhnp$ inferred from this fit without any additional assumptions is $9.8\%$. This is roughly comparable to what has been obtained under the additional model assumptions in the
LHC-only fit, cf.~\reffi{Fig:LHC_vs_ILCprospects_8dim_modeldependent}(b).
With the inclusion of the remaining ILC measurements from the baseline $250\gev$ run all
scale factors except $\ku$ and $\kg$ can be measured at the $\sim2.5\%$
level. $\ku$ and $\kg$ can be determined with a precision of $\sim 4.3\%$
and $\sim 3.3$, respectively. The only scale factor that is dominantly
constrained by the LHC data is that for the Higgs-photon coupling, $\kga$,
which remains the case even at the later ILC stages at $500\gev$ and
$1\tev$. With the ultimate ILC luminosity, including the upgrade, and combining all available
measurements from the HL--LHC and ILC, all Higgs coupling scale factors
are probed to at least a precision of $1.5\%$. The Higgs-weak gauge boson
couplings can even be probed at the per-mille level. At this level the estimated
accuracies are dominated by the assumed (reduced) theory
uncertainties. We find that our estimates for the later ILC stages have a slight tendency to be
more conservative than those of e.g.~\citeres{Dawson:2013bba,Peskin:2013xra},
since we include larger theoretical uncertainties for the ILC production
cross sections and their correlations.

\section{Conclusions}
\label{Sect:conclusions}

In this paper we have investigated in detail whether the coupling
properties of the discovered new particle show any significant deviations
from the predictions for a SM Higgs boson at the present level of accuracy.
We have further analyzed the room for potential coupling deviations, which remain consistent with the current measurements, and the associated parameter correlations.
The study has been carried out within a consistent statistical framework
using all available Higgs signal rate measurements from the LHC and Tevatron experiments by employing profile likelihood fits of Higgs coupling scale factors by means of the public program \HS. 
The fits have been done both for highly constrained and very generic
scale factor parametrizations of the Higgs couplings. 
All benchmark fits allow for additional Higgs decays to non-standard final states and various assumptions are discussed for constraining the total Higgs decay width at the LHC. In contrast to other investigations in the literature, we have paid
particular attention to the treatment of the general case where
no constraint on the total Higgs width --- or on the branching
fraction of Higgs decays to potentially undetectable final states of new physics --- is assumed.

We have employed the ``interim framework'' of Higgs coupling scale factors 
as a means to param\-etrize the relations between the physical collider observables (cross sections, branching ratios) and the possible deviations in the couplings of the new state from the predictions for a SM Higgs boson. While the scale factors probe different possible aspects of deviations from the SM predictions, their
inherent simplifications and restrictions make it non-trivial to directly map the results obtained in terms of Higgs coupling scale factors onto
realistic models of physics beyond the SM. The latter typically predict certain correlations that  differ from those assumed for the Higgs coupling scale factors. The investigation of particular models is therefore complementary to the analysis of Higgs coupling scale factors. The tool \HS, which has been used in the present analysis, has been specifically designed for this purpose,
and the statistical methods employed here can be directly taken over for fits of realistic new physics models.

The program \HS\ is a well-validated and accurate tool for the $\chi^2$
evaluation based on the signal rate predictions and the currently 80
included measurements from ATLAS, CMS, CDF and D\O. It takes into account
the correlations of luminosity, cross section and branching ratio
uncertainties among the signal rate measurements, as well as intrinsic
correlations among the cross section and branching ratio uncertainties
induced by common parametric uncertainties. For this study, we have further
included the correlations of the remaining major experimental systematics
for the ATLAS $\htotautau$ and CMS $\htogaga$ measurements. We validated the implementation against an official six-dimensional scale factor fit performed by CMS, yielding very good agreement. All these new developments, as documented here,
will be provided with \HS\ version \texttt{1.2.0}.

For all considered scale factor benchmark models we find very good agreement between the LHC and Tevatron
measurements and the signal rates predicted for the SM. For the SM itself, i.e.~all scale factors are set
to unity, we find a naive \pvalue\ of $\sim 35.0\%$, showing good agreement between data and theory. Thus, it is not surprising that the benchmark models
achieve similar \pvalue s, which we have found to be typically slightly lower than the SM \pvalue\ due to the smaller number of degrees of freedom at similar
minimal $\chi^2$. The lowest \pvalue\ of $\sim 27.8\%$ is obtained for the fit probing the Yukawa structure in Sect.~\ref{Sect:Vudl},
while the best \pvalue, excluding the SM \pvalue, is found with $\sim
33.9\%$ for the benchmark fit probing the loop-induced Higgs couplings to
gluons and photons, cf.~Sect.~\ref{Sect:gga}.

We find no indicative hint for deviations from the SM in any of the
fits. Indeed, all central values of the fitted Higgs coupling scale factors are
compatible with their SM values. The fitted values of an additional Higgs branching fraction, $\brhnp$,
are also compatible with zero. Uncertainties on the fitted scale factors range from around $10\%$
in the most constrained case, i.e.~a fit of only one
universal scaling parameter, up to $40\%$ for the top Yukawa scale factor, $\kappa_u$, 
in the seven-dimensional fit discussed in Sect.~\ref{Sect:7dim}. Comparing these results with the latest official scale factor determination performed by CMS for the Moriond 2013 conference, we find significant improvements in all scale factor precisions. This illustrates the power of a common interpretation of ATLAS and CMS (and Tevatron) measurements, as well as the importance of the recent measurements in the ATLAS $\htotautau$ and CMS $t\bar{t}H$-tagged searches.

The corresponding weakest observed limit from the fits on the invisible Higgs decay is $\brinv<17~[39]\%$ at the $68\%~[95\%]~\mathrm{C.L.}$, also taking into account direct searches for $\brinv$ at the LHC. We furthermore find for the total signal strength to known SM final states a lower limit of $\kappa^2\times (1-\brhnp)>81\%$ at the $95\%$~CL, employing the benchmark fit with one universal Higgs coupling scale factor $\kappa$.
This limit is independent of any further assumption, such as e.g.~$\kappa_{W,Z} \le 1$. Moreover, under the assumption that
$\kappa_{W,Z} \le 1$ holds, we find from the most general fit to the present data, which has seven free parameters, the limit $\brhnp<40\%$ at the $95\%$~C.L., where the final state(s) of such Higgs decay(s) may be undetectable to current LHC experiments.

Beyond the current measurements from the LHC and the Tevatron, we
have explored the capabilities of future Higgs coupling determinations
using projections of the signal rate measurements for the LHC with $300~\ifb$ (LHC~300) and $3000~\ifb$ (HL--LHC) at $14\tev$, as well as for various scenarios of an International Linear Collider (ILC). 
At the LHC~300 we find estimated precisions for the
determination of the Higgs coupling
scale factors within $\sim 5-18\%$ under the assumption
$\brhnp\equiv\brinv$. Possible improvements of theoretical uncertainties on the
cross sections and branching ratios turn out to have only a marginal effect on those estimated precisions. This changes at the HL--LHC, where the achievable precision of the Higgs-gluon coupling scale factor is significantly limited by the
theoretical uncertainty. The precision estimates of the remaining scale
factors, however, are hardly affected by varying assumptions on the
theoretical uncertainties. Overall, assuming $\brhnp\equiv\brinv$, we find
scale factor precisions of $\sim 3-10\%$ at the HL--LHC. If we make the
model assumption $\kV \le 1$ instead of the assumption that
additional non-standard Higgs decays result only in invisible final
states, then most of the estimated scale factor precisions marginally improve.

Concerning the prospects at the ILC, we have compared the ILC
capabilities of determining Higgs couplings with those of the HL--LHC
first for a model-dependent approach, i.e.~using the same assumptions as
for the HL--LHC analyses, namely assuming either $\brhnp\equiv\brinv$ or $\kV \le 1$ as a means to
constrain the total width. We find that already ILC measurements at
$250\gev$ for `baseline' assumptions on the integrated luminosity provide
significant improvements compared to the most optimistic scenario for the
HL--LHC along with complementary measurements that are of similar or
slightly worse accuracy compared with the projections for the HL--LHC. Starting from a center-of-mass
energy of $\sqrt{s}=500\gev$  for the corresponding `baseline' luminosity
assumption the ILC in fact has the potential to considerably improve upon
all measurements of the HL--LHC apart from the coupling of the Higgs to photons. At $\sqrt{s}=500\gev$, the $WW$ fusion
channel can be measured significantly better than at $250\gev$, which
leads to a significantly higher statistics for all considered
quantities and in particular to a further improvement in the determination 
of the total width. The further improvements from ILC running at $1 \tev$ and from exploiting
the ultimate ILC luminosity (LumiUp) turn out to be rather moderate for
the considered case of a model-dependent 8-parameter fit, which is related
to our fairly conservative estimates of the future theoretical
uncertainties.

The impact of the ILC on improving the determination of the Higgs
couplings becomes apparent most strikingly for the model-independent
analyses. Without employing additional theoretical assumptions the scale
factors at the LHC are essentially unconstrained from above. However,
taking into account a single measurement of the ILC --- the decay-mode independent
recoil analysis of the total Higgs production rate at $250\gev$ --- in
conjunction with the HL--LHC measurements already allows to perform a
significantly less model-dependent and more precise fit than with the HL--LHC alone. In particular, with this ILC measurement the
assumptions on the additional Higgs decay modes and on
 $\kV$ can be dropped.

From prospective measurements at the ILC up to $\sqrt{s}=1\tev$ 
with the `baseline' assumptions for the integrated luminosity together
with those from the HL--LHC, we find precision estimates for \textit{all} fitted Higgs coupling
scale factors of better than $2.5\%$. For some scale factors a precision better than $1\%$ is
  achieved. These estimates are obtained with the least amount of
model assumptions 
and 8 free fit parameters. With the ultimate ILC luminosity (LumiUp) this
precision would further increase significantly, reaching a level of better than $1.5\%$ for all scale factors. 

The Higgs coupling scale factor benchmark scenarios considered in this study typically have more
freedom to adjust the predicted signal rates to the measurements than realistic models.
Realistic model generally feature specific correlations among the predicted rates which depend non-trivially  on the model parameters. 
Moreover, limits from the electroweak precision data and other sectors, such as dark matter, collider searches, etc., may further restrict the allowed parameter space and thus the room for Higgs coupling deviations. The fact that the exploration of the Higgs couplings with those
rather general parametrizations does not improve the fit quality with
respect to the SM is a clear indication of the good agreement of the data
with the SM predictions. On the basis of this analysis one would not
expect a significant improvement in the description of the data from a
realistic model of physics beyond the SM. Thus, the full set of the
present public measurements from ATLAS, CMS, CDF and D\O\ in the Higgs
sector does not show any indications for physics beyond the SM.


Despite the lack of a concrete hint for any deviation from the SM in
the current measurements, there still is ample room for future
discoveries of deviations from the SM predictions for the Higgs couplings. 
In fact, the current uncertainties are still rather large and thus still allow for sizable deviations from the SM at the level of $\sim \mathcal{O}(10-40\%)$ at the $1\sigma$ level, even when making additional theory assumptions, namely $\brhnp\equiv\brinv$ or $\kV \le 1$.
Comparing those accuracies with the typical deviations
expected in realistic models of physics beyond the SM, a large improvement
in the experimental precision will be needed in order to sensitively probe
the parameter space of the most popular extensions of the SM. 
The measurements at an ILC-like machine, in conjunction with the HL--LHC,
will be crucial in this context for model-independent determinations of absolute Higgs couplings with
precisions at the percent level or better, offering great prospects for
identifying the underlying mechanism of electroweak symmetry breaking.

\acknowledgments{
We thank Andre David, Ansgar Denner, Klaus Desch, Manuel Drees,
Michael Duehrssen, Howie Haber, Alex Read, Bj\"orn Sarrazin, Daniel Schmeier and Tom Zirke for helpful discussions.
Part of the numerical calculations were performed using the computing infrastructure~\cite{Campos:2012vb} at the ``Instituto de F\'isica
de Cantabria''. This work was partially supported by the Helmholtz
Alliance ``Physics at the Terascale'',
the Collaborative Research Center SFB676 of the DFG, ``Particles, 
Strings and the early Universe'',
and by the European Commission through the
``HiggsTools'' Initial Training
Network PITN-GA-2012-316704.
The work of T.S.~was supported by the BMBF Grant No.\ 00160200 and the Bonn-Cologne Graduate School (BCGS).
S.H.~was supported by CICYT (Grant No.\ FPA 2010--22163-C02-01) and by 
the Spanish MICINN's Consolider-Ingenio 2010 Program under Grant MultiDark
No.\ CSD2009-00064. O.S.~is supported by the Swedish Research Council (VR)
through the Oskar Klein Centre.

Since one of the authors (P.B.) is also an ATLAS member, we would like to clarify that the work presented here is the responsibility of the individual
authors and does not represent an ATLAS result. This phenomenological analysis is purely based on public information.}

\appendix

\section{Experimental data}
\label{Sect:expdata}
\subsection{Implementation of current signal strength measurements}
\label{Sect:currentexpdata}

Tables~\ref{Tab:expdata1} and \ref{Tab:expdata2} list the signal strength measurements from ATLAS, CDF, CMS and D\O\ as implemented in \HSv{1.2.0}; there are 80 observables in total. The tables also provide numbers for the assumed signal composition of a SM Higgs boson for all measurements. Most of these results are used directly in the fits in Sect.~\ref{Sect:currentfits}, except for a few cases where a more careful treatment is required as described in detail below.

\begin{table}
\centering
\scalebox{0.86}{
\renewcommand{\arraystretch}{1.0}
\centering
\begin{threeparttable}[b]
\footnotesize
 \begin{tabular}{lccrrrrr}
\toprule
 Analysis &  energy $\sqrt{s}$ & $\muobs \pm \Delta\muobs$ & \multicolumn{5}{c}{SM signal composition [in \%]} \\
 & & & ggH & VBF & WH &  ZH & $t\bar{t}H$ \\
\midrule
ATL $(pp)\to h\to WW\to \ell\nu\ell\nu$~(0/1jet)~\cite{ATLAS:2013wla,Aad:2013wqa} & $7/8\tev$& $  0.82\substack{+  0.33\\ -  0.32}$ & $  97.2$ & $   1.6$ & $   0.7$ & $   0.4$ & $   0.1$\\ 
ATL $(pp)\to h\to WW\to \ell\nu\ell\nu$~(VBF)~\cite{ATLAS:2013wla,Aad:2013wqa} & $7/8\tev$& $  1.42\substack{+  0.70\\ -  0.56}$ & $  19.8$ & $  80.2$ & $   0.0$ & $   0.0$ & $   0.0$\\ 
ATL $(pp)\to h\to ZZ\to 4\ell$~(VBF/VH-like)~\cite{ATLAS:2013nma,Aad:2013wqa} & $7/8\tev$ & $  1.18\substack{+  1.64\\ -  0.90}$ & $  36.8$ & $  43.1$ & $  12.8$ & $   7.3$ & $   0.0$\\ 
ATL $(pp)\to h\to ZZ\to 4\ell$~(ggH-like)~\cite{ATLAS:2013nma,Aad:2013wqa} & $7/8\tev$& $  1.45\substack{+  0.43\\ -  0.37}$ & $  92.5$ & $   4.5$ & $   1.9$ & $   1.1$ & $   0.0$\\ 
ATL $(pp)\to h\to \gamma\gamma$~(unconv.-central-low $p_{Tt}$)~\cite{ATLAS:2012goa}  & $7\tev$& $  0.53\substack{+  1.41\\ -  1.48}$ & $  92.9$ & $   3.8$ & $   2.0$ & $   1.1$ & $   0.2$\\ 
ATL $(pp)\to h\to \gamma\gamma$~(unconv.-central-high $p_{Tt}$)~\cite{ATLAS:2012goa}  & $7\tev$& $  0.22\substack{+  1.94\\ -  1.95}$ & $  65.5$ & $  14.8$ & $  10.8$ & $   6.2$ & $   2.7$\\ 
ATL $(pp)\to h\to \gamma\gamma$~(unconv.-rest-low $p_{Tt}$)~\cite{ATLAS:2012goa} & $7\tev$ & $  2.52\substack{+  1.68\\ -  1.68}$ & $  92.6$ & $   3.7$ & $   2.2$ & $   1.2$ & $   0.2$\\ 
ATL $(pp)\to h\to \gamma\gamma$~(unconv.-rest-high $p_{Tt}$)~\cite{ATLAS:2012goa}  & $7\tev$& $ 10.44\substack{+  3.67\\ -  3.70}$ & $  64.4$ & $  15.2$ & $  11.8$ & $   6.6$ & $   2.0$\\ 
ATL $(pp)\to h\to \gamma\gamma$~(conv.-central-low $p_{Tt}$)~\cite{ATLAS:2012goa}  & $7\tev$& $  6.10\substack{+  2.63\\ -  2.62}$ & $  92.7$ & $   3.8$ & $   2.1$ & $   1.1$ & $   0.2$\\ 
ATL $(pp)\to h\to \gamma\gamma$~(conv.-central-high $p_{Tt}$)~\cite{ATLAS:2012goa}  & $7\tev$& $ -4.36\substack{+  1.80\\ -  1.81}$ & $  65.7$ & $  14.4$ & $  11.0$ & $   6.2$ & $   2.8$\\ 
ATL $(pp)\to h\to \gamma\gamma$~(conv.-rest-low $p_{Tt}$)~\cite{ATLAS:2012goa} & $7\tev$ & $  2.74\substack{+  1.98\\ -  2.01}$ & $  92.7$ & $   3.6$ & $   2.2$ & $   1.2$ & $   0.2$\\ 
ATL $(pp)\to h\to \gamma\gamma$~(conv.-rest-high $p_{Tt}$)~\cite{ATLAS:2012goa}  & $7\tev$& $ -1.59\substack{+  2.89\\ -  2.90}$ & $  64.4$ & $  15.1$ & $  12.1$ & $   6.4$ & $   2.0$\\ 
ATL $(pp)\to h\to \gamma\gamma$~(conv.-trans.)~\cite{ATLAS:2012goa}  & $7\tev$& $  0.37\substack{+  3.58\\ -  3.79}$ & $  89.2$ & $   5.0$ & $   3.7$ & $   1.9$ & $   0.3$\\ 
ATL $(pp)\to h\to \gamma\gamma$~(2 jet)~\cite{ATLAS:2012goa} & $7\tev$& $  2.72\substack{+  1.87\\ -  1.85}$ & $  23.3$ & $  75.9$ & $   0.5$ & $   0.2$ & $   0.1$\\ 
ATL $(pp)\to h\to \gamma\gamma$~(unconv.-central-low $p_{Tt}$)~\cite{ATLAS:2013oma} & $8\tev$& $  0.87\substack{+  0.73\\ -  0.70}$ & $  92.0$ & $   5.0$ & $   1.7$ & $   0.8$ & $   0.5$\\ 
ATL $(pp)\to h\to \gamma\gamma$~(unconv.-central-high $p_{Tt}$)~\cite{ATLAS:2013oma}& $8\tev$ & $  0.96\substack{+  1.07\\ -  0.95}$ & $  78.6$ & $  12.6$ & $   4.7$ & $   2.6$ & $   1.4$\\ 
ATL $(pp)\to h\to \gamma\gamma$~(unconv.-rest-low $p_{Tt}$)~\cite{ATLAS:2013oma} & $8\tev$& $  2.50\substack{+  0.92\\ -  0.77}$ & $  92.0$ & $   5.0$ & $   1.7$ & $   0.8$ & $   0.5$\\ 
ATL $(pp)\to h\to \gamma\gamma$~(unconv.-rest-high $p_{Tt}$)~\cite{ATLAS:2013oma} & $8\tev$& $  2.69\substack{+  1.35\\ -  1.17}$ & $  78.6$ & $  12.6$ & $   4.7$ & $   2.6$ & $   1.4$\\ 
ATL $(pp)\to h\to \gamma\gamma$~(conv.-central-low $p_{Tt}$)~\cite{ATLAS:2013oma} & $8\tev$& $  1.39\substack{+  1.01\\ -  0.95}$ & $  92.0$ & $   5.0$ & $   1.7$ & $   0.8$ & $   0.5$\\ 
ATL $(pp)\to h\to \gamma\gamma$~(conv.-central-high $p_{Tt}$)~\cite{ATLAS:2013oma} & $8\tev$& $  1.98\substack{+  1.54\\ -  1.26}$ & $  78.6$ & $  12.6$ & $   4.7$ & $   2.6$ & $   1.4$\\ 
ATL $(pp)\to h\to \gamma\gamma$~(conv.-rest-low $p_{Tt}$)~\cite{ATLAS:2013oma}& $8\tev$ & $  2.23\substack{+  1.14\\ -  1.01}$ & $  92.0$ & $   5.0$ & $   1.7$ & $   0.8$ & $   0.5$\\ 
ATL $(pp)\to h\to \gamma\gamma$~(conv.-rest-high $p_{Tt}$)~\cite{ATLAS:2013oma} & $8\tev$& $  1.27\substack{+  1.32\\ -  1.23}$ & $  78.6$ & $  12.6$ & $   4.7$ & $   2.6$ & $   1.4$\\ 
ATL $(pp)\to h\to \gamma\gamma$~(conv.-trans.)~\cite{ATLAS:2013oma}& $8\tev$ & $  2.78\substack{+  1.72\\ -  1.57}$ & $  92.0$ & $   5.0$ & $   1.7$ & $   0.8$ & $   0.5$\\ 
ATL $(pp)\to h\to \gamma\gamma$~(high mass, 2 jet, loose)~\cite{ATLAS:2013oma}  & $8\tev$ & $  2.75\substack{+  1.78\\ -  1.38}$ & $  45.3$ & $  53.7$ & $   0.5$ & $   0.3$ & $   0.2$\\ 
ATL $(pp)\to h\to \gamma\gamma$~(high mass, 2 jet, tight)~\cite{ATLAS:2013oma} & $8\tev$& $  1.61\substack{+  0.83\\ -  0.67}$ & $  27.1$ & $  72.5$ & $   0.3$ & $   0.1$ & $   0.0$\\ 
ATL $(pp)\to h\to \gamma\gamma$~(low mass, 2 jet)~\cite{ATLAS:2013oma} & $8\tev$& $  0.32\substack{+  1.72\\ -  1.44}$ & $  38.0$ & $   2.9$ & $  40.1$ & $  16.9$ & $   2.1$\\ 
ATL $(pp)\to h\to \gamma\gamma~(E_T^\mathrm{miss}~\mathrm{sign.})$~\cite{ATLAS:2013oma}& $8\tev$ & $  2.97\substack{+  2.71\\ -  2.15}$ & $   4.4$ & $   0.3$ & $  35.8$ & $  47.4$ & $  12.2$\\ 
ATL $(pp)\to h\to \gamma\gamma~(1\ell)$~\cite{ATLAS:2013oma} & $8\tev$& $  2.69\substack{+  1.97\\ -  1.66}$ & $   2.5$ & $   0.4$ & $  63.3$ & $  15.2$ & $  18.7$\\ 
ATL $(pp)\to h\to \tau\tau$~(VBF, had-had)~\cite{ATLAS-CONF-2013-108} & $8\tev$& $  1.03\substack{+  0.92\\ -  0.73}$ & $  25.1$ & $  74.9$ & $   0.0$ & $   0.0$ & $   0.0$\\ 
ATL $(pp)\to h\to \tau\tau$~(boosted, had-had)~\cite{ATLAS-CONF-2013-108}& $8\tev$ & $  0.77\substack{+  1.17\\ -  0.98}$ & $  65.1$ & $  16.1$ & $  12.5$ & $   6.3$ & $   0.0$\\ 
ATL $(pp)\to h\to \tau\tau$~(VBF, lep-had)~\cite{ATLAS-CONF-2013-108} & $8\tev$& $  1.61\substack{+  0.77\\ -  0.60}$ & $  13.9$ & $  86.1$ & $   0.0$ & $   0.0$ & $   0.0$\\ 
ATL $(pp)\to h\to \tau\tau$~(boosted, lep-had)~\cite{ATLAS-CONF-2013-108} & $8\tev$& $  1.21\substack{+  1.07\\ -  0.83}$ & $  68.8$ & $  16.1$ & $  10.1$ & $   5.0$ & $   0.0$\\ 
ATL $(pp)\to h\to \tau\tau$(VBF, lep-lep)~\cite{ATLAS-CONF-2013-108} & $8\tev$& $  2.19\substack{+  1.23\\ -  1.10}$ & $  12.4$ & $  87.6$ & $   0.0$ & $   0.0$ & $   0.0$\\ 
ATL $(pp)\to h\to \tau\tau$~(boosted, lep-lep)~\cite{ATLAS-CONF-2013-108} & $8\tev$& $  2.03\substack{+  1.80\\ -  1.45}$ & $  66.0$ & $  25.6$ & $   6.2$ & $   2.2$ & $   0.0$\\ 
%
ATL $(pp)\to Vh\to V(bb)~(0\ell)$~\cite{TheATLAScollaboration:2013lia}  & $7/8\tev$& $  0.46\substack{+  0.88\\ -  0.86}$ & $   0.0$ & $   0.0$ & $  21.2$ & $  78.8$ & $   0.0$\\ 
ATL $(pp)\to Vh\to V(bb)~(1\ell)$~\cite{TheATLAScollaboration:2013lia}  & $7/8\tev$& $  0.09\substack{+  1.01\\ -  1.00}$ & $   0.0$ & $   0.0$ & $  96.7$ & $   3.3$ & $   0.0$\\ 
ATL $(pp)\to Vh\to V(bb)~(2\ell)$~\cite{TheATLAScollaboration:2013lia}  & $7/8\tev$& $ -0.36\substack{+  1.48\\ -  1.38}$ & $   0.0$ & $   0.0$ & $   0.0$ & $ 100.0$ & $   0.0$\\ 
ATL $(pp)\to Vh\to V(WW)$~\cite{TheATLAScollaboration:2013hia}  & $7/8\tev$& $  3.70\substack{+  1.90\\ -  2.00}$ & $   0.0$ & $   0.0$ & $  63.8$ & $  36.2$ & $   0.0$\\ 
\midrule
CDF $(p\bar{p})\to h\to WW$~\cite{Aaltonen:2013ipa}  & $1.96\tev$& $  0.00\substack{+  1.78\\ -  1.78}$ & $  77.5$ & $   5.4$ & $  10.6$ & $   6.5$ & $   0.0$\\ 
CDF $(p\bar{p})\to h\to \gamma\gamma$~\cite{Aaltonen:2013ipa} & $1.96\tev$& $  7.81\substack{+  4.61\\ -  4.42}$ & $  77.5$ & $   5.4$ & $  10.6$ & $   6.5$ & $   0.0$\\ 
CDF $(p\bar{p})\to h\to \tau\tau$~\cite{Aaltonen:2013ipa} & $1.96\tev$& $  0.00\substack{+  8.44\\ -  8.44}$ & $  77.5$ & $   5.4$ & $  10.6$ & $   6.5$ & $   0.0$\\ 
CDF $(p\bar{p})\to Vh\to Vbb$~\cite{Aaltonen:2013ipa} & $1.96\tev$& $  1.72\substack{+  0.92\\ -  0.87}$ & $   0.0$ & $   0.0$ & $  61.9$ & $  38.1$ & $   0.0$\\ 
CDF $(p\bar{p})\to tth\to ttbb$~\cite{Aaltonen:2013ipa}& $1.96\tev$ & $  9.49\substack{+  6.60\\ -  6.28}$ & $   0.0$ & $   0.0$ & $   0.0$ & $   0.0$ & $ 100.0$\\ 
\bottomrule
 \end{tabular}
 \end{threeparttable}
 }
  \caption{Signal strength measurements from ATLAS and CDF.}
  \label{Tab:expdata1}
\end{table}

\begin{table}
\centering
\scalebox{0.86}{
\renewcommand{\arraystretch}{1.0}
\begin{threeparttable}[b]
\footnotesize
 \begin{tabular}{lccrrrrr}
\toprule
 Analysis &  energy $\sqrt{s}$ & $\muobs \pm \Delta\muobs$ & \multicolumn{5}{c}{SM signal composition [in \%]} \\
 & & & ggH & VBF & WH &  ZH & $t\bar{t}H$ \\
\midrule
CMS $(pp)\to h\to WW\to 2\ell 2\nu$~(0/1 jet)~\cite{Chatrchyan:2013iaa}  & $7/8\tev$& $  0.74\substack{+  0.22\\ -  0.20}$ & $  83.0$ & $  11.1$ & $   3.8$ & $   2.2$ & $   0.0$\\ 
CMS $(pp)\to h\to WW\to 2\ell 2\nu$~(VBF)~\cite{Chatrchyan:2013iaa}  & $7/8\tev$& $  0.60\substack{+  0.57\\ -  0.46}$ & $  19.8$ & $  80.2$ & $   0.0$ & $   0.0$ & $   0.0$\\ 
CMS $(pp)\to h\to WW\to 2\ell 2\nu$~(VH)~\cite{Chatrchyan:2013iaa}  & $7/8\tev$& $  0.39\substack{+  1.97\\ -  1.87}$ & $  56.2$ & $   4.5$ & $  25.1$ & $  14.2$ & $   0.0$\\ 
CMS $(pp)\to h\to WW\to 3\ell 3\nu$~(WH)~\cite{Chatrchyan:2013iaa}  & $7/8\tev$& $  0.56\substack{+  1.27\\ -  0.95}$ & $   0.0$ & $   0.0$ & $ 100.0$\tnote{1} & $   0.0$ & $   0.0$\\ 
CMS $(pp)\to Vh \to V(WW)$ (hadronic $V$)~\cite{CMS:2013xda} &$7/8\tev$& $  1.00\substack{+  2.00\\ -  2.00}$ & $  59.8$ & $   4.0$ & $  24.2$ & $  12.0$ & $   0.0$\\ 
CMS $(pp)\to h\to ZZ\to 4\ell$~(0/1 jet)~\cite{CMS:xwa} & $7/8\tev$& $  0.86\substack{+  0.32\\ -  0.26}$ & $  89.8$ & $  10.2$ & $   0.0$ & $   0.0$ & $   0.0$\\ 
CMS $(pp)\to h\to ZZ\to 4\ell$~(2 jet)~\cite{CMS:xwa}& $7/8\tev$ & $  1.24\substack{+  0.85\\ -  0.58}$ & $  71.2$ & $  28.8$ & $   0.0$ & $   0.0$ & $   0.0$\\ 
CMS $(pp)\to h\to \gamma\gamma$~(untagged 0)~\cite{CMS:2012paa,CMS:ril} & $7\tev$& $  3.88\substack{+  2.00\\ -  1.68}$ & $  61.4$ & $  16.9$ & $  12.0$ & $   6.6$ & $   3.1$\\ 
CMS $(pp)\to h\to \gamma\gamma$~(untagged 1)~\cite{CMS:2012paa,CMS:ril} & $7\tev$& $  0.20\substack{+  1.01\\ -  0.93}$ & $  87.7$ & $   6.2$ & $   3.6$ & $   2.0$ & $   0.5$\\ 
CMS $(pp)\to h\to \gamma\gamma$~(untagged 2)~\cite{CMS:2012paa,CMS:ril} & $7\tev$& $  0.04\substack{+  1.25\\ -  1.24}$ & $  91.4$ & $   4.4$ & $   2.5$ & $   1.4$ & $   0.3$\\ 
CMS $(pp)\to h\to \gamma\gamma$~(untagged 3)~\cite{CMS:2012paa,CMS:ril} & $7\tev$& $  1.47\substack{+  1.68\\ -  2.47}$ & $  91.3$ & $   4.4$ & $   2.6$ & $   1.5$ & $   0.2$\\ 
CMS $(pp)\to h\to \gamma\gamma$~(2 jet)~\cite{CMS:2012paa,CMS:ril} & $7\tev$& $  4.18\substack{+  2.31\\ -  1.78}$ & $  26.7$ & $  72.6$ & $   0.4$ & $   0.2$ & $   0.0$\\ 
CMS $(pp)\to h\to \gamma\gamma$~(untagged 0)~\cite{CMS:ril} & $8\tev$& $  2.20\substack{+  0.95\\ -  0.78}$ & $  72.9$ & $  11.7$ & $   8.2$ & $   4.6$ & $   2.6$\\ 
CMS $(pp)\to h\to \gamma\gamma$~(untagged 1)~\cite{CMS:ril} & $8\tev$& $  0.06\substack{+  0.69\\ -  0.67}$ & $  83.5$ & $   8.5$ & $   4.5$ & $   2.6$ & $   1.0$\\ 
CMS $(pp)\to h\to \gamma\gamma$~(untagged 2)~\cite{CMS:ril} & $8\tev$& $  0.31\substack{+  0.50\\ -  0.47}$ & $  91.5$ & $   4.5$ & $   2.3$ & $   1.3$ & $   0.4$\\ 
CMS $(pp)\to h\to \gamma\gamma$~(untagged 3)~\cite{CMS:ril} & $8\tev$ & $ -0.36\substack{+  0.88\\ -  0.81}$ & $  92.5$ & $   3.9$ & $   2.1$ & $   1.2$ & $   0.3$\\ 
CMS $(pp)\to h\to \gamma\gamma$~(2 jet, tight)~\cite{CMS:ril} & $8\tev$& $  0.27\substack{+  0.69\\ -  0.58}$ & $  20.6$ & $  79.0$ & $   0.2$ & $   0.1$ & $   0.1$\\ 
CMS $(pp)\to h\to \gamma\gamma$~(2 jet, loose)~\cite{CMS:ril} & $8\tev$& $  0.78\substack{+  1.10\\ -  0.98}$ & $  46.8$ & $  51.1$ & $   1.1$ & $   0.6$ & $   0.5$\\ 
CMS $(pp)\to h\to \gamma\gamma~(\mu)$~\cite{CMS:ril} & $8\tev$& $  0.38\substack{+  1.84\\ -  1.36}$ & $   0.0$ & $   0.2$ & $  50.4$ & $  28.6$ & $  20.8$\\ 
CMS $(pp)\to h\to \gamma\gamma~(e)$~\cite{CMS:ril} & $8\tev$& $ -0.67\substack{+  2.78\\ -  1.95}$ & $   1.1$ & $   0.4$ & $  50.2$ & $  28.5$ & $  19.8$\\ 
CMS $(pp)\to h\to \gamma\gamma~(E_T^\mathrm{miss})$~\cite{CMS:ril} & $8\tev$& $  1.89\substack{+  2.62\\ -  2.28}$ & $  22.1$ & $   2.6$ & $  40.6$ & $  23.0$ & $  11.7$\\ 
CMS $(pp)\to h\to \mu\mu$~\cite{CMS:2013aga}& $7/8\tev$ & $  2.90\substack{+  2.80\\ -  2.70}$ & $  92.5$ & $   7.5$ & $   0.0$ & $   0.0$ & $   0.0$\\ 
CMS $(pp)\to h\to \tau\tau$~(0 jet)~\cite{CMS:utj,Chatrchyan:2014nva}  & $7/8\tev$& $  0.40\substack{+  0.73\\ -  1.13}$ & $  98.2$ & $   1.0$ & $   0.5$ & $   0.3$ & $   0.0$\\ 
CMS $(pp)\to h\to \tau\tau$~(1 jet)~\cite{CMS:utj,Chatrchyan:2014nva}  & $7/8\tev$& $  1.06\substack{+  0.47\\ -  0.47}$ & $  76.0$ & $  14.9$ & $   5.8$ & $   3.3$ & $   0.0$\\ 
CMS $(pp)\to h\to \tau\tau$~(VBF)~\cite{CMS:utj,Chatrchyan:2014nva}  & $7/8\tev$& $  0.93\substack{+  0.41\\ -  0.41}$ & $  17.1$ & $  82.9$ & $   0.0$ & $   0.0$ & $   0.0$\\ 
CMS $(pp)\to Vh\to V(\tau\tau)$~\cite{CMS:utj,Chatrchyan:2014nva}  & $7/8\tev$& $  0.98\substack{+  1.68\\ -  1.50}$ & $   0.0$ & $   0.0$ & $  48.6$\tnote{2} & $   26.4$\tnote{2} & $   0.0$\\ 
CMS $(pp)\to Vh\to V(bb)$~\cite{CMS:2013dda} & $7/8\tev$& $  1.00\substack{+  0.51\\ -  0.49}$ & $   0.0$ & $   0.0$ & $  63.8$ & $  36.2$ & $   0.0$\\ 
CMS $(pp)\to tth\to 2\ell$~(same-sign)~\cite{CMS:2013tfa} &$8\tev$& $  5.30\substack{+  2.20\\ -  1.80}$ & $   0.0$ & $   0.0$ & $   0.0$ & $   0.0$ & $  100.0$\tnote{3}\\ 
CMS $(pp)\to tth\to 3\ell$~\cite{CMS:2013tfa} & $8\tev$&$  2.70\substack{+  2.20\\ -  1.80}$ & $   0.0$ & $   0.0$ & $   0.0$ & $   0.0$ & $  100.0$\tnote{4}\\ 
CMS $(pp)\to tth\to 4\ell$~\cite{CMS:2013tfa} &$8\tev$& $ -4.80\substack{+  5.00\\ -  1.20}$ & $   0.0$ & $   0.0$ & $   0.0$ & $   0.0$ & $  100.0$\tnote{5}\\ 
CMS $(pp)\to tth\to tt(bb)$~\cite{CMS:2013sea} &$7/8\tev$& $  1.00\substack{+  1.90\\ -  2.00}$ & $   0.0$ & $   0.0$ & $   0.0$ & $   0.0$ & $ 100.0$\\ 
CMS $(pp)\to tth\to tt(\tau\tau)$~\cite{CMS:2013sea} &$8\tev$& $ -1.40\substack{+  6.30\\ -  5.50}$ & $   0.0$ & $   0.0$ & $   0.0$ & $   0.0$ & $ 100.0$\\ 
CMS $(pp)\to tth\to tt(\gamma\gamma)$~\cite{CMS:2013fda} &$8\tev$& $ -0.20\substack{+  2.40\\ -  1.90}$ & $   0.0$ & $   0.0$ & $   0.0$ & $   0.0$ & $ 100.0$\\ 
\midrule
D\O\  $(p\bar{p})\to h\to WW$~\cite{Abazov:2013gmz} & $1.96\tev$& $  1.90\substack{+  1.63\\ -  1.52}$ & $  77.5$ & $   5.4$ & $  10.6$ & $   6.5$ & $   0.0$\\ 
D\O\  $(p\bar{p})\to h\to bb$~\cite{Abazov:2013gmz} & $1.96\tev$& $  1.23\substack{+  1.24\\ -  1.17}$ & $   0.0$ & $   0.0$ & $  61.9$ & $  38.1$ & $   0.0$\\ 
D\O\  $(p\bar{p})\to h\to \gamma\gamma$~\cite{Abazov:2013gmz} & $1.96\tev$& $  4.20\substack{+  4.60\\ -  4.20}$ & $  77.5$ & $   5.4$ & $  10.6$ & $   6.5$ & $   0.0$\\ 
D\O\  $(p\bar{p})\to h\to \tau\tau$~\cite{Abazov:2013gmz}& $1.96\tev$ & $  3.96\substack{+  4.11\\ -  3.38}$ & $  77.5$ & $   5.4$ & $  10.6$ & $   6.5$ & $   0.0$\\ 
\bottomrule
 \end{tabular}
  \begin{tablenotes}
 \footnotesize
 \item[1] The signal is contaminated to $15.0\%$ by $WH\to W(\tau\tau)$ in the SM.
 \item[2] The signal is contaminated to $17.2\%$ [$9.8\%$] by $WH\to WWW$ [$ZH\to ZWW$] in the SM.
 \item[3] The $t\bar{t}h \to \ell^\pm\ell^\pm$ signal is comprised of the final states $WW$ ($74.5\%$), $ZZ$ ($3.7\%$) and $\tau\tau$ (21.7\%) in the SM.
 \item[4] The $t\bar{t}h \to 3\ell$ signal is comprised of the final states $WW$ ($73.0\%$), $ZZ$ ($4.6\%$) and $\tau\tau$ (22.5\%) in the SM.
 \item[5] The $t\bar{t}h \to 4\ell$ signal is comprised of the final states $WW$ ($54.1\%$), $ZZ$ ($17.4\%$) and $\tau\tau$ (28.5\%) in the SM.
 \end{tablenotes}
 \end{threeparttable}
}
 \caption{Signal strength measurements from CMS and D\O.}
  \label{Tab:expdata2}
\end{table}

For the six signal strength category measurements of the ATLAS SM $\htotautau$ search we implement additional correlations inspired by the information given in Ref.~\cite{ATLAS-CONF-2013-108}, following the procedure outlined in Ref.~\cite{CorrSystDoc}. This includes
\begin{itemize}
\item correlated uncertainties of $\sim 5-10\%~(20-30\%)$ in the VBF (boosted) categories of the gluon fusion signal component, mostly representing the uncertainties of the differential $p_T$ distribution of this signal process,
\item correlated normalization uncertainties of the top and $Z\to \ell \ell$ background of $\sim 10-15\%$ among the leptonic-leptonic and leptonic-hadronic $\tau\tau$ categories,
\item correlated uncertainties from hadronic $\tau$ identification of $\sim4\%~(12\%)$ in the leptonic-hadronic (hadronic-hadronic) $\tau\tau$ categories,
\item correlated di-hadronic $\tau$ trigger efficiency uncertainties of $7\%$ among the two hadronic-hadronic $\tau\tau$ channels,
\item correlated $Z\to \tau\tau$ background normalization uncertainties of $\sim 10-12\%$ among the hadronic-leptonic and leptonic-leptonic $\tau\tau$ categories.
\end{itemize}
The effect of including these correlations is shown in Fig.~\ref{Fig:ATLAS_H-tautau_2D_CSscaling} for a fit in a two-dimensional scaling model. Here the gluon fusion and $t\bar{t}H$ production cross sections are scaled by $\mu_\text{ggF+ttH}$ and the VBF, $WH$ and $ZH$ production cross sections by $\mu_\text{VBF+VH}$. Both the original ATLAS result and the likelihood reconstructed using \HS\ are shown. It can clearly be seen that the agreement between the reconstructed and official likelihood is significantly improved by including the additional correlations.

\begin{figure}
\centering
\subfigure[Without correlations of experimental systematic uncertainies.]{\includegraphics[width=0.4\textwidth]{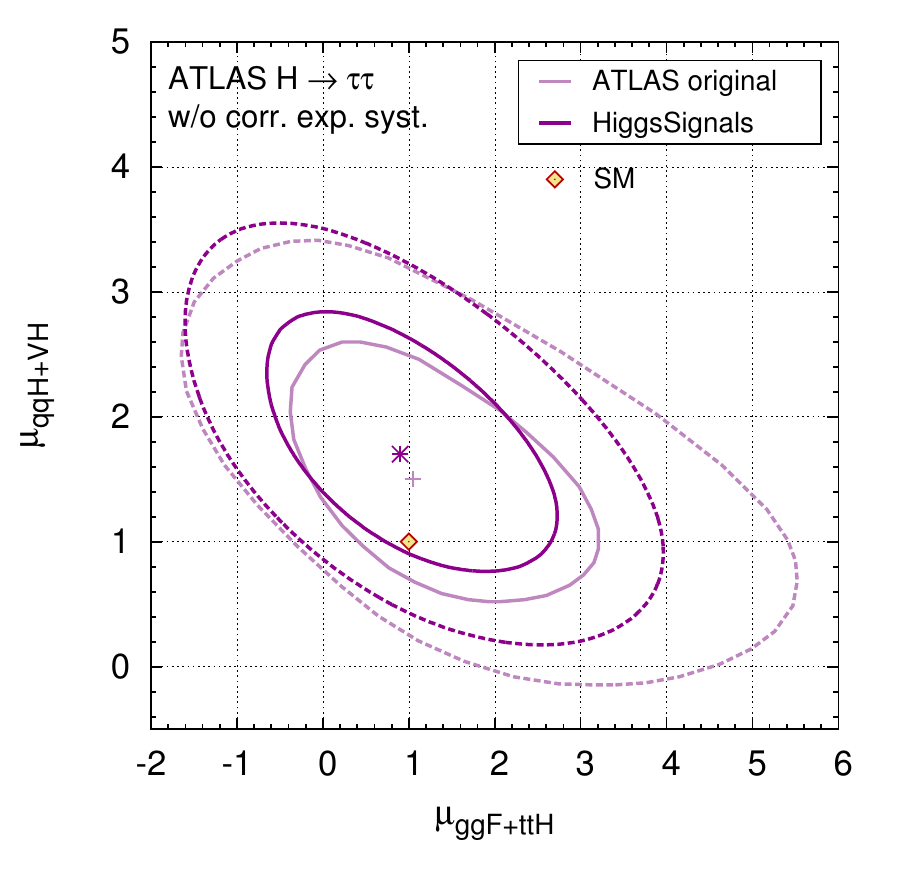}}\hspace{1cm}
\subfigure[With correlations of experimental systematic uncertainies.]{\includegraphics[width=0.4\textwidth]{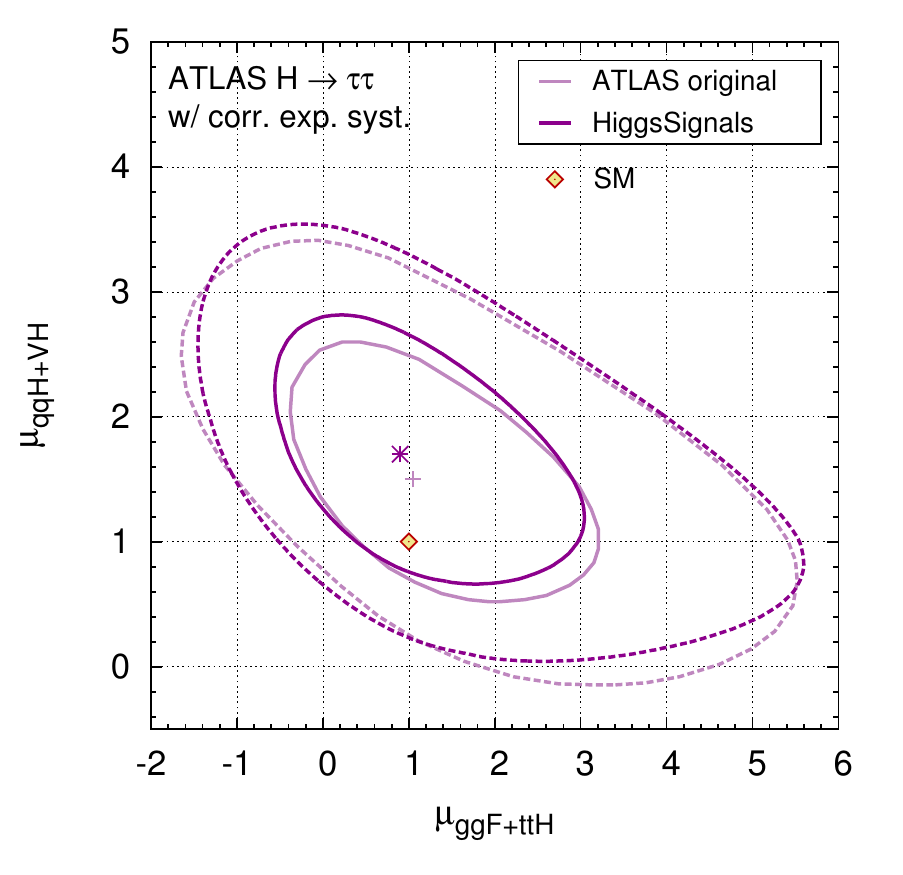}}
\caption{Comparison of our fit results with official ATLAS results for rescaled production cross sections of the gluon fusion (ggF) and $t\bar{t}H$ processes vs.~the vector boson fusion (qqH) and $VH$ ($V=W,Z$) processes using the ATLAS $\htotautau$ measurements~\cite{ATLAS-CONF-2013-108}. We compare the effects of neglecting or including correlations of known experimental systematic uncertainties in (a) and (b), respectively. The faint magenta curves indicates the original ATLAS results.}
\label{Fig:ATLAS_H-tautau_2D_CSscaling}
\end{figure}

\begin{figure}[h]
\centering
\subfigure[Using original measurements at $125.0\gev$ without correlations of experimental systematic uncertainies.]{\includegraphics[width=0.4\textwidth]{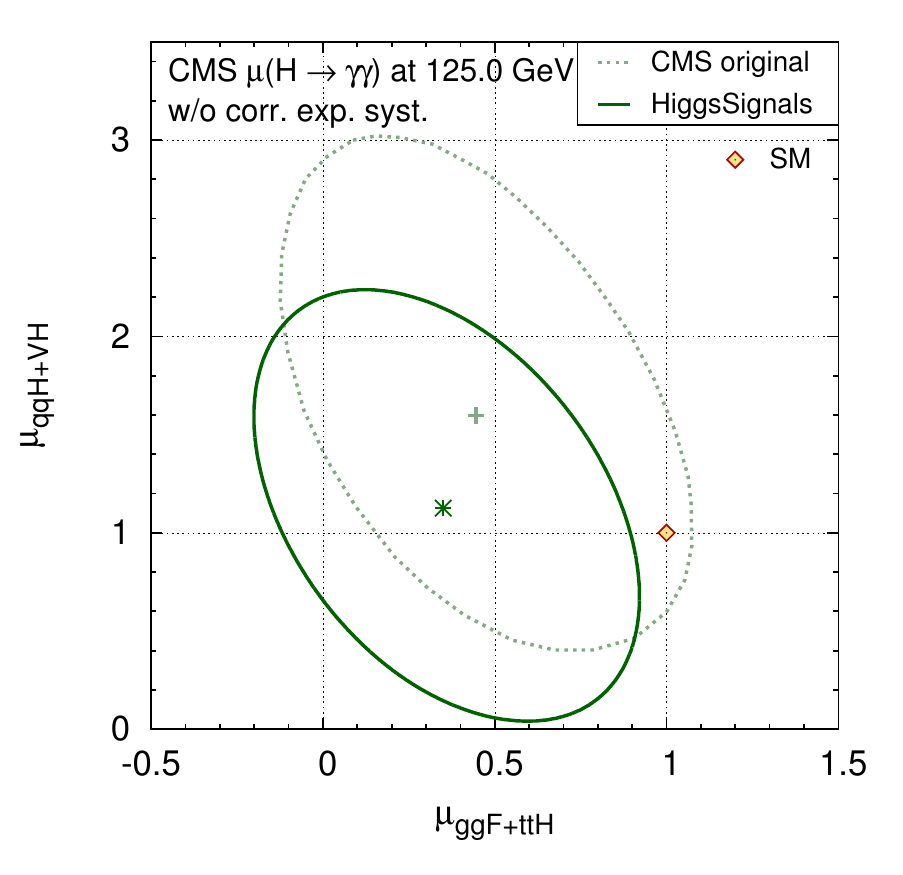}}\hspace{1cm}
\subfigure[Using approximated measurements at $125.7\gev$ without correlations of experimental systematic uncertainies.]{\includegraphics[width=0.4\textwidth]{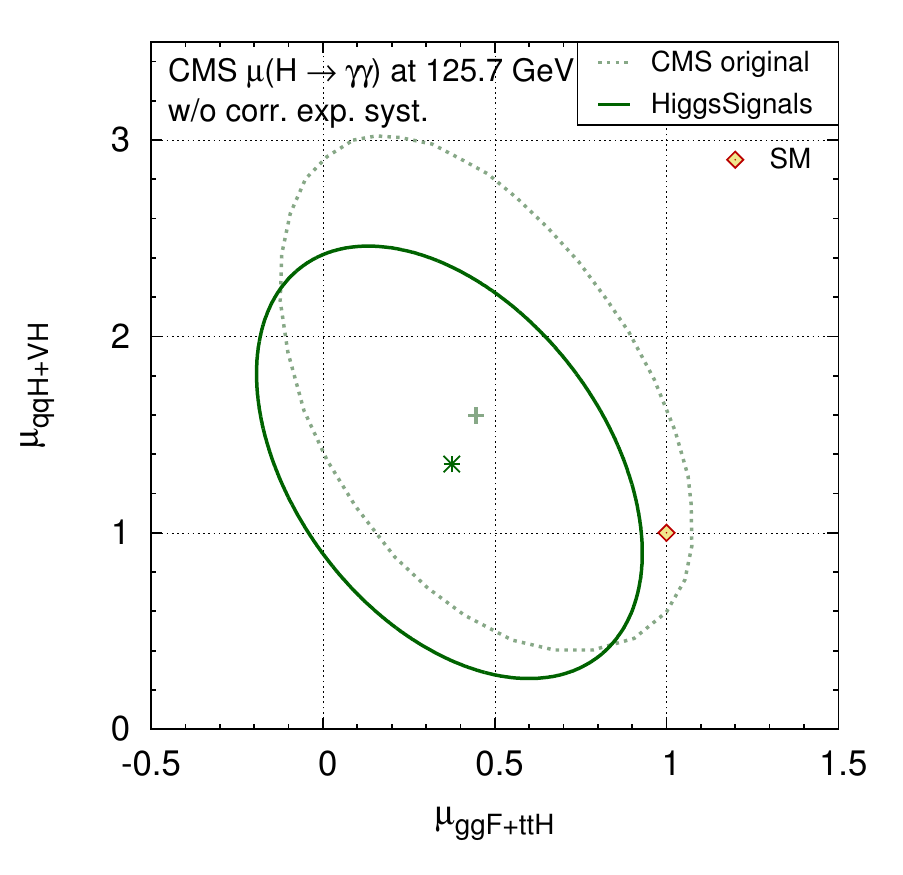}}
\subfigure[Using original measurements at $125.0\gev$ with correlations of experimental systematic uncertainies.]{\includegraphics[width=0.4\textwidth]{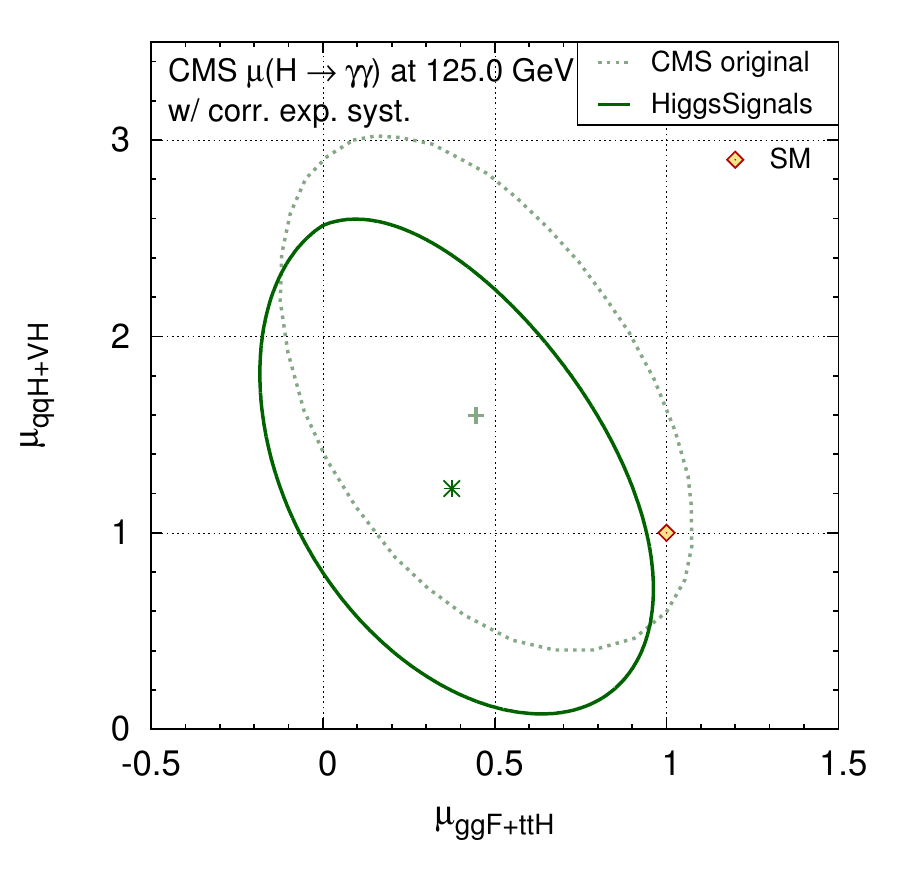}}\hspace{1cm}
\subfigure[Using approximated measurements at $125.7\gev$ with correlations of experimental systematic uncertainies.]{\includegraphics[width=0.4\textwidth]{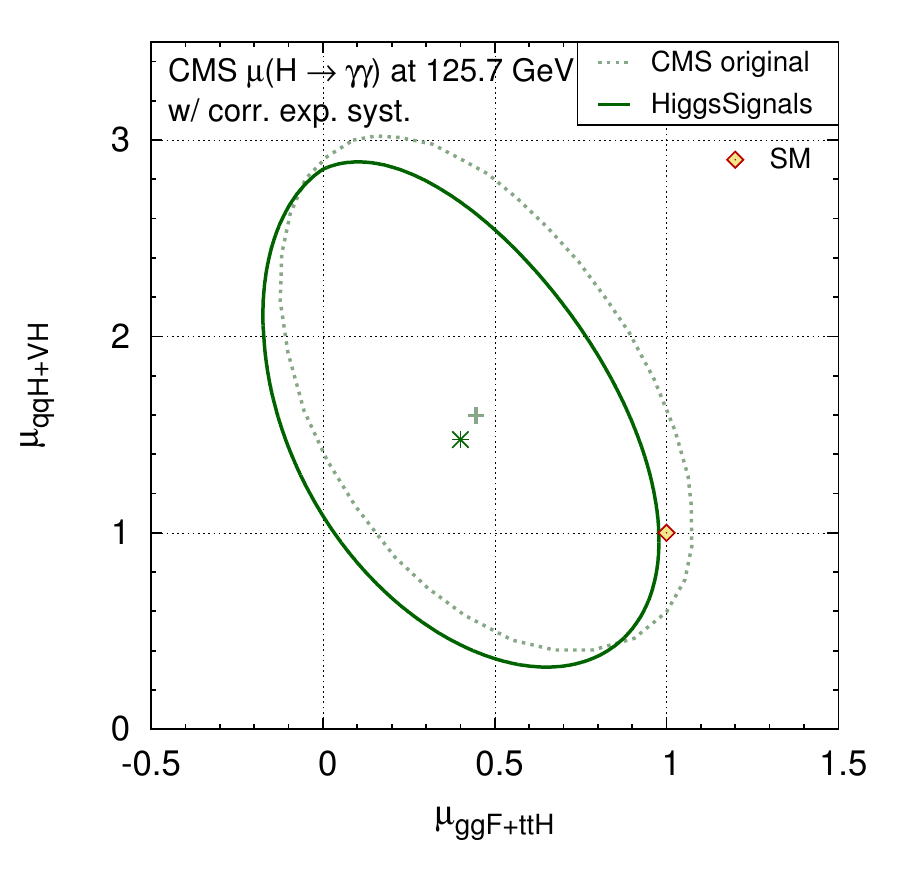}}
\caption{Comparison of our fit results with official CMS results for rescaled production cross sections of the gluon fusion (ggF) and $t\bar{t}H$ processes vs.~the vector boson fusion (qqH) and $VH$ ($V=W,Z$) processes using the CMS $\htogaga$ category measurements~\cite{CMS:2012paa,CMS:ril}. The results have been derived using either the original measurements given at a Higgs mass of $125.0\gev$, shown in (a,c), or approximated (rescaled) measurements at $125.7\gev$, shown in (b,d). We furthermore compare the effects of neglecting or including correlations of known experimental systematic uncertainties in (a,b) and (c,d), respectively. The dotted faint green curve indicates the original CMS results obtained for a Higgs boson mass of $125.7\gev$.}
\label{Fig:CMS_2D_CSscaling}
\end{figure}

In earlier validation fits~\cite{Bechtle:2013xfa} using the CMS $\htogaga$~\cite{CMS:ril,CMS:2012paa} results we found some discrepancies if only a simple $\chi^2$ test was performed. In this case the correlations among these observables introduced by common sources of experimental systematic uncertainties are non-negligible. Guided by the information given in Ref.~\cite{CMS:ril}, we therefore introduce the following correlations for the CMS $\htogaga$ category measurements:
\begin{itemize}
\item Event migration of $12.5\%$ between neighboring untagged categories for each $7\tev$ and $8\tev$,
\item Event migration of $15.0\%$ between the loose and tight dijet category at $8\tev$,
\item For the dijet categories, we include a dijet tagging efficiency uncertainty, corresponding to an anti-correlated uncertainty among the ggH and VBF channels, of $10-15\%$ and $30\%$, respectively.
\item $\etmiss$ cut efficiency uncertainty in the $\etmiss$ selection at $8\tev$ of $15\%$ for the ggH and VBF channels and $4\%$ for the $WH$, $ZH$, $t\bar{t}H$ channels, respectively.
\end{itemize}

One more complication arises because the signal rate measurements in the various categories of the $\htogaga$ analysis are only publicly available for a mass value of $m_H = 125.0\gev$. On the contrary, Ref.~\cite{CMS:yva} provides only fit results at $125.7\gev$ for the signal strengths
\begin{align}
\muobs(\htogaga,~\mathrm{untagged}) = 0.70 \substack{+0.33\\ -0.29},\\
\muobs(\htogaga,~\mathrm{VBF~tag}) = 1.01\substack{+0.63\\ -0.54},\\
\muobs(\htogaga,~\mathrm{VH~tag}) = 0.57 \substack{+1.34 \\ - 1.34},
\end{align}
combining the untagged, dijet and remaining leptonic/missing energy categories, respectively. Furthermore, the official scale factor fit results given by CMS, which can be used to validate our implementation, see Sect.~\ref{Sect:CMSvalidation}, assume a Higgs mass of $125.7\gev$~\cite{CMS:yva}. Given the category measurements at $125.0\gev$ (based on the MVA analysis), cf.~Tab.~\ref{Tab:expdata2}, we repeat these fits with \HS\ to obtain
\begin{align}
\muobs(\htogaga,~\mathrm{untagged}) = 0.64 \substack{+0.32\\ -0.30},\\
\muobs(\htogaga,~\mathrm{VBF~tag}) = 0.79\substack{+0.58\\ -0.54},\\
\muobs(\htogaga,~\mathrm{VH~tag}) = 0.63 \substack{+1.28 \\ - 1.14}.
\end{align}
We approximate the unknown category measurements at $125.7\gev$ by rescaling the category measurements at $125.0\gev$ by the ratio of the corresponding combined fit results.

\clearpage

In Fig.~\ref{Fig:CMS_2D_CSscaling} we show the effects of including the correlations of systematic experimental uncertainties and the rescaling of the category measurements to $m_H=125.7\gev$ for a 2D fit to common scale factors for the gluon fusion and $t\bar{t}H$ cross section, $\mu_{\mathrm{ggF}+\mathrm{ttH}}$, and for the vector boson fusion and $VH$ ($V=W,Z$) cross sections, $\mu_{\mathrm{qqH}+\mathrm{VH}}$, using only results from the CMS $\htogaga$ analysis~\cite{CMS:2012paa,CMS:ril}. The original CMS result obtained for $m_H=125.7\gev$ is overlaid in the figure. It can be seen that both effects have a sizable impact on the result. Acceptable agreement with the official CMS result can be obtained if both the correlations and the rescaling is taken into account, as shown in Fig.~\ref{Fig:CMS_2D_CSscaling}(d). We therefore use this setup of the CMS $\htogaga$ measurements for the fits presented in this paper.

\subsection{Validation fit using CMS data only}
\label{Sect:CMSvalidation}

We validate the fit procedure by performing a six-dimensional fit to the CMS Moriond 2013 data and comparing the results to the official fit results presented by CMS~\cite{CMS:yva}. The model parameters are identical to the scale factors of our general fit, i.e.~$\kV$, $\ku (\equiv \kt)$, $\kd (\equiv \kb)$, $\kl (\equiv \ktau)$, $\kg$ and $\kga$, but the total width is obtained from the rescaled effective couplings direcly (no additional Higgs decay modes). The CMS fit was performed assuming a Higgs boson mass of $125.7\gev$. The results are shown in Fig.~\ref{Fig:CMS_6dim}, where the blue curves indicate the original CMS results\cite{CMS:yva}. With the corrected implementation of the CMS $\htogaga$ measurements, as discussed above, the fit shows excellent agreement.

\begin{figure}
\centering
\includegraphics[width=0.55\textwidth]{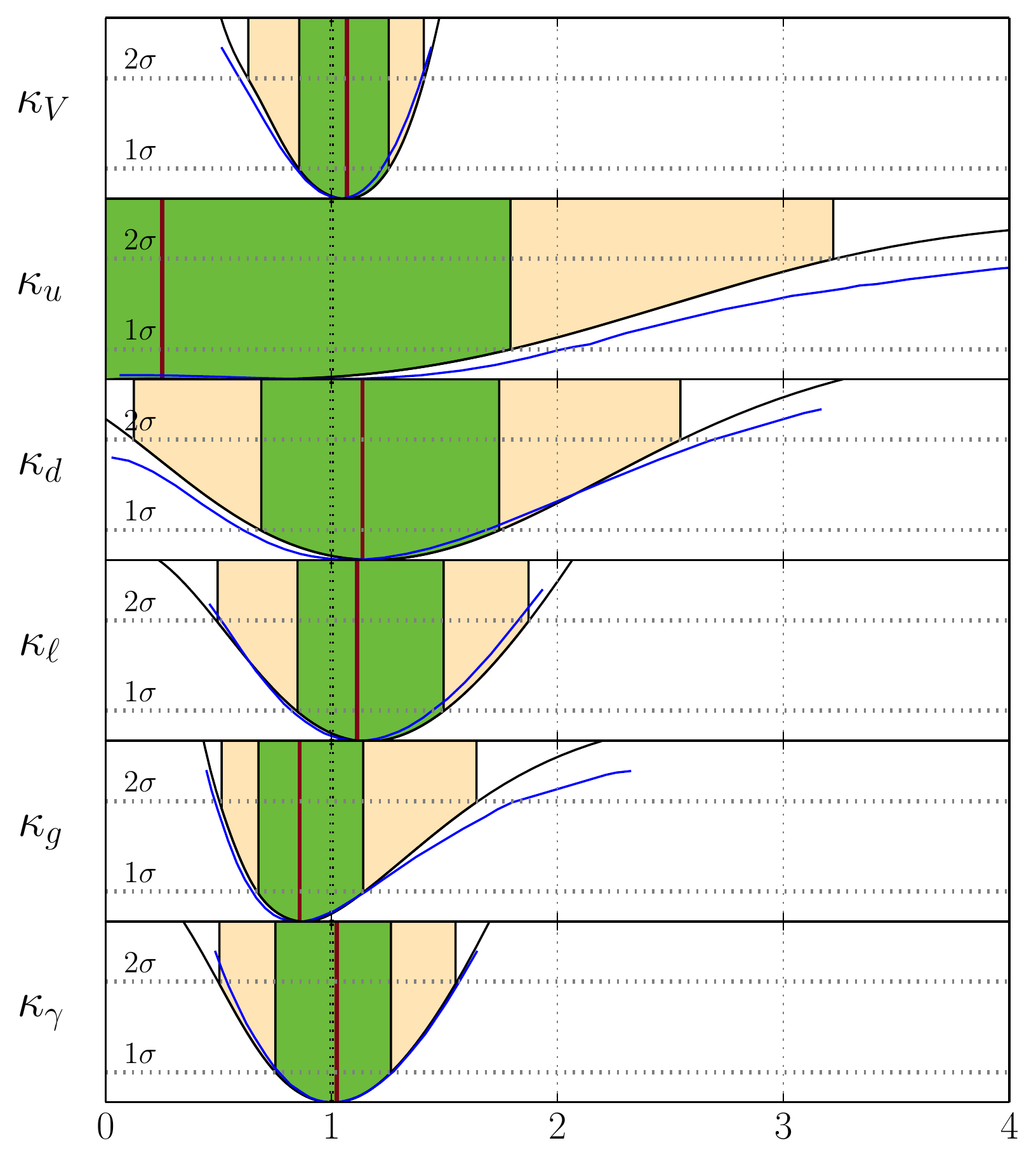}
\caption{One-dimensional $\chi^2$ profiles of the fitted Higgs coupling scale factors $\kV, \ku, \kd, \kl, \kg, \kga$ using only the CMS Moriond 2013 results~\cite{CMS:yva}. The CMS $\htogaga$ measurements were rescaled to a Higgs boson mass of $125.7\gev$ and include correlations of some experimental systematic uncertainties. The blue curves show the original fit result obtained by CMS~\cite{CMS:yva}.}
\label{Fig:CMS_6dim}
\end{figure}


\subsection{Projected sensitivity of future signal rate measurements}
\label{Sect:Appprojections}

The future estimates of signal strength measurements in various channels at the LHC for integrated luminosities of $300~\ifb$ and $3000~\ifb$ are given in Tab.~\ref{Tab:futureLHCmeasurements} for ATLAS~\cite{ATL-PHYS-PUB-2013-014} and CMS~\cite{CMS:2013xfa}. In Tab.~\ref{Tab:ILCobs} we list the estimated cross section and signal rate measurements at the ILC~\cite{Asner:2013psa}. These values are used for the study of the LHC and ILC capabilities of Higgs coupling determination presented in \refse{Sect:futurecouplings}.

\begin{table}
\centering
\scalebox{0.9}{
\renewcommand{\arraystretch}{1.0}
\centering
\begin{threeparttable}[b]
\footnotesize
 \begin{tabular}{lccrrrrr}
 \toprule
 Analysis & \multicolumn{2}{c}{$68\%$ C.L. precision} & \multicolumn{5}{c}{Assumed signal composition [in \%]} \\
 & 	$300~\ifb$& $3000~\ifb$	 & ggH & VBF & WH &  ZH & $t\bar{t}H$ \\
 \midrule
ATL $(pp)\to h\to \gamma\gamma~\mbox{(0jet)}$~\cite{ATL-PHYS-PUB-2013-014} & $   0.12$ & $0.05$	& $  91.6$ & $   2.7$ & $   3.2$ & $   1.8$ & $   0.6$\\ 
ATL $(pp)\to h\to \gamma\gamma~\mbox{(1jet)}$~\cite{ATL-PHYS-PUB-2013-014} & $   0.14$ & $0.05$	&$  81.8$ & $  13.2$ & $   2.9$ & $   1.6$ & $   0.5$\\ 
ATL $(pp)\to h\to \gamma\gamma~\mbox{(VBF-like)}$~\cite{ATL-PHYS-PUB-2013-014} & $   0.43$ & $0.16$&	 $  39.2$ & $  58.4$ & $   1.4$ & $   0.8$ & $   0.3$\\ 
ATL $(pp)\to h\to \gamma\gamma~(VH\mbox{-like)}$~\cite{EuropeanStrategyforParticlePhysicsPreparatoryGroup:2013fia} & $   0.77$ & $0.25$& $   2.5$ & $   0.4$ & $  63.3$ & $  15.2$ & $  18.7$\\ 
ATL $(pp)\to h\to \gamma\gamma~(t\bar{t}H\mbox{-like})$~\cite{EuropeanStrategyforParticlePhysicsPreparatoryGroup:2013fia} & $   0.54$ & $0.16$ &$   0.0$ & $   0.0$ & $   0.0$ & $   0.0$ & $ 100.0$\\ 
ATL $(pp)\to h\to WW~\mbox{(0jet)}$~\cite{ATL-PHYS-PUB-2013-014} & $   0.08$ &$0.05$& $  98.2$ & $   1.8$ & $   0.0$ & $   0.0$ & $   0.0$\\ 
ATL $(pp)\to h\to WW~\mbox{(1jet)}$~\cite{ATL-PHYS-PUB-2013-014} & $   0.17$ &$0.10$& $  88.4$ & $  11.6$ & $   0.0$ & $   0.0$ & $   0.0$\\ 
ATL $(pp)\to h\to WW~\mbox{(VBF-like)}$~\cite{EuropeanStrategyforParticlePhysicsPreparatoryGroup:2013fia} & $   0.20$ & $0.09$& $   8.1$ & $  91.9$ & $   0.0$ & $   0.0$ & $   0.0$\\ 
ATL $(pp)\to h\to ZZ~\mbox{(ggF-like)}$~\cite{ATL-PHYS-PUB-2013-014} & $   0.06$ & $0.04$ &$  88.7$ & $   7.2$ & $   2.0$ & $   1.4$ & $   0.7$\\ 
ATL $(pp)\to h\to ZZ~\mbox{(VBF-like)}$~\cite{ATL-PHYS-PUB-2013-014} & $   0.31$ &$0.16$ & $  44.7$ & $  53.2$ & $   0.7$ & $   0.4$ & $   1.0$\\ 
ATL $(pp)\to h\to ZZ~(VH\mbox{-like)}$~\cite{ATL-PHYS-PUB-2013-014} & $   0.31$ & $0.12$ &$  30.1$ & $   9.0$ & $  34.8$ & $  12.1$ & $  14.0$\\ 
ATL $(pp)\to h\to ZZ~(t\bar{t}H\mbox{-like)}$~\cite{ATL-PHYS-PUB-2013-014} & $   0.44$ & $0.16$& $   8.7$ & $   1.7$ & $   1.7$ & $   3.1$ & $  84.8$\\ 
ATL $(pp)\to h\to Z\gamma$~\cite{ATL-PHYS-PUB-2013-014} & $   1.45$ & $0.54$&$  87.6$ & $   7.1$ & $   3.1$ & $   1.7$ & $   0.6$\\ 
ATL $(pp)\to h\to \mu\mu$~\cite{ATL-PHYS-PUB-2013-014} & $   0.45$ & $0.15$ &$  87.6$ & $   7.1$ & $   3.1$ & $   1.7$ & $   0.6$\\ 
ATL $(pp)\to h\to \mu\mu~(t\bar{t}H)$~\cite{EuropeanStrategyforParticlePhysicsPreparatoryGroup:2013fia} & $   0.72$ & $0.23$ & $   0.0$ & $   0.0$ & $   0.0$ & $   0.0$ & $ 100.0$\\ 
ATL $(pp)\to h\to \tau\tau~(\mbox{VBF-like})$~\cite{ATL-PHYS-PUB-2013-014} & $   0.16$ & $0.12$ & $   19.8$ & $ 80.2$ & $   0.0$ & $   0.0$ & $   0.0$\\ 
\midrule
CMS $(pp)\to h\to \gamma\gamma$~\cite{CMS:2013xfa} & $   0.06$ & $0.04$ & $  87.6$ & $   7.1$ & $   3.1$ & $   1.7$ & $   0.6$\\ 
CMS $(pp)\to h\to WW$~\cite{CMS:2013xfa} & $   0.06$ & $0.04$ & $  88.1$  &$   7.1$ & $   3.1$ & $   1.7$ & $   0.0$\\ 
CMS $(pp)\to h\to ZZ$~\cite{CMS:2013xfa} & $   0.07$ & $0.04$ &$  88.1$ & $   7.1$ & $   3.1$ & $   1.7$ & $   0.0$\\ 
CMS $(pp)\to h\to Z\gamma$~\cite{CMS:2013xfa} & $   0.62$ & $0.20$& $  87.6$ & $   7.1$ & $   3.1$ & $   1.7$ & $   0.6$\\ 
CMS $(pp)\to h\to bb$~\cite{CMS:2013xfa} & $   0.11$ & $0.05$ & $   0.0$ & $   0.0$ & $  57.0$ & $  32.3$ & $  10.7$\\ 
CMS $(pp)\to h\to \mu\mu$~\cite{CMS:2013xfa} & $   0.40$ & $0.20$ & $  87.6$ & $   7.1$ & $   3.1$ & $   1.7$ & $   0.6$\\ 
CMS $(pp)\to h\to \tau\tau$~\cite{CMS:2013xfa} & $   0.08$ & $0.05$ &$  68.6$ & $  27.7$ & $   2.4$ & $   1.4$ & $   0.0$\\ 
\bottomrule
 \end{tabular}
 \end{threeparttable}
 }
  \caption{Projected experimental precision (i.e.~without theory
uncertainty) of signal strength measurements from ATLAS and CMS at
$\sqrt{s}=14\tev$ for $300~\ifb$ and $3000~\ifb$ (HL--LHC). The numbers from CMS correspond to Scenario 2 of their projections, however, we treat them as purely experimental precisions (see discussion in Sect.~\ref{Sect:futureLHC}).}
  \label{Tab:futureLHCmeasurements}
\end{table}

\begin{table}
\centering
\footnotesize
 \renewcommand{\arraystretch}{1.1}
\begin{tabular}{l|cc|ccc|cc}
\toprule
$\mathcal{L}$ and $\sqrt{s}$ 	&	\multicolumn{2}{c|}{$250\invfb$ at $250\gev$} &\multicolumn{3}{c|}{$500\invfb$ at $500\gev$} & \multicolumn{2}{c}{$1\invab$ at $1\tev$} \\
\midrule
 			&	$ZH$		&	$\nu\bar{\nu}H$			&	$ZH$		&	$\nu\bar{\nu}H$ 	&	$t\bar{t}H$	&	$\nu\bar{\nu}H$ 	&	$t\bar{t}H$  \\
\midrule
$\Delta \sigma/\sigma$	&	$2.6\%$	&	-	&	$3.0\%$	&	-	&	-	&	-	&	-	\\
$\brinv$				&	$<0.9\%$	&	-	&	-		&	-	&	-	&	-	&	-	\\
\midrule
mode		&	\multicolumn{7}{c}{$\Delta(\sigma \cdot \mathrm{BR}) / (\sigma \cdot \mathrm{BR})$} \\ 
\midrule
$\htobb$ 	&	$1.2\%$		&	$10.5\%$ 	& 	$1.8\%$	&	$0.7\%$ 	&	$28\%$	&	$0.5\%$		&	$6.0\%$ \\
$\htocc$	&	$8.3\%$		&	-		&	$13.0\%$	&	$6.2\%$	&	-		&	$3.1\%$		&	-	  \\
$\htogg$	&	$7.0\%$		&	-		&	$11\%$	&	$4.1\%$	& 	-		&	$2.6\%$		&	-	 \\ 
$\htoWW$	&	$6.4\%$		&	-		&	$9.2\%$	&	$2.4\%$	& 	-		&	$1.6\%$		&	-	 \\	
$\htotautau$&	$4.2\%$		&	-		&	$5.4\%$	&	$9.0\%$	& 	-		&	$3.1\%$		&	-	\\ 
$\htoZZ$	&	$18\%$		&	-		&	$25\%$	&	$8.2\%$	& 	-		&	$4.1\%$		&	-		 \\ 
$\htogaga$&	$34\%$		&	-		&	$ 34\%$	&	$23\%$	& 	-		&	$8.5\%$		&	-	\\ 
$\htomumu$ &	$100\%$		&	-		&	-		&	-		& 	-		&	$31\%$		&	-		\\ 
\midrule
\midrule
$\mathcal{L}$ and $\sqrt{s}$ 	&	\multicolumn{2}{c|}{$1150\invfb$ at $250\gev$} &\multicolumn{3}{c|}{$1600\invfb$ at $500\gev$} & \multicolumn{2}{c}{$2.5\invab$ at $1\tev$} \\
\midrule
 			&	$ZH$		&	$\nu\bar{\nu}H$			&	$ZH$		&	$\nu\bar{\nu}H$ 	&	$t\bar{t}H$	&	$\nu\bar{\nu}H$ 	&	$t\bar{t}H$  \\
\midrule
$\Delta \sigma/\sigma$	&	$1.2\%$	&	-	&	$1.7\%$	&	-	&	-	&	-	&	-	\\
$\brinv$				&	$<0.4\%$	&	-	&	-		&	-	&	-	&	-	&	-	\\
\midrule
mode		&	\multicolumn{7}{c}{$\Delta(\sigma \cdot \mathrm{BR}) / (\sigma \cdot \mathrm{BR})$} \\ 
\midrule
$\htobb$ 	&	$0.6\%$		&	$4.9\%$ 	& 	$1.0\%$	&	$0.4\%$ 	&	$16\%$	&	$0.3\%$		&	$3.8\%$ 	\\
$\htocc$	&	$3.9\%$		&	-		&	$7.2\%$	&	$3.5\%$	&	-		&	$2.0\%$		&	-	  	\\
$\htogg$	&	$3.3\%$		&	-		&	$6.0\%$	&	$2.3\%$	& 	-		&	$1.4\%$		&	-	 	\\ 
$\htoWW$	&	$3.0\%$		&	-		&	$5.1\%$	&	$5.1\%$	& 	-		&	$1.0\%$		&	-	 	\\	
$\htotautau$&	$2.0\%$		&	-		&	$3.0\%$	&	$3.0\%$	& 	-		&	$2.0\%$		&	-		\\ 
$\htoZZ$	&	$8.4\%$		&	-		&	$14.0\%$	&	$14.0\%$	& 	-		&	$2.6\%$		&	-		 \\ 
$\htogaga$&	$16.0\%$		&	-		&	$19.0\%$ &	$13.0\%$ & 	-		&	$5.4\%$		&	-		\\ 
$\htomumu$ &	$46.6\%$		&	-		&	-		&	-		& 	-		&	$20.0\%$		&	-		\\ 
\bottomrule
\end{tabular}
\caption{Expected accuracies for the measurements of signal rates and absolute production cross sections at various ILC stages of the baseline program (\textit{above}) and after a luminosity upgrade (\textit{below}) for a Higgs boson with mass $m_H = 125\gev$. Upper limits on $\brinv$ are given at $95\%~\mathrm{C.L.}$. The numbers are taken from Ref.~\cite{Asner:2013psa}, cf.~also Ref.~\cite{Baer:2013cma}.} 
\label{Tab:ILCobs}
\end{table}



\section{Investigating the \pvalue\ of $\chi^2$ fits to measured Higgs signal rates}
\label{Sect:pvalue}


As outlined in \refse{Sect:HS} and explained in detail in
Ref.~\cite{Bechtle:2013xfa}, \HS\ employs a $\chi^2$ approximation to
allow for a very fast evaluation of the model compatibility
with public results from Higgs rate and mass measurements in arbitrary models.
Comparisons to the results from ATLAS and CMS show that this
implementation yields a good approximation to the official
results~\cite{Bechtle:2013xfa} (see also
Appendices~\ref{Sect:currentexpdata} and~\ref{Sect:CMSvalidation} above). This allows for a reliable
phenomenological analysis of a very large variety of models of new
physics against the Higgs search results. In such studies, the
\pvalue, i.e.~the statistical agreement of the measured results with
the predictions from a theory, is of high interest. This can be
evaluated using toy Monte Carlo techniques. In this section we study to what
extent the specific implementation of the $\chi^2$ evaluation in \HS\
impacts the \pvalue\ calculation. This is also of interest
for other implementations of $\chi^2$ tests against Higgs mass and
rate measurements~\citefits, which employ different levels of detail concerning the implementation of uncertainties (correlated/uncorrelated, relative/absolute, symmetric/asymmetric, etc.).
In order to evaluate the impact of the calculation of uncertainties
and correlations on the $\chi^2$, we investigate the \pvalue\ of a SM-like Higgs boson modified by a global scale
parameter $\kappa$.  It is tested against the latest rate measurements
from ATLAS, CMS, CDF and D\O, see Appendix~\ref{Sect:expdata} for
details. Using a toy Monte Carlo technique the \pvalue\ is then evaluated from the \HS\ calculated
$\chi^2$ for sets of pseudo-measurements thrown around the best fit
point and according to the covariance matrix, which we obtain at the best fit
point. The exact \pvalue\ based on
the full likelihood distribution can of course only be calculated by
the experimental collaborations. However, no combination of the
experiments at LHC and the Tevatron is available, such that an
approximate calculation is of interest.

The default treatment of uncertainties in \HS\ suggests a deviation
from the ideal $\chi^2$ distribution in both the signal strength part,
$\chi^2_\mu$, and the Higgs mass part, $\chi^2_m$. Therefore, the
\pvalue\ can only approximately be extracted from the
observed $\chi^2$ at the best fit point and the number of degrees of freedom (ndf) assuming an ideal $\chi^2$
distribution. Instead, toy measurements have to be employed to take
into account the following effects in the \pvalue\ evaluation:

\begin{enumerate}
\item The usage of asymmetric (upper and lower) uncertainties in the
  rate measurements instead of averaged (symmetric) uncertainties. The
  choice for the observed rate uncertainty entering the $\chi^2$
  evaluation, $\Delta \muobs$, is dependent on the relative position of the
  model-predicted signal rate $\mu$ with respect to the observed value $\muobs$:
  \begin{align}
    \Delta \muobs = \left\{ \begin{array}{lll}
        \Delta\muobs^\mathrm{up} & ,\quad\mbox{if}\quad & \mu >
        \muobs\\ \Delta\muobs^\mathrm{low} &,\quad \mbox{if}\quad &
        \mu < \muobs \end{array} \right. .
  \end{align}
\item The usage of relative instead of absolute rate
  uncertainties. The luminosity uncertainty is scaled with the
  observed $\muobs$ value, while the theoretical rate uncertainties are
  scaled with the predicted $\mu$ value in \HS. Where the experimental
  systematics can not be attributed to either signal or background,
  they are treated as background-related and kept constant. This
  combination generally provides a good approximation of the
  experimental results.
\end{enumerate}
In case that the mass is also fitted, two additional effects arise:
\begin{enumerate}
\item[3.] Theoretical mass uncertainties can be treated as
  (anti-)correlated Gaussian errors in the $\chi^2_m$ evaluation. The
  theory mass uncertainty of two mass observables, $\mobs_i$,
  $\mobs_j$, is anti-correlated if the predicted mass lies in between
  these measurements, $\mobs_i < m < \muobs_j$.
\item[4.] The automatic assignment of the Higgs boson to the observables
  introduces a highly non-trivial deviation from the ideal $\chi^2$
  shape in both $\chi^2_\mu$ and $\chi^2_m$. This procedure takes care
  that the comparison of the predicted signal rate $\mu$ (at mass $m$)
  with the measured signal strength $\muobs$ (at mass $\mobs$) is
  still approximately valid, or otherwise adds a $\chi^2$ penalty to
  $\chi^2_\mu$. In the latter case, the mass measurement associated
  with the unassigned observable does not enter $\chi^2_m$ anymore.
  This issue is of course only relevant if a model with more than one
  Higgs boson is studied. It is not further studied in the examples
  below.
\end{enumerate}
The items (1, 2) lead to a dependence of the covariance matrix $C_\mu$
in the $\chi^2_\mu$ calculation on both the observed signal rate
values, $\muobs$, and the model-predicted signal rate values,
$\mu$. Hence, it changes for each set of pseudo-measurements and
depends on the tested
model. 
The items (3, 4) are of relevance only in the case of a non-trivial
model prediction of the Higgs mass. Here, we choose a fixed Higgs mass
of $m_H=\MHexp\gev$. We ensure a full assignment of all observables within \HS, while the actual constant $\chi^2$
contribution from the Higgs mass measurements is of no further
relevance in this study. It should be noted, however, that we hereby
make the approximation/assumption, that all signal rates measured by
the experiments for the Higgs signal at various mass positions between
$124.3\gev$ and $126.8\gev$ can be compared with the hypothesized
Higgs state at $m_H=\MHexp\gev$.

\begin{figure}
\centering
\subfigure[Absolute and symmetrical rate uncertainties.]{\includegraphics[width=0.48\textwidth]{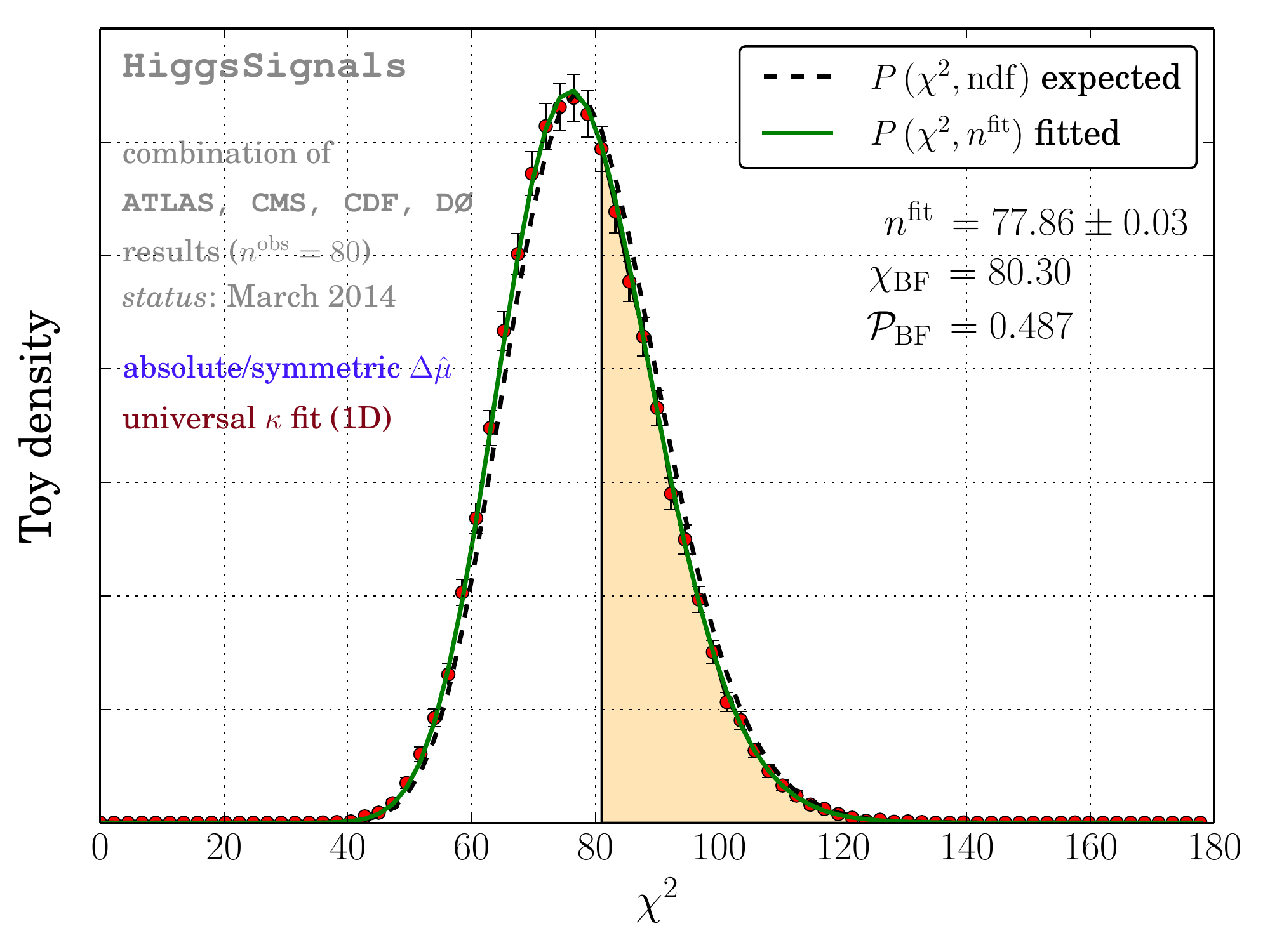}}
\subfigure[Relative and symmetrical rate uncertainties]{\includegraphics[width=0.48\textwidth]{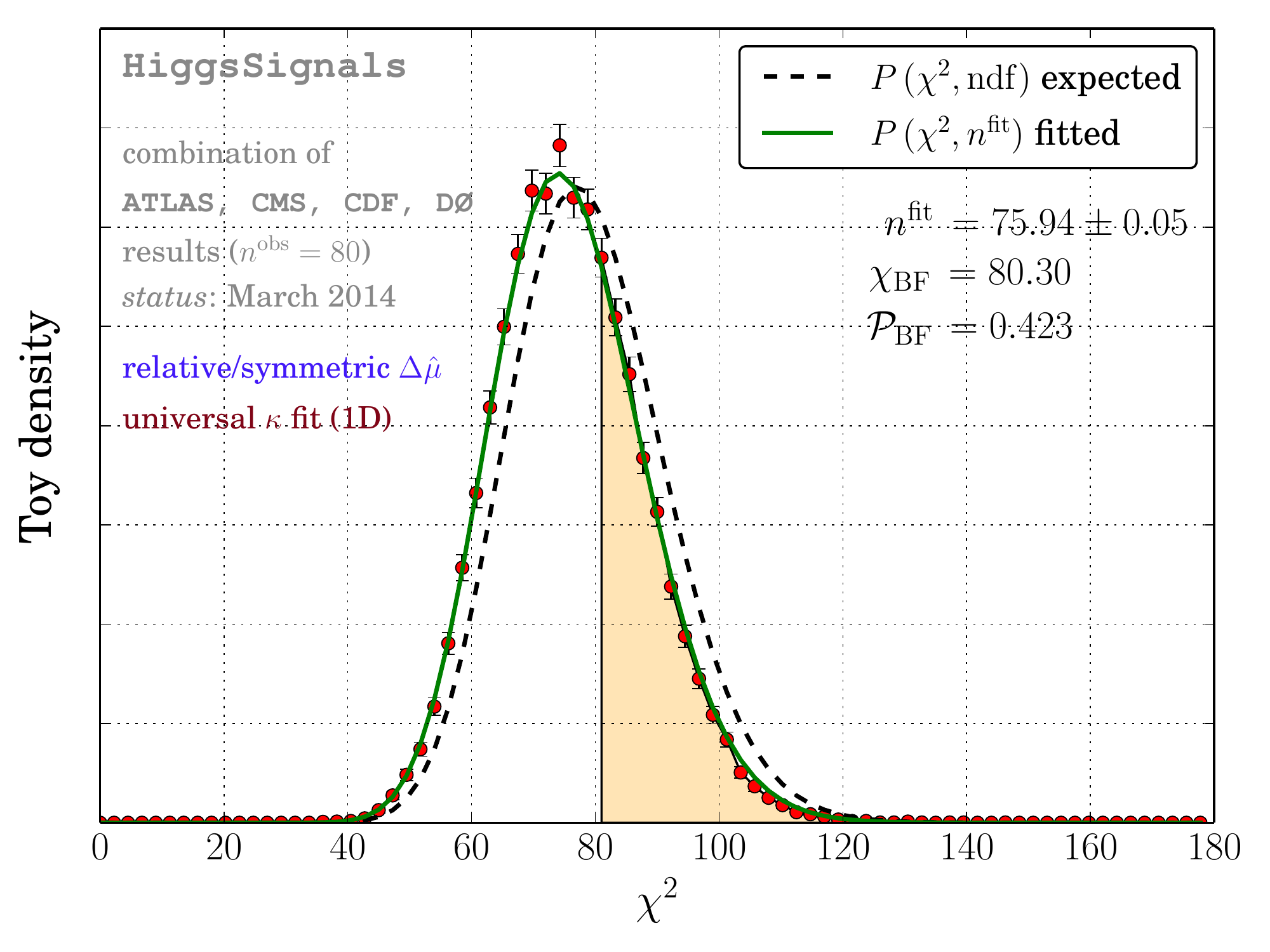}}
\subfigure[Absolute and asymmetric rate uncertainties.]{\includegraphics[width=0.48\textwidth]{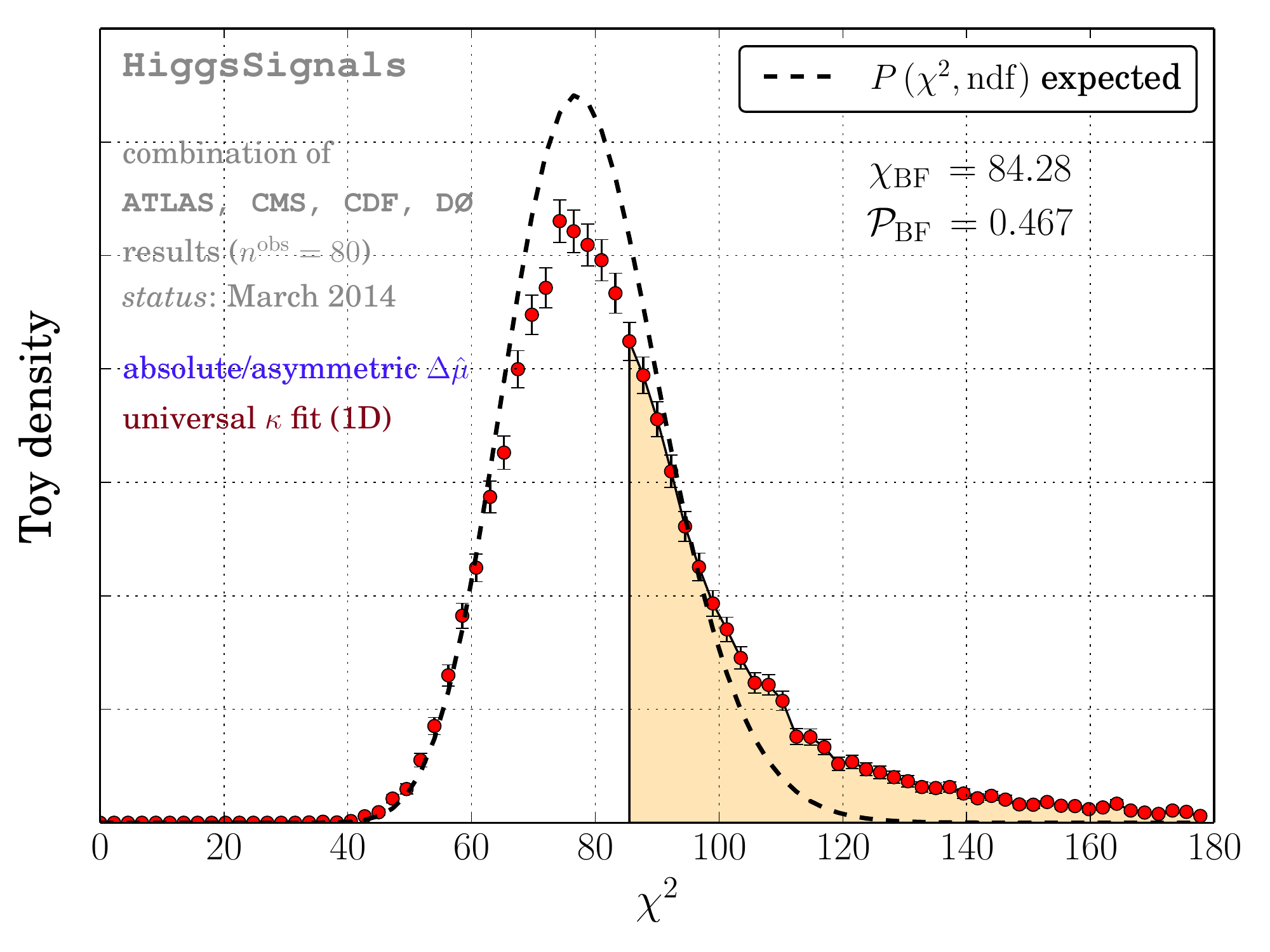}}
\subfigure[Relative and asymmetric rate uncertainties.]{\includegraphics[width=0.48\textwidth]{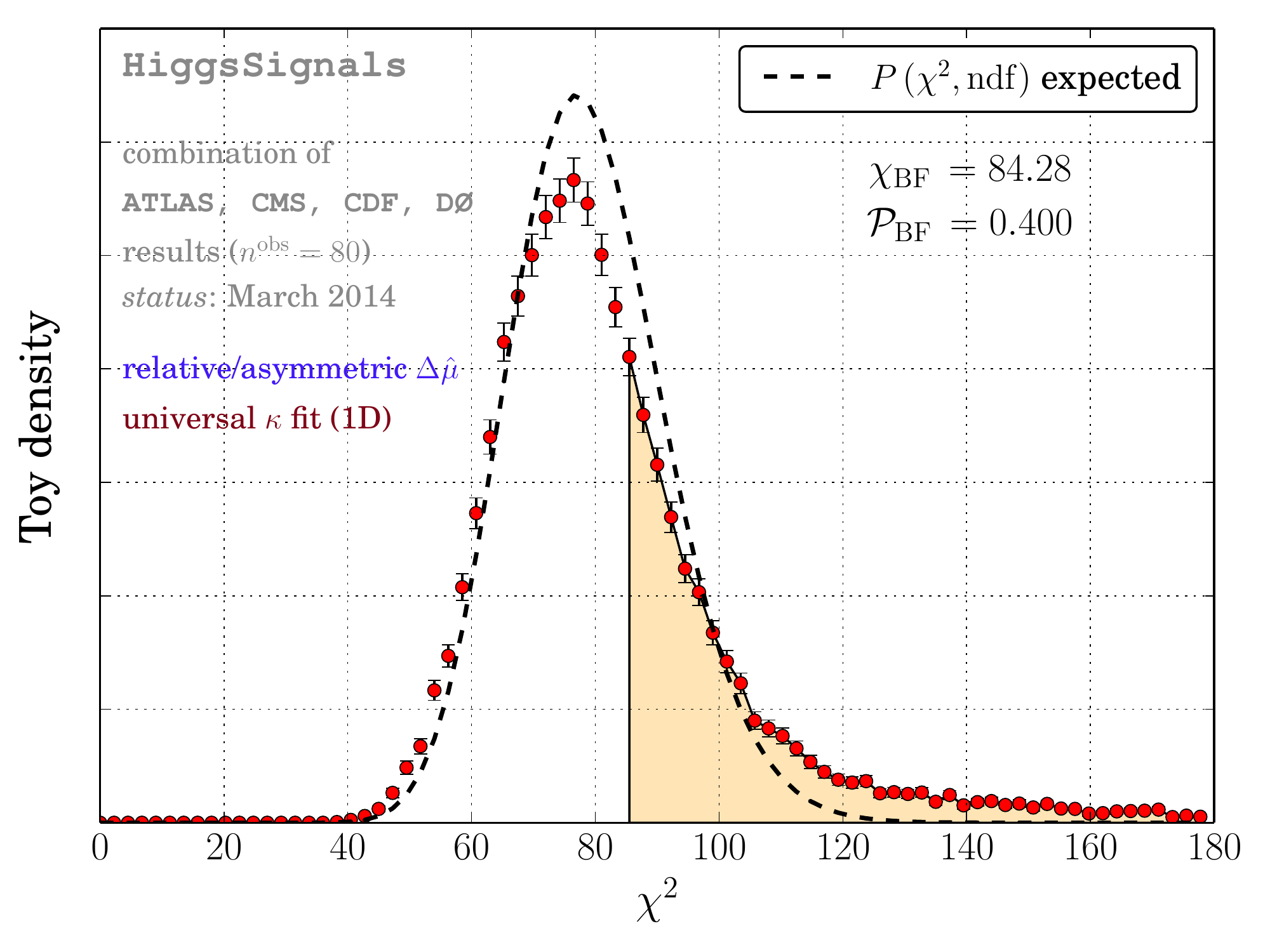}}
\caption{$\chi^2$ outcomes of the SM predicted Higgs rates tested
  against pseudo signal rate measurements in a fit setup with 80 rate
  measurements and one free parameter, a global scale factor $\kappa$
  for all {Higgs couplings}. The fits are performed with different \HS\ settings:
  In (\textit{a,c}) the luminosity and theory rate uncertainties are
  kept at their absolute values whereas in (\textit{b,d}) they are
  taken relative to the (pseudo-)measured signal rates as evaluated
  from the original measurements. In (\textit{a,b}) the signal rate
  uncertainties $\Delta \muobs$ are implemented as averaged
  (symmetrical) values, while (\textit{c,d}) asymmetrical upper and
  lower uncertainties as given in the original measurements are
  employed. The black dashed line shows the expected $\chi^2$
  distribution for $80$ signal rate observables and one parameter. The
  solid, green graph shows the best-fitting $\chi^2$ probability function to the toy outcomes. The
  yellow area underneath this curve as calculated from the observed
  best-fit $\chi^2$ value (obtained from the original measurements) to
  $\infty$ corresponds to the \pvalue.}
\label{fig:1Dfit_toy_chisq_dist}
\end{figure}

\begin{figure}
\centering
\subfigure[Absolute and symmetrical rate uncertainties.]{\includegraphics[width=0.48\textwidth]{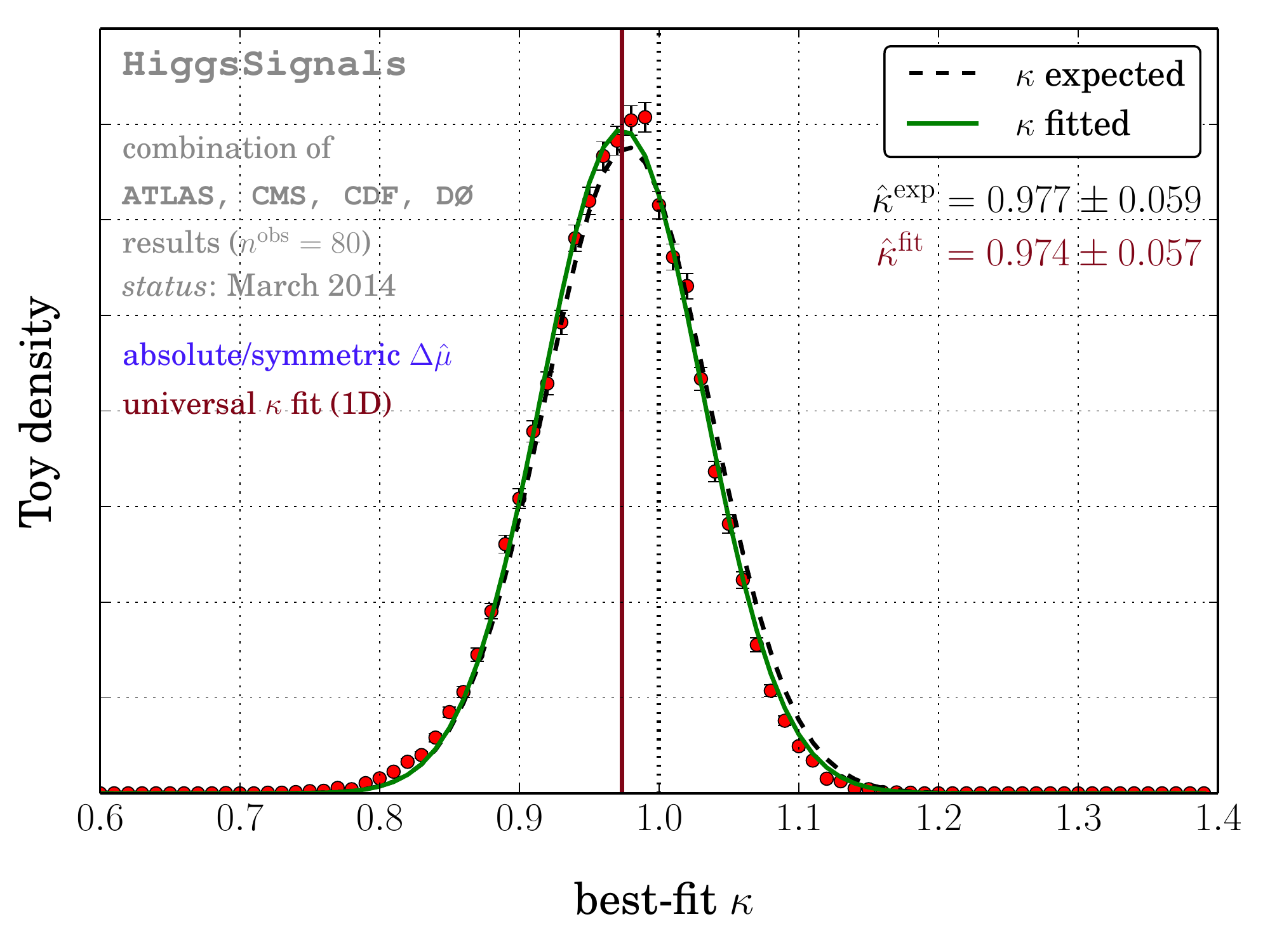}}
\subfigure[Relative and symmetrical rate uncertainties]{\includegraphics[width=0.48\textwidth]{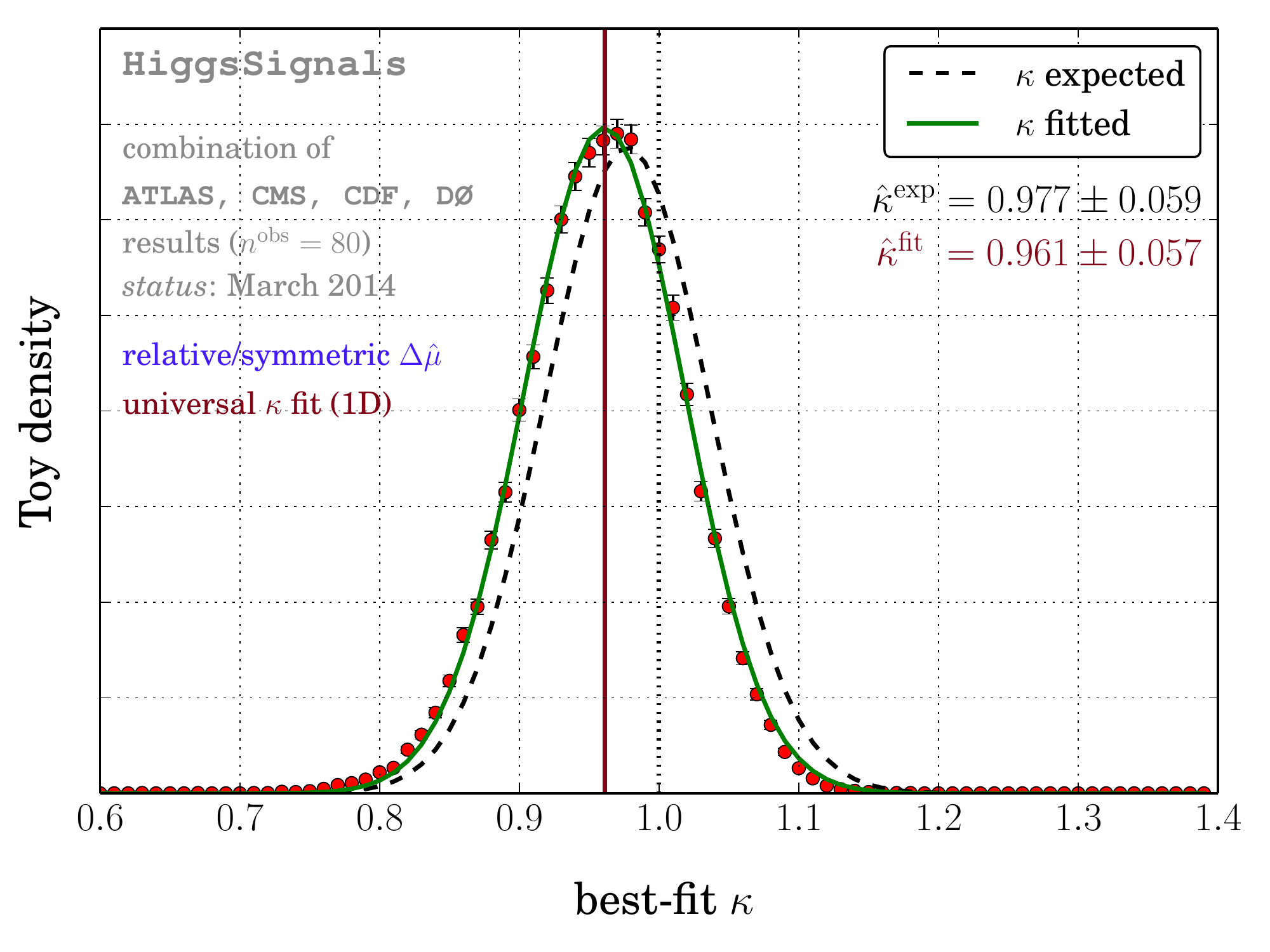}}
\subfigure[Absolute and asymmetric rate uncertainties.]{\includegraphics[width=0.48\textwidth]{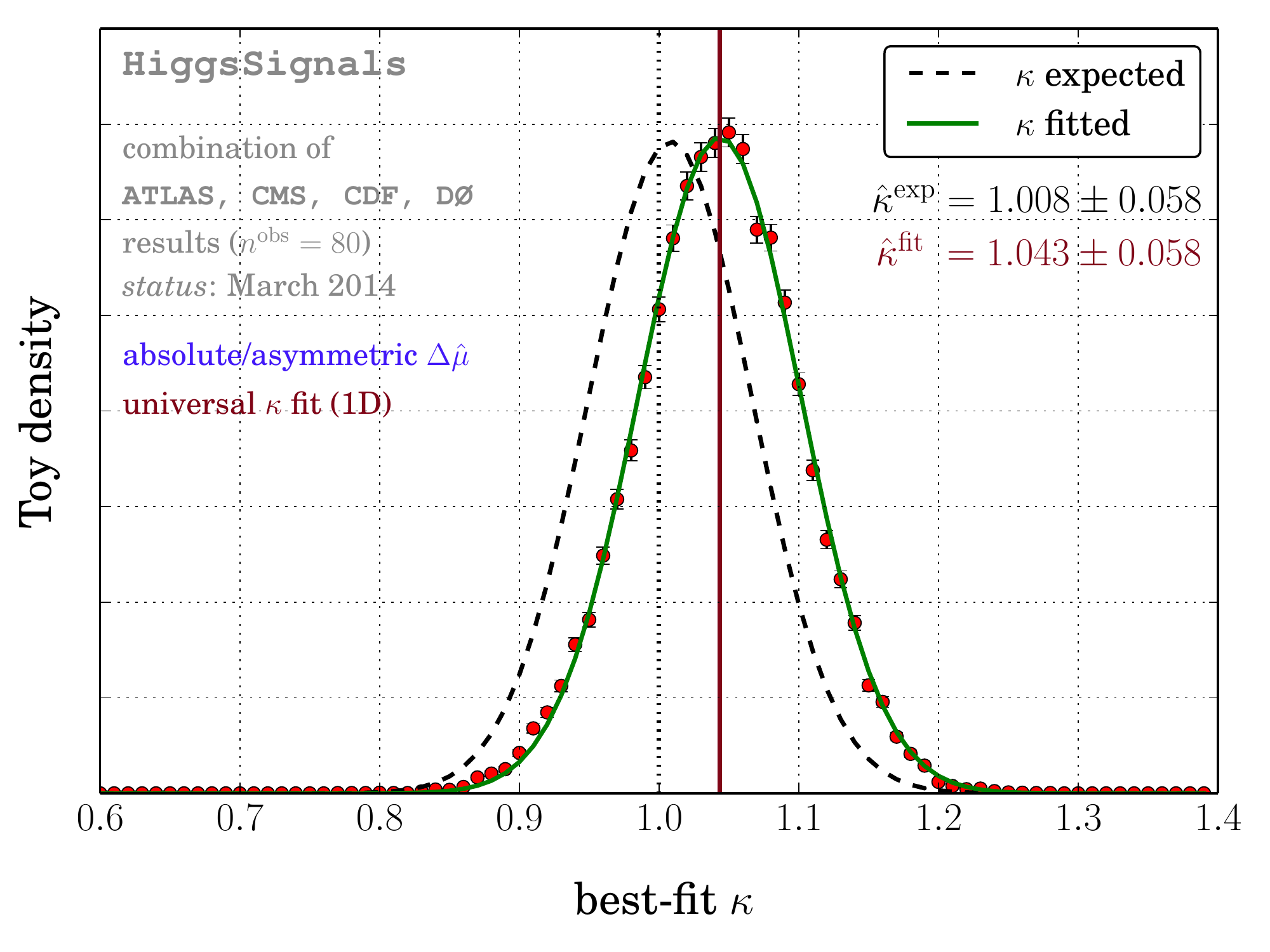}}
\subfigure[Relative and asymmetric rate uncertainties.]{\includegraphics[width=0.48\textwidth]{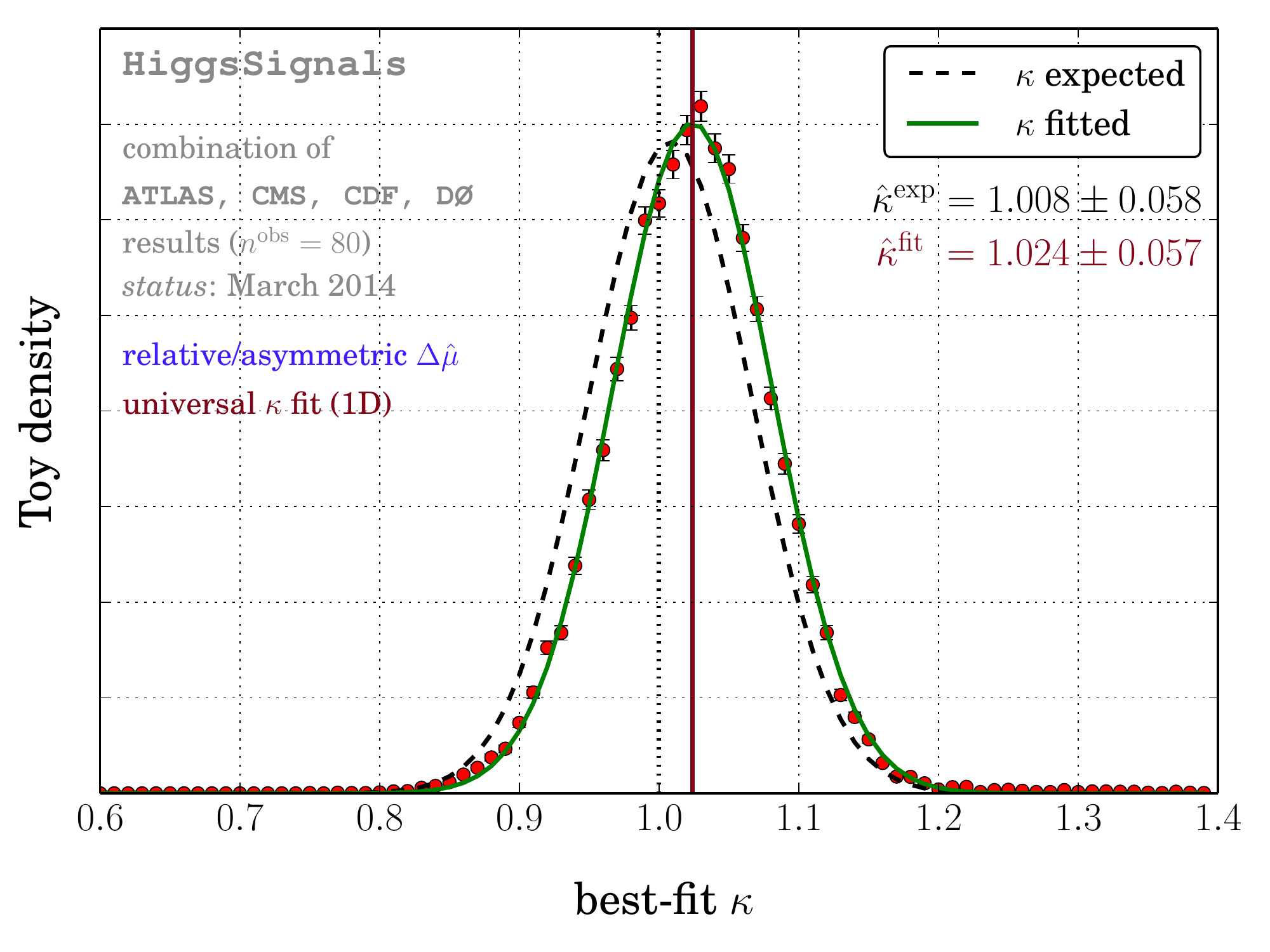}}
\caption{Best fit values $\mu$ of the same toy fits and \HS\ settings as discussed in Fig.~\ref{fig:1Dfit_toy_chisq_dist}.
  The black dashed line shows the expected Gaussian distribution for
  the original best fit point and $1\sigma$ uncertainties extracted at
  $\Delta \chi^2 = \chi^2 - \chi^2_\text{BF} =1$.  The solid, green
  curve shows the fit of a Gaussian to the toy outcomes.}
\label{fig:1Dfit_toy_mu_dist}
\end{figure}

As a simple generic toy model we employ a fit with only one free
parameter, namely a global Higgs coupling scale factor
$\kappa$ affecting all Higgs couplings to bosons and fermions in
  the same way, thus the SM predictions for the Higgs boson
  signal rates are universally scaled by $\kappa^2$.
The toy data is created
using the covariance matrix constructed under the principles outlined
above and evaluated at the best fit point. The resulting distributions
of the minimal $\chi^2$ from the toy experiments thrown around the
best fit point in $\mu$ is shown in
Fig.~\ref{fig:1Dfit_toy_chisq_dist}. In
Fig.~\ref{fig:1Dfit_toy_chisq_dist}~(a), the main effects leading to a
deviation from the naive $\chi^2$-distribution are deactivated:
Absolute rate uncertainties are used instead of relative ones for all
statistical and systematic errors, and the experimental uncertainties
are symmetrized. As expected, a nearly perfect $\chi^2$ shape is
obtained. The original best fit point is located at
$\kappa^\text{BF}=0.977$ with $\chi^2_\text{BF,abs/sym}=80.3$. The
\pvalue\ is given by the area under the obtained $\chi^2$ distribution
for $\chi^2 \ge \chi^2_\text{BF}$. In this treatment we obtain ${\cal
  P}^\text{BF}_\text{abs/sym}=48.7\%$, indicating very good
agreement of all Higgs rate measurements with the toy model chosen
here. Note, that the best fit point is extremely close to the SM (with
$\kappa=1$), which features a $\chi^2_\text{SM,abs/sym}=80.4$ in this
treatment and thus a very similar \pvalue.

The more realistic treatment of the uncertainties, however, has
significant impact on the \pvalue, as shown in
Fig.~\ref{fig:1Dfit_toy_chisq_dist}(d). The full model dependence of
the covariance matrix is used including relative errors and asymmetric
experimental uncertainties. This is the most accurate approximation to
the real likelihood distribution and thus provides a more accurate
guess of the \pvalue\ than the naive calculation above, where these
effects have been ignored. The result ${\cal
  P}^\text{BF}_\text{rel/asym}=40.0\%$ differs from the previously
obtained ${\cal P}^\text{BF}_\text{abs/sym}$. More importantly, the
shape of the histogram of the obtained $\chi^2_\text{min}$ values from the
toy fits does not follow an ideal $\chi^2$ distribution anymore.
More toy outcomes accumulate in the tail of the distribution at larger $\chi^2$ values, thus leading to a slightly improved \pvalue\ of the original best fit point than expected when assuming an ideal $\chi^2$ shape.
Toy MC studies like this will be of greater importance once the data
is more precise, and in particular if significant deviations between the SM and the data emerge. $\chi^2$ analyses that do not take into account the
effects described above might thus lead to conclusion significantly deviating from the full results. 

In order to show the origin of the deviation of the \pvalue\ from the
idealized implementation, Fig.~\ref{fig:1Dfit_toy_chisq_dist}(a), the
two major effects yielding deviations from the naive expectation are
singled out in Fig.~\ref{fig:1Dfit_toy_chisq_dist}(b) and (c). In
Fig.~\ref{fig:1Dfit_toy_chisq_dist}(b), only the effect of relative
errors, cf. item (2) above, is applied while the uncertainties are
kept symmetrized. It can be seen that the treatment of relative
uncertainties by itself has rather small effects. This is because the
preferred range of the global scale factor $\kappa$ is with
$\Delta\kappa \sim 6\,\%$ already quite narrow. Hence, $\kappa$ varies only in a
small range and the impact from uncertainties varying with $\kappa$ is
rather insignificant. However, the picture will change in more complex
models with more freedom in the variation of individual rates,
including some of the benchmark scale factor fits that are
discussed in Sect.~\ref{Sect:currentfits}.

In Fig.~\ref{fig:1Dfit_toy_chisq_dist}(c) the effect of asymmetric
errors, cf.~item (1) above, is studied. In this case we hold the
values of the uncertainties fixed for every toy measurement (absolute
uncertainties). It can be seen that for the \pvalue\ this effect fully
dominates the full implementation in
Fig.~\ref{fig:1Dfit_toy_chisq_dist}(d) and should not be omitted in
any implementation, since it could have a significant effect on the
conclusion.

In Fig.~\ref{fig:1Dfit_toy_mu_dist} we show the toy distribution
  of the best fit global scale factor $\kappa$ for the four different
  settings discussed above.
Again, Fig.~\ref{fig:1Dfit_toy_mu_dist}(a) shows the idealized
result with absolute and symmetrized uncertainties and (d) shows
the result from the full implementation of relative and
  asymmetric uncertainties. The same variations as explained for the
\pvalue\ can also be observed in the distribution of the best fit
points.
A small negative bias of about $-1.6\%$ in the universal coupling scale factor estimator
$\kappa$ is introduced by the relative uncertainties, as can be seen
in Fig.~\ref{fig:1Dfit_toy_mu_dist}(b). A much larger positive bias of the order of $3.5\%$,
however, results from the correct treatment of asymmetric errors,
cf.~Fig.~\ref{fig:1Dfit_toy_mu_dist}~(c). This stems from the fact
that the experimental uncertainties are typically larger for
variations in the upward direction as a direct consequence of the
likelihood shape. As expected the full result in
Fig.~\ref{fig:1Dfit_toy_mu_dist}(d) is in between (b) and (c) since
both biases apply, leading to an upward shift between expected and fitted universal scale factor of $\sim 1.6\%$. Note also, that the best-fit $\mu$ distribution
happens to be systematically slightly narrower than what is expected
from the naive $\chi^2$ comparison, but in this case by only
$\sigma_\text{fit}/\sigma_\text{exp}=0.057/0.058$, which corresponds to a change
of only $\lesssim 2\%$. 
We thus conclude that the Gaussian shape of the uncertainties is
approximately preserved, and that the uncertainties derived from the profile likelihood in the
main part of this paper are expected to be reliable estimates of the
uncertainties obtained in a full MC toy based treatment, or even the full
likelihood analysis in the experimental collaborations.

In summary, this simple toy model study shows that there are
potentially significant effects affecting the evaluation of \pvalue s
of arbitrary Higgs models tested against the signal rate
measurements. These effects stem from non-Gaussian likelihood effects
such as asymmetric uncertainties as well as different scaling behavior
of systematic uncertainties with either the measured or predicted
rates. Both effects are approximately accounted for in the $\chi^2$
evaluation in \HS, leading to an outcome that does not strictly follow
the naive expectation of an ideal $\chi^2$ probability distribution
with $n^\text{dof} = n^\text{obs} - n^\text{par}$ due to visible
changes in the $\chi^2_\text{min}$ probability density function. In a
detailed evaluation of the \pvalue\ we therefore advice to take these
effects into account by using toy experiments.


\section{Theoretical uncertainties of Higgs production and decay modes}
\label{Sect:AppTHU}

The (correlated) uncertainties of the Higgs production and decay rates induced by the dependence on (common) parameters are evaluated as follows. We introduce a random variable $x_i$ following a Normal distribution,
\begin{align}
P_i(x_i; \alpha) = \frac{1}{\sqrt{2\pi} \alpha} \cdot e^{-\frac{x_i^2}{2\alpha^2}},
\label{Eq:GaussianDist}
\end{align}
for each common parametric dependence $i$. In particular, the following common parametric dependencies are of relevance:
\begin{itemize}
\item $i \in \{\alpha_s, m_c, m_b, m_t\}$ for the partial width uncertainties of \textit{all} Higgs decay modes,
\item $i = \mathrm{PDF+\alpha_s}$ for the ggH and $t\bar{t}H$ cross section uncertainties,
\end{itemize}
The smearing of the common parameter, described by $x_i$, thus affects the resulting uncertainties of the corresponding production or decay modes in a fully correlated way (see also below).
For the remaining parametric dependencies $j$, individual Normal-distributed random variables $x_j^a$ are introduced per production or decay mode $a$, thus these uncertainty sources are regarded as uncorrelated. Similarly, the theoretical uncertainties corresponding to estimates of the missing higher-order corrections are described by individual (and thus uncorrelated) Normal-distributed variations, $x_\mathrm{th}^a$, except in the case of $WH$ and $ZH$ production which are treated as fully correlated.

In Eq.~\eqref{Eq:GaussianDist}, $\alpha$ is introduced as an artificial scale factor of the standard deviation of the parametric uncertainties. Usually, we choose $\alpha=1$, corresponding to a $68\%~\mathrm{C.L.}$ interpretation of the quoted uncertainties. For comparison, however, we define the setting `LHCHXSWG-matched', where we adjust $\alpha = 1.5~[1.7]$ for the cross section [partial width] uncertainties in order to approximately match with the uncertainty estimates given by the LHCHXSWG. Note that all (theoretical, correlated or uncorrelated parametric) variations are described by Eq.~\eqref{Eq:GaussianDist}, hence, for simplicity, the scale factor $\alpha$ affects all variations in the same way.

\begin{table}
\small
\centering
\begin{threeparttable}[b]
\begin{tabular}{l | c | c | ccc}
\toprule
Mode	&	LHCHXSWG\tnote{1}	&	LHCHXSWG	&	LHC-S1	&	LHC-S2		&	ILC	\\
		&	from Ref.~\cite{Heinemeyer:2013tqa}		&	 matched\tnote{2}	&			&			&		\\
\midrule
$\sigma(gg\to H)$ (ggH)		&	$15.3\%$	&	$15.6\%$	&	$10.4\%$	&	$5.2\%$	&	-	\\	
$\sigma(qq \to qqH)$ (VBF)	&	$6.9\%$	&	$5.6\%$	&	$3.7\%$	&	$1.9\%$	&	-	\\	
$\sigma(pp\to WH)$			&	$3.3\%$	&	$4.0\%$	&	$2.7\%$	&	$1.3\%$	&	-	\\	
$\sigma(\pptoZH)$			&	$5.7\%$	&	$6.3\%$	&	$4.2\%$	&	$2.1\%$	&	-	\\	
$\sigma(pp\to t\bar{t}H)$		&	$17.4\%$	&	$15.6\%$	&	$10.4\%$	&	$5.2\%$	&	-	\\	
\midrule
$\sigma(\epemtoZH)$			&	-	&	-	&	-	&	-		&	$0.5\%$	\\
$\sigma(\epem\to \nu\bar{\nu}H)$	&	-	&	-	&	-	&	-		&	$1.0\%$	\\
$\sigma(\epem\to t\bar{t}H)$		&	-	&	-	&	-	&	-		&	$1.0\%$	\\
\midrule
					\multicolumn{6}{c}{Using a Gaussian-shaped parameter variation}	\\
\midrule				
$\brhgaga$		&	$4.9\%$	&	$4.5\%$	&	$2.7\%$	&	$2.3\%$	&	$1.3\%$	\\
$\brhWW$		&	$4.2\%$	&	$4.3\%$	&	$2.5\%$	&	$2.3\%$	&	$1.3\%$	\\
$\brhZZ$			&	$4.1\%$	&	$4.3\%$	&	$2.5\%$	&	$2.3\%$	&	$1.3\%$	\\
$\brhtautau$ 		&	$5.7\%$	&	$5.3\%$	&	$3.1\%$	&	$2.4\%$	&	$1.6\%$	\\
$\brhbb$			&	$3.3\%$	&	$3.6\%$	&	$2.1\%$	&	$1.9\%$	&	$1.1\%$	\\
$\brhZga$			&	$8.9\%$	&	$9.5\%$	&	$5.6\%$	&	$3.4\%$	&	$2.8\%$	\\
$\brhcc$			&	$12.2\%$	&	$15.3\%$	&	$9.0\%$	&	$8.8\%$	&	$4.5\%$	\\
$\brhmumu$		&	$5.9\%$	&	$5.4\%$	&	$3.2\%$	&	$2.5\%$	&	$1.6\%$	\\
$\brhgg$			&	$10.1\%$	&	$10.9\%$	&	$6.4\%$	&	$5.9\%$	&	$3.2\%$	\\
\bottomrule
\end{tabular}
\begin{tablenotes}
\footnotesize
\item[1] Taken from Ref.~\cite{Heinemeyer:2013tqa}, using (always the larger) uncertainty estimates for $\sqrt{s}= 8\tev$, $m_H = \MHexp\gev$. Theoretical and parametric uncertainties are added linearly. In our \textit{naive} fit, we use these numbers as maximum error estimates and neglect all correlations of common parametric uncertainty sources, total width, etc..
\item[2] Using an artificially enlarged range for the parametric variation of $\alpha =1.5$ and $1.7$ for the production cross section and partial width uncertainties, respectively.
\end{tablenotes}
\end{threeparttable}
\caption{Relative theoretical uncertainties of LHC and ILC production cross sections and Higgs branching ratios for various implementations and future scenarios discussed in Tab.~\ref{Tab:scenarios}.}
\label{Tab:relativeuncertainties}
\end{table}

We now employ a Monte-Carlo (MC) calculation, where each iteration $k$ (also called ``toy'') is defined by throwing random numbers for the parametric and theoretical uncertainty variations, $k \equiv \{x_i, x_j^a, x_\mathrm{th}^a\}_k$. Then, the production cross sections and partial widths predicted for this toy are evaluated as
\begin{align}
\sigma^a_k &= \overline\sigma^a + \sum_{i} x_i \Delta\sigma^a_i + \sum_{j} x_j^a \Delta\sigma^a_j + x^a_\mathrm{th} \Delta\sigma^a_\mathrm{th},\\
\Gamma^a_k &= \overline\Gamma^a + \sum_{i} x_i \Delta\Gamma^a_i + \sum_{j} x_j^a \Delta\Gamma^a_j + x^a_\mathrm{th} \Delta\Gamma^a_\mathrm{th},
\end{align}
where $\overline\sigma^a$ and $\overline\Gamma^a$ are the central values of the production cross sections and partial widths in the SM, respectively, and $\Delta \sigma$ and $\Delta \Gamma$ their parametric or theoretical uncertainties as given in Ref.~\cite{Heinemeyer:2013tqa}. We take into account possibly asymmetric uncertainties:
\begin{align}
\Delta\sigma, \Delta\Gamma  = \left\{ \begin{array}{ll} \Delta\sigma^\mathrm{upper}, \Delta\Gamma^\mathrm{upper} & \mbox{for} \quad x > 0, \\ \Delta\sigma^\mathrm{lower}, \Delta\Gamma^\mathrm{lower} & \mbox{for} \quad x < 0. \end{array} \right.
\end{align}
Note, that $\Delta \sigma$ and $\Delta \Gamma$ can also be negative, depending on the response of the calculated quantity to the parameter variation. From the partial widths we can simply derive for each toy $k$ the total decay width, $\Gamma_k^\mathrm{tot} = \sum_a \Gamma_k^a$, and branching ratios, $\mathrm{BR}_k^a = \Gamma_k^a/\Gamma_k^\mathrm{tot}$. The covariance matrices are then given by
\begin{align}
\mathrm{cov}(X)_{ab} = \langle X_a X_b \rangle - \langle X_a \rangle \langle X_b \rangle,
\end{align}
where $\langle \cdot \rangle$ denotes the arithmetic mean for the full toy MC sample and $X = \sigma, \Gamma$ or $\mathrm{BR}$. 

In Tab.~\ref{Tab:relativeuncertainties} we give the relative uncorrelated uncertainties, given by $\mathrm{cov}(X)_{aa} / \overline{X_a}^2$, for the LHC production cross sections and Higgs branching ratios for the future scenarios `LHC-S1', `LHC-S2' and `ILC' discussed in Tab.~\ref{Tab:scenarios}. These are compared with the uncertainty estimates given by the LHCHXSWG~\cite{Heinemeyer:2013tqa}, where the parametric and theoretical uncertainties for a Higgs mass of $m_H = \MHexp\gev$ and a $(pp)$ center-of-mass energy of $\sqrt{s}=8\tev$ are added linearly. For the scenario `LHCHXSWG-matched' we employ the toy MC procedure with the artificial scale factor $\alpha = 1.5$ and $1.7$ for the uncertainties of the production cross sections and branching ratios, respectively. We furthermore compared these uncertainty estimates with those obtained when using a uniform (box-shaped) smearing of the parametric and theoretical uncertainties instead of \refeq{Eq:GaussianDist}. The deviations found are rather small, being typically $\lesssim\mathcal{O}(10\%)$.

\begin{figure}
\centering
\includegraphics[width=0.6\textwidth]{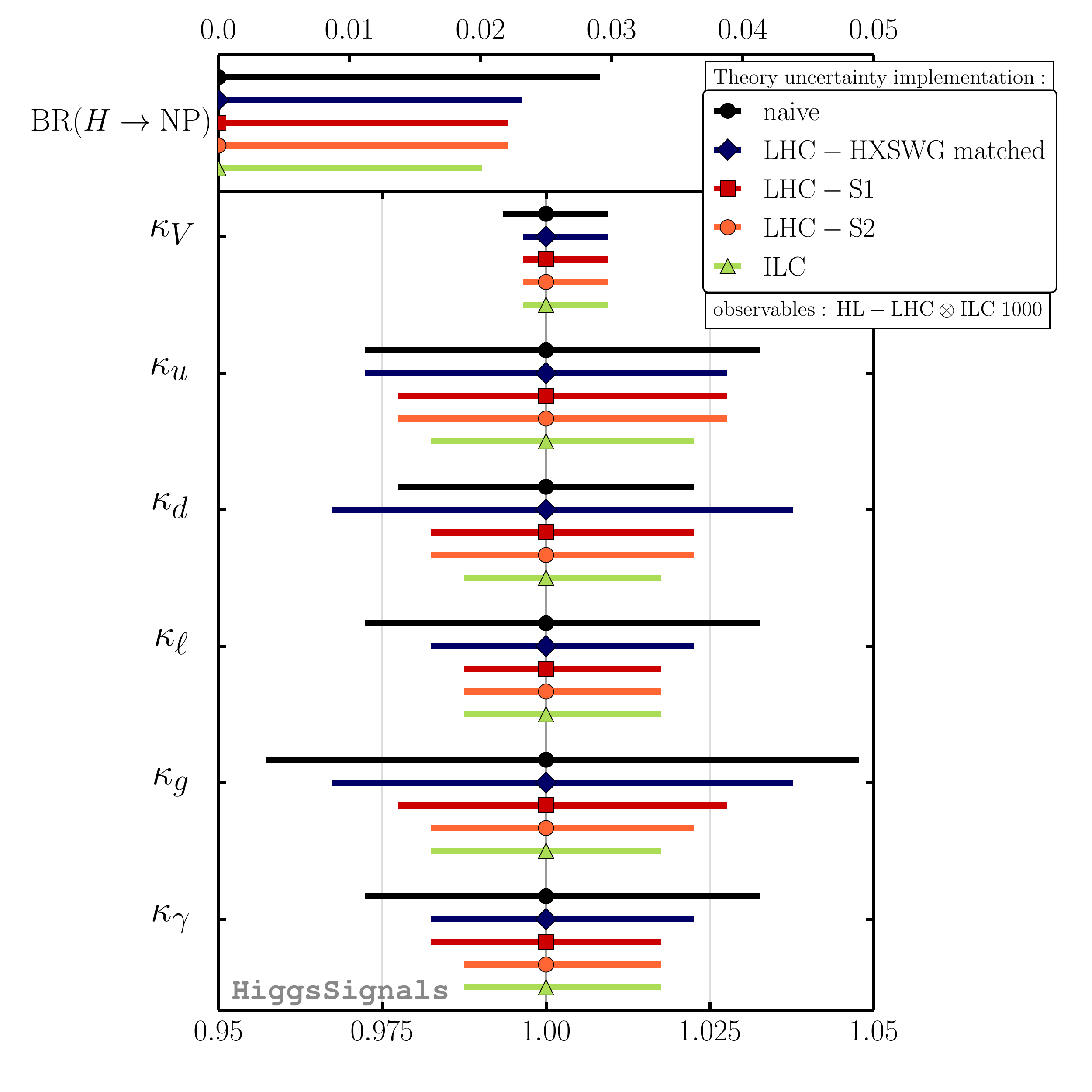}
\caption{Comparison of Higgs coupling precision estimates 
obtained for various implementations of theoretical rate
  uncertainties. The comparison uses all   available measurements from the HL--LHC and the ultimate ILC stage at
  $1\tev$ with $1\invab$ of data (and including measurements of previous
  ILC stages).}
\label{Fig:theo_err_comparison}
\end{figure}

In order to investigate the impact of the different theoretical
  uncertainty implementations on the precision estimates of the Higgs coupling scale factors we
  perform the seven-dimensional scale factor fit, cf.~\refse{Sect:7dim}
  and \ref{Sect:futureLHC}, to the same combined future projections
  for the high-luminosity LHC and all baseline ILC stages up to $1\tev$,
  $1\invab$, for all implementations. The result is shown in
  \reffi{Fig:theo_err_comparison}. Comparing the `naive' implementation,
  where simply the estimates from the LHCHXSWG are taken (cf.~Tab.~\ref{Tab:relativeuncertainties}) and all correlations among the cross section and branching ratio
  predictions are neglected, with the `LHC-HXSWG matched'
  implementation, we see that for the latter, $\brhnp$ and all scale
  factors except $\kd$ can be determined more precisely. Note, however,
  that the major effect causing these differences is actually the
  remaining mismatch of the uncertainty estimates,
  cf.~\refta{Tab:relativeuncertainties}, and not the inclusion of
  correlations. Nevertheless, as we have argued in this work, we find it
  more consistent to evaluate the covariances of the cross section and
  branching ratio predictions directly via the toy MC outlined above, leading to the
  uncertainty estimates denoted by `LHC-S1'. Here, we find the largest
  differences to the `naive' implementation in the achievable precisions
  of $\kd$, $\kl$, $\kg$ and $\kga$, being $\mathcal{O}(1-2\%)$. 
As expected, in the scenarios with improved theoretical uncertainties
the Higgs coupling precision is further improved, indicating that in
this high-statistics scenario the theoretical uncertainties are a
dominant limiting factor for the achievable precision.

\clearpage
\bibliographystyle{JHEP}
\bibliography{HiggsCouplings}

\providecommand{\href}[2]{#2}\begingroup\raggedright\begin{thebibliography}{100}

\bibitem{Aad:2012tfa}
{ATLAS Collaboration}, {\it {Observation of a new particle in the search for
  the Standard Model Higgs boson with the ATLAS detector at the LHC}},  {\em
  Phys.~Lett.~B} {\bf 716} (2012) 1,
  [\href{http://xxx.lanl.gov/abs/1207.7214}{{\tt arXiv:1207.7214}}].

\bibitem{Chatrchyan:2012ufa}
{CMS Collaboration}, {\it {Observation of a new boson at a mass of 125 GeV with
  the CMS experiment at the LHC}},  {\em Phys.~Lett.~B} {\bf 716} (2012) 30,
  [\href{http://xxx.lanl.gov/abs/1207.7235}{{\tt arXiv:1207.7235}}].

\bibitem{ATLAS:2013sla}
{ATLAS Collaboration}, {\it {Combined coupling measurements of the Higgs-like
  boson with the ATLAS detector using up to 25 fb$^{-1}$ of proton-proton
  collision data}},  \href{http://xxx.lanl.gov/abs/ATLAS-CONF-2013-034,
  ATLAS-COM-CONF-2013-035}{{\tt ATLAS-CONF-2013-034, ATLAS-COM-CONF-2013-035}}.

\bibitem{ATLAS:2013mla}
{ATLAS Collaboration}, {\it {Study of the spin of the new boson with up to
  25~fb$^{-1}$ of ATLAS data}},
  \href{http://xxx.lanl.gov/abs/ATLAS-CONF-2013-040,
  ATLAS-COM-CONF-2013-048}{{\tt ATLAS-CONF-2013-040, ATLAS-COM-CONF-2013-048}}.

\bibitem{CMS:yva}
{CMS Collaboration}, {\it {Combination of standard model Higgs boson searches
  and measurements of the properties of the new boson with a mass near 125
  GeV}},  \href{http://xxx.lanl.gov/abs/CMS-PAS-HIG-13-005}{{\tt
  CMS-PAS-HIG-13-005}}.

\bibitem{CMS:2013wda}
{CMS Collaboration}, {\it {Properties of the observed Higgs-like resonance
  using the diphoton channel}},
  \href{http://xxx.lanl.gov/abs/CMS-PAS-HIG-13-016}{{\tt CMS-PAS-HIG-13-016}}.

\bibitem{Aaltonen:2013kxa}
{CDF and D{\O} Collaborations}, {\it {Higgs Boson Studies at the Tevatron}},
  {\em Phys.~Rev.~D} {\bf 88} (2013) 052014,
  [\href{http://xxx.lanl.gov/abs/1303.6346}{{\tt arXiv:1303.6346}}].

\bibitem{LHCHiggsCrossSectionWorkingGroup:2012nn}
A.~David, A.~Denner, M.~Duehrssen, M.~Grazzini, et~al., {\it {LHC HXSWG interim
  recommendations to explore the coupling structure of a Higgs-like particle}},
   \href{http://xxx.lanl.gov/abs/1209.0040}{{\tt arXiv:1209.0040}}.

\bibitem{Heinemeyer:2013tqa}
S.~Heinemeyer et~al., {\it {Handbook of LHC Higgs Cross Sections: 3. Higgs
  Properties}},  \href{http://xxx.lanl.gov/abs/1307.1347}{{\tt
  arXiv:1307.1347}}.

\bibitem{Zeppenfeld:2000td}
D.~Zeppenfeld, R.~Kinnunen, A.~Nikitenko, and E.~Richter-Was, {\it {Measuring
  Higgs boson couplings at the CERN LHC}},  {\em Phys.~Rev.~D} {\bf 62} (2000)
  013009, [\href{http://xxx.lanl.gov/abs/hep-ph/0002036}{{\tt
  hep-ph/0002036}}].

\bibitem{Duhrssen:2003tba}
M.~Duhrssen, {\it {Prospects for the measurement of Higgs boson coupling
  parameters in the mass range from 110--190 GeV}},
  \href{http://xxx.lanl.gov/abs/ATL-PHYS-2003-030}{{\tt ATL-PHYS-2003-030}}.

\bibitem{Duhrssen:2004cv}
M.~Duhrssen, S.~Heinemeyer, H.~Logan, D.~Rainwater, G.~Weiglein, et~al., {\it
  {Extracting Higgs boson couplings from CERN LHC data}},  {\em Phys.~Rev.~D}
  {\bf 70} (2004) 113009, [\href{http://xxx.lanl.gov/abs/hep-ph/0406323}{{\tt
  hep-ph/0406323}}].

\bibitem{Duhrssen:2004uu}
M.~Duhrssen, S.~Heinemeyer, H.~Logan, D.~Rainwater, G.~Weiglein, et~al., {\it
  {Determination of Higgs-boson couplings at the LHC}},
  \href{http://xxx.lanl.gov/abs/hep-ph/0407190}{{\tt hep-ph/0407190}}.

\bibitem{Lafaye:2009vr}
R.~Lafaye, T.~Plehn, M.~Rauch, D.~Zerwas, and M.~Duhrssen, {\it {Measuring the
  Higgs Sector}},  {\em JHEP} {\bf 0908} (2009) 009,
  [\href{http://xxx.lanl.gov/abs/0904.3866}{{\tt arXiv:0904.3866}}].

\bibitem{Klute:2012pu}
M.~Klute, R.~Lafaye, T.~Plehn, M.~Rauch, and D.~Zerwas, {\it {Measuring Higgs
  Couplings from LHC Data}},  {\em Phys.~Rev.~Lett.} {\bf 109} (2012) 101801,
  [\href{http://xxx.lanl.gov/abs/1205.2699}{{\tt arXiv:1205.2699}}].

\bibitem{Plehn:2012iz}
T.~Plehn and M.~Rauch, {\it {Higgs Couplings after the Discovery}},  {\em
  Europhys.~Lett.} {\bf 100} (2012) 11002,
  [\href{http://xxx.lanl.gov/abs/1207.6108}{{\tt arXiv:1207.6108}}].

\bibitem{Klute:2013cx}
M.~Klute, R.~Lafaye, T.~Plehn, M.~Rauch, and D.~Zerwas, {\it {Measuring Higgs
  Couplings at a Linear Collider}},  {\em Europhys.~Lett.} {\bf 101} (2013)
  51001, [\href{http://xxx.lanl.gov/abs/1301.1322}{{\tt arXiv:1301.1322}}].

\bibitem{Dobrescu:2012td}
B.~A. Dobrescu and J.~D. Lykken, {\it {Coupling spans of the Higgs-like
  boson}},  {\em JHEP} {\bf 1302} (2013) 073,
  [\href{http://xxx.lanl.gov/abs/1210.3342}{{\tt arXiv:1210.3342}}].

\bibitem{Espinosa:2012im}
J.~Espinosa, C.~Grojean, M.~M{\"u}hlleitner, and M.~Trott, {\it {First Glimpses
  at Higgs' face}},  {\em JHEP} {\bf 1212} (2012) 045,
  [\href{http://xxx.lanl.gov/abs/1207.1717}{{\tt arXiv:1207.1717}}].

\bibitem{Cacciapaglia:2012wb}
G.~Cacciapaglia, A.~Deandrea, G.~D. La~Rochelle, and J.-B. Flament, {\it {Higgs
  couplings beyond the Standard Model}},  {\em JHEP} {\bf 1303} (2013) 029,
  [\href{http://xxx.lanl.gov/abs/1210.8120}{{\tt arXiv:1210.8120}}].

\bibitem{Belanger:2012gc}
G.~Belanger, B.~Dumont, U.~Ellwanger, J.~Gunion, and S.~Kraml, {\it {Higgs
  Couplings at the End of 2012}},  {\em JHEP} {\bf 1302} (2013) 053,
  [\href{http://xxx.lanl.gov/abs/1212.5244}{{\tt arXiv:1212.5244}}].

\bibitem{Ellis:2013lra}
J.~Ellis and T.~You, {\it {Updated Global Analysis of Higgs Couplings}},  {\em
  JHEP} {\bf 1306} (2013) 103, [\href{http://xxx.lanl.gov/abs/1303.3879}{{\tt
  arXiv:1303.3879}}].

\bibitem{Djouadi:2013qya}
A.~Djouadi and G.~Moreau, {\it {The couplings of the Higgs boson and its CP
  properties from fits of the signal strengths and their ratios at the 7+8 TeV
  LHC}},  \href{http://xxx.lanl.gov/abs/1303.6591}{{\tt arXiv:1303.6591}}.

\bibitem{Cheung:2013kla}
K.~Cheung, J.~S. Lee, and P.-Y. Tseng, {\it {Higgs Precision (Higgcision) Era
  begins}},  {\em JHEP} {\bf 1305} (2013) 134,
  [\href{http://xxx.lanl.gov/abs/1302.3794}{{\tt arXiv:1302.3794}}].

\bibitem{Holdom:2013axa}
B.~Holdom, {\it {Far from standard Higgs couplings}},
  \href{http://xxx.lanl.gov/abs/1306.1564}{{\tt arXiv:1306.1564}}.

\bibitem{Chpoi:2013wga}
S.~Choi, S.~Jung, and P.~Ko, {\it {Implications of LHC data on 125 GeV
  Higgs-like boson for the Standard Model and its various extensions}},  {\em
  JHEP} {\bf 1310} (2013) 225, [\href{http://xxx.lanl.gov/abs/1307.3948}{{\tt
  arXiv:1307.3948}}].

\bibitem{Bechtle:2013xfa}
P.~Bechtle, S.~Heinemeyer, O.~St{\aa}l, T.~Stefaniak, and G.~Weiglein, {\it
  {HiggsSignals: Confronting arbitrary Higgs sectors with measurements at the
  Tevatron and the LHC}},  {\em Eur.~Phys.~J.~C} {\bf 74} (2013) 2711,
  [\href{http://xxx.lanl.gov/abs/1305.1933}{{\tt arXiv:1305.1933}}].

\bibitem{Belanger:2013xza}
G.~Belanger, B.~Dumont, U.~Ellwanger, J.~Gunion, and S.~Kraml, {\it {Global fit
  to Higgs signal strengths and couplings and implications for extended Higgs
  sectors}},  {\em Phys.~Rev.~D} {\bf 88} (2013) 075008,
  [\href{http://xxx.lanl.gov/abs/1306.2941}{{\tt arXiv:1306.2941}}].

\bibitem{Caola:2013yja}
F.~Caola and K.~Melnikov, {\it {Constraining the Higgs boson width with ZZ
  production at the LHC}},  {\em Phys.~Rev.~D} {\bf 88} (2013) 054024,
  [\href{http://xxx.lanl.gov/abs/1307.4935}{{\tt arXiv:1307.4935}}].

\bibitem{Campbell:2011cu}
J.~M. Campbell, R.~K. Ellis, and C.~Williams, {\it {Gluon-Gluon Contributions
  to W+ W- Production and Higgs Interference Effects}},  {\em JHEP} {\bf 1110}
  (2011) 005, [\href{http://xxx.lanl.gov/abs/1107.5569}{{\tt
  arXiv:1107.5569}}].

\bibitem{Campbell:2013wga}
J.~M. Campbell, R.~K. Ellis, and C.~Williams, {\it {Bounding the Higgs width at
  the LHC: complementary results from $H \to WW$}},
  \href{http://xxx.lanl.gov/abs/1312.1628}{{\tt arXiv:1312.1628}}.

\bibitem{Campbell:2013una}
J.~M. Campbell, R.~K. Ellis, and C.~Williams, {\it {Bounding the Higgs width at
  the LHC using full analytic results for $gg \to 2e 2\mu$}},
  \href{http://xxx.lanl.gov/abs/1311.3589}{{\tt arXiv:1311.3589}}.

\bibitem{Dixon:2013haa}
L.~J. Dixon and Y.~Li, {\it {Bounding the Higgs Boson Width Through
  Interferometry}},  {\em Phys.~Rev.~Lett.} {\bf 111} (2013) 111802,
  [\href{http://xxx.lanl.gov/abs/1305.3854}{{\tt arXiv:1305.3854}}].

\bibitem{Buckley:2012em}
M.~R. Buckley and D.~Hooper, {\it {Are There Hints of Light Stops in Recent
  Higgs Search Results?}},  {\em Phys.~Rev.~D} {\bf 86} (2012) 075008,
  [\href{http://xxx.lanl.gov/abs/1207.1445}{{\tt arXiv:1207.1445}}].

\bibitem{Arbey:2012na}
A.~Arbey, M.~Battaglia, and F.~Mahmoudi, {\it {Light Neutralino Dark Matter in
  the pMSSM: Implications of LEP, LHC and Dark Matter Searches on SUSY Particle
  Spectra}},  {\em Eur.~Phys.~J.~C} {\bf 72} (2012) 2169,
  [\href{http://xxx.lanl.gov/abs/1205.2557}{{\tt arXiv:1205.2557}}].

\bibitem{Arbey:2012dq}
A.~Arbey, M.~Battaglia, A.~Djouadi, and F.~Mahmoudi, {\it {The Higgs sector of
  the phenomenological MSSM in the light of the Higgs boson discovery}},  {\em
  JHEP} {\bf 1209} (2012) 107, [\href{http://xxx.lanl.gov/abs/1207.1348}{{\tt
  arXiv:1207.1348}}].

\bibitem{Akula:2012kk}
S.~Akula, P.~Nath, and G.~Peim, {\it {Implications of the Higgs Boson Discovery
  for mSUGRA}},  {\em Phys.~Lett.~B} {\bf 717} (2012) 188,
  [\href{http://xxx.lanl.gov/abs/1207.1839}{{\tt arXiv:1207.1839}}].

\bibitem{Cao:2012yn}
J.~Cao, Z.~Heng, J.~M. Yang, and J.~Zhu, {\it {Status of low energy SUSY models
  confronted with the LHC 125 GeV Higgs data}},  {\em JHEP} {\bf 1210} (2012)
  079, [\href{http://xxx.lanl.gov/abs/1207.3698}{{\tt arXiv:1207.3698}}].

\bibitem{Howe:2012xe}
K.~Howe and P.~Saraswat, {\it {Excess Higgs Production in Neutralino Decays}},
  {\em JHEP} {\bf 1210} (2012) 065,
  [\href{http://xxx.lanl.gov/abs/1208.1542}{{\tt arXiv:1208.1542}}].

\bibitem{Drees:2012fb}
M.~Drees, {\it {A Supersymmetric Explanation of the Excess of Higgs--Like
  Events at the LHC and at LEP}},  {\em Phys.~Rev.~D} {\bf 86} (2012) 115018,
  [\href{http://xxx.lanl.gov/abs/1210.6507}{{\tt arXiv:1210.6507}}].

\bibitem{Haisch:2012re}
U.~Haisch and F.~Mahmoudi, {\it {MSSM: Cornered and Correlated}},  {\em JHEP}
  {\bf 1301} (2013) 061, [\href{http://xxx.lanl.gov/abs/1210.7806}{{\tt
  arXiv:1210.7806}}].

\bibitem{Bechtle:2012jw}
P.~Bechtle, S.~Heinemeyer, O.~St{\aa}l, T.~Stefaniak, G.~Weiglein, and
  L.~Zeune, {\it {MSSM Interpretations of the LHC Discovery: Light or Heavy
  Higgs?}},  {\em Eur.~Phys.~J.~C} {\bf 73} (2013) 2354,
  [\href{http://xxx.lanl.gov/abs/1211.1955}{{\tt arXiv:1211.1955}}].

\bibitem{Arbey:2012bp}
A.~Arbey, M.~Battaglia, A.~Djouadi, and F.~Mahmoudi, {\it {An update on the
  constraints on the phenomenological MSSM from the new LHC Higgs results}},
  {\em Phys.~Lett.~B} {\bf 720} (2013) 153,
  [\href{http://xxx.lanl.gov/abs/1211.4004}{{\tt arXiv:1211.4004}}].

\bibitem{Ke:2012zq}
J.~Ke, H.~Luo, M.-x. Luo, K.~Wang, L.~Wang, et~al., {\it {Revisit to
  Non-decoupling MSSM}},  {\em Phys.~Lett.~B} {\bf 723} (2013) 113,
  [\href{http://xxx.lanl.gov/abs/1211.2427}{{\tt arXiv:1211.2427}}].

\bibitem{Chakraborty:2013si}
A.~Chakraborty, B.~Das, J.~L. Diaz-Cruz, D.~K. Ghosh, S.~Moretti, et~al., {\it
  {The 125 GeV Higgs signal at the LHC in the CP Violating MSSM}},
  \href{http://xxx.lanl.gov/abs/1301.2745}{{\tt arXiv:1301.2745}}.

\bibitem{Carmona:2013cq}
A.~Carmona and F.~Goertz, {\it {Custodial Leptons and Higgs Decays}},  {\em
  JHEP} {\bf 1304} (2013) 163, [\href{http://xxx.lanl.gov/abs/1301.5856}{{\tt
  arXiv:1301.5856}}].

\bibitem{Arbey:2013jla}
A.~Arbey, M.~Battaglia, and F.~Mahmoudi, {\it {Supersymmetric Heavy Higgs
  Bosons at the LHC}},  {\em Phys.~Rev.~D} {\bf 88} (2013) 015007,
  [\href{http://xxx.lanl.gov/abs/1303.7450}{{\tt arXiv:1303.7450}}].

\bibitem{Cao:2013wqa}
J.~Cao, P.~Wan, J.~M. Yang, and J.~Zhu, {\it {The SM extension with color-octet
  scalars: diphoton enhancement and global fit of LHC Higgs data}},  {\em JHEP}
  {\bf 1308} (2013) 009, [\href{http://xxx.lanl.gov/abs/1303.2426}{{\tt
  arXiv:1303.2426}}].

\bibitem{Bhattacherjee:2013vga}
B.~Bhattacherjee, M.~Chakraborti, A.~Chakraborty, U.~Chattopadhyay, D.~Das,
  et~al., {\it {Implications of 98 GeV and 125 GeV Higgs scenario in
  non-decoupling SUSY with updated ATLAS, CMS and PLANCK data}},  {\em
  Phys.~Rev.~D} {\bf 88} (2013) 035011,
  [\href{http://xxx.lanl.gov/abs/1305.4020}{{\tt arXiv:1305.4020}}].

\bibitem{Lopez-Val:2013yba}
D.~Lopez-Val, T.~Plehn, and M.~Rauch, {\it {Measuring extended Higgs sectors as
  a consistent free couplings model}},  {\em JHEP} {\bf 1310} (2013) 134,
  [\href{http://xxx.lanl.gov/abs/1308.1979}{{\tt arXiv:1308.1979}}].

\bibitem{Belyaev:2013ida}
A.~Belyaev, M.~S. Brown, R.~Foadi, and M.~T. Frandsen, {\it {The Technicolor
  Higgs in the Light of LHC Data}},
  \href{http://xxx.lanl.gov/abs/1309.2097}{{\tt arXiv:1309.2097}}.

\bibitem{Cao:2013gba}
J.~Cao, F.~Ding, C.~Han, J.~M. Yang, and J.~Zhu, {\it {A light Higgs scalar in
  the NMSSM confronted with the latest LHC Higgs data}},  {\em JHEP} {\bf 1311}
  (2013) 018, [\href{http://xxx.lanl.gov/abs/1309.4939}{{\tt
  arXiv:1309.4939}}].

\bibitem{Bharucha:2013ela}
A.~Bharucha, A.~Goudelis, and M.~McGarrie, {\it {En-gauging Naturalness}},
  \href{http://xxx.lanl.gov/abs/1310.4500}{{\tt arXiv:1310.4500}}.

\bibitem{Cheung:2013rva}
K.~Cheung, J.~S. Lee, and P.-Y. Tseng, {\it {Higgcision in the Two-Higgs
  Doublet Models}},  {\em JHEP} {\bf 1401} (2014) 085,
  [\href{http://xxx.lanl.gov/abs/1310.3937}{{\tt arXiv:1310.3937}}].

\bibitem{Bechtle:2013mda}
P.~Bechtle, K.~Desch, H.~K. Dreiner, M.~Hamer, M.~Kr{\"a}mer, et~al., {\it
  {Constrained Supersymmetry after the Higgs Boson Discovery: A global analysis
  with Fittino}},  \href{http://xxx.lanl.gov/abs/1310.3045}{{\tt
  arXiv:1310.3045}}.

\bibitem{Cao:2013mqa}
J.~Cao, C.~Han, L.~Wu, P.~Wu, and J.~M. Yang, {\it {A light SUSY dark matter
  after CDMS-II, LUX and LHC Higgs data}},
  \href{http://xxx.lanl.gov/abs/1311.0678}{{\tt arXiv:1311.0678}}.

\bibitem{Djouadi:2013lra}
A.~Djouadi, {\it {Implications of the Higgs discovery for the MSSM}},
  \href{http://xxx.lanl.gov/abs/1311.0720}{{\tt arXiv:1311.0720}}.

\bibitem{Enberg:2013jba}
R.~Enberg, J.~Rathsman, and G.~Wouda, {\it {Higgs phenomenology in the Stealth
  Doublet Model}},  \href{http://xxx.lanl.gov/abs/1311.4367}{{\tt
  arXiv:1311.4367}}.

\bibitem{Cao:2013cfa}
J.~Cao, Y.~He, P.~Wu, M.~Zhang, and J.~Zhu, {\it {Higgs Phenomenology in the
  Minimal Dilaton Model after Run I of the LHC}},  {\em JHEP} {\bf 1401} (2014)
  150, [\href{http://xxx.lanl.gov/abs/1311.6661}{{\tt arXiv:1311.6661}}].

\bibitem{Wang:2013sha}
L.~Wang and X.-F. Han, {\it {Status of the aligned two-Higgs-doublet model
  confronted with the Higgs data}},
  \href{http://xxx.lanl.gov/abs/1312.4759}{{\tt arXiv:1312.4759}}.

\bibitem{Fan:2014txa}
J.~Fan and M.~Reece, {\it {A New Look at Higgs Constraints on Stops}},
  \href{http://xxx.lanl.gov/abs/1401.7671}{{\tt arXiv:1401.7671}}.

\bibitem{Belanger:2014roa}
G.~Belanger, V.~Bizouard, and G.~Chalons, {\it {Boosting Higgs decays into
  gamma and a Z in the NMSSM}},  \href{http://xxx.lanl.gov/abs/1402.3522}{{\tt
  arXiv:1402.3522}}.

\bibitem{Stal:2014sua}
O.~St{\aa}l, {\it {Prospects for Higgs boson scenarios beyond the Standard
  Model}},  \href{http://xxx.lanl.gov/abs/1402.6732}{{\tt arXiv:1402.6732}}.

\bibitem{Stal:2013hwa}
O.~St{\aa}l and T.~Stefaniak, {\it {Constraining extended Higgs sectors with
  HiggsSignals}},  {\em PoS} {\bf EPS-HEP} (2013) 314,
  [\href{http://xxx.lanl.gov/abs/1310.4039}{{\tt arXiv:1310.4039}}].

\bibitem{Cornwall:1973tb}
J.~M. Cornwall, D.~N. Levin, and G.~Tiktopoulos, {\it {Uniqueness of
  spontaneously broken gauge theories}},  {\em Phys.~Rev.~Lett.} {\bf 30}
  (1973) 1268.

\bibitem{LlewellynSmith:1973ey}
C.~Llewellyn~Smith, {\it {High-Energy Behavior and Gauge Symmetry}},  {\em
  Phys.~Lett.~B} {\bf 46} (1973) 233.

\bibitem{Dittmaier:2011ti}
S.~Dittmaier et~al., {\it {Handbook of LHC Higgs Cross Sections: 1. Inclusive
  Observables}},  \href{http://xxx.lanl.gov/abs/1101.0593}{{\tt
  arXiv:1101.0593}}.

\bibitem{Dittmaier:2012vm}
S.~Dittmaier et~al., {\it {Handbook of LHC Higgs Cross Sections: 2.
  Differential Distributions}},  \href{http://xxx.lanl.gov/abs/1201.3084}{{\tt
  arXiv:1201.3084}}.

\bibitem{Harlander:2013mla}
R.~V. Harlander, S.~Liebler, and T.~Zirke, {\it {Higgs Strahlung at the Large
  Hadron Collider in the 2-Higgs-Doublet Model}},
  \href{http://xxx.lanl.gov/abs/1307.8122}{{\tt arXiv:1307.8122}}.

\bibitem{Eboli:2000ze}
O.~J. Eboli and D.~Zeppenfeld, {\it {Observing an invisible Higgs boson}},
  {\em Phys.Lett.} {\bf B495} (2000) 147--154,
  [\href{http://xxx.lanl.gov/abs/hep-ph/0009158}{{\tt hep-ph/0009158}}].

\bibitem{Bechtle:2008jh}
P.~Bechtle, O.~Brein, S.~Heinemeyer, G.~Weiglein, and K.~E. Williams, {\it
  {HiggsBounds: Confronting Arbitrary Higgs Sectors with Exclusion Bounds from
  LEP and the Tevatron}},  {\em Comput.~Phys.~Commun.} {\bf 181} (2010) 138,
  [\href{http://xxx.lanl.gov/abs/0811.4169}{{\tt arXiv:0811.4169}}].

\bibitem{Bechtle:2011sb}
P.~Bechtle, O.~Brein, S.~Heinemeyer, G.~Weiglein, and K.~E. Williams, {\it
  {HiggsBounds 2.0.0: Confronting Neutral and Charged Higgs Sector Predictions
  with Exclusion Bounds from LEP and the Tevatron}},  {\em
  Comput.~Phys.~Commun.} {\bf 182} (2011) 2605,
  [\href{http://xxx.lanl.gov/abs/1102.1898}{{\tt arXiv:1102.1898}}].

\bibitem{Bechtle:2013gu}
P.~Bechtle, O.~Brein, S.~Heinemeyer, O.~St{\aa}l, T.~Stefaniak, et~al., {\it
  {Recent Developments in HiggsBounds and a Preview of HiggsSignals}},  {\em
  PoS} {\bf CHARGED2012} (2012) 024,
  [\href{http://xxx.lanl.gov/abs/1301.2345}{{\tt arXiv:1301.2345}}].

\bibitem{Bechtle:2013wla}
P.~Bechtle, O.~Brein, S.~Heinemeyer, O.~St{\aa}l, T.~Stefaniak, et~al., {\it
  {HiggsBounds-4: Improved Tests of Extended Higgs Sectors against Exclusion
  Bounds from LEP, the Tevatron and the LHC}},  {\em Eur.~Phys.~J.~C} {\bf 74}
  (2013) 2693, [\href{http://xxx.lanl.gov/abs/1311.0055}{{\tt
  arXiv:1311.0055}}].

\bibitem{Banerjee:2013apa}
S.~Banerjee, S.~Mukhopadhyay, and B.~Mukhopadhyaya, {\it {Higher dimensional
  operators and LHC Higgs data : the role of modified kinematics}},
  \href{http://xxx.lanl.gov/abs/1308.4860}{{\tt arXiv:1308.4860}}.

\bibitem{Anderson:2013afp}
I.~Anderson, S.~Bolognesi, F.~Caola, Y.~Gao, A.~V. Gritsan, et~al., {\it
  {Constraining anomalous HVV interactions at proton and lepton colliders}},
  \href{http://xxx.lanl.gov/abs/1309.4819}{{\tt arXiv:1309.4819}}.

\bibitem{Azatov:2013xha}
A.~Azatov and A.~Paul, {\it {Probing Higgs couplings with high $p_T$ Higgs
  production}},  \href{http://xxx.lanl.gov/abs/1309.5273}{{\tt
  arXiv:1309.5273}}.

\bibitem{Boos:2013mqa}
E.~Boos, V.~Bunichev, M.~Dubinin, and Y.~Kurihara, {\it {Higgs boson signal at
  complete tree level in the SM extension by dimension-six operators}},
  \href{http://xxx.lanl.gov/abs/1309.5410}{{\tt arXiv:1309.5410}}.

\bibitem{Chen:2013waa}
M.~Chen, T.~Cheng, J.~S. Gainer, A.~Korytov, K.~T. Matchev, et~al., {\it {The
  role of interference in unraveling the ZZ-couplings of the newly discovered
  boson at the LHC}},  \href{http://xxx.lanl.gov/abs/1310.1397}{{\tt
  arXiv:1310.1397}}.

\bibitem{Buchalla:2013mpa}
G.~Buchalla, O.~Cata, and G.~D'Ambrosio, {\it {Nonstandard Higgs Couplings from
  Angular Distributions in $h\to Z \ell^+\ell^-$}},
  \href{http://xxx.lanl.gov/abs/1310.2574}{{\tt arXiv:1310.2574}}.

\bibitem{Dumont:2013wma}
B.~Dumont, S.~Fichet, and G.~von Gersdorff, {\it {A Bayesian view of the Higgs
  sector with higher dimensional operators}},  {\em JHEP} {\bf 1307} (2013)
  065, [\href{http://xxx.lanl.gov/abs/1304.3369}{{\tt arXiv:1304.3369}}].

\bibitem{Buchmuller:1985jz}
W.~Buchm{\"u}ller and D.~Wyler, {\it {Effective Lagrangian Analysis of New
  Interactions and Flavor Conservation}},  {\em Nucl.~Phys.~B} {\bf 268} (1986)
  621.

\bibitem{Contino:2013kra}
R.~Contino, M.~Ghezzi, C.~Grojean, M.~M{\"u}hlleitner, and M.~Spira, {\it
  {Effective Lagrangian for a light Higgs-like scalar}},  {\em JHEP} {\bf 1307}
  (2013) 035, [\href{http://xxx.lanl.gov/abs/1303.3876}{{\tt
  arXiv:1303.3876}}].

\bibitem{Pomarol:2013zra}
A.~Pomarol and F.~Riva, {\it {Towards the Ultimate SM Fit to Close in on Higgs
  Physics}},  {\em JHEP} {\bf 1401} (2014) 151,
  [\href{http://xxx.lanl.gov/abs/1308.2803}{{\tt arXiv:1308.2803}}].

\bibitem{Boudjema:2013qla}
F.~Boudjema, G.~Cacciapaglia, K.~Cranmer, G.~Dissertori, A.~Deandrea, et~al.,
  {\it {On the presentation of the LHC Higgs Results}},
  \href{http://xxx.lanl.gov/abs/1307.5865}{{\tt arXiv:1307.5865}}.

\bibitem{CMS:ril}
{CMS Collaboration}, {\it {Updated measurements of the Higgs boson at 125 GeV
  in the two photon decay channel}},
  \href{http://xxx.lanl.gov/abs/CMS-PAS-HIG-13-001}{{\tt CMS-PAS-HIG-13-001}}.

\bibitem{ATLAS-CONF-2013-108}
{ATLAS Collaboration}, {\it {Evidence for Higgs Boson Decays to the
  $\tau^+\tau^-$ Final State with the ATLAS Detector}},
  \href{http://xxx.lanl.gov/abs/ATLAS-CONF-2013-108,
  ATLAS-COM-CONF-2013-095}{{\tt ATLAS-CONF-2013-108, ATLAS-COM-CONF-2013-095}}.

\bibitem{CorrSystDoc}
P.~Bechtle and T.~Stefaniak, ``{On the presentation of correlated systematic
  uncertainties in Higgs boson rate measurements}.'' \emph{available online at}
  \url{http://higgsbounds.hepforge.org}.

\bibitem{Cranmer:2013hia}
K.~Cranmer, S.~Kreiss, D.~Lopez-Val, and T.~Plehn, {\it {A Novel Approach to
  Higgs Coupling Measurements}},  \href{http://xxx.lanl.gov/abs/1401.0080}{{\tt
  arXiv:1401.0080}}.

\bibitem{haario2001ama}
H.~Haario, E.~Saksman, and J.~Tamminen, {\it {An adaptive Metropolis
  algorithm}},  {\em Bernoulli} {\bf 7} (2001), no.~2 223.

\bibitem{Patil:Huard:Fonnesbeck:2010}
A.~Patil, D.~Huard, and C.~J. Fonnesbeck, {\it {PyMC: Bayesian Stochastic
  Modelling in Python}},  {\em Journal of Statistical Software} {\bf 35}
  (2010), no.~4 1.

\bibitem{HSreleasenote}
P.~Bechtle, S.~Heinemeyer, O.~St{\aa}l, T.~Stefaniak, and G.~Weiglein,
  ``{Release Note for HiggsSignals-1.1}.'' \emph{available online at}
  \url{http://higgsbounds.hepforge.org}.

\bibitem{Denner:2011mq}
A.~Denner, S.~Heinemeyer, I.~Puljak, D.~Rebuzzi, and M.~Spira, {\it {Standard
  Model Higgs-Boson Branching Ratios with Uncertainties}},  {\em
  Eur.~Phys.~J.~C} {\bf 71} (2011) 1753,
  [\href{http://xxx.lanl.gov/abs/1107.5909}{{\tt arXiv:1107.5909}}].

\bibitem{Chatrchyan:2013mxa}
{CMS Collaboration}, {\it {Measurement of the properties of a Higgs boson in
  the four-lepton final state}},  \href{http://xxx.lanl.gov/abs/1312.5353}{{\tt
  arXiv:1312.5353}}.

\bibitem{Belanger:2013kya}
G.~Belanger, B.~Dumont, U.~Ellwanger, J.~Gunion, and S.~Kraml, {\it {Status of
  invisible Higgs decays}},  {\em Phys.~Lett.~B} {\bf 723} (2013) 340--347,
  [\href{http://xxx.lanl.gov/abs/1302.5694}{{\tt arXiv:1302.5694}}].

\bibitem{Cline:2013gha}
J.~M. Cline, K.~Kainulainen, P.~Scott, and C.~Weniger, {\it {Update on scalar
  singlet dark matter}},  {\em Phys.~Rev.~D} {\bf 88} (2013) 055025,
  [\href{http://xxx.lanl.gov/abs/1306.4710}{{\tt arXiv:1306.4710}}].

\bibitem{Dreiner:1997uz}
H.~K. Dreiner, {\it {An Introduction to explicit R-parity violation}},
  \href{http://xxx.lanl.gov/abs/hep-ph/9707435}{{\tt hep-ph/9707435}}.

\bibitem{Barbier:2004ez}
R.~Barbier, C.~Berat, M.~Besancon, M.~Chemtob, A.~Deandrea, et~al., {\it
  {R-parity violating supersymmetry}},  {\em Phys.~Rept.} {\bf 420} (2005) 1,
  [\href{http://xxx.lanl.gov/abs/hep-ph/0406039}{{\tt hep-ph/0406039}}].

\bibitem{Georgi1985}
H.~Georgi and M.~Machacek, {\it {DOUBLY CHARGED HIGGS BOSONS}},  {\em
  Nucl.~Phys.~B} {\bf 262} (1985) 463.

\bibitem{Chanowitz1985}
M.~S. Chanowitz and M.~Golden, {\it {Higgs Boson Triplets With M ($W$) = M
  ($Z$) $\cos \theta \omega$}},  {\em Phys.~Lett.~B} {\bf 165} (1985) 105.

\bibitem{Gunion1990}
J.~Gunion, R.~Vega, and J.~Wudka, {\it {Higgs triplets in the standard model}},
   {\em Phys.~Rev.~D} {\bf 42} (1990) 1673.

\bibitem{oai:arXiv.org:hep-ph/0306034}
S.~Chang, {\it {A 'Littlest Higgs' model with custodial SU(2) symmetry}},  {\em
  JHEP} {\bf 0312} (2003) 057,
  [\href{http://xxx.lanl.gov/abs/hep-ph/0306034}{{\tt hep-ph/0306034}}].

\bibitem{oai:arXiv.org:1202.1532}
A.~Falkowski, S.~Rychkov, and A.~Urbano, {\it {What if the Higgs couplings to W
  and Z bosons are larger than in the Standard Model?}},  {\em JHEP} {\bf 1204}
  (2012) 073, [\href{http://xxx.lanl.gov/abs/1202.1532}{{\tt
  arXiv:1202.1532}}].

\bibitem{Chang2012}
S.~Chang, C.~A. Newby, N.~Raj, and C.~Wanotayaroj, {\it {Revisiting Theories
  with Enhanced Higgs Couplings to Weak Gauge Bosons}},  {\em Phys.~Rev.~D}
  {\bf 86} (2012) 095015, [\href{http://xxx.lanl.gov/abs/1207.0493}{{\tt
  arXiv:1207.0493}}].

\bibitem{Hisano:2013sn}
J.~Hisano and K.~Tsumura, {\it {Higgs boson mixes with an SU(2) septet
  representation}},  {\em Phys.~Rev.~D} {\bf 87} (2013) 053004,
  [\href{http://xxx.lanl.gov/abs/1301.6455}{{\tt arXiv:1301.6455}}].

\bibitem{Kanemura:2013mc}
S.~Kanemura, M.~Kikuchi, and K.~Yagyu, {\it {Probing exotic Higgs sectors from
  the precise measurement of Higgs boson couplings}},  {\em Phys.~Rev.~D} {\bf
  88} (2013) 015020, [\href{http://xxx.lanl.gov/abs/1301.7303}{{\tt
  arXiv:1301.7303}}].

\bibitem{ATLAS:2013pma}
{ATLAS Collaboration}, {\it {Search for invisible decays of a Higgs boson
  produced in association with a Z boson in ATLAS}},
  \href{http://xxx.lanl.gov/abs/ATLAS-CONF-2013-011,
  ATLAS-COM-CONF-2013-013}{{\tt ATLAS-CONF-2013-011, ATLAS-COM-CONF-2013-013}}.

\bibitem{CMS:2013yda}
{CMS Collaboration}, {\it {Search for invisible Higgs produced in association
  with a Z boson}},  \href{http://xxx.lanl.gov/abs/CMS-PAS-HIG-13-018}{{\tt
  CMS-PAS-HIG-13-018}}.

\bibitem{CMS-PAS-HIG-13-028}
{CMS Collaboration}, {\it {Search for the Higgs boson decaying to invisible
  particles produced in association with Z bosons decaying to bottom quarks}},
  \href{http://xxx.lanl.gov/abs/CMS-PAS-HIG-13-028}{{\tt CMS-PAS-HIG-13-028}}.

\bibitem{Heinemeyer:1998yj}
S.~Heinemeyer, W.~Hollik, and G.~Weiglein, {\it {FeynHiggs: A Program for the
  calculation of the masses of the neutral CP even Higgs bosons in the MSSM}},
  {\em Comput.~Phys.~Commun.} {\bf 124} (2000) 76,
  [\href{http://xxx.lanl.gov/abs/hep-ph/9812320}{{\tt hep-ph/9812320}}].

\bibitem{Hahn:2009zz}
T.~Hahn, S.~Heinemeyer, W.~Hollik, H.~Rzehak, and G.~Weiglein, {\it {FeynHiggs:
  A program for the calculation of MSSM Higgs-boson observables - Version
  2.6.5}},  {\em Comput.~Phys.~Commun.} {\bf 180} (2009) 1426.

\bibitem{Aglietti:2006tp}
U.~Aglietti, R.~Bonciani, G.~Degrassi, and A.~Vicini, {\it {Analytic Results
  for Virtual QCD Corrections to Higgs Production and Decay}},  {\em JHEP} {\bf
  0701} (2007) 021, [\href{http://xxx.lanl.gov/abs/hep-ph/0611266}{{\tt
  hep-ph/0611266}}].

\bibitem{Bonciani:2007ex}
R.~Bonciani, G.~Degrassi, and A.~Vicini, {\it {Scalar particle contribution to
  Higgs production via gluon fusion at NLO}},  {\em JHEP} {\bf 0711} (2007)
  095, [\href{http://xxx.lanl.gov/abs/0709.4227}{{\tt arXiv:0709.4227}}].

\bibitem{Djouadi:1997yw}
A.~Djouadi, J.~Kalinowski, and M.~Spira, {\it {HDECAY: A Program for Higgs
  boson decays in the standard model and its supersymmetric extension}},  {\em
  Comput.~Phys.~Commun.} {\bf 108} (1998) 56,
  [\href{http://xxx.lanl.gov/abs/hep-ph/9704448}{{\tt hep-ph/9704448}}].

\bibitem{Spira:1997dg}
M.~Spira, {\it {QCD effects in Higgs physics}},  {\em Fortsch.~Phys.} {\bf 46}
  (1998) 203, [\href{http://xxx.lanl.gov/abs/hep-ph/9705337}{{\tt
  hep-ph/9705337}}].

\bibitem{Schabinger:2005ei}
R.~Schabinger and J.~D. Wells, {\it {A Minimal spontaneously broken hidden
  sector and its impact on Higgs boson physics at the large hadron collider}},
  {\em Phys.~Rev.~D} {\bf 72} (2005) 093007,
  [\href{http://xxx.lanl.gov/abs/hep-ph/0509209}{{\tt hep-ph/0509209}}].

\bibitem{Patt:2006fw}
B.~Patt and F.~Wilczek, {\it {Higgs-field portal into hidden sectors}},
  \href{http://xxx.lanl.gov/abs/hep-ph/0605188}{{\tt hep-ph/0605188}}.

\bibitem{Barger:2007im}
V.~Barger, P.~Langacker, M.~McCaskey, M.~J. Ramsey-Musolf, and G.~Shaughnessy,
  {\it {LHC Phenomenology of an Extended Standard Model with a Real Scalar
  Singlet}},  {\em Phys.~Rev.~D} {\bf 77} (2008) 035005,
  [\href{http://xxx.lanl.gov/abs/0706.4311}{{\tt arXiv:0706.4311}}].

\bibitem{Barger:2008jx}
V.~Barger, P.~Langacker, M.~McCaskey, M.~Ramsey-Musolf, and G.~Shaughnessy,
  {\it {Complex Singlet Extension of the Standard Model}},  {\em Phys.~Rev.~D}
  {\bf 79} (2009) 015018, [\href{http://xxx.lanl.gov/abs/0811.0393}{{\tt
  arXiv:0811.0393}}].

\bibitem{Bhattacharyya:2007pb}
G.~Bhattacharyya, G.~C. Branco, and S.~Nandi, {\it {Universal Doublet-Singlet
  Higgs Couplings and phenomenology at the CERN Large Hadron Collider}},  {\em
  Phys.~Rev.~D} {\bf 77} (2008) 117701,
  [\href{http://xxx.lanl.gov/abs/0712.2693}{{\tt arXiv:0712.2693}}].

\bibitem{Bock:2010nz}
S.~Bock, R.~Lafaye, T.~Plehn, M.~Rauch, D.~Zerwas, et~al., {\it {Measuring
  Hidden Higgs and Strongly-Interacting Higgs Scenarios}},  {\em Phys.~Lett.~B}
  {\bf 694} (2010) 44, [\href{http://xxx.lanl.gov/abs/1007.2645}{{\tt
  arXiv:1007.2645}}].

\bibitem{Englert:2011yb}
C.~Englert, T.~Plehn, D.~Zerwas, and P.~M. Zerwas, {\it {Exploring the Higgs
  portal}},  {\em Phys.~Lett.~B} {\bf 703} (2011) 298,
  [\href{http://xxx.lanl.gov/abs/1106.3097}{{\tt arXiv:1106.3097}}].

\bibitem{Englert:2011aa}
C.~Englert, T.~Plehn, M.~Rauch, D.~Zerwas, and P.~M. Zerwas, {\it {LHC:
  Standard Higgs and Hidden Higgs}},  {\em Phys.~Lett.~B} {\bf 707} (2012) 512,
  [\href{http://xxx.lanl.gov/abs/1112.3007}{{\tt arXiv:1112.3007}}].

\bibitem{Pruna:2013bma}
G.~M. Pruna and T.~Robens, {\it {The Higgs Singlet extension parameter space in
  the light of the LHC discovery}},  {\em Phys.~Rev.~D} {\bf 88} (2013) 115012,
  [\href{http://xxx.lanl.gov/abs/1303.1150}{{\tt arXiv:1303.1150}}].

\bibitem{Giudice:2007fh}
G.~Giudice, C.~Grojean, A.~Pomarol, and R.~Rattazzi, {\it {The
  Strongly-Interacting Light Higgs}},  {\em JHEP} {\bf 0706} (2007) 045,
  [\href{http://xxx.lanl.gov/abs/hep-ph/0703164}{{\tt hep-ph/0703164}}].

\bibitem{Agashe:2004rs}
K.~Agashe, R.~Contino, and A.~Pomarol, {\it {The Minimal composite Higgs
  model}},  {\em Nucl.~Phys.~B} {\bf 719} (2005) 165,
  [\href{http://xxx.lanl.gov/abs/hep-ph/0412089}{{\tt hep-ph/0412089}}].

\bibitem{Contino:2006qr}
R.~Contino, L.~Da~Rold, and A.~Pomarol, {\it {Light custodians in natural
  composite Higgs models}},  {\em Phys.~Rev.~D} {\bf 75} (2007) 055014,
  [\href{http://xxx.lanl.gov/abs/hep-ph/0612048}{{\tt hep-ph/0612048}}].

\bibitem{Altarelli:1990zd}
G.~Altarelli and R.~Barbieri, {\it {Vacuum polarization effects of new physics
  on electroweak processes}},  {\em Phys.~Lett.~B} {\bf 253} (1991) 161.

\bibitem{Peskin:1991sw}
M.~E. Peskin and T.~Takeuchi, {\it {Estimation of oblique electroweak
  corrections}},  {\em Phys.~Rev.~D} {\bf 46} (1992) 381.

\bibitem{Baak:2012kk}
M.~Baak, M.~Goebel, J.~Haller, A.~Hoecker, D.~Kennedy, et~al., {\it {The
  Electroweak Fit of the Standard Model after the Discovery of a New Boson at
  the LHC}},  {\em Eur.~Phys.~J.~C} {\bf 72} (2012) 2205,
  [\href{http://xxx.lanl.gov/abs/1209.2716}{{\tt arXiv:1209.2716}}].

\bibitem{Ciuchini:2013pca}
M.~Ciuchini, E.~Franco, S.~Mishima, and L.~Silvestrini, {\it {Electroweak
  Precision Observables, New Physics and the Nature of a 126 GeV Higgs Boson}},
   \href{http://xxx.lanl.gov/abs/1306.4644}{{\tt arXiv:1306.4644}}.

\bibitem{Lee:1973iz}
T.~Lee, {\it {A Theory of Spontaneous T Violation}},  {\em Phys.~Rev.~D} {\bf
  8} (1973) 1226.

\bibitem{Glashow:1976nt}
S.~L. Glashow and S.~Weinberg, {\it {Natural Conservation Laws for Neutral
  Currents}},  {\em Phys.~Rev.~D} {\bf 15} (1977) 1958.

\bibitem{Deshpande:1977rw}
N.~G. Deshpande and E.~Ma, {\it {Pattern of Symmetry Breaking with Two Higgs
  Doublets}},  {\em Phys.~Rev.~D} {\bf 18} (1978) 2574.

\bibitem{Donoghue:1978cj}
J.~F. Donoghue and L.~F. Li, {\it {Properties of Charged Higgs Bosons}},  {\em
  Phys.~Rev.~D} {\bf 19} (1979) 945.

\bibitem{Haber:1978jt}
H.~Haber, G.~L. Kane, and T.~Sterling, {\it {The Fermion Mass Scale and
  Possible Effects of Higgs Bosons on Experimental Observables}},  {\em
  Nucl.~Phys.~B} {\bf 161} (1979) 493.

\bibitem{Hall:1981bc}
L.~J. Hall and M.~B. Wise, {\it {FLAVOR CHANGING HIGGS - BOSON COUPLINGS}},
  {\em Nucl.~Phys.~B} {\bf 187} (1981) 397.

\bibitem{Gunion:2002zf}
J.~F. Gunion and H.~E. Haber, {\it {The CP conserving two Higgs doublet model:
  The Approach to the decoupling limit}},  {\em Phys.~Rev.~D} {\bf 67} (2003)
  075019, [\href{http://xxx.lanl.gov/abs/hep-ph/0207010}{{\tt
  hep-ph/0207010}}].

\bibitem{Branco:2011iw}
G.~Branco, P.~Ferreira, L.~Lavoura, M.~Rebelo, M.~Sher, et~al., {\it {Theory
  and phenomenology of two-Higgs-doublet models}},  {\em Phys.~Rept.} {\bf 516}
  (2012) 1, [\href{http://xxx.lanl.gov/abs/1106.0034}{{\tt arXiv:1106.0034}}].

\bibitem{Cheng:1987rs}
T.~Cheng and M.~Sher, {\it {Mass Matrix Ansatz and Flavor Nonconservation in
  Models with Multiple Higgs Doublets}},  {\em Phys.~Rev.~D} {\bf 35} (1987)
  3484.

\bibitem{Haber:1984rc}
H.~E. Haber and G.~L. Kane, {\it {The Search for Supersymmetry: Probing Physics
  Beyond the Standard Model}},  {\em Phys.~Rept.} {\bf 117} (1985) 75--263.

\bibitem{Gunion:1984yn}
J.~Gunion and H.~E. Haber, {\it {Higgs Bosons in Supersymmetric Models. 1.}},
  {\em Nucl.~Phys.~B} {\bf 272} (1986) 1.

\bibitem{Gunion:1986nh}
J.~Gunion and H.~E. Haber, {\it {Higgs Bosons in Supersymmetric Models. 2.
  Implications for Phenomenology}},  {\em Nucl.~Phys.~} {\bf 278} (1986) 449.

\bibitem{Gunion:1989we}
J.~F. Gunion, H.~E. Haber, G.~L. Kane, and S.~Dawson, {\it {The Higgs Hunter's
  Guide}},  {\em Front.~Phys.} {\bf 80} (2000) 1.

\bibitem{Mahmoudi:2009zx}
F.~Mahmoudi and O.~St{\aa}l, {\it {Flavor constraints on the two-Higgs-doublet
  model with general Yukawa couplings}},  {\em Phys.~Rev.~D} {\bf 81} (2010)
  035016, [\href{http://xxx.lanl.gov/abs/0907.1791}{{\tt arXiv:0907.1791}}].

\bibitem{Crivellin:2012ye}
A.~Crivellin, C.~Greub, and A.~Kokulu, {\it {Explaining $B\to D\tau\nu$, $B\to
  D^*\tau\nu$ and $B\to \tau\nu$ in a 2HDM of type III}},  {\em Phys.~Rev.~D}
  {\bf 86} (2012) 054014, [\href{http://xxx.lanl.gov/abs/1206.2634}{{\tt
  arXiv:1206.2634}}].

\bibitem{Crivellin:2013wna}
A.~Crivellin, A.~Kokulu, and C.~Greub, {\it {Flavor-phenomenology of
  two-Higgs-doublet models with generic Yukawa structure}},  {\em Phys.~Rev.~D}
  {\bf 87} (2013) 094031, [\href{http://xxx.lanl.gov/abs/1303.5877}{{\tt
  arXiv:1303.5877}}].

\bibitem{Hempfling:1993kv}
R.~Hempfling, {\it {Yukawa coupling unification with supersymmetric threshold
  corrections}},  {\em Phys.~Rev.~D} {\bf 49} (1994) 6168.

\bibitem{Hall:1993gn}
L.~J. Hall, R.~Rattazzi, and U.~Sarid, {\it {The Top quark mass in
  supersymmetric SO(10) unification}},  {\em Phys.~Rev.~D} {\bf 50} (1994)
  7048, [\href{http://xxx.lanl.gov/abs/hep-ph/9306309}{{\tt hep-ph/9306309}}].

\bibitem{Carena:1994bv}
M.~S. Carena, M.~Olechowski, S.~Pokorski, and C.~Wagner, {\it {Electroweak
  symmetry breaking and bottom - top Yukawa unification}},  {\em Nucl.~Phys.~B}
  {\bf 426} (1994) 269, [\href{http://xxx.lanl.gov/abs/hep-ph/9402253}{{\tt
  hep-ph/9402253}}].

\bibitem{CMS:utj}
{CMS Collaboration}, {\it {Search for the Standard-Model Higgs boson decaying
  to tau pairs in proton-proton collisions at sqrt(s) = 7 and 8 TeV}},
  \href{http://xxx.lanl.gov/abs/CMS-PAS-HIG-13-004}{{\tt CMS-PAS-HIG-13-004}}.
  Updated results (dated Dec 2013) taken from TWiki page:
  \url{https://twiki.cern.ch/twiki/bin/view/CMSPublic/Hig13004TWikiUpdate}.

\bibitem{Chatrchyan:2014nva}
{CMS Collaboration}, {\it {Evidence for the 125 GeV Higgs boson decaying to a
  pair of $\tau$ leptons}},  \href{http://xxx.lanl.gov/abs/1401.5041}{{\tt
  arXiv:1401.5041}}.

\bibitem{Chatrchyan:2014vua}
{CMS Collaboration}, {\it {Evidence for the direct decay of the 125 GeV Higgs
  boson to fermions}},  \href{http://xxx.lanl.gov/abs/1401.6527}{{\tt
  arXiv:1401.6527}}.

\bibitem{Carena:2011aa}
M.~Carena, S.~Gori, N.~R. Shah, and C.~E. Wagner, {\it {A 125 GeV SM-like Higgs
  in the MSSM and the $\gamma \gamma$ rate}},  {\em JHEP} {\bf 1203} (2012)
  014, [\href{http://xxx.lanl.gov/abs/1112.3336}{{\tt arXiv:1112.3336}}].

\bibitem{Carena:2012gp}
M.~Carena, S.~Gori, N.~R. Shah, C.~E. Wagner, and L.-T. Wang, {\it {Light Stau
  Phenomenology and the Higgs $\gamma\gamma$ Rate}},  {\em JHEP} {\bf 1207}
  (2012) 175, [\href{http://xxx.lanl.gov/abs/1205.5842}{{\tt
  arXiv:1205.5842}}].

\bibitem{Basso:2012tr}
L.~Basso and F.~Staub, {\it {Enhancing $h \to \gamma \gamma$ with staus in SUSY
  models with extended gauge sector}},  {\em Phys.~Rev.~D} {\bf 87} (2013)
  015011, [\href{http://xxx.lanl.gov/abs/1210.7946}{{\tt arXiv:1210.7946}}].

\bibitem{Carena:2013qia}
M.~Carena, S.~Heinemeyer, O.~St{\aa}l, C.~Wagner, and G.~Weiglein, {\it {MSSM
  Higgs Boson Searches at the LHC: Benchmark Scenarios after the Discovery of a
  Higgs-like Particle}},  {\em Eur.~Phys.~J.~C} {\bf 73} (2013) 2552,
  [\href{http://xxx.lanl.gov/abs/1302.7033}{{\tt arXiv:1302.7033}}].

\bibitem{SchmidtHoberg:2012yy}
K.~Schmidt-Hoberg and F.~Staub, {\it {Enhanced $h\rightarrow \gamma \gamma$
  rate in MSSM singlet extensions}},  {\em JHEP} {\bf 1210} (2012) 195,
  [\href{http://xxx.lanl.gov/abs/1208.1683}{{\tt arXiv:1208.1683}}].

\bibitem{Hemeda:2013hha}
M.~Hemeda, S.~Khalil, and S.~Moretti, {\it {Light Chargino Effects onto $H\to
  \gamma \gamma$ in the MSSM}},  \href{http://xxx.lanl.gov/abs/1312.2504}{{\tt
  arXiv:1312.2504}}.

\bibitem{Posch:2010hx}
P.~Posch, {\it {Enhancement of $h \to \gamma \gamma$ in the Two Higgs Doublet
  Model Type I}},  {\em Phys.~Lett.~B} {\bf 696} (2011) 447,
  [\href{http://xxx.lanl.gov/abs/1001.1759}{{\tt arXiv:1001.1759}}].

\bibitem{Drozd:2012vf}
A.~Drozd, B.~Grzadkowski, J.~F. Gunion, and Y.~Jiang, {\it {Two-Higgs-Doublet
  Models and Enhanced Rates for a 125 GeV Higgs}},  {\em JHEP} {\bf 1305}
  (2013) 072, [\href{http://xxx.lanl.gov/abs/1211.3580}{{\tt
  arXiv:1211.3580}}].

\bibitem{Cordero-Cid:2013sxa}
A.~Cordero-Cid, J.~Hernandez-Sanchez, C.~Honorato, S.~Moretti, M.~Perez,
  et~al., {\it {Impact of a four-zero Yukawa texture on $h\to \gamma \gamma$
  and $\gamma Z$ in the framework of the 2-Higgs Doublet Model Type III}},
  \href{http://xxx.lanl.gov/abs/1312.5614}{{\tt arXiv:1312.5614}}.

\bibitem{Arhrib:2012ia}
A.~Arhrib, R.~Benbrik, and N.~Gaur, {\it {$H\to \gamma \gamma$ in Inert Higgs
  Doublet Model}},  {\em Phys.~Rev.~D} {\bf 85} (2012) 095021,
  [\href{http://xxx.lanl.gov/abs/1201.2644}{{\tt arXiv:1201.2644}}].

\bibitem{Goudelis:2013uca}
A.~Goudelis, B.~Herrmann, and O.~St{\aa}l, {\it {Dark matter in the Inert
  Doublet Model after the discovery of a Higgs-like boson at the LHC}},  {\em
  JHEP} {\bf 1309} (2013) 106, [\href{http://xxx.lanl.gov/abs/1303.3010}{{\tt
  arXiv:1303.3010}}].

\bibitem{Krawczyk:2013jta}
M.~Krawczyk, D.~Sokolowska, P.~Swaczyna, and B.~Swiezewska, {\it {Constraining
  Inert Dark Matter by $R_{\gamma\gamma}$ and WMAP data}},  {\em JHEP} {\bf
  1309} (2013) 055, [\href{http://xxx.lanl.gov/abs/1305.6266}{{\tt
  arXiv:1305.6266}}].

\bibitem{Krawczyk:2013pea}
M.~Krawczyk, D.~Sokolowska, P.~Swaczyna, and B.~Swiezewska, {\it {Higgs $\to
  \gamma \gamma $, $Z\gamma $ in the Inert Doublet Model}},  {\em Acta
  Phys.~Polon.~B} {\bf 44} (2013) 2163,
  [\href{http://xxx.lanl.gov/abs/1309.7880}{{\tt arXiv:1309.7880}}].

\bibitem{oai:arXiv.org:hep-ph/9806315}
A.~Djouadi, {\it {Squark effects on Higgs boson production and decay at the
  LHC}},  {\em Phys.~Lett.~B} {\bf 435} (1998) 101,
  [\href{http://xxx.lanl.gov/abs/hep-ph/9806315}{{\tt hep-ph/9806315}}].

\bibitem{oai:arXiv.org:hep-ph/0202167}
M.~S. Carena, S.~Heinemeyer, C.~Wagner, and G.~Weiglein, {\it {Suggestions for
  benchmark scenarios for MSSM Higgs boson searches at hadron colliders}},
  {\em Eur.~Phys.~J.~C} {\bf 26} (2003) 601,
  [\href{http://xxx.lanl.gov/abs/hep-ph/0202167}{{\tt hep-ph/0202167}}].

\bibitem{CMS:2013tfa}
{CMS Collaboration}, {\it {Search for the standard model Higgs boson produced
  in association with top quarks in multilepton final states}},
  \href{http://xxx.lanl.gov/abs/CMS-PAS-HIG-13-020}{{\tt CMS-PAS-HIG-13-020}}.

\bibitem{CMS:2013sea}
{CMS Collaboration}, {\it {Search for Higgs Boson Production in Association
  with a Top-Quark Pair and Decaying to Bottom Quarks or Tau Leptons}},
  \href{http://xxx.lanl.gov/abs/CMS-PAS-HIG-13-019}{{\tt CMS-PAS-HIG-13-019}}.

\bibitem{CMS:2013fda}
{CMS Collaboration}, {\it {Search for ttH production in events where H decays
  to photons at 8 TeV collisions}},
  \href{http://xxx.lanl.gov/abs/CMS-PAS-HIG-13-015}{{\tt CMS-PAS-HIG-13-015}}.

\bibitem{CMStthcombination}
{CMS Collaboration}, ``Combination of search results for higgs boson production
  in association with a top-quark pair.''
  \url{https://twiki.cern.ch/twiki/bin/view/CMSPublic/ttHCombinationTWiki}.

\bibitem{EuropeanStrategyforParticlePhysicsPreparatoryGroup:2013fia}
R.~Aleksan et~al., {\it {Physics Briefing Book: Input for the Strategy Group to
  draft the update of the European Strategy for Particle Physics}}, .

\bibitem{ATLAS:2013hta}
{ATLAS Collaboration}, {\it {Physics at a High-Luminosity LHC with ATLAS}},
  \href{http://xxx.lanl.gov/abs/1307.7292}{{\tt arXiv:1307.7292}}.

\bibitem{ATL-PHYS-PUB-2013-014}
{ATLAS Collaboration}, {\it {Projections for measurements of Higgs boson cross
  sections, branching ratios and coupling parameters with the ATLAS detector at
  a HL-LHC}},  \href{http://xxx.lanl.gov/abs/ATL-PHYS-PUB-2013-014}{{\tt
  ATL-PHYS-PUB-2013-014}}.

\bibitem{CMS:2013xfa}
{CMS Collaboration}, {\it {Projected Performance of an Upgraded CMS Detector at
  the LHC and HL-LHC: Contribution to the Snowmass Process}},
  \href{http://xxx.lanl.gov/abs/1307.7135}{{\tt arXiv:1307.7135}}.

\bibitem{Dawson:2013bba}
S.~Dawson, A.~Gritsan, H.~Logan, J.~Qian, C.~Tully, et~al., {\it {Higgs Working
  Group Report of the Snowmass 2013 Community Planning Study}},
  \href{http://xxx.lanl.gov/abs/1310.8361}{{\tt arXiv:1310.8361}}.

\bibitem{Peskin:2013xra}
M.~E. Peskin, {\it {Estimation of LHC and ILC Capabilities for Precision Higgs
  Boson Coupling Measurements}},  \href{http://xxx.lanl.gov/abs/1312.4974}{{\tt
  arXiv:1312.4974}}.

\bibitem{Asner:2013psa}
D.~Asner, T.~Barklow, C.~Calancha, K.~Fujii, N.~Graf, et~al., {\it {ILC Higgs
  White Paper}},  \href{http://xxx.lanl.gov/abs/1310.0763}{{\tt
  arXiv:1310.0763}}.

\bibitem{Campbell:2013qaa}
J.~Campbell, K.~Hatakeyama, J.~Huston, F.~Petriello, J.~R. Andersen, et~al.,
  {\it {Report of the Snowmass 2013 energy frontier QCD working group}},
  \href{http://xxx.lanl.gov/abs/1310.5189}{{\tt arXiv:1310.5189}}.

\bibitem{Baer:2013cma}
H.~Baer, T.~Barklow, K.~Fujii, Y.~Gao, A.~Hoang, et~al., {\it {The
  International Linear Collider Technical Design Report - Volume 2: Physics}},
  \href{http://xxx.lanl.gov/abs/1306.6352}{{\tt arXiv:1306.6352}}.

\bibitem{Han:2013kya}
T.~Han, Z.~Liu, and J.~Sayre, {\it {Potential Precision on Higgs Couplings and
  Total Width at the ILC}},  \href{http://xxx.lanl.gov/abs/1311.7155}{{\tt
  arXiv:1311.7155}}.

\bibitem{Campos:2012vb}
I.~Campos et~al., {\it {Phenomenology Tools on Cloud Infrastructures using
  OpenStack}},  {\em Eur.~Phys.~J.~C} {\bf 73} (2013) 2375,
  [\href{http://xxx.lanl.gov/abs/1212.4784}{{\tt arXiv:1212.4784}}].

\bibitem{ATLAS:2013wla}
{ATLAS Collaboration}, {\it {Measurements of the properties of the Higgs-like
  boson in the $WW^{(\ast)} \to \ell \nu \ell \nu$ decay channel with the ATLAS
  detector using 25 fb$^{-1}$ of proton-proton collision data}},
  \href{http://xxx.lanl.gov/abs/ATLAS-CONF-2013-030,
  ATLAS-COM-CONF-2013-028}{{\tt ATLAS-CONF-2013-030, ATLAS-COM-CONF-2013-028}}.

\bibitem{Aad:2013wqa}
{ATLAS Collaboration}, {\it {Measurements of Higgs boson production and
  couplings in diboson final states with the ATLAS detector at the LHC}},  {\em
  Phys.~Lett.~B} {\bf 726} (2013) 88,
  [\href{http://xxx.lanl.gov/abs/1307.1427}{{\tt arXiv:1307.1427}}].

\bibitem{ATLAS:2013nma}
{ATLAS Collaboration}, {\it {Measurements of the properties of the Higgs-like
  boson in the four lepton decay channel with the ATLAS detector using 25 fb?1
  of proton-proton collision data}},
  \href{http://xxx.lanl.gov/abs/ATLAS-CONF-2013-013,
  ATLAS-COM-CONF-2013-018}{{\tt ATLAS-CONF-2013-013, ATLAS-COM-CONF-2013-018}}.

\bibitem{ATLAS:2012goa}
{ATLAS Collaboration}, {\it {Observation of an excess of events in the search
  for the Standard Model Higgs boson in the gamma-gamma channel with the ATLAS
  detector}},  \href{http://xxx.lanl.gov/abs/ATLAS-CONF-2012-091,
  ATLAS-COM-CONF-2012-109}{{\tt ATLAS-CONF-2012-091, ATLAS-COM-CONF-2012-109}}.

\bibitem{ATLAS:2013oma}
{ATLAS Collaboration}, {\it {Measurements of the properties of the Higgs-like
  boson in the two photon decay channel with the ATLAS detector using 25
  $\mathrm{fb}^{-1}$ of proton-proton collision data}},
  \href{http://xxx.lanl.gov/abs/ATLAS-CONF-2013-012,
  ATLAS-COM-CONF-2013-015}{{\tt ATLAS-CONF-2013-012, ATLAS-COM-CONF-2013-015}}.

\bibitem{TheATLAScollaboration:2013lia}
{ATLAS collaboration}, {\it {Search for the bb decay of the Standard Model
  Higgs boson in associated W/ZH production with the ATLAS detector}},
  \href{http://xxx.lanl.gov/abs/ATLAS-CONF-2013-079,
  ATLAS-COM-CONF-2013-080}{{\tt ATLAS-CONF-2013-079, ATLAS-COM-CONF-2013-080}}.

\bibitem{TheATLAScollaboration:2013hia}
{ATLAS Collaboration}, {\it {Search for associated production of the Higgs
  boson in the $WH\to WWW^{(*)} \to \ell\nu\ell\nu\ell\nu$ and $ZH\to
  ZWW^{(*)}\to \ell\ell\ell\nu\ell\nu$ channels with the ATLAS detector at the
  LHC}},  \href{http://xxx.lanl.gov/abs/ATLAS-CONF-2013-075,
  ATLAS-COM-CONF-2013-069}{{\tt ATLAS-CONF-2013-075, ATLAS-COM-CONF-2013-069}}.

\bibitem{Aaltonen:2013ipa}
{CDF Collaboration}, {\it {Combination of searches for the Higgs boson using
  the full CDF data set}},  {\em Phys.~Rev.~D} {\bf 88} (2013) 052013,
  [\href{http://xxx.lanl.gov/abs/1301.6668}{{\tt arXiv:1301.6668}}].

\bibitem{Chatrchyan:2013iaa}
{CMS Collaboration}, {\it {Measurement of Higgs boson production and properties
  in the WW decay channel with leptonic final states}},
  \href{http://xxx.lanl.gov/abs/1312.1129}{{\tt arXiv:1312.1129}}.

\bibitem{CMS:2013xda}
{CMS Collaboration}, {\it {VH with H$\rightarrow$WW$\rightarrow\ell\nu\ell\nu$
  and V$\rightarrow$jj}},
  \href{http://xxx.lanl.gov/abs/CMS-PAS-HIG-13-017}{{\tt CMS-PAS-HIG-13-017}}.

\bibitem{CMS:xwa}
{CMS Collaboration}, {\it {Properties of the Higgs-like boson in the decay $H
  \to ZZ \to 4\ell$ in $pp$ collisions at $\sqrt{s} = 7$ and $8$ TeV}},
  \href{http://xxx.lanl.gov/abs/CMS-PAS-HIG-13-002}{{\tt CMS-PAS-HIG-13-002}}.

\bibitem{CMS:2012paa}
{CMS Collaboration}, {\it {Evidence for a new state decaying into two photons
  in the search for the standard model Higgs boson in pp collisions}},
  \href{http://xxx.lanl.gov/abs/CMS-PAS-HIG-12-015}{{\tt CMS-PAS-HIG-12-015}}.

\bibitem{CMS:2013aga}
{CMS Collaboration}, {\it {Search for the standard model Higgs boson in the
  dimuon decay channel in pp collisions at sqrt(s)= 7 and 8 TeV}},
  \href{http://xxx.lanl.gov/abs/CMS-PAS-HIG-13-007}{{\tt CMS-PAS-HIG-13-007}}.

\bibitem{CMS:2013dda}
{CMS Collaboration}, {\it {Search for the standard model Higgs boson produced
  in association with W or Z bosons, and decaying to bottom quarks for LHCp
  2013}},  \href{http://xxx.lanl.gov/abs/CMS-PAS-HIG-13-012}{{\tt
  CMS-PAS-HIG-13-012}}.

\bibitem{Abazov:2013gmz}
{D{\O} Collaboration}, {\it {Combined search for the Higgs boson with the D0
  experiment}},  {\em Phys.~Rev.~D} {\bf 88} (2013) 052011,
  [\href{http://xxx.lanl.gov/abs/1303.0823}{{\tt arXiv:1303.0823}}].

\end{thebibliography}\endgroup

\end{document}